\documentclass[twocolumn,aps,prd,showpacs,showkeys,superscriptaddress,groupedaddress,floatfix,letterpaper,10pt]{revtex4-1}
\pdfoutput=1
\usepackage{ifthen}
\usepackage{tikz}
\usetikzlibrary{calc}
\usepackage{bchart}
\usepackage{bigdelim}
\usepackage{array}
\usepackage{dcolumn}
\newcolumntype{d}[1]{D{.}{.}{#1}}
\usepackage{longtable}
\usepackage{color}
\usepackage{enumitem}
\usepackage{etex}
\usepackage{graphicx}
\DeclareGraphicsExtensions{.pdf}
\usepackage{diagbox}
\usepackage{multirow}
\usepackage{rotating}
\usepackage{hyperref}
\hypersetup{
  colorlinks,
  linkcolor=blue!60!black,
  citecolor=blue!60!black,
   urlcolor=blue!60!black
}
\usepackage{tabb}
\usepackage{upgreek}
\usepackage{relsize}
\usepackage{epstopdf}
\usepackage{helvet}
\usepackage{overpic}
\usepackage{afterpage}
\usepackage{preprintcover}

\newcommand\HwwPrintOne[1]{
  {\PM}#1
}

\newcommand\HwwPrintTwo[2]{
  {\PM}#1~\textrm{(stat)}
  {\PM}#2~\textrm{(syst)}
}

\newcommand\HwwPrintFour[4]{
  {\PM}#1\,(\textrm{stat})
  {\PM}#2\,\Big(\!\begin{tabular}{c}{\rm\footnotesize expt}\\ \noalign{\vskip -0.10truecm}{\rm\footnotesize syst}\end{tabular}\!\Big)
  {\PM}#3\,\Big(\!\begin{tabular}{c}{\rm\footnotesize theo}\\ \noalign{\vskip -0.10truecm}{\rm\footnotesize syst}\end{tabular}\!\Big)
  {\PM}#4\,\Big(\!\begin{tabular}{c}{\rm\footnotesize lumi}\\ \noalign{\vskip -0.10truecm}{\rm\footnotesize syst}\end{tabular}\!\Big)
}

\newcommand\HwwPrintFive[5]{
  {\PM}#1\,\Big(\!\begin{tabular}{c}{\sc\footnotesize sr  }\\ \noalign{\vskip -0.10truecm}{\rm\footnotesize stat}\end{tabular}\!\Big)
  {\PM}#2\,\Big(\!\begin{tabular}{c}{\sc\footnotesize cr  }\\ \noalign{\vskip -0.10truecm}{\rm\footnotesize stat}\end{tabular}\!\Big)
  {\PM}#3\,\Big(\!\begin{tabular}{c}{\rm\footnotesize expt}\\ \noalign{\vskip -0.10truecm}{\rm\footnotesize syst}\end{tabular}\!\Big)
  {\PM}#4\,\Big(\!\begin{tabular}{c}{\rm\footnotesize theo}\\ \noalign{\vskip -0.10truecm}{\rm\footnotesize syst}\end{tabular}\!\Big)
  {\PM}#5\,\Big(\!\begin{tabular}{c}{\rm\footnotesize lumi}\\ \noalign{\vskip -0.10truecm}{\rm\footnotesize syst}\end{tabular}\!\Big)
}

\newcommand\HwwAtlasEtaJetCtr       {\ABS{\etajet}{\LT}2.4}
\newcommand\HwwAtlasEtaJetFwd       {2.4{\LE}\ABS{\etajet}{\LT}4.5}
\newcommand\HwwLumi[2]{
  \ifthenelse{\equal{#1}{0}}{
    \ifthenelse{\equal{#1}{0}}{25}{}
    \ifthenelse{\equal{#1}{8}}{20}{}
    \ifthenelse{\equal{#1}{7}}{5}{}
  }{}
  \ifthenelse{\equal{#1}{1}}{
    \ifthenelse{\equal{#1}{0}}{24.8}{}
    \ifthenelse{\equal{#1}{8}}{20.3}{}
    \ifthenelse{\equal{#1}{7}}{4.5}{}
  }{}
}

\newcommand\HwwLumiAvgMuPeak[1]{
  \ifthenelse{\equal{#1}{8}}{\AvgMu{\EQ}35}{}
  \ifthenelse{\equal{#1}{7}}{\AvgMu{\EQ}10}{}
}
\newcommand\HwwHiggsMass[1]{
  \ifthenelse{\equal{#1}{HWW}}   {128.X}{}
  \ifthenelse{\equal{#1}{ATLAS}} {125.36}{}
  \ifthenelse{\equal{#1}{CMS}}   {125.03}{}
  \ifthenelse{\equal{#1}{minp0}} {130}{}
}
\newcommand\HwwHiggsMassError[1]{
  \ifthenelse{\equal{#1}{ATLAS}} {0.41}{}
  \ifthenelse{\equal{#1}{CMSERR}}{0.30}{}
}
\newcommand\HwwSignif[2]{
  \ifthenelse{\equal{#1}{obs}}{
    \ifthenelse{\equal{#2}{all}}{6.1}{}
    \ifthenelse{\equal{#2}{ggF}}{4.3}{}
    \ifthenelse{\equal{#2}{VX}} {3.2}{}
    \ifthenelse{\equal{#2}{VXggFratio}} {3.2}{}
    \ifthenelse{\equal{#2}{WH}} {1.xx}{}
    \ifthenelse{\equal{#2}{minp0}}{6.1}{}
  }{}
  \ifthenelse{\equal{#1}{exp}}{
    \ifthenelse{\equal{#2}{all}}{5.8}{}
    \ifthenelse{\equal{#2}{ggF}}{4.3}{}
    \ifthenelse{\equal{#2}{VX}} {2.7}{}
    \ifthenelse{\equal{#2}{VXggFratio}} {2.7}{}
    \ifthenelse{\equal{#2}{WH}} {1.xy}{}
  }{}
}
\newcommand\HwwSigmu[3]{
  \ifthenelse{\equal{#1}{obs}}{
    \ifthenelse{\equal{#2}{0}}{
      \ifthenelse{\equal{#3}{all}}{1.08}{}
      \ifthenelse{\equal{#3}{ggF}}{1.01}{}
      \ifthenelse{\equal{#3}{VX}} {1.27}{}
      \ifthenelse{\equal{#3}{WH}} {1.xx}{}
    }{}
    \ifthenelse{\equal{#2}{1}}{
      \ifthenelse{\equal{#3}{all}}{\HwwSigmu{#1}{0}{#3}\HwwPrintOne{0.23}}{}
      \ifthenelse{\equal{#3}{ggF}}{\HwwSigmu{#1}{0}{#3}\HwwPrintOne{0.2x}}{}
      \ifthenelse{\equal{#3}{VX}} {\HwwSigmu{#1}{0}{#3}\HwwPrintOne{0.4x}}{}
      \ifthenelse{\equal{#3}{WH}} {\HwwSigmu{#1}{0}{#3}\HwwPrintOne{0.8x}}{}
    }{}
    \ifthenelse{\equal{#2}{2}}{
      \ifthenelse{\equal{#3}{all}}{\HwwSigmu{#1}{0}{#3}\HwwPrintTwo{0.16}{0.17}}{}
      \ifthenelse{\equal{#3}{ggF}}{\HwwSigmu{#1}{0}{#3}\HwwPrintTwo{0.2x}{0.2x}}{}
      \ifthenelse{\equal{#3}{VX}} {\HwwSigmu{#1}{0}{#3}\HwwPrintTwo{0.2x}{0.2x}}{}
      \ifthenelse{\equal{#3}{WH}} {\HwwSigmu{#1}{0}{#3}\HwwPrintTwo{0.2x}{0.2x}}{}
    }{}
   \ifthenelse{\equal{#2}{4}}{
      \ifthenelse{\equal{#3}{all}}{\HwwSigmu{#1}{0}{#3}\HwwPrintFour{0.16}{0.13}{0.10}{0.1x}}{}
      \ifthenelse{\equal{#3}{ggF}}{\HwwSigmu{#1}{0}{#3}\HwwPrintFour{0.1x}{0.2x}{0.2x}{0.1x}}{}
      \ifthenelse{\equal{#3}{VX}} {\HwwSigmu{#1}{0}{#3}\HwwPrintFour{0.1x}{0.2x}{0.2x}{0.1x}}{}
      \ifthenelse{\equal{#3}{WH}} {\HwwSigmu{#1}{0}{#3}\HwwPrintFour{0.1x}{0.2x}{0.2x}{0.1x}}{}
    }{}
    \ifthenelse{\equal{#2}{5}}{
      \ifthenelse{\equal{#3}{all}}{\HwwSigmu{#1}{0}{#3}\HwwPrintFive{0.13}{0.09}{0.13}{0.10}{0.1x}}{}
      \ifthenelse{\equal{#3}{ggF}}{\HwwSigmu{#1}{0}{#3}\HwwPrintFive{0.2x}{0.1x}{0.2x}{0.2x}{0.1x}}{}
      \ifthenelse{\equal{#3}{VX}} {\HwwSigmu{#1}{0}{#3}\HwwPrintFive{0.2x}{0.1x}{0.2x}{0.2x}{0.1x}}{}
      \ifthenelse{\equal{#3}{WH}} {\HwwSigmu{#1}{0}{#3}\HwwPrintFive{0.2x}{0.1x}{0.2x}{0.2x}{0.1x}}{}
    }{}
  }{}
  \ifthenelse{\equal{#1}{exp}}{
    \ifthenelse{\equal{#2}{0}}{
      \ifthenelse{\equal{#3}{all}}{1.xx}{}
      \ifthenelse{\equal{#3}{ggF}}{1.xx}{}
      \ifthenelse{\equal{#3}{VX}} {1.xx}{}
      \ifthenelse{\equal{#3}{WH}} {1.xx}{}
    }{}
    \ifthenelse{\equal{#2}{1}}{
      \ifthenelse{\equal{#3}{all}}{\HwwSigmu{#1}{0}{#3}\HwwPrintOne{0.3x}}{}
      \ifthenelse{\equal{#3}{ggF}}{\HwwSigmu{#1}{0}{#3}\HwwPrintOne{0.3x}}{}
      \ifthenelse{\equal{#3}{VX}} {\HwwSigmu{#1}{0}{#3}\HwwPrintOne{0.3x}}{}
      \ifthenelse{\equal{#3}{WH}} {\HwwSigmu{#1}{0}{#3}\HwwPrintOne{0.3x}}{}
    }{}
    \ifthenelse{\equal{#2}{2}}{
      \ifthenelse{\equal{#3}{all}}{\HwwSigmu{#1}{0}{#3}\HwwPrintTwo{0.2x}{0.2x}}{}
      \ifthenelse{\equal{#3}{ggF}}{\HwwSigmu{#1}{0}{#3}\HwwPrintTwo{0.2x}{0.2x}}{}
      \ifthenelse{\equal{#3}{VX}} {\HwwSigmu{#1}{0}{#3}\HwwPrintTwo{0.2x}{0.2x}}{}
      \ifthenelse{\equal{#3}{WH}} {\HwwSigmu{#1}{0}{#3}\HwwPrintTwo{0.2x}{0.2x}}{}
    }{}
    \ifthenelse{\equal{#2}{5}}{
      \ifthenelse{\equal{#3}{all}}{\HwwSigmu{#1}{0}{#3}\HwwPrintFive{0.2x}{0.1x}{0.2x}{0.2x}{0.1x}}{}
      \ifthenelse{\equal{#3}{ggF}}{\HwwSigmu{#1}{0}{#3}\HwwPrintFive{0.2x}{0.1x}{0.2x}{0.2x}{0.1x}}{}
      \ifthenelse{\equal{#3}{VX}} {\HwwSigmu{#1}{0}{#3}\HwwPrintFive{0.2x}{0.1x}{0.2x}{0.2x}{0.1x}}{}
      \ifthenelse{\equal{#3}{WH}} {\HwwSigmu{#1}{0}{#3}\HwwPrintFive{0.2x}{0.1x}{0.2x}{0.2x}{0.1x}}{}
    }{}
  }{}
}
\newcommand\HwwXsecEight[3]{
  \ifthenelse{\equal{#1}{obs}}{
    \ifthenelse{\equal{#2}{0}}{
      \ifthenelse{\equal{#3}{all}}{25.xx}{}
      \ifthenelse{\equal{#3}{ggF}}{20.xx}{}
      \ifthenelse{\equal{#3}{VX}} { 5.xx}{}
      \ifthenelse{\equal{#3}{WH}} { 1.xx}{}
    }{}
    \ifthenelse{\equal{#2}{1}}{
      \ifthenelse{\equal{#3}{all}}{\HwwSigmu{#1}{0}{#3}\HwwPrintOne{0.3x}\pb}{}
      \ifthenelse{\equal{#3}{ggF}}{\HwwSigmu{#1}{0}{#3}\HwwPrintOne{0.3x}\pb}{}
      \ifthenelse{\equal{#3}{VX}} {\HwwSigmu{#1}{0}{#3}\HwwPrintOne{0.3x}\pb}{}
      \ifthenelse{\equal{#3}{WH}} {\HwwSigmu{#1}{0}{#3}\HwwPrintOne{0.3x}\pb}{}
    }{}
    \ifthenelse{\equal{#2}{2}}{
      \ifthenelse{\equal{#3}{all}}{\HwwSigmu{#1}{0}{#3}\HwwPrintTwo{0.2x}{0.2x}\pb}{}
      \ifthenelse{\equal{#3}{ggF}}{\HwwSigmu{#1}{0}{#3}\HwwPrintTwo{0.2x}{0.2x}\pb}{}
      \ifthenelse{\equal{#3}{VX}} {\HwwSigmu{#1}{0}{#3}\HwwPrintTwo{0.2x}{0.2x}\pb}{}
      \ifthenelse{\equal{#3}{WH}} {\HwwSigmu{#1}{0}{#3}\HwwPrintTwo{0.2x}{0.2x}\pb}{}
    }{}
    \ifthenelse{\equal{#2}{5}}{
      \ifthenelse{\equal{#3}{all}}{\HwwSigmu{#1}{0}{#3}\HwwPrintFive{0.2x}{0.1x}{0.2x}{0.2x}{0.1x}\pb}{}
      \ifthenelse{\equal{#3}{ggF}}{\HwwSigmu{#1}{0}{#3}\HwwPrintFive{0.2x}{0.1x}{0.2x}{0.2x}{0.1x}\pb}{}
      \ifthenelse{\equal{#3}{VX}} {\HwwSigmu{#1}{0}{#3}\HwwPrintFive{0.2x}{0.1x}{0.2x}{0.2x}{0.1x}\pb}{}
      \ifthenelse{\equal{#3}{WH}} {\HwwSigmu{#1}{0}{#3}\HwwPrintFive{0.2x}{0.1x}{0.2x}{0.2x}{0.1x}\pb}{}
    }{}
  }{}
  \ifthenelse{\equal{#1}{exp}}{
    \ifthenelse{\equal{#2}{0}}{
      \ifthenelse{\equal{#3}{all}}{1.xx\pb}{}
      \ifthenelse{\equal{#3}{ggF}}{1.xx\pb}{}
      \ifthenelse{\equal{#3}{VX}} {1.xx\pb}{}
      \ifthenelse{\equal{#3}{WH}} {1.xx\pb}{}
    }{}
    \ifthenelse{\equal{#2}{1}}{
      \ifthenelse{\equal{#2}{all}}{\HwwSigmu{#1}{0}{#3}\HwwPrintOne{0.3x\pb}}{}
      \ifthenelse{\equal{#2}{ggF}}{\HwwSigmu{#1}{0}{#3}\HwwPrintOne{0.3x\pb}}{}
      \ifthenelse{\equal{#2}{VX}} {\HwwSigmu{#1}{0}{#3}\HwwPrintOne{0.3x\pb}}{}
      \ifthenelse{\equal{#2}{WH}} {\HwwSigmu{#1}{0}{#3}\HwwPrintOne{0.3x\pb}}{}
    }{}
    \ifthenelse{\equal{#2}{2}}{
      \ifthenelse{\equal{#2}{all}}{\HwwSigmu{#1}{0}{#3}\HwwPrintTwo{0.2x}{0.2x}\pb}{}
      \ifthenelse{\equal{#2}{ggF}}{\HwwSigmu{#1}{0}{#3}\HwwPrintTwo{0.2x}{0.2x}\pb}{}
      \ifthenelse{\equal{#2}{VX}} {\HwwSigmu{#1}{0}{#3}\HwwPrintTwo{0.2x}{0.2x}\pb}{}
      \ifthenelse{\equal{#2}{WH}} {\HwwSigmu{#1}{0}{#3}\HwwPrintTwo{0.2x}{0.2x}\pb}{}
    }{}
    \ifthenelse{\equal{#2}{5}}{
      \ifthenelse{\equal{#1}{all}}{\HwwSigmu{#1}{0}{#3}\HwwPrintFive{0.2x}{0.1x}{0.2x}{0.2x}{0.1x}\pb}{}
      \ifthenelse{\equal{#1}{ggF}}{\HwwSigmu{#1}{0}{#3}\HwwPrintFive{0.2x}{0.1x}{0.2x}{0.2x}{0.1x}\pb}{}
      \ifthenelse{\equal{#1}{VX}} {\HwwSigmu{#1}{0}{#3}\HwwPrintFive{0.2x}{0.1x}{0.2x}{0.2x}{0.1x}\pb}{}
      \ifthenelse{\equal{#1}{WH}} {\HwwSigmu{#1}{0}{#3}\HwwPrintFive{0.2x}{0.1x}{0.2x}{0.2x}{0.1x}\pb}{}
    }{}
  }{}
}
\newcommand\HwwXsecEightFiducial[3]{
  \ifthenelse{\equal{#1}{obs}}{
    \ifthenelse{\equal{#2}{0}}{
      \ifthenelse{\equal{#3}{all}}{25.xx}{}
      \ifthenelse{\equal{#3}{ggF}}{20.xx}{}
    }{}
    \ifthenelse{\equal{#2}{1}}{
      \ifthenelse{\equal{#3}{all}}{\HwwSigmu{#1}{0}{#3}\HwwPrintOne{0.3x}\pb}{}
      \ifthenelse{\equal{#3}{ggF}}{\HwwSigmu{#1}{0}{#3}\HwwPrintOne{0.3x}\pb}{}
    }{}
    \ifthenelse{\equal{#2}{2}}{
      \ifthenelse{\equal{#3}{all}}{\HwwSigmu{#1}{0}{#3}\HwwPrintTwo{0.2x}{0.2x}\pb}{}
      \ifthenelse{\equal{#3}{ggF}}{\HwwSigmu{#1}{0}{#3}\HwwPrintTwo{0.2x}{0.2x}\pb}{}
    }{}
    \ifthenelse{\equal{#2}{5}}{
      \ifthenelse{\equal{#3}{all}}{\HwwSigmu{#1}{0}{#3}\HwwPrintFive{0.2x}{0.1x}{0.2x}{0.2x}{0.1x}\pb}{}
      \ifthenelse{\equal{#3}{ggF}}{\HwwSigmu{#1}{0}{#3}\HwwPrintFive{0.2x}{0.1x}{0.2x}{0.2x}{0.1x}\pb}{}
    }{}
  }{}
  \ifthenelse{\equal{#1}{exp}}{
    \ifthenelse{\equal{#2}{0}}{
      \ifthenelse{\equal{#3}{all}}{1.xx\pb}{}
      \ifthenelse{\equal{#3}{ggF}}{1.xx\pb}{}
    }{}
    \ifthenelse{\equal{#2}{1}}{
      \ifthenelse{\equal{#2}{all}}{\HwwSigmu{#1}{0}{#3}\HwwPrintOne{0.3x\pb}}{}
      \ifthenelse{\equal{#2}{ggF}}{\HwwSigmu{#1}{0}{#3}\HwwPrintOne{0.3x\pb}}{}
    }{}
    \ifthenelse{\equal{#2}{2}}{
      \ifthenelse{\equal{#2}{all}}{\HwwSigmu{#1}{0}{#3}\HwwPrintTwo{0.2x}{0.2x}\pb}{}
      \ifthenelse{\equal{#2}{ggF}}{\HwwSigmu{#1}{0}{#3}\HwwPrintTwo{0.2x}{0.2x}\pb}{}
    }{}
    \ifthenelse{\equal{#2}{5}}{
      \ifthenelse{\equal{#1}{all}}{\HwwSigmu{#1}{0}{#3}\HwwPrintFive{0.2x}{0.1x}{0.2x}{0.2x}{0.1x}\pb}{}
      \ifthenelse{\equal{#1}{ggF}}{\HwwSigmu{#1}{0}{#3}\HwwPrintFive{0.2x}{0.1x}{0.2x}{0.2x}{0.1x}\pb}{}
    }{}
  }{}
}
\newcommand\HwwLimit[1]{
  \ifthenelse{\equal{#1}{obs}}{132{\LT}\mH{\LT}200}{}
  \ifthenelse{\equal{#1}{exp}}{114{\LT}\mH{\LT}200}{}
}

\newcommand\ABS  [1]{|\,{#1}\,|}
\newcommand\CEN  [1]{C_{#1}}
\newcommand\MBF  [1]{\hbox{\boldmath $#1$}}
\newcommand\SPACE[1]{\,{#1}\,}
\newcommand\MBFr[1]{\multicolumn{1}{r}{\MBF{#1}}}
\newcommand\MCOL[2]{\multicolumn{#1}{l}{#2}}

\newcommand\MROW[2]{\multirow{#1}{*}{#2}}
\newcommand\mcolz{\multicolumn{2}{c}{~-}}

\newcommand\TO     {\SPACE{\rightarrow}}
\newcommand\MINUS  {\SPACE{-}}
\newcommand\PLUS   {\SPACE{+}}

\newcommand\CDOT   {\SPACE{\cdot}}

\newcommand\PM     {\SPACE{\pm}}

\newcommand\EQ     {\SPACE{=}}
\newcommand\APPROX {\SPACE{\approx}}
\newcommand\GT     {\SPACE{>}}
\newcommand\LT     {\SPACE{<}}
\newcommand\GE     {\SPACE{\ge}}
\newcommand\LE     {\SPACE{\le}}

\newcommand\TIMES  {\SPACE{\times}}
\newcommand\data   {\rm data}

\newcommand\myphi  {\protect\raisebox{0.5px}{\ensuremath{\phi}}}
\newcommand\myeta  {\protect\raisebox{0.5px}{\ensuremath{\eta}}}
\newcommand\myy    {\protect\raisebox{0.5px}{\ensuremath{y}}}
\newcommand\mynu   {\protect\raisebox{0.5px}{\ensuremath{\nu}}}

\newcommand\mm     {\mbox{\,mm}}

\newcommand\mb     {\mbox{\,mb}}
\newcommand\pb     {\mbox{\,pb}}
\newcommand\fb     {\mbox{\,fb}}
\newcommand\iab    {\mbox{\,ab$^{-1}$}}
\newcommand\ifb    {\mbox{\,fb$^{-1}$}}

\newcommand\Tesla  {\mbox{\,T}}
\newcommand\Teslam {\mbox{\,T\,m}}
\newcommand\Hz     {\ifmmode{\,\mathrm{Hz}}\else
                      \,\textrm{Hz}\fi}
\newcommand\TeV    {\ifmmode{\,\mathrm{Te\kern -0.1em V}}\else
                      \,\textrm{Te\kern -0.1em V}\fi}
\newcommand\GeV    {\ifmmode{\,\mathrm{Ge\kern -0.1em V}}\else
                      \,\textrm{Ge\kern -0.1em V}\fi}
\newcommand\MeV    {\ifmmode{\,\mathrm{Me\kern -0.1em V}}\else
                      \,\textrm{Me\kern -0.1em V}\fi}

\newcommand\RMS          {rms}
\newcommand\CLs          {\mathrm{CL}_{\tiny S}}
\newcommand\likelihood   {\mathcal{L}}
\newcommand\Lmax         {\likelihood_\mathrm{max}}
\newcommand\BF           {\mathcal{B}}
\newcommand\Acc          {\mathcal{A}}
\newcommand\Cor          {\mathcal{C}}
\newcommand\Lint         {\mbox{\raisebox{2pt}{\footnotesize$\int$}} L\,{\rm{d}}t}
\newcommand\sigmu        {\upmu}

\newcommand\qmu          {q_\sigmu}
\newcommand\qzero        {q_{\mbox{\tiny $0$}}}
\newcommand\pzero        {p_{\mbox{\tiny $0$}}}
\newcommand\hatsigmu     {\hat{\sigmu}}
\newcommand\responsefn   {\mynu}
\newcommand\nuipar       {\theta}
\newcommand\nuipars      {\MBF{\nuipar}}

\newcommand\hatnuipar    {\hat{\nuipar}}
\newcommand\hatnuiparsmu {\hat{\nuipars}_\sigmu}
\newcommand\Nuipar       {\vartheta}

\newcommand\kF           {\kappa_{\mbox{\tiny $F$}}}
\newcommand\kV           {\kappa_{\mbox{\tiny $V$}}}

\newcommand\JETVHETO     {{\sc jetvheto}}
\newcommand\PROPHECY     {{\sc prophecy}}
\newcommand\MCFM         {{\sc mcfm}}
\newcommand\HAWK         {{\sc hawk}}
\newcommand\POWHEG       {{\sc powheg}}
\newcommand\PYTHIA       {{\sc pythia}}

\newcommand\VBFATNNLO    {{\sc vbf@nnlo}}
\newcommand\HERWIG       {{\sc herwig}}
\newcommand\ALPGEN       {{\sc alpgen}}
\newcommand\MADGRAPH     {{\sc madgraph}}
\newcommand\MCATNLO      {{\sc mc@nlo}}
\newcommand\aMCATNLO     {{\sc amc@nlo}}

\newcommand\GGTOVV       {{\sc gg2vv}}
\newcommand\SHERPA       {{\sc sherpa}}
\newcommand\ACERMC       {{\sc acermc}}
\newcommand\JIMMY        {{\sc jimmy}}
\newcommand\HDECAY       {{\sc hdecay}}
\newcommand\GEANT        {{\sc geant}}
\newcommand\HRES         {{\sc hres}}
\newcommand\TOPPP        {{\sc top}{\raisebox{1.5pt}{\tiny\bf ++}}}
\newcommand\CT           {{\sc ct}}
\newcommand\CTEQ         {{\sc cteq}}

\newcommand\MRST         {{\sc mrst}}

\newcommand\DYNNLO       {{\sc dynnlo}}

\newcommand\mH           {\M_H}
\newcommand\mW           {\M_W}
\newcommand\mZ           {\M_Z}

\newcommand\alphaS       {\alpha_{\tiny\textrm{S}}}
\newcommand\bjet         {\mbox{$b$-jet}}

\newcommand\emu          {e\mu}

\renewcommand\ll         {\ell\ell}
\newcommand\lj           {\ell j}
\newcommand\llj          {\ell\lj}

\newcommand\DFchan       {\emu}
\newcommand\SFchan       {ee/\mu\mu}

\newcommand\Jpsi         {J/\psi}
\newcommand\WW           {WW}
\newcommand\WWs          {WW^\ast}
\newcommand\ZZs          {ZZ^\ast}

\newcommand\VV           {VV}
\newcommand\Wj           {W\!j}

\newcommand\jj           {jj}
\newcommand\qq           {q\bar{q}}
\newcommand\Wg           {W\gamma}
\newcommand\Zg           {Z\gamma}
\newcommand\Zgs          {Z\gstar}
\newcommand\gstar        {\gamma^{\ast}}
\newcommand\Wgs          {W\gstar}

\newcommand\ZDY          {Z/\gstar}
\newcommand\ttbar        {t\bar{t}}
\newcommand\tautau       {\tau\tau}

\newcommand\WZ           {W\!Z}
\newcommand\ZZ           {ZZ}

\newcommand\nonDY        {{\rm non\textrm{-}DY}}

\newcommand\DY           {{\rm DY}}
\newcommand\scMC         {{\textsc{\scriptsize mc}}}
\newcommand\scggF        {{\tiny\rm gg}{\textsc{\scriptsize f}}}
\newcommand\scVBF        {{\textsc{\scriptsize vbf}}}
\newcommand\scVX         {{\textsc{\scriptsize vbf}}}
\newcommand\scVBFVH   {{\textsc{\scriptsize vbf+vh}}}
\newcommand\scVH         {{\textsc{\scriptsize vh}}}
\newcommand\ggF          {{\rm ggF}}
\newcommand\VBF          {{\rm VBF}}
\newcommand\VH           {{\rm VH}}

\newcommand\VX           {{\rm VBF}}

\newcommand\all          {{\rm all}}
\newcommand\obs          {{\rm obs}}
\renewcommand\exp        {{\rm exp}}
\newcommand\stat         {{\rm stat}}
\newcommand\syst         {{\rm syst}}

\newcommand\est          {{\rm est}}

\renewcommand\top        {{\rm top}}
\newcommand\fakes        {{\rm misid}}

\newcommand\btag         {b{\textrm{\tiny -}{\rm tag}}}

\newcommand\Fakes        {{\rm Misid{.}}}
\newcommand\Wjets        {\ensuremath{W}{+}{\rm jets}}
\newcommand\Zjets        {\ensuremath{Z}{+}{\rm jets}}

\newcommand\scnonDY      {\rm non\textrm{\footnotesize -}{\textsc{\footnotesize dy}}}
\newcommand\scDY         {{\textsc{\footnotesize dy}}}

\newcommand\scWW         {\mbox{\tiny{$WW$}}}

\newcommand\Njet         {n_j}

\newcommand\Nextrajet    {n_{{\rm extra\textrm{\footnotesize -}}j}}
\newcommand\Nbjet        {n_b}
\newcommand\NbjetEQzero  {n_b{\EQ}0}
\newcommand\NbjetEQone   {n_b{\EQ}1}

\newcommand\NbjetGEone   {n_b{\GE}1}
\newcommand\NjetEQzero   {n_j{\EQ}0}
\newcommand\NjetEQone    {n_j{\EQ}1}
\newcommand\NjetEQtwo    {n_j{\EQ}2}
\newcommand\NjetGEtwo    {n_j{\GE}2}
\newcommand\NjetGEthree  {n_j{\GE}3}

\newcommand\NjetLEone    {n_j{\LE}1}

\newcommand\NjetEQzeroone{\NjetEQzero{\rm\ and\ }\NjetEQone}

\newcommand\Nobs         {N_{\rm obs}}
\newcommand\Nbkg         {N_{\rm bkg}}
\newcommand\Nsig         {N_{\rm sig}}

\newcommand\NggF         {N_{\rm ggF}}
\newcommand\NVBF         {N_{\rm VBF}}
\newcommand\NVH          {N_{\rm VH}}
\newcommand\NVBFVH       {N_{\rm VBF}}
\newcommand\NWW          {N_{\WW}}

\newcommand\NVV          {N_{\VV}}
\newcommand\NWj          {N_{\Wj}}
\newcommand\Njj          {N_{\jj}}
\newcommand\Nttbar       {N_{\ttbar}}
\newcommand\Nt           {N_{t}}
\newcommand\Nll          {N_{\SFchan}}
\newcommand\Ntautau      {N_{\tautau}}

\newcommand\Ntop         {N_{\rm top}}
\newcommand\Nfakes       {N_{\rm misid}}
\newcommand\Ndy          {N_{\rm DY}}
\newcommand\Ndrellyan    {N_{\rm Drell\textrm{\footnotesize -}Yan}}

\newcommand\NWWqcd       {\NWW^\textrm{\tiny QCD}}
\newcommand\NWWew        {\NWW^\textrm{\tiny EW}}
\newcommand\Ntautauqcd   {\Ntautau^\textrm{\tiny QCD}}
\newcommand\Ntautauew    {\Ntautau^\textrm{\tiny EW}}

\newcommand\SR           {\textrm{\footnotesize\sc sr}}
\newcommand\CR           {\textrm{\footnotesize\sc cr}}
\newcommand\MC           {\textrm{\footnotesize\sc mc}}
\newcommand\pass         {{\rm pass}}
\newcommand\fail         {{\rm fail}}

\newcommand\NMC          {N_{\MC}}
\newcommand\BSR          {B_{\SR}}
\newcommand\BCR          {B_{\CR}}

\newcommand\NCR          {N_{\CR}}
\newcommand\ZCR          {Z\CR}

\newcommand\Npass          {N_{\pass}}
\newcommand\Nfail          {N_{\fail}}

\newcommand\lvlv         {\ell\nu\ell\nu}

\newcommand\Wlv          {W{\TO}\ell\nu}

\newcommand\WWlvlv       {\WWs{\TO}\lvlv}

\newcommand\Htt          {H{\TO}\tautau}
\newcommand\Hff          {H{\TO}f\bar{f}}
\newcommand\Hgg          {H{\TO} gg}
\newcommand\Hgamgam      {H{\TO} \gamma\gamma}
\newcommand\HVV          {H{\TO} VV}
\newcommand\HWW          {H{\TO}\WWs}
\newcommand\HZZ          {H{\TO}\ZZs}

\newcommand\HWWlvlv      {H{\TO}\WWlvlv}

\newcommand\WWevmv       {\WWs{\TO}{e\nu\mu\nu}}
\newcommand\HWWevmv      {H{\TO}\WWevmv}

\newcommand\Zmumug       {Z{\TO}\mu\mu\gamma}
\newcommand\ZDYll        {\ZDY{\TO}ee,\,\mu\mu}
\newcommand\ZDYtt        {\ZDY{\TO}\tau\tau}

\newcommand\Ztt          {\ZDYtt}

\newcommand\M            {m}
\newcommand\E            {\MBF{E}}
\renewcommand\P          {\MBF{p}}
\newcommand\p            {\MBF{p}}
\newcommand\T            {\textrm{\footnotesize\sc t}}
\newcommand\Trel         {{\T},{\rm rel}}
\newcommand\tot          {\rm sum}
\newcommand\miss         {\rm miss}
\newcommand\missnojet    {\rm miss\,(trk)}

\newcommand\eT           {E_{\T}}
\newcommand\pT           {p_{\T}}

\newcommand\vpT          {\P_{\T}}

\newcommand\DR           {{\Delta}R}
\newcommand\DeltaR       {\DR}

\newcommand\AvgMu        {\langle\mu\rangle}

\newcommand\bdt          {O_{\rm BDT}}

\newcommand\frecoil      {f_{\rm recoil}}
\newcommand\jvf          {\mbox{\sc jvf}}
\newcommand\fBvarsignif  {\zeta}
\newcommand\fBsignif     {\xi}

\newcommand\fAlpha       {\alpha}
\newcommand\fNorm        {\beta}
\newcommand\fbtag        {\gamma}
\newcommand\fEff         {\varepsilon}

\newcommand\fcor	       {\gamma_{1j}}
\newcommand\effbest	     {\epsilon_{1j}^{\rm est}}
\newcommand\effbonej	   {\epsilon_{1j}}
\newcommand\effbtwoj	   {\epsilon_{2j}}
\newcommand\effbtwojdata {\epsilon_{2j}^{\rm data}}
\newcommand\effrest	     {\epsilon_{\rm rest}}

\newcommand\met          {\textrm{missing transverse momentum}}
\newcommand\Met          {\textrm{Missing transverse momentum}}
\newcommand\METsc        {\textrm{\sc met}}

\newcommand\calMet       {E}
\newcommand\trkMet       {p}
\newcommand\vcalMet      {\E}
\newcommand\vtrkMet      {\P}
\newcommand\MET          {\calMet_{\T}^{\miss}}
\newcommand\MPT          {\trkMet_{\T}^{\missnojet}}
\newcommand\MPTj         {\trkMet_{\T}^{\miss}}
\newcommand\METrel       {\calMet_{\Trel}^{\miss}}
\newcommand\MPTrel       {\trkMet_{\Trel}^{\missnojet}}
\newcommand\MPTjrel      {\trkMet_{\Trel}^{\miss}}
\newcommand\vMET         {\vcalMet_{\T}^{\miss}}
\newcommand\vMPT         {\vtrkMet_{\T}^{\missnojet}}
\newcommand\vMPTj        {\vtrkMet_{\T}^{\miss}}

\newcommand\mTlep        {\M_{\T}^{\ell}}
\newcommand\mTl          {\mTlep}
\newcommand\mTlepi       {\M_{\T}^{\ell_{i}}}
\newcommand\mTlead       {\M_{\T}^{\ell1}}
\newcommand\mTsublead    {\M_{\T}^{\ell2}}

\newcommand\pTnu         {p_{\T}^{\,\nu\nu}}

\newcommand\pTlepi       {p_{\T}^{\,\ell_{i}}}
\newcommand\pTlead       {p_{\T}^{\,\ell1}}
\newcommand\pTsublead    {p_{\T}^{\,\ell2}}

\newcommand\vpTnu        {\p_{\T}^{\,\nu\nu}}

\newcommand\etalead      {\myeta_{\ell1}}
\newcommand\etasublead   {\myeta_{\ell2}}

\newcommand\vpTjet       {\p_{\T}^{\,j}}
\newcommand\pTjet        {p_{\T}^{\,j}}
\newcommand\vpTj         {\vpTjet}
\newcommand\pTj          {\pTjet}

\newcommand\pTleadjet    {p_{\T}^{\,j1}}
\newcommand\pTsubleadjet {p_{\T}^{\,j2}}
\newcommand\pTthirdjet   {p_{\T}^{\,j3}}
\newcommand\etajet       {\myeta_{j}}

\newcommand\yleadjet     {y_{j1}}
\newcommand\ysubleadjet  {y_{j2}}
\newcommand\etaleadjet   {\myeta_{j1}}
\newcommand\etasubleadjet{\myeta_{j2}}
\newcommand\etathirdjet  {\myeta_{j3}}

\newcommand\sumeT        {\Sigma\,\eT}
\newcommand\sumpT        {\Sigma\,\pT}
\newcommand\sumpTsq      {\Sigma\,(\pT)^2}

\newcommand\eTll         {E_{\T}^{\,\ll}}
\newcommand\pTll         {p_{\T}^{\,\ll}}
\newcommand\pTllj        {p_{\T}^{\,\llj}}

\newcommand\pTZDY        {p_{\T}^{\ZDY}}
\newcommand\pTtot        {p_{\T}^{\,\tot}}
\newcommand\vpTll        {\p_{\T}^{\,\ll}}

\newcommand\vpTllj       {\p_{\T}^{\,\llj}}

\newcommand\vpTtot       {\p_{\T}^{\,\tot}}

\newcommand\mTH          {\M_{\T}}

\newcommand\mTtwo        {\M_{\T2}}

\newcommand\mTHcut       {\frac{3}{4}\,\mH{\LT}\mTH{\LT}\mH}

\providecommand{\mtt}    {\M_{\tau\tau}}
\renewcommand\mtt        {\M_{\tau\tau}}
\newcommand\mll          {\M_{\ell\ell}}
\newcommand\dphi         {\Delta\myphi}

\newcommand\dy           {\Delta\myy}

\newcommand\dphill       {\dphi_{\ell\ell}}

\newcommand\dphillMET    {\Delta{\myphi_{\ell\ell, \textrm{\footnotesize \sc met}}}}
\newcommand\dphiNear     {\dphi_{\rm near}}

\newcommand\cjv          {\CEN{j3}}
\newcommand\olv          {\CEN{\ell}}
\newcommand\olvlead      {\CEN{\ell1}}
\newcommand\olvsublead   {\CEN{\ell2}}
\newcommand\contolv      {\Sigma\,\CEN{\ell}}
\newcommand\mlj          {\Sigma\,\M_{\lj}}
\newcommand\mjj          {\M_{\jj}}
\newcommand\detajj       {\Delta\myeta_{jj}}

\newcommand\sumetajj     {\Sigma\,\myeta_{jj}}

\newcommand\dyjj         {\dy_{jj}}

\definecolor{myWW}{rgb}{0,0,255}
\definecolor{myVV}{rgb}{305,69,94}
\definecolor{myMisid}{rgb}{180,81,92}
\definecolor{myDY}{rgb}{120,89,72}
\definecolor{myTop}{rgb}{60,70,92}
\definecolor{myr}{rgb}{0.584, 0, 0.102}
\definecolor{myb}{rgb}{0.004, 0.145, 0.431}
\newcommand\myb{\color{myb}}
\newcommand\myr{\color{myr}}
\newcommand{\slice}[6]{
  \pgfmathparse{0.5*#1+0.5*#2}
  \let\midangle\pgfmathresult

  \ifthenelse{\equal{#2}{0}}{}{
    \draw[fill=#6] (0,0) -- (#1:1) arc (#1:#2:1) -- cycle;}

  \node[label=\midangle:#5] at (\midangle:0.95) {};

  \ifthenelse{#3 < 5}{}{
    \pgfmathparse{min((#2-#1-20)/110*(-0.3),0)}
    \let\temp\pgfmathresult
    \pgfmathparse{max(\temp,-0.4) + 0.8}
    \let\innerpos\pgfmathresult
      \node at (\midangle:\innerpos){
        \ifthenelse{\equal{blue}{#6}}{\color{white} #4}{
          \ifthenelse{\equal{magenta}{#6}}{\color{white} #4}{
            \ifthenelse{\equal{cyan}{#6}}{\color{white} #4}{
              \ifthenelse{\equal{black}{#6}}{\color{white} #4}{
                #4
              }
            }
          }
        }
      };
  }
}

\makeatletter
\newcommand*{\balancecolsandclearpage}{
  \close@column@grid
  \clearpage
  \twocolumngrid
}
\makeatother

\newcommand\HwwPlotDetail[1]{#1 Fig.~\ref{fig:MET} for plotting details}
\newcommand\HwwPlotSelectionDetail[1]{#1 Figs.~\ref{fig:MET} and~\ref{fig:0j} for plotting details}
\newcommand\HwwPlotFitDetailShort[1]{#1 Figs.~\ref{fig:MET} and~\ref{fig:mT_fitDF} for plotting details}
\newcommand\paper        {paper}

\newcommand\no           {\!\!}
\newcommand\np           {\no\no}
\newcommand\nq           {\np\np}
\newcommand\nqq          {\nq\nq\nq}

\newcommand\zz           {\phantom{..}}
\newcommand\z            {\phantom{0}}
\newcommand\Z            {\phantom{.0}}

\newcommand\CDF          {T.~Aaltonen \etal\ (CDF Collaboration)}
\newcommand\DZ           {V.~M.~Abazov \etal\ (D0 Collaboration)}
\newcommand\CDFDZ        {T.~Aaltonen \etal\ (CDF and D0 Collaborations)}
\newcommand\ATLAS        {ATLAS Collaboration}
\newcommand\CMS          {CMS Collaboration}
\newcommand\ATLASCMS     {ATLAS and CMS Collaborations}
\newcommand\LEP          {ALEPH, DELPHI, L3, OPAL, and SLD Collaborations; LEP Electroweak Working Group; SLD Electroweak and Heavy Flavor Groups}

\newcommand\Note[2]      {Report No.\ #2, \href{#1}{#1}}
\newcommand\CPC[1]       {Comput.~Phys.~Commun.~{\bf #1}}
\newcommand\EPJC[1]      {Eur.~Phys.~J.~C {\bf #1}}
\newcommand\JHEP[1]      {J.~High Energy Phys.~#1}
\newcommand\JINST[1]     {JINST {\bf #1}}
\newcommand\JPHYSG[1]    {J.~Phys.~{\bf G~#1}}
\newcommand\NIMA         {Nucl.~Instrum.~Methods~Phys.~Res.,~Sect.~A~}
\newcommand\NPhys[1]     {Nucl.~Phys.\ {\bf #1}}
\newcommand\PLB[1]       {Phys.~Lett.~B {\bf #1}}
\newcommand\PLett[1]     {Phys.~Lett.\ {\bf #1}}
\newcommand\PRep[1]      {Phys.~Rep.\ {\bf #1}}
\newcommand\PRev[1]      {Phys.~Rev.\ {\bf #1}}
\newcommand\PRD[1]       {Phys.~Rev.~D {\bf #1}}
\newcommand\PRL[1]       {Phys.~Rev.~Lett.\ {\bf #1}}
\newcommand\ZPC[1]       {Z.~Phys.~C {\bf #1}}
\newcommand\arxiv[1]     {\href{http://arxiv.org/abs/#1}{arXiv:#1}}

\newcommand\ie           {i.\,e{.}}
\newcommand\eg           {e.\,g{.}}
\newcommand\etal         {{\it et al{.}}}
\newcommand\ibid         {{\it ibid{.}}}
\newcommand\insitu       {{\it in situ}}

\newcommand\dbline{\noalign{\vskip 0.10truecm\hrule\vskip 0.05truecm\hrule\vskip 0.10truecm}}
\newcommand\sgline{\noalign{\vskip 0.10truecm\hrule\vskip 0.10truecm}}
\newcommand\clineskip{\noalign{\vskip 0.10truecm}}
\PreprintCoverPaperTitle{Observation and measurement of Higgs boson decays to $\WWs$ with the ATLAS detector}
\PreprintIdNumber{CERN-PH-EP-2014-270}
\PreprintCoverAbstract{%
We report the observation of Higgs boson decays to $\WWs$ based on an excess over background
of $\HwwSignif{obs}{all}$ standard deviations in the dilepton final state, where the
Standard Model expectation is $\HwwSignif{exp}{all}$ standard deviations.  Evidence for the
vector-boson fusion (VBF) production process is obtained with a significance of
$\HwwSignif{obs}{VX}$ standard deviations.  The results are obtained from a data sample
corresponding to an integrated luminosity of $\HwwLumi{0}{0}\ifb$ from $\sqrt{s}{\EQ}7$ and
$8\TeV$ $pp$ collisions recorded by the ATLAS detector at the LHC.  For a Higgs boson mass
of $\HwwHiggsMass{ATLAS}\GeV$, the ratio of the measured value to the expected value of the
total production cross section times branching fraction is
$1.09^{+0.16}_{-0.15}\,\textrm{(stat)}^{+0.17}_{-0.14}\,\textrm{(syst)}$.  The
corresponding ratios for the gluon fusion and vector-boson fusion production mechanisms are
$1.02{\PM}0.19\,\textrm{(stat)}\,^{+0.22}_{-0.18}\,\textrm{(syst)}$ and
$1.27\,^{+0.44}_{-0.40}\,\textrm{(stat)}\,^{+0.30}_{-0.21}\,\textrm{(syst)}$, respectively.
At $\sqrt{s}{\EQ}8\TeV$, the total production cross sections are measured to be
$\sigma(gg{\TO}\HWW) = 4.6{\PM}0.9\,\textrm{(stat)}\,^{+0.8}_{-0.7}\,\textrm{(syst)}\pb$ and
$\sigma(\VBF~\HWW){\EQ}0.51\,^{+0.17}_{-0.15}\,\textrm{(stat)}\,^{+0.13}_{-0.08}\,\textrm{(syst)}\pb$.
The fiducial cross section is determined for the gluon-fusion
process in exclusive final states with zero or one associated jet.}

\PreprintJournalName{Phys.~Rev.~D}

\begin{document}

\title{\bf \boldmath Observation~and~measurement~of~Higgs~boson~decays~to~$\WWs$~with~the~ATLAS~detector}
\author{G.~Aad \etal}
\email{Full author list given at the end of the article}
\homepage{}
\affiliation{\rm (ATLAS Collaboration)}
\date{\today}

\begin{abstract}
We report the observation of Higgs boson decays to $\WWs$ based on an excess over background
of $\HwwSignif{obs}{all}$ standard deviations in the dilepton final state, where the
Standard Model expectation is $\HwwSignif{exp}{all}$ standard deviations.  Evidence for the
vector-boson fusion (VBF) production process is obtained with a significance of
$\HwwSignif{obs}{VX}$ standard deviations.  The results are obtained from a data sample
corresponding to an integrated luminosity of $\HwwLumi{0}{0}\ifb$ from $\sqrt{s}{\EQ}7$ and
$8\TeV$ $pp$ collisions recorded by the ATLAS detector at the LHC.  For a Higgs boson mass
of $\HwwHiggsMass{ATLAS}\GeV$, the ratio of the measured value to the expected value of the
total production cross section times branching fraction is
$1.09^{+0.16}_{-0.15}\,\textrm{(stat)}^{+0.17}_{-0.14}\,\textrm{(syst)}$.  The
corresponding ratios for the gluon fusion and vector-boson fusion production mechanisms are
$1.02{\PM}0.19\,\textrm{(stat)}\,^{+0.22}_{-0.18}\,\textrm{(syst)}$ and
$1.27\,^{+0.44}_{-0.40}\,\textrm{(stat)}\,^{+0.30}_{-0.21}\,\textrm{(syst)}$, respectively.
At $\sqrt{s}{\EQ}8\TeV$, the total production cross sections are measured to be
$\sigma(gg{\TO}\HWW) = 4.6{\PM}0.9\,\textrm{(stat)}\,^{+0.8}_{-0.7}\,\textrm{(syst)}\pb$ and
$\sigma(\VBF~\HWW){\EQ}0.51\,^{+0.17}_{-0.15}\,\textrm{(stat)}\,^{+0.13}_{-0.08}\,\textrm{(syst)}\pb$.
The fiducial cross section is determined for the gluon-fusion
process in exclusive final states with zero or one associated jet.
\end{abstract}
\maketitle

\section{\boldmath Introduction \label{sec:introduction}}

In the Standard Model of particle physics (SM), the Higgs boson results from
the Brout-Englert-Higgs mechanism~\cite{Englert:1964xx} that breaks the
electroweak symmetry~\cite{Glashow:1961xx} and gives mass to the $W$ and $Z$
gauge bosons~\cite{WZmasses}.  It has a spin-parity of $0^+$, with couplings
to massive particles that are precisely determined by their measured masses.
A new particle compatible with the spin and gauge-boson couplings of the SM
Higgs boson was discovered in 2012 by the ATLAS and CMS experiments at the
LHC using the $\ZZs$, $\gamma\gamma$, and $\WWs$ final
states~\cite{discovery, couplings, spin, cmscouplingsspin, cmsMassPaper}.
Measurements of the particle's mass~\cite{cmsMassPaper, atlasMassPaper}
yield a value of approximately $125\GeV$, consistent with the mass of the SM
Higgs boson provided by a global fit to electroweak measurements~\cite{ewkfit}.
Evidence for production of this boson at the Tevatron~\cite{tevfermions} and
for its decay to fermions at the LHC~\cite{lhcfermions} are also consistent
with the properties of the SM Higgs boson.

The direct observation of the Higgs boson in individual decay channels provides
an essential confirmation of the SM predictions.  For a Higgs boson with a mass
of $125\GeV$, the $\HWW$ decay has the second largest branching fraction ($22\%$)
and is a good candidate for observation.  The sequential decay $\HWWlvlv$, where
$\ell$ is an electron or muon, is a sensitive experimental signature.  Searches
for this decay produced the first direct limits on the mass of the Higgs boson
at a $pp$ collider~\cite{TeVlimits, LHClimits}, and measurements following
the boson discovery are among the most precise in determining its couplings and
spin~\cite{couplings, spin, cmscouplingsspin}.

\begin{figure}[tb!]
\hspace{0pt}\includegraphics{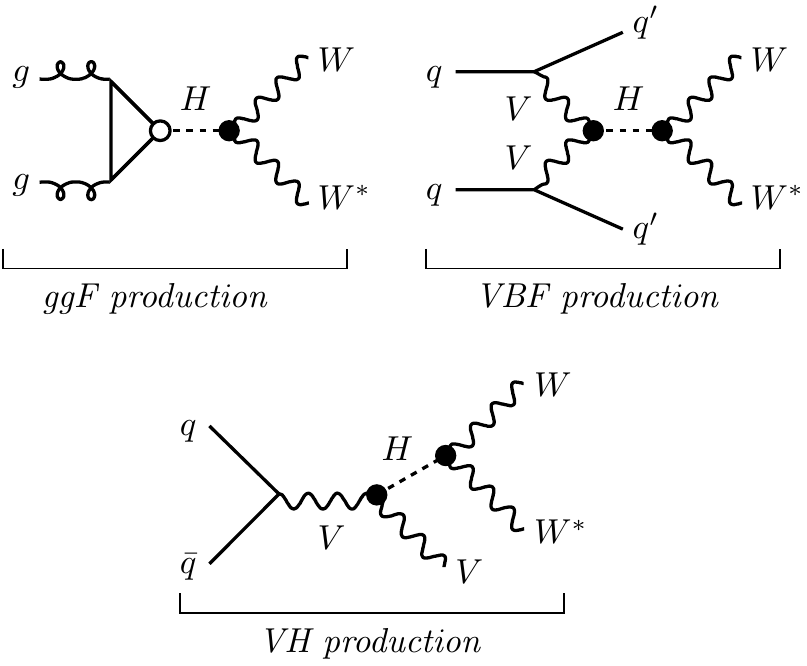}
\caption{
  Feynman diagrams for the leading production modes (ggF, VBF, and $\VH$), where the
  $VV\!H$ and $qqH$ coupling vertices are marked by $\bullet$ and $\circ$, respectively.
  The $V$ represents a $W$ or $Z$ vector boson.
}
\label{fig:production}
\end{figure}

The dominant Higgs boson production mode in high-energy hadron collisions is
gluon fusion (ggF), where the interacting gluons produce a Higgs boson
predominantly through a top-quark loop.  The next most abundant production
mechanism, with a factor of $12$ reduction in rate, is the fusion of vector
bosons radiated by the interacting quarks into a Higgs boson (vector-boson
fusion or VBF).  At a further reduced rate, a Higgs boson can be produced in
association with a $W$ or $Z$ boson (vector and Higgs boson production or $\VH$).  The leading-order
production processes are depicted in Fig.~\ref{fig:production}.

This \paper\ describes the observation and measurement of the Higgs boson in its
decay to a pair of $W$ bosons, with the Higgs boson produced by the ggF and VBF
processes at center-of-mass energies of $7$ and $8\TeV$.  The ggF production process
probes Higgs boson couplings to heavy quarks, while the VBF and $\VH$ processes
probe its couplings to $W$ and $Z$ bosons.  The branching fraction $\mathcal{B}_{\HWW}$
is sensitive to Higgs boson couplings to the fermions and bosons through the total
width.  To constrain these couplings, the rates of the ggF and VBF $\HWW$ processes
are measured---individually and combined---and normalized by the SM predictions for
a Higgs boson with mass $\HwwHiggsMass{ATLAS}\GeV$~\cite{atlasMassPaper}
to obtain the ``signal strength'' parameters $\sigmu$, $\sigmu_{\scggF}$, and
$\sigmu_{\scVX}$.  The total cross section for each process is also measured, along
with fiducial cross sections for the ggF process.

A prior measurement of these processes with the same data set yielded a combined
result of $\sigmu{\EQ}1.0{\PM}0.3$~\cite{couplings}.  The results presented
here supersede this measurement and contain improvements in signal acceptance,
background determination and rejection, and signal yield extraction.  Together,
these improvements increase the expected significance of an excess of $\HWW$ decays
over background from $3.7$ to $\HwwSignif{exp}{all}$ standard deviations, and
reduce the expected relative uncertainty on the corresponding $\sigmu$ measurement
by $30\%$.

The \paper\ is organized as follows. Section~\ref{sec:analysis} provides
an overview of the signal and backgrounds, and of the data analysis strategy.
Section~\ref{sec:atlas} describes the ATLAS detector and data, and the
event reconstruction. The selection of events in the different
final states is given in Sec.~\ref{sec:selection}.  Sections~\ref{sec:signal}
and~\ref{sec:bkg} discuss the modeling of the signal and the background
processes, respectively. The signal yield extraction and the various
sources of systematic uncertainty are described in Sec.~\ref{sec:systematics}.
Section~\ref{sec:yields} provides the event yields and the distributions of
the final discriminating variables; the differences with respect to previous
ATLAS measurements in this channel \cite{couplings} are given in Sec.~\ref{sec:yields_differences}.
The results are presented in
Sec.~\ref{sec:results}, and the conclusions given in
Sec.~\ref{sec:conclusions}.

\section{\boldmath Analysis overview \label{sec:analysis}}

\begin{figure}[bt!]
\hspace{-10pt}\includegraphics{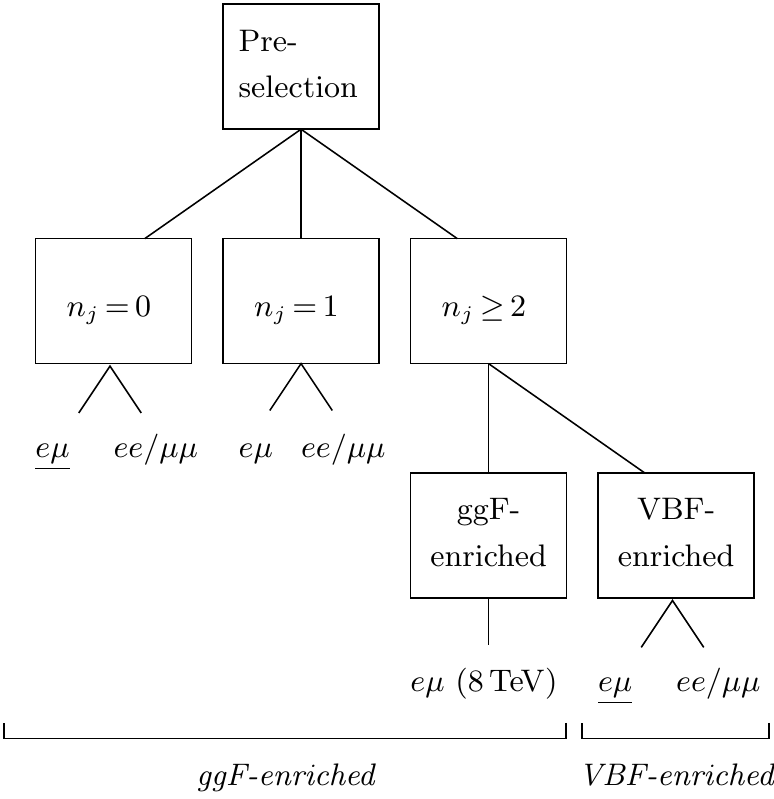}
\caption{
  Analysis divisions in categories based on jet multiplicity ($\Njet$) and
  lepton-flavor samples ($\emu$ and $ee/\mu\mu$).  The most sensitive
  signal region for ggF production is $\NjetEQzero$ in $\emu$, while for VBF
  production it is $\NjetGEtwo$ in $\emu$. These two samples are underlined.
  The $\emu$ samples with $\NjetLEone$ are further subdivided as described
  in the text.
}
\label{fig:flowchart}
\end{figure}

The $\HWW$ final state with the highest purity at the LHC occurs when
each $W$ boson decays leptonically, $W{\TO}\ell\nu$, where $\ell$ is an
electron or muon.  The analysis therefore selects events consistent with
a final state containing neutrinos and a pair of opposite-charge leptons.
The pair can be an electron and a muon, two electrons, or two muons.  The
relevant backgrounds to these final states are shown in Table~\ref{tab:process}
and are categorized as $\WW$, top quarks, misidentified leptons, other
dibosons, and Drell-Yan.  The distinguishing features of these backgrounds,
discussed in detail below, motivate the definition of event categories
based on lepton flavor and jet multiplicity, as illustrated in
Fig.~\ref{fig:flowchart}.  In the final step of the analysis, a profile
likelihood fit is simultaneously performed on all categories in order to
extract the signal from the backgrounds and measure its yield.

\begin{table}[bt!]
\caption{
  Backgrounds to the $\HWW$ measurement in the final state with two charged leptons
  ($\ell{\EQ}e$ or $\mu$) and neutrinos, and no jet that contains a $b$-quark.
  Irreducible backgrounds have the same final state; other backgrounds are shown with
  the features that lead to this final state.  Quarks from the first or second
  generation are denoted as $q$, and $j$ represents a jet of any flavor.
}
\label{tab:process}
\begin{tabular*}{0.480\textwidth}{p{0.075\textwidth} p{0.180\textwidth} l}
\dbline
Name             & Process                                  & Feature(s) \\
\sgline
\multicolumn{1}{l}{$WW$} & $\WW$                            & Irreducible \\
\clineskip
\clineskip
\multicolumn{2}{l}{Top quarks} \\
\quad$\ttbar$    & $\ttbar{\TO}Wb\,W\bar{b}$                & Unidentified $b$-quarks \\
\multirow{2}{*}{\quad$t$}
                 & \hspace{-12.5pt}\raisebox{2pt}{\ldelim\{{2}{0pt}[]}\hspace{9.5pt}
                   $tW$                                     & Unidentified $b$-quark \\
                 & $t\bar{b}$, $tq\bar{b}$                  & $q$ or $b$ misidentified as $\ell$; \\
                 &                                          & unidentified $b$-quarks \\
\clineskip
\multicolumn{2}{l}{Misidentified leptons (Misid)} \\
\quad$\Wj$       & $W{\PLUS}\textrm{jet(s)}$                & $j$ misidentified as $\ell$ \\
\quad$\jj$       & Multijet production                      & $jj$ misidentified as $\ell\ell$; \\
                 &                                          & misidentified neutrinos \\
\clineskip
\multicolumn{2}{l}{Other dibosons} \\
 & \hspace{-13.5pt}\ldelim\{{4}{0pt}[]\hspace{10.5pt} $\Wg$ & $\gamma$ misidentified as $e$ \\
\multirow{2}{*}{\quad$VV$}
                 & $\Wgs$, $\WZ$, $\ZZ{\TO}\ll\,\ll$        & Unidentified lepton(s) \\
                 & $\ZZ{\TO}\ll\,\nu\nu$                    & Irreducible \\
                 & $Z\gamma$                                & $\gamma$ misidentified as $e$; \\
                 &                                          & unidentified lepton \\
\clineskip
\multicolumn{2}{l}{Drell-Yan (DY)} \\
\quad$\SFchan$   & $\ZDYll$                                 & Misidentified neutrinos \\
\quad$\tautau$   & $\ZDYtt{\TO}\ell\nu\nu\,\ell\nu\nu$      & Irreducible \\
\dbline
\end{tabular*}
\end{table}

The Drell-Yan (DY) process is the dominant source of events with two
identified leptons, and contributes to the signal final state when there
is a mismeasurement of the net particle momentum in the direction
transverse to the beam (individual particle momentum in this direction is
denoted $\vpT$).  The DY background is strongly reduced in events with
different-flavor leptons ($\DFchan$), as these arise through fully leptonic
decays of $\tau$-lepton pairs with a small branching fraction and reduced
lepton momenta.  The analysis thus separates $\DFchan$ events from those with
same-flavor leptons ($\SFchan$) in the event selection and the likelihood
fit.

Pairs of top quarks are also a prolific source of lepton pairs, which are
typically accompanied by high-momentum jets.  Events are removed if they have
a jet identified to contain a $b$-hadron decay ($\bjet$), but the $t\bar{t}$
background remains large due to inefficiencies in the $b$-jet identification
algorithm.  Events are therefore categorized by the number of jets.  The
top-quark background provides a small contribution to the zero-jet category
but represents a significant fraction of the total background in categories
with one or more jets.

In events with two or more jets, the sample is separated by signal production
process (``VBF-enriched'' and ``ggF-enriched'').  The VBF process is
characterized by two quarks scattered at a small angle, leading to two
well-separated jets with a large invariant mass~\cite{vbf}.  These and
other event properties are inputs to a boosted decision tree (BDT)
algorithm~\cite{BDT} that yields a single-valued discriminant to isolate
the VBF process.  A separate analysis based on a sequence of individual
selection criteria provides a cross-check of the BDT analysis.  The
ggF-enriched sample contains all events with two or more jets that do
not pass either of the VBF selections.

\begin{figure}[bt!]
\includegraphics{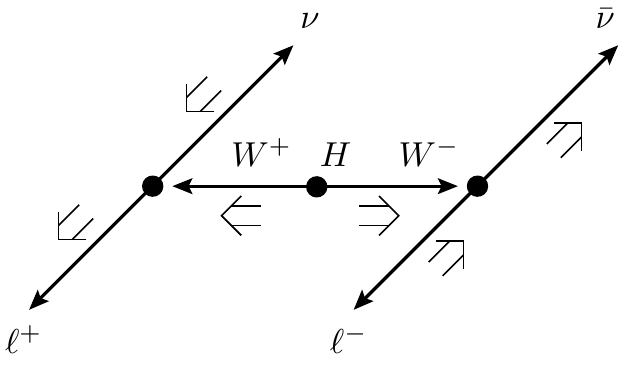}
\caption{
  Illustration of the $H{\TO}WW$ decay.  The small arrows indicate the particles'
  directions of motion and the large double arrows indicate their spin projections.
  The spin-$0$ Higgs boson decays to $W$ bosons with opposite spins, and the spin-$1$
  $W$ bosons decay into leptons with aligned spins.  The $H$ and $W$ boson decays are shown
  in the decaying particle's rest frame.  Because of the $V{\MINUS}A$ decay of the
  $W$ bosons, the charged leptons have a small opening angle in the laboratory frame.
  This feature is also present when one $W$ boson is off shell.
}
\label{fig:decay}
\end{figure}

Due to the large Drell-Yan and top-quark backgrounds in events with
same-flavor leptons or with jets, the most sensitive signal region is in the
$\emu$ zero-jet final state.  The dominant background to this category is $WW$
production, which is effectively suppressed by exploiting the properties of
$W$ boson decays and the spin-0 nature of the Higgs boson
(Fig.~\ref{fig:decay}).  This property generally leads to a lepton pair with
a small opening angle~\cite{spintheory} and a correspondingly low invariant mass
$\mll$, broadly distributed in the range below $\mH/2$.  The dilepton invariant
mass is used to select signal events, and the signal likelihood fit is
performed in two ranges of $\mll$ in $\emu$ final states with $\NjetLEone$.

Other background components are distinguished by $\pTsublead$, the magnitude
of the transverse momentum of the lower-$\pT$ lepton in the event (the
``subleading'' lepton).  In the signal process, one of the $W$ bosons from the
Higgs boson decay is off shell, resulting in relatively low subleading lepton
$\pT$ (peaking near 22~GeV, half the difference between the Higgs and $W$ boson
masses). In the background from $W$ bosons produced in association with a jet or
photon (misreconstructed as a lepton) or an off-shell photon producing a low-mass
lepton pair (where one lepton is not reconstructed), the $\pTsublead$ distribution
falls rapidly with increasing $\pT$.  The $\emu$ sample is therefore subdivided
into three regions of subleading lepton $\pT$ for $\NjetLEone$.  The jet and
photon misidentification rates differ for electrons and muons, so this sample is
further split by subleading lepton flavor.

Because of the neutrinos produced in the signal process, it is not possible
to fully reconstruct the invariant mass of the final state.  However, a
``transverse mass'' $\mTH$~\cite{mTpub} can be calculated without the
unknown longitudinal neutrino momenta:
\begin{equation}
  \mTH = \sqrt{\big(\eTll + \pTnu\big)^2 - \big|\,\vpTll + \vpTnu\,\big|^2},
\label{eqn:mTH}
\end{equation}
where $\eTll{\EQ}\sqrt{(\pTll)^2+(\mll)^2}$, $\vpTnu$ ($\vpTll$) is the
vector sum of the neutrino (lepton) transverse momenta, and $\pTnu$
($\pTll$) is its modulus.  The distribution has a kinematic upper bound at
the Higgs boson mass, effectively separating Higgs boson production from the
dominant nonresonant $\WW$ and top-quark backgrounds.  For the VBF analysis,
the transverse mass is one of the inputs to the BDT distribution used to fit
for the signal yield.  In the ggF and cross-check VBF analyses, the signal
yield is obtained from a direct fit to the $\mTH$ distribution for each
category.

Most of the backgrounds are modeled using Monte Carlo samples normalized to
data, and include theoretical uncertainties on the extrapolation from the
normalization region to the signal region, and on the shape of the distribution
used in the likelihood fit.  For the $W$+jet(s) and multijet backgrounds, the
high rates and the uncertainties in modeling misidentified leptons motivate a
model of the kinematic distributions based on data.  For a few minor backgrounds,
the process cross sections are taken from theoretical calculations.  Details of
the background modeling strategy are given in Sec.~\ref{sec:bkg}.

The analyses of the $7$ and $8\TeV$ data sets are separate, but use common methods
where possible; differences arise primarily because of the lower instantaneous
and integrated luminosities in the $7\TeV$ data set.  As an example, the
categorization of $7\TeV$ data does not include a ggF-enriched category for events
with at least two jets, since the expected significance of such a category is
very low.  Other differences are described in the text or in dedicated
subsections.

\section{\boldmath Data samples and reconstruction \label{sec:atlas}}

This section begins with a description of the ATLAS detector,
the criteria used to select events during data-taking (triggers)
and the data sample used for this analysis.
A description of the event reconstruction follows.
The Monte Carlo simulation samples used in this analysis are described
next, and then differences between the 2012 and 2011 analyses are summarized.

\subsection{\boldmath Detector and data samples \label{sec:atlas_samples}}

The ATLAS detector~\cite{Aad:2008zzm} is a multipurpose particle detector with approximately
forward-backward symmetric cylindrical geometry.
The experiment uses a right-handed coordinate system with the origin at
the nominal $pp$ interaction point at the center of the detector.  The
positive $x$ axis is defined by the direction from the origin to the center
of the LHC ring, the positive $y$ axis points upwards, and the $z$ axis is
along the beam direction.  Cylindrical coordinates $(r,\phi)$ are used in the
plane transverse to the beam; $\myphi$ is the azimuthal angle around the beam
axis.  Transverse components of vectors are indicated by the subscript T.
The pseudorapidity is defined in terms of the polar angle $\theta$ as
$\myeta{\EQ}{-}\ln\tan(\theta/2)$.

The inner tracking detector (ID) consists of a silicon-pixel
detector, which is closest to the interaction point,
a silicon-microstrip detector
surrounding the pixel detector---both covering
$\ABS{\myeta}{\LT}2.5$---and an outer transition-radiation straw-tube tracker (TRT) covering $\ABS{\myeta}{\LT}2$.
The TRT provides substantial discriminating power between electrons and pions over a wide energy range.
The ID is surrounded by a thin superconducting
solenoid providing a $2\Tesla$ axial magnetic field.

A highly segmented lead/liquid-argon (LAr)
sampling electromagnetic calorimeter measures the energy and the position of electromagnetic showers
with $\ABS{\myeta}{\LT}3.2$.  The LAr calorimeter includes a presampler (for $\ABS{\myeta}{\LT}1.8$) and
three sampling layers, longitudinal in shower depth, up to $\ABS{\myeta}{\LT}2.5$.  The LAr sampling calorimeters
are also used to measure hadronic showers in the endcap ($1.5{\LT}\ABS{\myeta}{\LT}3.2$) and
electromagnetic and hadronic showers in the forward regions ($3.1{\LT}\ABS{\myeta}{\LT}4.9$), while a
steel/scintillator tile calorimeter measures hadronic showers in the central region
($\ABS{\myeta}{\LT}1.7$).

The muon spectrometer (MS) surrounds the calorimeters and is designed to detect muons in
the pseudorapidity range $\ABS{\myeta}{\LT}2.7$. The MS consists of one barrel
($\ABS{\myeta}{\LT}1.05$) and two endcap regions. A system of three large superconducting air-core
toroid magnets, each with eight coils, provides a magnetic field with a bending integral of about
$2.5\Teslam$ in the barrel and up to $6\Teslam$ in the endcaps.
Monitored drift tube chambers
in both the barrel and endcap regions and
cathode strip chambers
covering $2.0{\LT}\ABS{\myeta}{\LT}2.7$ are used as precision-measurement chambers,
whereas resistive plate chambers
in the barrel
and thin gap chambers
in the endcaps are used as trigger chambers,
covering $\ABS{\myeta}{\LT}2.4$.
The chambers are arranged in three layers,
so high-$\pT$ particles traverse at least three stations with a lever
arm of several meters.

A three-level trigger system selects events to be recorded for offline analysis.
The first level (level-1 trigger) is hardware based, and the second two levels
(high-level trigger) are software based.
This analysis uses events selected by triggers that required either a single lepton or two leptons (dilepton).
The single-lepton triggers had more restrictive lepton identification requirements and higher
$\pT$~thresholds than the dilepton triggers.
The specific triggers used for the $8\TeV$ data with the corresponding thresholds
at the hardware and software levels are listed in Table~\ref{tab:trigger}.
Offline, two leptons---either $ee$, $\mu\mu$ or $e\mu$---with opposite charge are required.
The leading lepton ($\ell_1$) is required to have $\pT{\GE}22\GeV$ and the
subleading lepton ($\ell_2$) is required to have $\pT{\GE}10\GeV$.

\begin{table}[b!]
\caption{
  Summary of the
  minimum lepton $\pT$ trigger requirements (in $\!\GeV$)
  during the $8\TeV$ data-taking.
  For single-electron triggers, the hardware and software thresholds are either
  $18$ and $24$i or $30$ and $60$, respectively.
  The ``i'' denotes an isolation requirement that is less restrictive than the isolation
  requirement imposed in the offline selection.
  For dilepton triggers, the pair of thresholds corresponds to the leading and subleading lepton, respectively;
  the ``$\mu, \mu$'' dilepton trigger requires only a single muon at level-1.
  The ``and'' and ``or'' are logical.
}
\label{tab:trigger}
{\small
  \centering
\begin{tabular*}{0.480\textwidth}{
  p{0.135\textwidth}
  p{0.165\textwidth}
  l
  @{\extracolsep{\fill}}*{1}{l}
}
\dbline
  Name
& Level-1 trigger
& High-level trigger
\\
\sgline
\multicolumn{3}{l}{Single lepton} \\
\quad$e$          & $18$ or $30$  & $24$i or $60$ \\
\quad$\mu$        & $15$          & $24$i or $36$ \\
\clineskip
\clineskip
\multicolumn{3}{l}{Dilepton} \\
\quad$e$, $e$     & $10$ and $10$ & $12$ and $12$ \\
\quad$\mu$, $\mu$ & $15$          & $18$ and  $8$ \\
\quad$e$, $\mu$   & $10$ and $6$  & $12$ and  $8$ \\
\dbline
\end{tabular*}
}
\end{table}

The efficiency of the trigger selection is measured using a tag-and-probe method with
a data sample of $\ZDYll$ candidates.
For muons, the single-lepton trigger efficiency varies with $\myeta$ and is
approximately $70\%$ for $\ABS{\myeta}{\LT}1.05$ and $90\%$ for $\ABS{\myeta}{\GT}1.05$.
For electrons, the single-lepton trigger efficiency increases with
$\pT$, and its average is approximately $90\%$.
These trigger efficiencies are for leptons that satisfy the analysis
selection criteria described below.
Dilepton triggers increase the signal acceptance by allowing lower leading-lepton $\pT$
thresholds to be applied offline while still remaining in the kinematic range
that is in the plateau of the trigger efficiency.
The trigger efficiencies for signal events satisfying the selection criteria described in
Sec.~\ref{sec:selection}
are $95\%$ for events with a leading electron and a subleading muon,
$81\%$ for events with a leading muon and subleading electron,
$89\%$ for $\mu\mu$ events and $97\%$ for $ee$ events.
These efficiencies are for the $\NjetEQzero$ category;
the efficiencies are slightly larger for categories with higher jet multiplicity.

The data are subjected to quality requirements: events recorded when the relevant detector
components were not operating correctly are rejected.
The resulting integrated luminosity is $20.3\ifb$ taken at $\sqrt{s}{\EQ}8\TeV$ in 2012
and $4.5\ifb$ at $7\TeV$ in 2011.
The mean number of inelastic collisions per bunch crossing
had an average value of $20$ in 2012 and $9$ in 2011.
Overlapping signals in the detector due to these multiple interactions---as well as signals due to interactions
occurring in other nearby bunch crossings---are referred to as ``pile-up.''

\subsection{\boldmath Event reconstruction \label{sec:atlas_detector}}

The primary vertex of each event
must have at least three tracks with $\pT{\GE}400\MeV$
and is selected as the vertex with the largest value of $\sumpTsq$,
where the sum is over all the tracks associated with that particular vertex.

Muon candidates are identified by matching a reconstructed ID track
with a reconstructed MS track~\cite{MCPpaper2014}.
The MS track is required to have a track segment in at least two layers of the MS.
The ID tracks are required to have at least a minimum number of associated hits
in each of the ID subdetectors to ensure good track reconstruction.
This analysis uses muon candidates referred to as ``combined muons'' in Ref.~\cite{MCPpaper2014},
in which the track parameters of the MS track and the ID track are combined statistically.
Muon candidates are required to have $\ABS{\myeta}{\LT}2.50$.
The efficiencies for reconstructing and identifying combined muons are provided in Ref.~\cite{MCPpaper2014}.

Electron candidates are clusters of energy deposited in the electromagnetic calorimeter that are associated
with ID tracks~\cite{Aad:2011mk}.  All candidate electron tracks are fitted using a Gaussian sum
filter~\cite{GSFConf} (GSF) to account for bremsstrahlung energy losses.
The GSF fit reduces the difference between the energy measured in the calorimeter
and the momentum measured in the ID and improves the measured electron direction
and impact parameter resolutions.
The impact parameter is the lepton track's distance of closest approach in the transverse plane
to the reconstructed position of the primary vertex.
The electron transverse energy is computed from the cluster energy and the track direction
at the interaction point.

Electron identification is performed in the range $\ABS{\myeta}{\LT}2.47$,
excluding the transition region between the barrel and endcap EM calorimeters, $1.37{\LT}\ABS{\myeta}{\LT}1.52$.
The identification is based on criteria that require the longitudinal and transverse shower
profiles to be consistent with those expected for electromagnetic showers, the track and cluster
positions to match in $\myeta$ and $\myphi$, and
signals of transition radiation in the TRT.
The electron identification has been improved relative to that described in Ref.~\cite{couplings}
by adding a likelihood-based method in addition to the selection-based method.
The likelihood allows the inclusion of discriminating variables that are difficult to use with explicit
requirements without incurring significant efficiency losses.
Detailed discussions of the likelihood identification and selection-based identification
and the corresponding efficiency measurements can be found in Ref.~\cite{ElectronEff2012}.
Electrons with $10{\LT}\eT{\LT}25$\,GeV must satisfy the  ``very tight'' likelihood requirement,
which reduces backgrounds from light-flavor jets and photon conversions by $35\%$ relative
to the selection-based identification with the same signal efficiency.
For $\eT{\GT}25\GeV$, where misidentification backgrounds are less important,
electrons must satisfy the ``medium'' selection-based requirement.
The single-lepton trigger applies the medium selection-based requirements.
Using a likelihood-based selection criterion in addition to this selection-based requirement would result in
a loss of signal efficiency without sufficient compensation in background rejection.
Finally, additional requirements reduce the contribution of electrons from photon conversions by rejecting
electron candidates that have an ID track that is part of a conversion vertex or that do not
have a hit in the innermost layer of the pixel detector.

To further reduce backgrounds from misidentified leptons,
additional requirements are imposed on the lepton impact parameter and isolation.
The significance of the transverse impact parameter, defined as the measured transverse impact parameter $d_0$
divided by its estimated uncertainty $\sigma_{d_0}$, is required to satisfy $\ABS{d_0}/\sigma_{d_0}{\LT}3.0$;
the longitudinal impact parameter $z_0$ must satisfy the requirement
$\ABS{z_0\sin\theta}{\LT}0.4\mm$ for electrons and $1.0\mm$ for muons.

Lepton isolation is defined using track-based and calorimeter-based quantities.
Details about the definition of
electron isolation can be found in Ref.~\cite{ElectronEff2012}.
The track isolation is based on the scalar sum $\sumpT$ of all tracks
with $\pT{\GT}400\MeV$ for electrons ($\pT{\GT}1\GeV$ for muons)
that are found in a cone in \mbox{$\myeta$-$\myphi$} space around
the lepton, excluding the lepton track.
Tracks used in this scalar sum are required to be consistent with
coming from the primary vertex.
The cone size is $\DeltaR{\EQ}0.4$ for leptons with $\pT{\LT}15\GeV$,
where $\DeltaR{\EQ}\sqrt{(\Delta\myphi)^2{\PLUS}(\Delta\myeta)^2}$,
and $\DeltaR{\EQ}0.3$ for $\pT{\GT}15\GeV$.
The track isolation selection criterion uses the ratio of the $\sumpT$ divided by the electron $\eT$ (muon $\pT$).
This ratio is required to be less than $0.06$ for leptons with $10{\LT}\pT{\LT}15\GeV$,
and this requirement increases monotonically to $0.10$ for electrons ($0.12$ for muons) for $\pT{\GT}25\GeV$.

The calorimeter isolation selection criterion---like the track isolation---is based on a ratio.
The relative calorimetric isolation for electrons is computed as the sum of the
cluster transverse energies $\sumeT$ of surrounding energy deposits
in the electromagnetic and hadronic calorimeters
inside a cone of $\DeltaR{\EQ}0.3$ around the candidate electron cluster,
divided by the electron $\eT$.
The cells within $0.125{\TIMES}0.175$ in $\myeta{\TIMES}\myphi$ around the electron
cluster barycenter are excluded.
The pile-up and underlying-event contributions to the calorimeter isolation
are estimated and subtracted event by event.
The electron relative calorimetric isolation upper bound varies monotonically with electron $\eT$:
it is
$0.20$ for $10{\LT}\eT{\LT}15\GeV$, increasing to
$0.28$ for $\eT{\GT}25\GeV$.
In the case of muons, the relative calorimetric isolation discriminant is defined
as the $\sumeT$ calculated from calorimeter cells within $\DeltaR{\EQ}0.3$ of the
muon candidate, and with energy above a noise threshold, divided by the
muon $\pT$.
All calorimeter cells within the range $\DeltaR{\LT}0.05$ around the muon candidate are
excluded from $\sumeT$.
A correction based on the number of reconstructed primary vertices in the event
is made to $\sumeT$ to compensate for extra energy due to pile-up.
The muon relative calorimetric isolation upper bound also varies monotonically with muon $\pT$;
it is
$0.06$ for $10{\LT}\pT{\LT}15\GeV$, increasing to
$0.28$ for $\pT{\GT}25\GeV$.
The signal efficiencies of the impact parameter and isolation requirements
are measured using a tag-and-probe method with a data sample of $\ZDYll$ candidates.
The efficiencies of the combined impact parameter and isolation requirements range from
$68\%$ ($60\%$) for electrons (muons) with $10{\LT}\pT{\LT}15\GeV$ to
greater than $90\%$ ($96\%$) for electrons (muons) with $\pT{\GT}25\GeV$.

Jets are reconstructed using the anti-$k_t$ sequential recombination clustering
algorithm~\cite{Cacciari:2005hq} with a radius parameter $R{\EQ}0.4$.
The inputs to the reconstruction are three-dimensional
clusters of energy~\cite{Lampl:1099735,Aad:2011he} in the calorimeter.
The algorithm for this clustering suppresses noise by keeping only
cells with a significant energy deposit and their neighboring cells.
To take into account the differences in calorimeter response to
electrons and photons and hadrons, each cluster is classified,
prior to the jet reconstruction, as coming from an
electromagnetic or hadronic shower using information from its shape.
Based on this classification,
the local cell signal weighting calibration method~\cite{Aad:2014bia}
applies dedicated corrections for the effects of calorimeter noncompensation,
signal losses due to noise threshold effects and energy lost in regions that are not instrumented.
Jets are corrected for contributions from in-time and out-of-time
pile-up~\cite{ATLAS:2012lla}, and the position of the primary interaction vertex.
Subsequently, the jets are calibrated to the hadronic energy scale using $\pT$- and
$\myeta$-dependent correction factors determined in a first pass from simulation
and then refined in a second pass from data~\cite{Aad:2011he,Aad:2014bia}.
The systematic uncertainties on these correction factors are determined
from the same control samples in data.

To reduce the number of jet candidates originating from pile-up vertices,
a requirement is imposed on the jet vertex fraction, denoted $\jvf$:
for jets with $\pT{\LT}50\GeV$ and $\ABS{\myeta}{\LT}2.4$, more than $50\%$
of the summed scalar $\pT$ of tracks within $\DeltaR{\EQ}0.4$ of the jet axis must be from tracks
associated with the primary vertex
($\jvf{\GT}0.50$)~\cite{TheATLAScollaboration:2013pia}.
No $\jvf$ selection requirement is applied to jets that have no associated tracks.

For the purposes of classifying an event in terms of jet multiplicity $\Njet$,
a jet is required to have $\pTjet{\GT}25\GeV$ for $\HwwAtlasEtaJetCtr$,
and $\pTjet{\GT}30\GeV$ if $\HwwAtlasEtaJetFwd$.
The increased threshold in the higher-$\ABS{\myeta}$ region suppresses jets
from pile-up.
The two highest-$\pT$ jets ($j_1$, $j_2$, ordered in $\pT$) are the ``VBF jets'' used to compute dijet variables in the VBF-enhanced $\NjetGEtwo$ category.

Additional jets not counted in $\Njet$ have lower thresholds in three scenarios.
First, those used to reject events because they lie in the $\myeta$ range
spanned by the two leading jets
in the VBF-enriched selection (see Sec.~\ref{sec:selection_2jvbf}) are considered if they have $\pTjet{\GT}20\GeV$.
Second, the jets for $b$-jet identification---described below---are required to have $\pTjet{\GT}20\GeV$ and
$\ABS{\etajet}{\LT}2.4$.
Third, the jets used for the calculation of soft hadronic recoil (see Sec.~\ref{sec:selection_0j}
and the $\frecoil$ definition therein)
are required to have $\pTjet{\GT}10\GeV$ and have no $\jvf$ requirement.
The calibration procedure described above is applied only to jets with $\pTjet{\GT}20\GeV$.  Jets with
$10\GeV{\LT}\pTjet{\LT}20\GeV$ are used only in the $\frecoil$ definition, and the efficiency for the
requirements on this quantity are measured directly from the data, so the analysis is not sensitive to
the modeling of the energy scale of these soft jets in the Monte Carlo simulation.

The identification of $b$-quark jets ($b$-jets) is limited to the acceptance of the ID ($\ABS{\myeta}{\LT}2.5$).
The $b$-jets are identified with a multivariate
technique---the MV1 algorithm~\cite{ATLAS-CONF-2014-046}---that is based on quantities that
separate $b$ and $c$ jets from ``light jets'' arising from light-flavor quarks and gluons.
The inputs~\cite{ATLAS-CONF-2011-102} to this algorithm use quantities such as the presence of secondary
vertices, the impact parameters of tracks, and the
topologies of weak heavy-quark decays.
The efficiency for identifying $b$-jets is measured~\cite{ATLAS-CONF-2014-004}
in a large data sample of dilepton $\ttbar$~pair candidates.
An operating point that is $85\%$ efficient for identifying $b$-jets is adopted.
At this operating point, the probability of misidentifying a light jet
as a $b$-jet is $10.3\%$.

Two leptons or a lepton and a jet may be close in \mbox{$\myeta$-$\myphi$}~space.
The following procedure is adopted in the case of overlapping objects.
Electron candidates that have tracks that extend to the MS are removed.
If a muon candidate and an electron candidate are separated by
$\DeltaR{\LT}0.1$,
then the muon is retained, and the electron is removed.
These cases usually indicate a muon that has undergone bremsstrahlung in the ID material
or calorimeter.
A high-$\pT$ electron is always also reconstructed as a jet,
so if an electron and the nearest jet are separated by less than $\DeltaR{\EQ}0.3$,
the jet is removed.
In contrast, if a muon and a jet are separated by less than $\DeltaR{\EQ}0.3$,
the muon candidate is removed,
as it is more likely to be a nonprompt muon from heavy-flavor decay.
Finally, due to early bremsstrahlung, a prompt electron may produce more than one electron candidate
in its vicinity.
In the case of two electrons separated by less than $\DeltaR{\EQ}0.1$,
the electron candidate with larger $\eT$ is retained.

The signature of a high-momentum neutrino is a momentum imbalance in the transverse
plane.  The reconstruction of this ``missing'' transverse momentum~\cite{met-perf}
is calculated as the negative vector sum of the momentum of objects selected according
to ATLAS identification algorithms, such as leptons, photons, and jets, and of the
remaining ``soft'' objects that typically have low values of $\pT$.  The calculation
can thus be summarized as
\begin{equation}
\vMET = -\raisebox{-3.0pt}{\bigg(}
     \displaystyle
     \sum_{\rm selected}\no\!\vpT
   + \sum_{\rm soft}\,\vpT
   \raisebox{-3.0pt}{\bigg)},
\label{eqn:MET}
\end{equation}
\noindent
where the reconstruction of soft objects and the choice of selected objects differ
between different methods of evaluating the missing transverse momentum.  Three
methods of reconstruction are used in this analysis; $\vMET$ is used to represent
one particular method, as described below.

The large coverage in rapidity ($y$) of the calorimeter and its sensitivity to neutral
particles motivate a calorimeter-based reconstruction of the missing transverse
momentum.  Selected objects are defined as the leptons selected by the analysis,
and photons and jets with $\eT{\GT}20\GeV$.  The transverse momenta of these
objects are added vectorially using object-specific calibrations.  For the
remaining soft objects, calibrated calorimeter cluster energy measurements are
used to determine their net transverse momentum.  The resulting missing transverse
momentum is denoted $\vMET$.

\begin{figure}[bt!]
\includegraphics[width=0.45\textwidth]{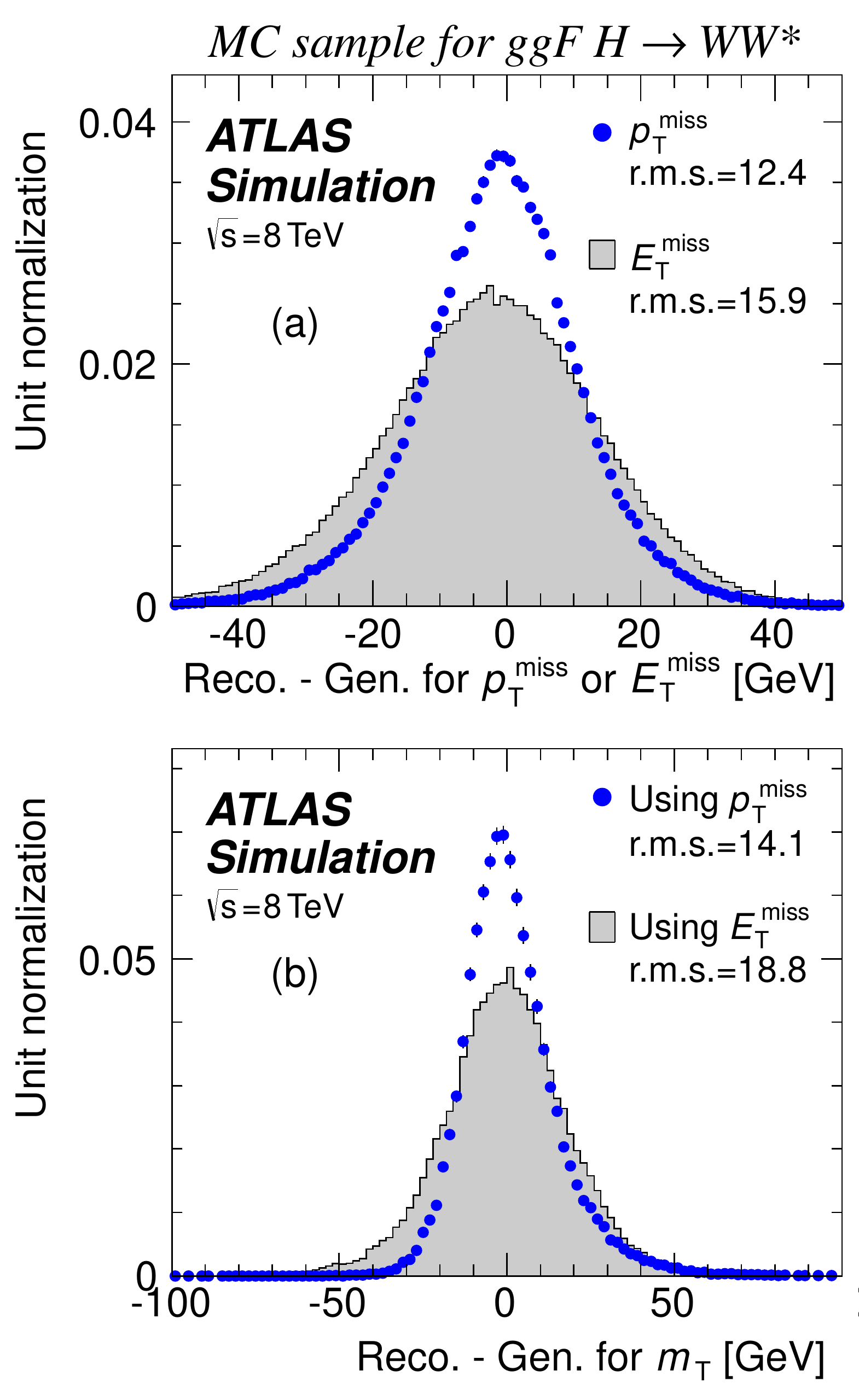}
\caption{
  Simulated resolutions of (a) $\met$ and (b) $\mTH$ for the ggF signal MC in the
  $\NjetEQzero$ category.  The comparisons are made between the calorimeter-based
  reconstruction ($\MET$) and the track-based reconstruction ($\MPTj$) of the soft objects
  [see Eq.~(\ref{eqn:MET})].
  The resolution is measured as the difference of the reconstructed (Reco) and generated (Gen) quantities;
  the \RMS\ values of the distributions are given with the legends in units of $\!\GeV$.
}
\label{fig:metres}
\end{figure}

The significant pile-up present in the data degrades the resolution of the
calorimeter-based measurement of missing transverse momentum.  An ${\cal{O}}(20\%)$
improvement in resolution is obtained using a track-based measurement of the soft
objects, where the tracks are required to have $\pT{\GT}0.5\GeV$ and originate from
the primary vertex.  Tracks associated with identified leptons or jets are not
included, as these selected objects are added separately to the calculation of the
missing transverse momentum.  This reconstruction of missing transverse momentum,
denoted $\vMPTj$, is used in the final fit to the $\mTH$ distribution and improves
the signal resolution relative to the $\vMET$ used for the previous
measurement~\cite{couplings}.  Figure~\ref{fig:metres} shows the expected resolution
for the magnitude of $\vMET$ and $\vMPTj$ ($\MET$ and $\MPTj$ respectively), and for
$\mTH$ in the $\NjetEQzero$ category, all evaluated by subtracting the reconstructed
quantity from the corresponding quantity obtained using generated leptons and
neutrinos in ggF $\HWW$ events.  The \RMS\ of the $\mTH$ difference decreases from
$19\GeV$ to $14\GeV$ when using $\MPTj$ instead of $\MET$ in the reconstruction.  The
improved resolution significantly increases the discrimination between signal and
certain background processes (such as $\Wg$).

A simplified version of $\vMPTj$ is used to suppress the Drell-Yan background in events
with same-flavor leptons.  This definition, denoted $\vMPT$, differs from $\vMPTj$ in
that the tracks associated with jets are also used, replacing the calorimeter-based jet
measurement.  This tends to align $\vMPT$ with the jet(s) in Drell-Yan events, while in
signal events $\vMPT$ generally remains in the direction of the neutrinos.  Incorporating
the direction of $\vMPT$ relative to the jet directions in the event selection thus
improves Drell-Yan rejection.

The direction of $\vMET$ relative to the lepton and jet directions is also used to
reject Drell-Yan, particularly the case of $\tau\tau$ production where $\vMET$ tends to
align with a final-state lepton.  A relative quantity $\METrel$ is defined as follows:
\begin{equation}
  \begin{array}{ll}
  \multirow{2}{*}{$\METrel$ =\ \bigg\{ }
    &\!\!\!\!\MET\ \sin\dphiNear\quad\textrm{if $\dphiNear<\pi/2$}\\
    &\!\!\!\!\MET\ \phantom{\sin\dphiNear}\quad\textrm{otherwise,}
  \end{array}
\label{eqn:METrel}
\end{equation}
\noindent
where $\dphiNear$ is the azimuthal separation of the $\vMET$ and the nearest
high-$\pT$ lepton or jet.  A similar calculation defines $\MPTjrel$ and $\MPTrel$.

\subsection{\boldmath Monte Carlo samples \label{sec:atlas_samples_mc}}

Given the large number of background contributions to the signal region and the
broadly peaking signal $\mTH$ distribution, Monte Carlo modeling is an important
aspect of the analysis.  Dedicated samples are generated to evaluate all but the
$\Wjets$~and multijet backgrounds, which are estimated using data (see
Sec.~\ref{sec:bkg_misid}).  Most samples use the {\POWHEG}~\cite{Nason:2004rx}
generator to include corrections at next-to-leading order (NLO) in $\alphaS$.  In
cases where higher parton multiplicities are important, {\ALPGEN}~\cite{alpgen}
or {\SHERPA}~\cite{Gleisberg:2008ta}~provide merged calculations at tree level
for up to five additional partons.  In a few cases, only leading-order generators
(such as \ACERMC~\cite{Kersevan:2004yg}~or \GGTOVV~\cite{Kauer:2013qba}) are
available.  Table~\ref{tab:mc} shows the generator and cross section used for
each process.

\begin{table}[b!]
\caption{
  Monte Carlo samples used to model the signal and background processes.  The
  corresponding cross sections times branching fractions, $\sigma{\CDOT}\mathcal{B}$,
  are quoted at $\sqrt{s}{\EQ}8\TeV$.  The branching fractions include the decays
  $t{\TO}Wb$, $\Wlv$, and $Z{\TO}\ell\ell$ (except for the process
  $\ZZ{\TO}\ell\ell\,\nu\nu$).  Here $\ell$
  refers to $e$, $\mu$, or $\tau$ for signal and background processes.  The neutral
  current $Z/\gamma^\ast{\TO}\ell\ell$ process is denoted $Z$ or $\gamma^\ast$,
  depending on the mass
  of the produced lepton pair.  Vector-boson scattering (VBS) and vector-boson
  fusion (VBF) background processes include all leading-order diagrams with zero
  QCD vertices for the given final state (except for diagrams with Higgs bosons,
  which only appear in the signal processes).
}
\label{tab:mc}
\begin{tabular*}{0.480\textwidth}{
    lll p{0.150\textwidth} d{3}
}
\dbline
\multicolumn{3}{l}{Process}
& \multicolumn{1}{l}{MC generator$\nq$}
& \multicolumn{1}{r}{$\sigma{\CDOT}\mathcal{B}$~(pb)}
\\
\sgline
\multicolumn{2}{l}{Signal }& & \\
\quad ggF    &$\HWW$                                                             && \POWHEG+\PYTHIA8      & 0.435 \\
\quad VBF    &$\HWW$                                                             && \POWHEG+\PYTHIA8      & 0.0356 \\
\quad $\VH$  &$\HWW$                                                             && \PYTHIA8              & 0.0253 \\
\clineskip\clineskip
\multicolumn{3}{l}{$\WW$ }& & \\
\multicolumn{3}{l}{\quad $\qq{\TO}\WW$ and $qg{\TO}\WW$                          }& \POWHEG+\PYTHIA6      & 5.68 \\
\multicolumn{3}{l}{\quad $gg{\TO}\WW$                                            }& \GGTOVV+\HERWIG       & 0.196 \\
\multicolumn{3}{l}{\quad $(\qq{\TO}W){\PLUS}(\qq{\TO}W)$                         }& \PYTHIA8              & 0.480 \\
\multicolumn{3}{l}{\quad $\qq{\TO}\WW$                                           }& \SHERPA               & 5.68 \\
\multicolumn{3}{l}{\quad VBS $\WW{+\,}2\,\textrm{jets}$                          }& \SHERPA               & 0.0397 \\
\clineskip\clineskip
\multicolumn{3}{l}{Top quarks }& & \\
\multicolumn{3}{l}{\quad $\ttbar$                                                }& \POWHEG+\PYTHIA6      & 26.6 \\
\multicolumn{3}{l}{\quad $Wt$                                                    }& \POWHEG+\PYTHIA6      & 2.35 \\
\multicolumn{3}{l}{\quad $tq\bar{b}$                                             }& \ACERMC+\PYTHIA6      & 28.4 \\
\multicolumn{3}{l}{\quad $t\bar{b}$                                              }& \POWHEG+\PYTHIA6      & 1.82 \\
\clineskip\clineskip
\multicolumn{3}{l}{Other dibosons ($VV$)}& & \\
\multicolumn{1}{l}{\quad $\Wg$  } &\multicolumn{2}{l}{($\pT^{\gamma}{\GT}8\GeV$) }& \ALPGEN+\HERWIG       & 369 \\
\multicolumn{1}{l}{\quad $\Wgs$ } &\multicolumn{2}{l}{($\mll{\LE}7\GeV$)         }& \SHERPA               & 12.2 \\
\multicolumn{1}{l}{\quad $\WZ$  } &\multicolumn{2}{l}{($\mll{\GT}7\GeV$)         }& \POWHEG+\PYTHIA8      & 12.7 \\
\multicolumn{3}{l}{\quad VBS $\WZ{\PLUS}2\,\textrm{jets}$                        }& \SHERPA               & 0.0126 \\
\multicolumn{1}{l}{\quad        } & ($\mll{\GT}7\GeV$)                            &                       & \\
\multicolumn{1}{l}{\quad $\Zg$  } &\multicolumn{2}{l}{($\pT^{\gamma}{\GT}8\GeV$) }& \SHERPA               & 163 \\
\multicolumn{1}{l}{\quad $\Zgs$ } &\multicolumn{2}{l}{(min.\ $\mll{\LE}4\GeV$)   }& \SHERPA               & 7.31 \\
\multicolumn{1}{l}{\quad $\ZZ$  } &\multicolumn{2}{l}{($\mll{\GT}4\GeV$)         }& \POWHEG+\PYTHIA8      & 0.733 \\
\multicolumn{3}{l}{\quad $\ZZ{\TO}\ell\ell\,\nu\nu$ ($\mll{\GT}4\GeV$)           }& \POWHEG+\PYTHIA8      & 0.504 \\
\clineskip\clineskip
\multicolumn{3}{l}{Drell-Yan }& & \\
\multicolumn{1}{l}{\quad $Z$   } &\multicolumn{2}{l}{($\mll{\GT}10\GeV$)         }& \ALPGEN+\HERWIG  $\np$& 16500 \\
\multicolumn{3}{l}{\quad VBF $Z{\PLUS}2\,\textrm{jets}$                          }& \SHERPA               & 5.36 \\
\multicolumn{1}{l}{\quad        } & ($\mll{\GT}7\GeV$)                            &                       & \\
\dbline
\end{tabular*}
\end{table}

The matrix-element-level Monte Carlo calculations are matched to a model of the
parton shower, underlying event and hadronization, using either
\PYTHIA6~\cite{Sjostrand:2006za}, \PYTHIA8~\cite{Sjostrand:2007gs},
\HERWIG~\cite{Corcella:2000bw} (with the underlying event modeled by
{\JIMMY}~\cite{jimmy}), or \SHERPA.  Input parton distribution functions (PDFs)
are taken from \CT10~\cite{Lai:2010vv} for the \POWHEG~and \SHERPA~samples and
\CTEQ6L1~\cite{cteq6} for \ALPGEN+\HERWIG~and \ACERMC~samples.  The $\ZDY$ sample
is reweighted to the {\MRST}mcal PDF set~\cite{mrst}.

Pile-up interactions are modeled with \PYTHIA8, and the ATLAS detector response
is simulated~\cite{atlassim} using either \GEANT4~\cite{GEANT4} or \GEANT4 combined
with a parametrized \GEANT4-based calorimeter simulation~\cite{AFII}.  Events are
filtered during generation where necessary, allowing up to $2\iab$ of equivalent
luminosity for high cross section processes such as $\ZDY$ in the {\VBF} category.

The ggF and VBF production modes for the $\HWW$ signal are modeled with
\POWHEG+\PYTHIA8 \cite{Bagnaschi:2011tu,Nason:2009ai} at $\mH{\EQ}125\GeV$, and the corresponding cross sections are
shown in Table~\ref{tab:mc}.  A detailed description of these processes and
their modeling uncertainties is given in Sec.~\ref{sec:signal}.  The smaller
contribution from the $\VH$ process, with subsequent $\HWW$ decay, is also
shown in Table~\ref{tab:mc}.  Not shown are the $H\rightarrow\tau\tau$ MC
samples, which have an even smaller contribution but are included in the
signal modeling for completeness using the same generators as for the $\HWW$
decay.  The $\HZZ$ decay contributes negligibly after event selection and is not included
in the analysis.

Cross sections are calculated for the dominant diboson and top-quark processes
as follows: the inclusive $\WW$ cross section is calculated to NLO in $\alphaS$ with
\MCFM~\cite{mcfm6}; nonresonant gluon fusion is calculated and modeled to
leading order in $\alphaS$ (LO) with~\GGTOVV, including both $\WW$ and $\ZZ$
production and their interference; $\ttbar$ production is normalized to the
calculation at next-to-next-to-leading order in $\alphaS$ (NNLO) with resummation
of higher-order terms to the next-to-next-to-leading logarithms (NNLL), evaluated
with \TOPPP2.0~\cite{Czakon:2011xx}; and single-top processes are normalized to
NNLL following the calculations from
Refs.~\cite{Kidonakis:2010tc,Kidonakis:2011wy,Kidonakis:2010ux}
for the $s$-channel, $t$-channel, and $Wt$ processes, respectively.
The $\ttbar$, $Wt$, and single-top $s$-channel kinematics are modeled with
\POWHEG+\PYTHIA6 \cite{Frixione:2007nw,Re:2010bp,Alioli:2009je}, while
\ACERMC~\cite{Kersevan:2004yg} is used for single-top $t$-channel process.
The $\WW$
kinematics are modeled using the {\POWHEG+\PYTHIA6}~\cite{Melia:2011tj} sample for the $\NjetLEone$
categories and the merged multileg {\SHERPA}~sample for the $\NjetGEtwo$
categories.  Section~\ref{sec:bkg_ww} describes this modeling and the normalization
of the double parton interaction process $(\qq{\TO}W){\PLUS}(\qq{\TO}W)$, which is
modeled using the \PYTHIA8~generator.  For $\WW$, $\WZ$, and $\ZZ$ production via
nonresonant vector-boson scattering, the \SHERPA~generator provides the LO cross
section and is used for event modeling.  The negligible vector-boson scattering $\ZZ$ process is not
shown in the table but is included in the background modeling for completeness.

The process $\Wgs$ is defined as associated $W$+$\ZDY$ production, where there
is an opposite-charge same-flavor lepton pair with invariant mass $\mll$ less
than $7\GeV$.  This process is modeled using \SHERPA~with up to one additional
parton.  The range $\mll{\GT}7\GeV$ is simulated with \POWHEG+\PYTHIA8~\cite{Melia:2011tj} and
normalized to the {\POWHEG} cross section. The use of {\SHERPA} for $\Wgs$
is due to the inability of \POWHEG+\PYTHIA8 to model invariant masses down to
the dielectron production threshold.  The \SHERPA~sample requires two leptons
with $\pT{\GT}5\GeV$ and $\ABS{\myeta}{\LT}3$.  The jet multiplicity is
corrected using a {\SHERPA} sample generated with $0.5{\LT}\mll{\LT}7\GeV$
and up to two additional partons,
while the total cross section is corrected using the ratio of the {\MCFM} NLO
to \SHERPA~LO calculations in the same restricted mass range.  A similar
procedure is used to model $Z\gamma^*$, defined as $\ZDY$ pair production
with one same-flavor opposite-charge lepton pair having $\mll{\LE}4\GeV$ and
the other having $\mll{\GT}4\GeV$.

The $\Wg$ and DY processes are modeled using \ALPGEN +\HERWIG~with merged
tree-level calculations of up to five jets.  The merged samples are normalized
to the NLO calculation of \MCFM~(for $\Wg$) or the NNLO calculation of
\DYNNLO~\cite{DYNNLO} (for $\ZDY$).  The $\Wg$ sample is generated with the
requirements $\pT^{\gamma}{\GT}8\GeV$ and $\DeltaR(\gamma, \ell){\GT}0.25$.
A $\Wg$ calculation at NNLO~\cite{WgammaNNLO} finds a correction of less than $8\%$
in the modeled phase space, which falls within the uncertainty of the NLO calculation.

A \SHERPA~sample is used to accurately model the $Z(\to\ll) \gamma$
background.  The photon is required to have $\pT^{\gamma}{\GT}8\GeV$ and
$\DeltaR(\gamma, \ell){\GT}0.1$; the lepton pair must satisfy $\mll{\GT}10\GeV$.
The cross section is normalized to NLO using \MCFM.  Events are removed from
the \ALPGEN +\HERWIG~DY samples if they overlap with the kinematics
defining the \SHERPA~$Z(\to\ll) \gamma$ sample.

The uncertainties are discussed for each specific background in
Sec.~\ref{sec:bkg}, and their treatment in the likelihood fit is
summarized in Sec.~\ref{sec:systematics}.

\subsection{\boldmath Modifications for $7\TeV$ data \label{sec:atlas_7tev}}

The $7\TeV$ data are selected using single-lepton triggers with a
muon $\pT$ threshold of $18\GeV$ and with varying electron $\pT$ thresholds
($20$ or $22\GeV$ depending on the data-taking period).
The identification of the electrons uses the ``tight'' selection-based requirement
described in Ref.~\cite{Aad:2014fxa} over the entire $\eT$ range,
and the GSF fit is not used.
Muons are identified with the same selection used for the analysis of the $8\TeV$ data.
The lepton isolation requirements are tighter than in the $8\TeV$ analysis due to a
statistically and systematically less precise estimation of the backgrounds with misidentified leptons.
The jet $\pT$ thresholds are the same as in the $8\TeV$ analysis,
but due to less severe pile-up conditions, the requirement on the jet vertex fraction $\jvf{\GT}0.75$
can be stricter without loss in signal efficiency.

The MC samples used for the analysis of the $7\TeV$ data have been
chosen to reflect closely the samples used for the $8\TeV$ data (see Table~\ref{tab:mc}).
The same matrix-element calculations and parton-shower models are used for all samples except for the $\WZ$
and $\ZZ$ backgrounds where {\POWHEG+\PYTHIA6} is used instead of
{\POWHEG+\PYTHIA8}. The pile-up events are simulated  with {\PYTHIA6}
instead of {\PYTHIA8}. The samples are normalized to inclusive cross
sections computed following the same prescriptions described in Sec.~\ref{sec:atlas_samples_mc}.

\section{\boldmath Event selection \label{sec:selection}}

The initial sample of events is based on the data quality, trigger, lepton $\pT$ threshold, and
two identified leptons discussed in the previous section.
Events with more than two identified leptons with $\pT{\GT}10\GeV$ are rejected.

After the leptons are required to have opposite charge and pass the $\pT$-threshold selections,
the $\DFchan$ sample of approximately $1.33{\TIMES}10^5$ events is composed primarily of
contributions from $\ZDYtt$ and $\ttbar$, with
approximately $800$ expected signal events. The $\SFchan$ sample of $1.6{\TIMES}10^7$
events is dominated by $\ZDYll$ production, which is largely
reduced (by approximately $90\%$) by requiring $\ABS{\mll{\MINUS}\mZ}{\GT}15\GeV$.
Low-mass meson resonances and $\ZDY$ (Drell-Yan or DY) events are removed with the $\mll{\GT}10\GeV$ $(12\GeV)$ selection for the
$\DFchan$ ($\SFchan$) samples. The DY, and $\Wjets$ and multijets
events are further reduced with requirements on the \met\ distributions. Figure~\ref{fig:MET}(a)
shows the $\METrel$ distribution in the $\NjetLEone$ $\SFchan$ sample, where the dominant $\ZDYll$
contribution is suppressed by the $\METrel{\GT}40\GeV$ requirement. In the $\NjetLEone$
and $\NjetGEtwo$ ggF-enriched $\DFchan$ samples, a $\MPTj{\GT}20\GeV$ selection is applied to significantly
reduce the $\ZDYtt$ background and the multijet backgrounds with misidentified leptons
[see Figs.~\ref{fig:MET}(b) and~\ref{fig:MET}(c) for the $\NjetLEone$ categories].
The $\NjetGEtwo$ VBF-enriched $\DFchan$ sample requires no $\met$ selection,
and thus recovers signal acceptance for the statistically limited VBF measurement.
In the $\SFchan$ sample, more stringent selections are applied: $\MET{\GT}45\GeV$ and $\MPTj{\GT}40\GeV$.
Table~\ref{tab:selection} lists these so-called preselection criteria.

The different background composition as a function of jet multiplicity
motivates the division of the data sample
into the various $\Njet$ categories.
Figures~\ref{fig:Njet}(a) and~\ref{fig:Njet}(b)
show the jet multiplicity distributions in the $\SFchan$
and $\DFchan$ samples, respectively.
The $\ZDYll$ background dominates the $\NjetLEone$ $\SFchan$ samples
even after the above-mentioned \met\ requirements.
The top-quark background
becomes more significant at higher jet multiplicities.  Its suppression is primarily based on
the $b$-jet multiplicity; the distribution is shown in Fig.~\ref{fig:Njet}(c) for the $\DFchan$ sample.

In each of the $\Njet$ and lepton-flavor categories,
further criteria are applied to increase the precision
of the signal measurement.
Sections~\ref{sec:selection_0j} to \ref{sec:selection_2jggf} present
the discriminating distributions and the resulting event yields.
The selections are also listed in Table~\ref{tab:selection} along with the preselection.
Section~\ref{sec:selection_7tev} details the selection modifications for the $7\TeV$ data analysis.
Section~\ref{sec:selection_summary} concludes with the distributions after all
requirements are applied.

In this section, the background processes are normalized using control regions
(see Sec.~\ref{sec:bkg}).
The distributions in the figures and the rates in the tables
for the signal contribution correspond to the expectations for an SM Higgs boson with $\mH{\EQ}125\GeV$.
The $\VBF$ contribution includes the small contribution from $\VH$ production,
unless stated otherwise.

\subsection{\boldmath $\NjetEQzero$ category \label{sec:selection_0j}}

Events with a significant mismeasurement of the missing transverse momentum are
suppressed by requiring $\vMPTj$ to point away from the dilepton transverse momentum
($\dphillMET{\GT}\pi/2$). In the absence of a reconstructed jet to balance the
dilepton system, the magnitude of the dilepton momentum $\pTll$ is
expected to be small in DY events.  A requirement of
$\pTll{\GT}30\GeV$ further reduces the DY contribution while retaining the majority of the
signal events, as shown for the $\DFchan$ sample in Fig.~\ref{fig:0j}(a).
At this stage, the DY background is sufficiently reduced in the $\DFchan$ sample,
but still dominates in the $\SFchan$ one. In this latter sample, a requirement of
$\MPTrel{\GT}40\GeV$ is applied to provide further rejection against
DY events.

The continuum $\WW$ production and the resonant Higgs boson production
processes can be separated by exploiting the spin-$0$ property of the
Higgs boson, which, when combined with the $V{\MINUS}A$ nature of the $W$ boson
decay, leads to a small opening angle between the charged leptons (see
Sec.~\ref{sec:analysis}).
A requirement of $\dphill{\LT}1.8$ reduces both the $WW$ and DY backgrounds while retaining
$90\%$ of the signal.  A related requirement of $\mll{\LT}55\GeV$ combines the small lepton
opening angle with the kinematics of a low-mass Higgs boson
($\mH{\EQ}125\GeV$).  The $\mll$ and $\dphill$ distributions are shown
for the $\DFchan$ sample in Figs.~\ref{fig:0j}(b) and~\ref{fig:0j}(c), respectively.

\begin{figure}[t!]
\hspace{-3pt}\includegraphics[width=0.49\textwidth]{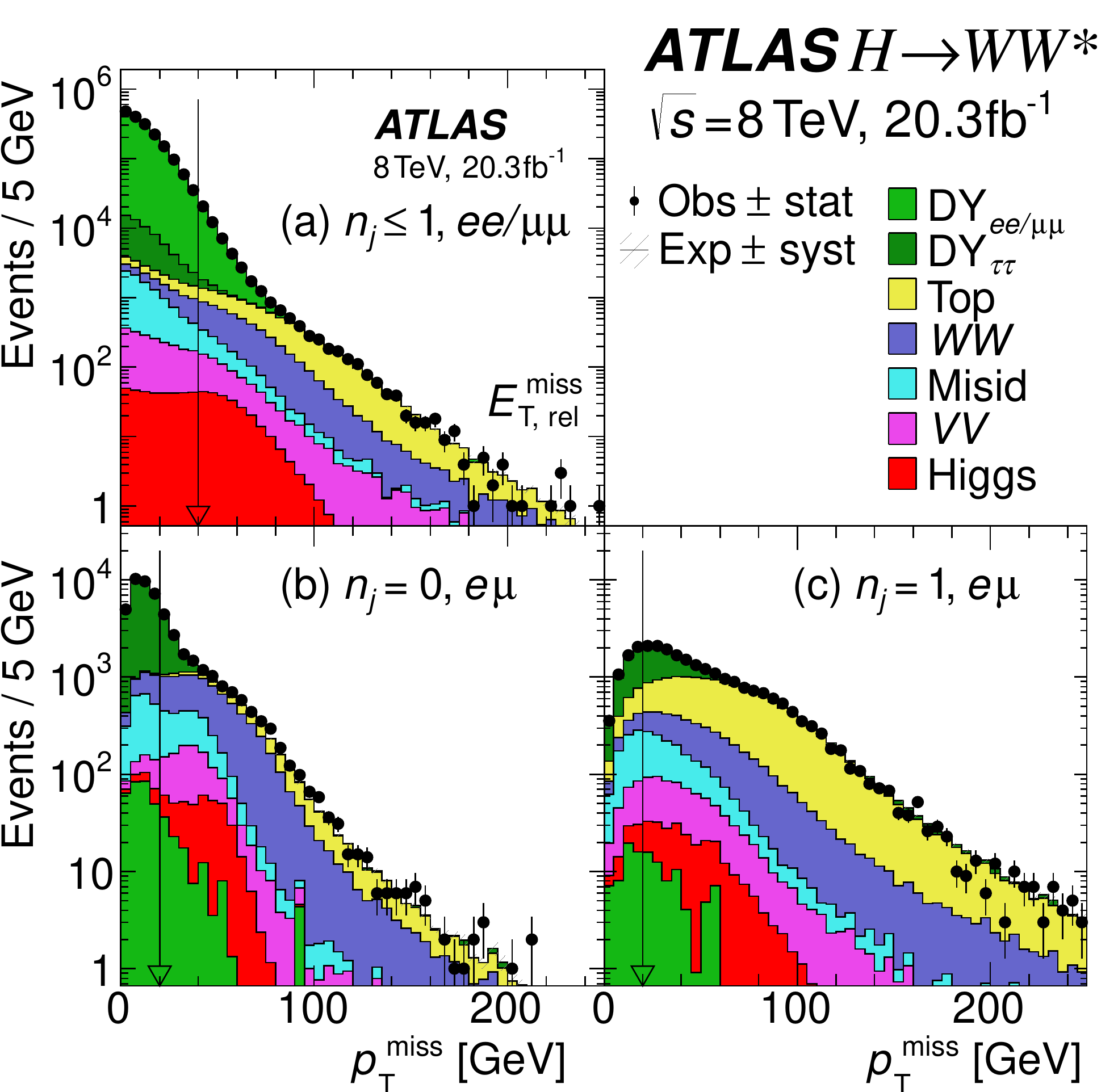}
\caption{
  \Met\ distributions.
  The plots for $\MET$ and $\MPTj$ [see Eq.~(\ref{eqn:MET})]
  are made after applying the preselection criteria common to all $\Njet$ categories (see Table~\ref{tab:selection}).
  The observed data points (Obs, $\bullet$) with their statistical
  uncertainty (stat) are compared
  with the histograms representing the cumulative expected contributions (Exp, --),
  for which the systematic uncertainty (syst) is represented by the
  shaded band. The band accounts for experimental uncertainties and for theoretical
  uncertainties on the acceptance for background and signal and is only visible in the tails of the distributions.
  Of the listed contributions (see Table~\ref{tab:process}), the dominant DY backgrounds peak at low values.
  The legend order follows the histogram stacking order of the
  plots with the exception of DY$_{\SFchan}$; it is at the top for (a) and at the bottom for the others.
  The arrows mark the threshold of the selection requirements.
}
\label{fig:MET}
\end{figure}

\begin{table*}[p!]
\caption{
  Event selection summary.
  Selection requirements specific to the $\DFchan$ and $\SFchan$ lepton-flavor samples are noted as such
  (otherwise, they apply to both); a dash (-) indicates no selection.
  For the $\NjetGEtwo$ VBF-enriched category, $\METsc$ denotes all types of missing transverse momentum observables.
  Values are given for the analysis of $8\TeV$ data for $\mH{\EQ}125\GeV$;
  the modifications for $7\TeV$ are given in Sec.~\ref{sec:selection_7tev}.
  All energy-related values are in $\!\GeV$.
}
\label{tab:selection}
{
\small
  \centering
\begin{tabular*}{1\textwidth}{
  p{0.180\textwidth}
  lll
  p{0.005\textwidth}
  p{0.040\textwidth}
  l
}
\dbline
  \multirow{2}*{Objective}
& \multicolumn{3}{c}{ggF-enriched}
&
& \multicolumn{2}{c}{VBF-enriched}
\\
\clineskip\cline{2-4}\cline{6-7}\clineskip
& \multicolumn{1}{p{0.190\textwidth}}{$\NjetEQzero$}
& \multicolumn{1}{p{0.190\textwidth}}{$\NjetEQone$}
& \multicolumn{1}{p{0.160\textwidth}}{$\NjetGEtwo$ ggF}
&
& \multicolumn{2}{p{0.225\textwidth}}{$\NjetGEtwo$ VBF}
\\
\sgline
Preselection
                    &
                     \multicolumn{6}{l}{
                       \hspace{-38.5pt}\ldelim\{{6}{20pt}[All $\Njet$\,]
                       \hspace{15.5pt}
                         $\pTlead{\GT}22$    for the leading lepton $\ell_1$} \\
                    &\multicolumn{6}{l}{$\pTsublead{\GT}10$ for the subleading lepton $\ell_2$} \\
                    &\multicolumn{6}{l}{Opposite-charge leptons} \\
                    &\multicolumn{6}{l}{$\mll{\GT}10$                  for the $\DFchan$ sample} \\
                    &\multicolumn{6}{l}{$\mll{\GT}12$                  for the $\SFchan$ sample} \\
                    &\multicolumn{6}{l}{$\ABS{\mll{\MINUS}\mZ}{\GT}15$ for the $\SFchan$ sample}
                    \\
                    &$\MPTj{\,\GT}20$ for $\DFchan$
                    &$\MPTj{\,\GT}20$ for $\DFchan$
                    &$\MPTj{\,\GT}20$ for $\DFchan$
                    &
                    &\multicolumn{2}{l}{\,No $\METsc$ requirement for $\DFchan\nqq$}
                    \\
                    &$\METrel{\GT}40$ for $\SFchan$
                    &$\METrel{\GT}40$ for $\SFchan$
                    &\quad-
                    &
                    &\multicolumn{2}{l}{\,\quad-}
                    \\
\sgline
Reject backgrounds
                    &\hspace{-28.5pt}\ldelim\{{4}{20pt}[DY~]\hspace{8.5pt}$\MPTrel{>}40$ for $\SFchan$
                    &$\MPTrel{>}35$ for $\SFchan\np\no$
                    &\quad-
                    &
                    &\multicolumn{2}{l}{$\MPTj{\GT}40$ for $\SFchan$}
                    \\
                    &$\frecoil{\LT}0.1$ for $\SFchan$
                    &$\frecoil{\LT}0.1$ for $\SFchan$
                    &\quad-
                    &
                    &\multicolumn{2}{l}{$\MET{\GT}45$ for $\SFchan$}
                    \\
                    &\multicolumn{1}{l}{$\pTll{\GT}30$}
                    &\multicolumn{1}{l}{$\mtt{\LT}\mZ{\MINUS}25$}
                    &\multicolumn{1}{l}{$\mtt{\LT}\mZ{\MINUS}25$}
                    &
                    &\multicolumn{2}{l}{$\mtt{\LT}\mZ{\MINUS}25$}
                    \\
                    &\multicolumn{1}{l}{$\dphillMET{\GT}\pi/2$}
                    &\quad-
                    &\quad-
                    &
                    &\multicolumn{2}{l}{\quad-}
                    \\
\multicolumn{1}{r}{Misid.}
                    &\quad-
                    &$\mTlep{\GT}50$ for $\DFchan$
                    &\quad-
                    &
                    &\multicolumn{2}{l}{\quad-}
                    \\
                    &\multicolumn{1}{l}{\hspace{-29pt}\ldelim\{{3}{21pt}[Top~]\hspace{8pt}
                        $\NjetEQzero$}
                    &\multicolumn{1}{l}{$\Nbjet{\EQ}0$}
                    &\multicolumn{1}{l}{$\Nbjet{\EQ}0$}
                    &
                    &\multicolumn{2}{l}{$\Nbjet{\EQ}0$}
                    \\
                    &\multicolumn{1}{l}{\quad-}
                    &\quad-
                    &\quad-
                    &
                    &$\pTtot$ &inputs to BDT
                    \\
                    &\multicolumn{1}{l}{\quad-}
                    &\quad-
                    &\quad-
                    &
                    &$\mlj\nq$ &inputs to BDT
                    \\
\sgline
VBF topology
                    &\multirow{6}{*}{ \hspace{-5pt}\begin{tabular}{l} \quad- \end{tabular} }
                    &\multirow{6}{*}{ \hspace{-5pt}\begin{tabular}{l} \quad- \end{tabular} }
                    &\multirow{6}{*}{ \hspace{-5pt}
                        \begin{tabular}{l}
                        See Sec.~\ref{sec:selection_2jggf} for \\
                        rejection of VBF \&$\nq$ \\
                        VH ($W,Z{\TO}jj$), \\
                        where $\HWW$ \\
                        \end{tabular}
                    }
                    &
                    &$\mjj$ &inputs to BDT
                    \\
                    &
                    &
                    &
                    &
                    &$\dyjj\no$ &inputs to BDT
                    \\
                    &
                    &
                    &
                    &
                    &$\contolv$ &inputs to BDT
                    \\
                    &
                    &
                    &
                    &
                    &\multicolumn{2}{l}{$\olvlead{\LT}1$ and $\olvsublead{\LT}1$}
                    \\
                    &
                    &
                    &
                    &
                    &\multicolumn{2}{l}{$\cjv{\GT}1$ for $j_3$ with $\pTthirdjet{\GT}20\nqq$}
                    \\
                    &
                    &
                    &
                    &
                    &\multicolumn{2}{l}{$\bdt{\GE}-0.48$}
                    \\
\sgline
$\HWWlvlv$
                    &\multicolumn{1}{l}{$\mll{\LT}55$}
                    &\multicolumn{1}{l}{$\mll{\LT}55$}
                    &\multicolumn{1}{l}{$\mll{\LT}55$}
                    &
                    &$\mll$ &inputs to BDT
                    \\
decay topology
                    &\multicolumn{1}{l}{$\dphill{\LT}1.8$}
                    &\multicolumn{1}{l}{$\dphill{\LT}1.8$}
                    &\multicolumn{1}{l}{$\dphill{\LT}1.8$}
                    &
                    &$\dphill\no$ &inputs to BDT
                    \\
                    &\multicolumn{1}{l}{No $\mTH$ requirement}
                    &\multicolumn{1}{l}{No $\mTH$ requirement}
                    &\multicolumn{1}{l}{No $\mTH$ requirement}
                    &
                    &$\mTH$ &inputs to BDT
                    \\
\dbline
\end{tabular*}
}
\end{table*}

An additional discriminant, $\frecoil$, based on soft jets, is defined
to reduce the remaining DY contribution in the $\SFchan$ sample.
This residual DY background satisfies the event selection primarily
when the measurement of the energy associated with partons from initial-state radiation is underestimated,
resulting in an apparent imbalance of transverse momentum in the event.
To further suppress such mismeasured DY events,
jets with $\pTjet{\GT}10\GeV$, within a $\pi/2$ wedge in $\phi$ (noted as $\wedge$) centered
on $-\vpTll$, are used to define a fractional jet recoil relative to
the dilepton transverse momentum:
\begin{equation}
\frecoil = \raisebox{-3pt}{\bigg|} \sum_{{\rm jets\,}j{\rm\,in\,}\wedge}
           \no\jvf_{\,j}\cdot\vpTjet
           ~\raisebox{-3pt}{\bigg|}
           ~\raisebox{-3pt}{\bigg/} \pTll.
\label{eqn:frecoil}
\end{equation}
The jet transverse momenta are weighted by their associated $\jvf$
value to suppress the contribution from jets originating from pile-up
interactions. Jets with no associated tracks are assigned a weight of 1.
The $\frecoil$ distribution is shown in Fig.~\ref{fig:0j}(d); a requirement of $\frecoil{\LT}0.1$
reduces the residual DY background in the $\SFchan$ sample by a factor
of $7$.

The expected signal and background yields at each stage of selection are shown in
Table~\ref{tab:sr_0j}, together with the observed yields.
At the final stage, the table also shows the event yields
in the range $\frac{3}{4}\mH{\LT}\mTH{\LT}\mH$ where most of the
signal resides.  This $\mTH$ selection is not used to extract the final results,
but nicely illustrates the expected signal-to-background ratios in the different categories.

\subsection{\boldmath $\NjetEQone$ category \label{sec:selection_1j}}

The one-jet requirement significantly increases the top-quark background.
Since top quarks decay to $Wb$, jets with
$\pT{\GT}20\GeV$ are rejected  if they are identified as containing a $b$-quark [$\NbjetEQzero$,
see Fig.~\ref{fig:Njet}(c)].  After this requirement, the $WW$ and the DY
background processes are dominant in the sample; as shown in Table~\ref{tab:sr_1j}.

In the case of the $\DFchan$ sample, a requirement is applied to the
transverse mass defined for a single lepton $\ell_i$:
\begin{equation}
\mTlepi = \sqrt{2\,\pTlepi{\CDOT}\MPTj{\CDOT}\big(1 - \cos\Delta\phi\big)},
\label{eqn:mTlep}
\end{equation}
\noindent
where $\Delta\phi$ is the angle between the lepton transverse momentum and $\vMPTj$.
This quantity tends to have small values for the DY background and large values
for the signal process.  It also has small values for multijet production,
where misidentified leptons are frequently measured with energy lower
than the jets from which they originate.
The $\mTlep$ distribution, chosen to be the larger of $\mTlead$ or $\mTsublead$, is presented in
Fig.~\ref{fig:1j}(a), and shows a clear difference in shape between the
DY and multijet backgrounds, which lie mostly at low values of
$\mTlep$, and the other background processes.
Thus, both the DY and multijet processes are substantially reduced with a requirement of $\mTlep{\GT}50\GeV$
in the $\DFchan$ sample.

The requirement of a jet allows for improved rejection of the
$\ZDYtt$ background.
Using the direction of the measured missing transverse momentum,
the mass of the $\tau$-lepton pair can be reconstructed using the
so-called collinear approximation~\cite{collinear}.
A requirement of $\mtt{\LT}\mZ{\MINUS}25\GeV$ significantly reduces the
remaining DY contribution in the $\DFchan$ sample,
as can be seen in Fig.~\ref{fig:1j}(b).

\begin{figure}[t!]
\hspace{-3pt}\includegraphics[width=0.49\textwidth]{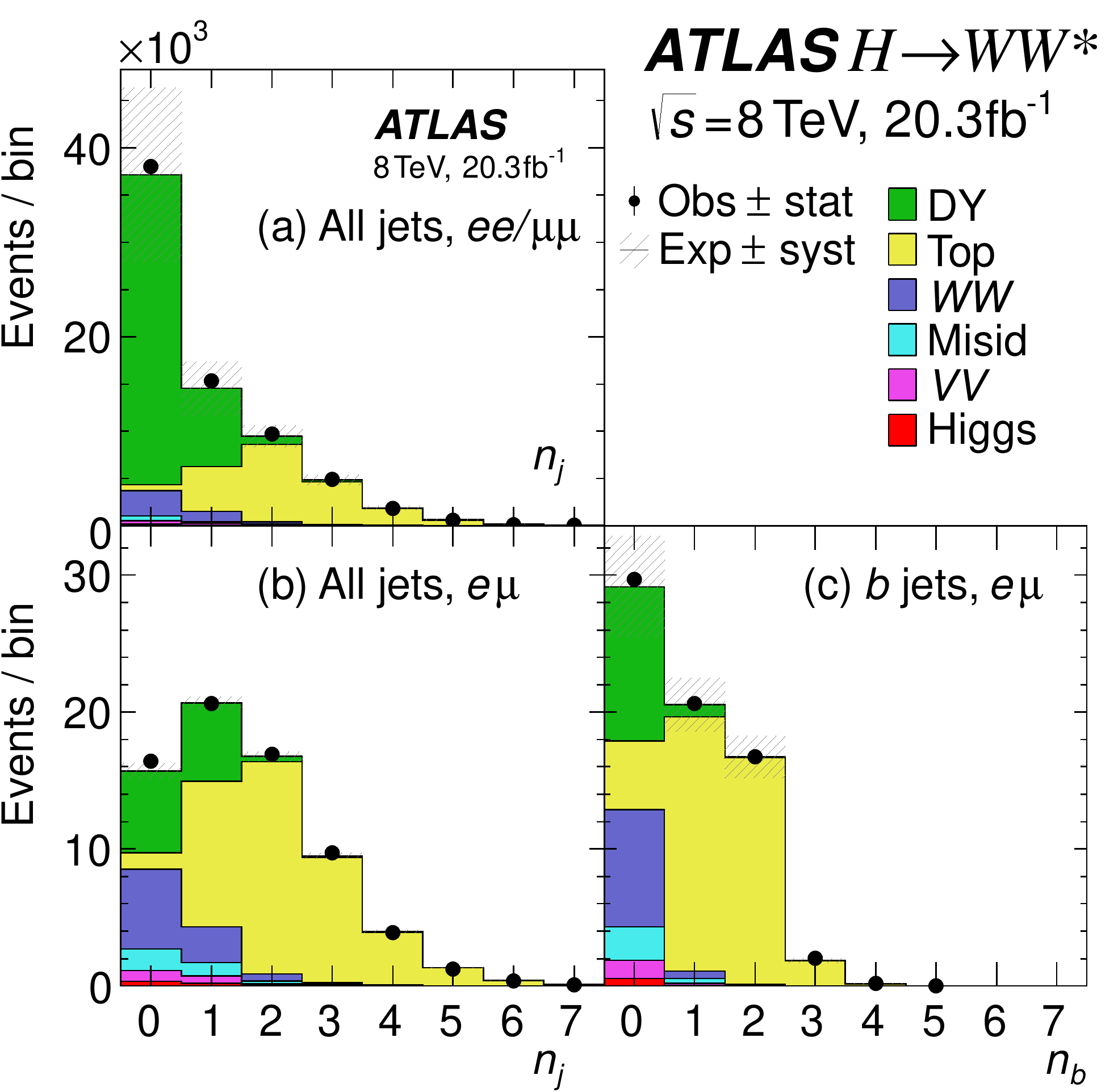}
\caption{
  Jet multiplicity distributions for all jets ($\Njet$) and $b$-tagged jets ($\Nbjet$).
  The plots are made after applying the preselection criteria common
  to all $\Njet$ categories (see Table~\ref{tab:selection}).
  \HwwPlotDetail{See}.
}
\label{fig:Njet}
\end{figure}

\begin{figure*}[t!]
\includegraphics[width=1.00\textwidth]{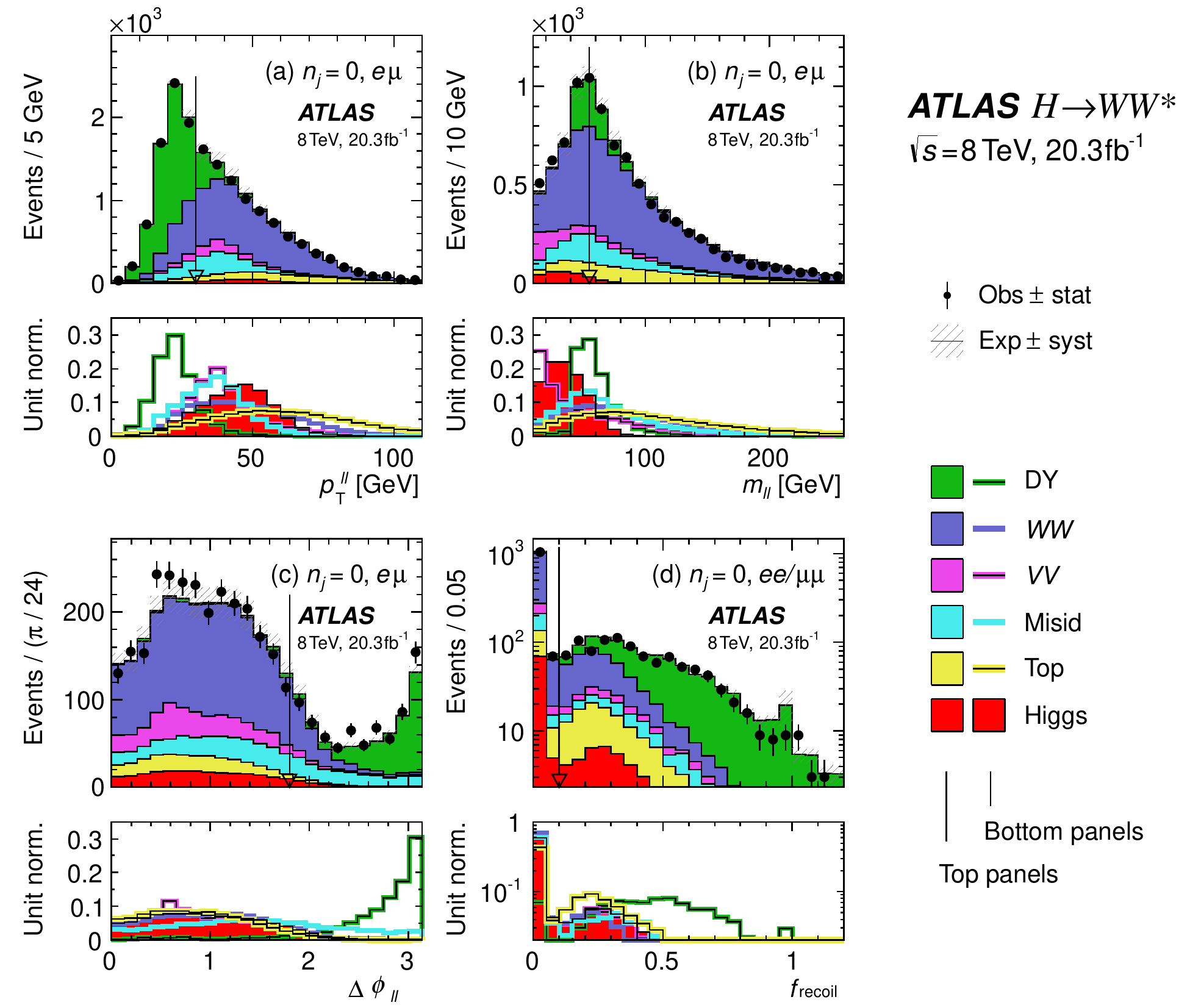}
\caption{
  Distributions of
  (a) $\pTll$,
  (b) $\mll$,
  (c) $\dphill$, and
  (d) $\frecoil$,
  for the $\NjetEQzero$ category.
  The plot in (a) is made after requiring all selections up to $\pTll$,
  (b) up to $\mll$, (c) up to $\dphill$ and (d) up to $\frecoil$ (see Table~\ref{tab:sr_0j}).
  For each variable, the top panel compares the observed and the cumulative expected distributions;
  the bottom panel shows the overlay of the distributions of the
  individual expected contributions, normalized to unit area, to
  emphasize shape differences.
  \HwwPlotDetail{See}.
}
\label{fig:0j}
\end{figure*}

\begin{table*}[t!]
\caption{
  Event selection for the $\NjetEQzero$ category in the $8\TeV$ data analysis.
  The selection is presented separately for the $\DFchan$ and $\SFchan$ samples.
  The summary columns give
  the observed yields ($\Nobs$),
  the expected background yields ($\Nbkg$),
  their ratios, and
  the expected signal yields ($\Nsig$).
  For the dominant backgrounds, the expected yields are normalized
  using control regions, as described in Sec.~\ref{sec:bkg}.
  The $\Nsig$ values are given for $\mH{\EQ}125\GeV$ and
  are subdivided into the $\NggF$ and $\NVBFVH$ contributions.
  The composition columns give the contributions to $\Nbkg$ (see Sec.~\ref{sec:bkg}).
  The requirements are imposed sequentially from top to bottom;
  entries are shown as $0.0$ (-) if they are less than $0.1$ ($0.01$) events.
  The entries are rounded to a precision commensurate with the statistical
  uncertainties due to the random error associated with the central value of the yield ($\textrm{stat}_{\obs}{\EQ}\sqrt{\Nobs}$) and
  the sampling error associated with the finite sample size used for the prediction for background type $k$ ($\textrm{stat}_{{\rm bkg},k}$).
  The errors on $\Nobs/\Nbkg$ are due to the combined statistical uncertainty on $\textrm{stat}_{\obs}$ and $\textrm{stat}_{\rm bkg}$.
  Energy-related quantities are in $\!\GeV$.
}
\label{tab:sr_0j}
{\small
  \centering
\begin{tabular*}{1\textwidth}{
  l
  r@{$\PM$}l
  d{0}d{1}d{1}d{1}
  p{0.010\textwidth}
  d{1}d{1}d{1}d{1}
  d{1}d{1}d{1}d{1}
}
\dbline
&\multicolumn{6}{c}{Summary}
&&\multicolumn{8}{c}{Composition of $\Nbkg$}
\\
\clineskip\cline{2-7}\cline{9-16}\clineskip
\multicolumn{1}{p{0.185\textwidth}}{Selection}
& \multicolumn{2}{p{0.090\textwidth}}{$\Nobs/\Nbkg$}
& \multicolumn{1}{p{0.040\textwidth}}{$\Nobs\nq$}
& \multicolumn{1}{p{0.040\textwidth}}{$\Nbkg$}
& \multicolumn{2}{p{0.100\textwidth}}{~~~~$\Nsig$}
&
& \multicolumn{1}{p{0.060\textwidth}}{$\NWW$}
& \multicolumn{2}{l}{~~~$\Ntop$}
& \multicolumn{2}{l}{~~~$\Nfakes$}
& \multicolumn{1}{p{0.044\textwidth}}{$\NVV$}
& \multicolumn{2}{l}{~~~~~~$\Ndy$}
\\
\multicolumn{2}{l}{}
&
&
&
& \multicolumn{1}{l}{$\NggF$}
& \multicolumn{1}{l}{$\NVBFVH$}
&
&
& \multicolumn{1}{p{0.038\textwidth}}{$\Nttbar$}
& \multicolumn{1}{p{0.038\textwidth}}{$\Nt$}
& \multicolumn{1}{p{0.050\textwidth}}{$\NWj$}
& \multicolumn{1}{p{0.050\textwidth}}{$\Njj$}
&
& \multicolumn{1}{p{0.045\textwidth}}{$\Nll$}
& \multicolumn{1}{p{0.045\textwidth}}{$\Ntautau$}
\\
\sgline
$\DFchan$ sample            & 1.01 & 0.01 & 16423 & 16330 & 290 & 12.1 && 7110 & 820 & 407 & 1330 & 237   & 739 &   115   & 5570   \\
\quad$\dphillMET{\GT}\pi/2$ & 1.00 & 0.01 & 16339 & 16270 & 290 & 12.1 && 7110 & 812 & 405 & 1330 & 230   & 736 &   114   & 5530   \\
\quad$\pTll{\GT}30$         & 1.00 & 0.01 &  9339 &  9280 & 256 & 10.3 && 5690 & 730 & 363 & 1054 &  28   & 571 &    60   &  783   \\
\quad$\mll{\LT}55$          & 1.11 & 0.02 &  3411 &  3060 & 224 &  6.3 && 1670 & 141 &  79 &  427 &  12   & 353 &    27   &  350   \\
\quad$\dphill{\LT}1.8$      & 1.12 & 0.02 &  2642 &  2350 & 203 &  5.9 && 1500 & 132 &  75 &  278 &   9.2 & 324 &    19   &   12   \\
\quad$\mTHcut\nq$           & 1.20 & 0.04 &  1129 &   940 & 131 &  2.2 &&  660 &  40 &  21 &  133 &   0.8 &  78 &     4.3 &    2.3 \\
\clineskip\clineskip
$\SFchan$ sample            & 1.04 & 0.01 & 38040 & 36520 & 163 &  7.2 && 3260 & 418 & 211 &  504 &  29   & 358 & 31060   &  685   \\
\quad$\dphillMET{\GT}\pi/2$ & 1.05 & 0.01 & 35445 & 33890 & 163 &  7.1 && 3250 & 416 & 211 &  493 &  26   & 355 & 28520   &  622   \\
\quad$\pTll{\GT}30$         & 1.06 & 0.01 & 11660 & 11040 & 154 &  6.8 && 3010 & 394 & 201 &  396 &   2.6 & 309 &  6700   &   21   \\
\quad$\mll{\LT}55$          & 1.01 & 0.01 &  6786 &  6710 & 142 &  5.0 && 1260 & 109 &  64 &  251 &   2.0 & 179 &  4840   &    8.7 \\
\quad$\MPTrel{\GT}40$       & 1.02 & 0.02 &  2197 &  2160 & 117 &  4.3 && 1097 &  99 &  59 &  133 &   0.5 & 106 &   660   &    0.3 \\
\quad$\dphill{\LT}1.8$      & 1.01 & 0.02 &  2127 &  2100 & 113 &  4.2 && 1068 &  96 &  57 &  122 &   0.5 & 104 &   649   &    0.3 \\
\quad$\frecoil{\LT}0.1$     & 1.01 & 0.03 &  1108 &  1096 &  72 &  2.7 &&  786 &  41 &  31 &   79 &   0.0 &  69 &    91   &    0.1 \\
\quad$\mTHcut\nq$           & 0.99 & 0.05 &   510 &   517 &  57 &  1.3 &&  349 &  11 &   8 &   53 &   $-$ &  31 &    64   &    0.1 \\
\dbline
\end{tabular*}
}
\end{table*}

\begin{table*}[t!]
\caption{
  Event selection for the $\NjetEQone$ category in the $8\TeV$ data analysis (see Table~\ref{tab:sr_0j} for presentation details).
}
\label{tab:sr_1j}
{
\small
  \centering
\begin{tabular*}{1\textwidth}{ l r@{$\PM$}l d{0}d{1}d{1}d{1} p{0.010\textwidth} d{0}d{0}d{1}d{1}d{1}d{1}d{1}d{1} }
\dbline
&\multicolumn{6}{c}{Summary}
&&\multicolumn{8}{c}{Composition of $\Nbkg$}
\\
\clineskip\cline{2-7}\cline{9-16}\clineskip
\multicolumn{1}{p{0.185\textwidth}}{Selection}
& \multicolumn{2}{p{0.090\textwidth}}{$\Nobs/\Nbkg$}
& \multicolumn{1}{p{0.040\textwidth}}{$\Nobs\nq$}
& \multicolumn{1}{p{0.040\textwidth}}{$\Nbkg$}
& \multicolumn{2}{p{0.100\textwidth}}{~~~~$\Nsig$}
&
& \multicolumn{1}{p{0.060\textwidth}}{$\NWW$}
& \multicolumn{2}{l}{~~~$\Ntop$}
& \multicolumn{2}{l}{~~~$\Nfakes$}
& \multicolumn{1}{p{0.040\textwidth}}{$\NVV$}
& \multicolumn{2}{l}{~~~~~$\Ndy$}
\\
\multicolumn{2}{l}{}
&
&
&
& \multicolumn{1}{l}{$\NggF$}
& \multicolumn{1}{l}{$\NVBFVH$}
&
&
& \multicolumn{1}{p{0.040\textwidth}}{$\Nttbar$}
& \multicolumn{1}{p{0.040\textwidth}}{$\Nt$}
& \multicolumn{1}{p{0.050\textwidth}}{$\NWj$}
& \multicolumn{1}{p{0.050\textwidth}}{$\Njj$}
&
& \multicolumn{1}{p{0.045\textwidth}}{$\Nll$}
& \multicolumn{1}{p{0.045\textwidth}}{$\Ntautau$}
\\
\sgline
$\DFchan$ sample              & 1.00 & 0.01 & 20607 & 20700 & 131 & 32   && 2750 & 8410 & 2310 & 663 & 334   & 496 &   66   & 5660 \\
$\quad\Nbjet{\EQ}0$           & 1.01 & 0.01 & 10859 & 10790 & 114 & 26   && 2410 & 1610 &  554 & 535 & 268   & 423 &   56   & 4940 \\
$\quad\mTlep{\GT}50$          & 1.01 & 0.01 &  7368 &  7280 & 103 & 23   && 2260 & 1540 &  530 & 477 &  62   & 366 &   43   & 1990 \\
$\quad\mtt{\LT}\mZ{\MINUS}25$ & 1.02 & 0.02 &  4574 &  4490 &  96 & 20   && 1670 & 1106 &  390 & 311 &  32   & 275 &   21   &  692 \\
$\quad\mll{\LT}55$            & 1.05 & 0.02 &  1656 &  1570 &  84 & 15   &&  486 &  297 &  111 & 129 &  19   & 139 &    6.4 &  383 \\
$\quad\dphill{\LT}1.8$        & 1.10 & 0.03 &  1129 &  1030 &  74 & 13   &&  418 &  269 &  102 &  88 &   6.1 & 119 &    5.0 &   22 \\
$\quad\mTHcut\nq$             & 1.21 & 0.06 &   407 &   335 &  42 &  6.6 &&  143 &   76 &   30 &  40 &   0.5 &  42 &    1.1 &    2 \\
\clineskip\clineskip
$\SFchan$ sample              & 1.05 & 0.01 & 15344 & 14640 &  61 & 15   && 1111 & 3770 &  999 & 178 &  13   & 192 & 8100   &  280 \\
$\quad\Nbjet{\EQ}0$           & 1.08 & 0.02 &  9897 &  9140 &  53 & 12.1 &&  972 &  725 &  245 & 137 &  10   & 163 & 6640   &  241 \\
$\quad\mll{\LT}55$            & 1.16 & 0.02 &  5127 &  4410 &  48 &  9.4 &&  351 &  226 &   85 &  73 &   7.8 &  79 & 3420   &  168 \\
$\quad\MPTrel{\GT}35$         & 1.14 & 0.04 &   960 &   842 &  36 &  6.9 &&  292 &  193 &   73 &  38 &   0.2 &  49 &  194   &    2 \\
$\quad\dphill{\LT}1.8$        & 1.14 & 0.04 &   889 &   783 &  32 &  6.3 &&  265 &  179 &   68 &  30 &   0.2 &  44 &  194   &    2 \\
$\quad\frecoil{\LT}0.1$       & 1.16 & 0.05 &   467 &   404 &  20 &  3.6 &&  188 &   98 &   44 &  17 &   $-$ &  29 &   26   &    1 \\
$\quad\mTHcut\nq$             & 1.11 & 0.10 &   143 &   129 &  14 &  2.0 &&   59 &   23 &   11 &  11 &   $-$ &  11 &   14   &   $-$\\
\dbline
\end{tabular*}
}
\end{table*}

\begin{figure*}[t!]
\includegraphics[width=1.00\textwidth]{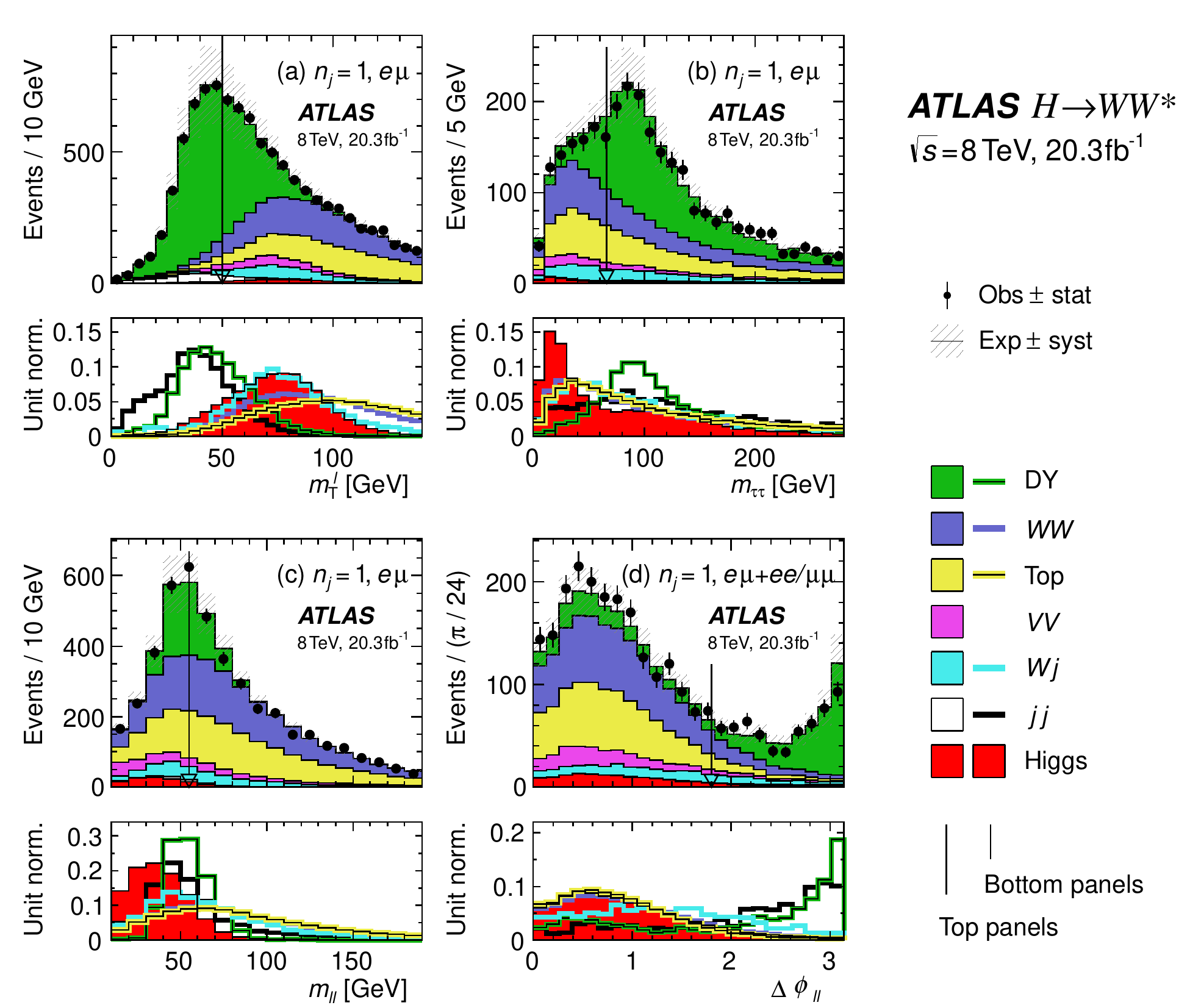}
\caption{
  Distributions of
  (a) $\mTlep$,
  (b) $\mtt$,
  (c) $\mll$, and
  (d) $\dphill$,
  for the $\NjetEQone$ category.
  The plot in (a) is made after requiring all selections up to $\mtt$,
  (b) up to $\mTlep$, (c) up to $\mll$ and (d) up to $\dphill$ (see Table~\ref{tab:sr_1j}).
  \HwwPlotSelectionDetail{See}
  (the sum of the $\jj$ and $\Wj$ contributions corresponds to the label ``Misid.'' in Fig.~\ref{fig:MET}).
}
\label{fig:1j}
\end{figure*}

The remaining selection criteria ($\MPTrel$, $\frecoil$, $\mll$,
$\dphill$) are the same as in the
$\NjetEQzero$ category, except that $\vpTll$ is
replaced with the magnitude of $\vpTllj{\EQ}\vpTll{\PLUS}\vpTjet$ in the calculation of
$\frecoil$, and the $\MPTrel$ threshold is reduced to $35\GeV$. The
$\mll$ and $\dphill$ distributions are shown in Figs.~\ref{fig:1j}(c)
and~\ref{fig:1j}(d), respectively. Differences between the shapes of the signal or $\WW$ processes and
the $\ZDY$ background processes are more apparent in the $\dphill$
distribution of the $\DFchan{\PLUS}\SFchan$ events than of the $\DFchan$ events.

\subsection{\boldmath VBF-enriched $\NjetGEtwo$ category \label{sec:selection_2jvbf}}

\begin{table*}[t!]
\caption{
  Event selection for the $\NjetGEtwo$ VBF-enriched category in
  the $8\TeV$ cross-check data analysis (see Table~\ref{tab:sr_0j} for presentation details).
  The $\NggF$, $\NVBF$, and $\NVH$ expected yields are shown separately.
  The expected yields for $\WW$ and $\ZDYtt$ are divided into QCD and electroweak (EW)
  processes, where the latter includes VBF production.
}
\label{tab:sr_2jvbf}
{\small
  \centering
\begin{tabular*}{1\textwidth}{ l r@{$\PM$}l d{0}d{0}d{0}d{1}d{1} p{0.001\textwidth} d{0}d{0}d{0}d{0}d{0}d{0} d{1}d{1}d{0} d{2}d{1}d{1} }
\dbline
&\multicolumn{7}{c}{Summary}
&&\multicolumn{10}{c}{Composition of $\Nbkg$}
\\
\clineskip\cline{2-8}\cline{10-19}\clineskip
\multicolumn{1}{p{0.165\textwidth}}{Selection}
& \multicolumn{2}{p{0.050\textwidth}}{$\Nobs/\Nbkg\nq$}
& \multicolumn{1}{p{0.040\textwidth}}{$\Nobs\nq\no$}
& \multicolumn{1}{p{0.040\textwidth}}{$\Nbkg\np$}
& \multicolumn{3}{p{0.125\textwidth}}{~~~~~~$N_{\rm signal}$}
&
& \multicolumn{2}{l}{~~~~~$\NWW$}
& \multicolumn{2}{l}{~~~~~$\Ntop$}
& \multicolumn{2}{l}{~\,$\Nfakes$}
& \multicolumn{1}{l}{$\NVV$}
& \multicolumn{3}{l}{~~~~~~~$\Ndrellyan$}
\\
\multicolumn{2}{l}{}
&
&
&
& \multicolumn{1}{l}{$\NggF\no$}
& \multicolumn{1}{l}{$\NVBF\no$}
& \multicolumn{1}{l}{$\NVH\no$}
&
& \multicolumn{1}{p{0.020\textwidth}}{$\NWWqcd\np$}
& \multicolumn{1}{p{0.030\textwidth}}{$\NWWew$}
& \multicolumn{1}{p{0.020\textwidth}}{~~$\Nttbar\nq$}
& \multicolumn{1}{p{0.015\textwidth}}{$\Nt\nq$}
& \multicolumn{1}{p{0.020\textwidth}}{$\NWj\nq$}
& \multicolumn{1}{p{0.020\textwidth}}{$\Njj\np$}
& \multicolumn{1}{p{0.020\textwidth}}{$\!\Nll\nq$}
& \multicolumn{1}{p{0.020\textwidth}}{$~\,\Ntautauqcd\nq$}
& \multicolumn{1}{p{0.020\textwidth}}{$~\,\Ntautauew\nq$}
\\
\sgline
$\DFchan$ sample                           &1.00 &0.00 &61434 &61180   &85   &32   &26   &&1350   &68   &51810   &2970   &847   &308   &380       & 51   &3260   &46   \\
$\quad\Nbjet{\EQ}0$                        &1.02 &0.01 & 7818 & 7700   &63   &26   &16   && 993   &43   & 3000   & 367   &313   &193   &273       & 35   &2400   &29   \\
$\quad\pTtot{\LT}15$                       &1.03 &0.01 & 5787 & 5630   &46   &23   &13   && 781   &38   & 1910   & 270   &216   &107   &201       & 27   &2010   &23   \\
$\quad\mtt{\LT}\mZ{\MINUS}25\nq$           &1.05 &0.02 & 3129 & 2970   &40   &20   & 9.9 && 484   &22   & 1270   & 177   &141   & 66   &132       &  7.6 & 627   & 5.8 \\
$\quad\mjj{\GT}600$                        &1.31 &0.12 &  131 &  100   & 2.3 & 8.2 & $-$ &&  18   & 8.9 &   40   &   5.3 &  1.8 &  2.4 &  5.1     &  0.1 &  15   & 1.0 \\
$\quad\dyjj{\GT}3.6$                       &1.33 &0.13 &  107 &   80   & 2.1 & 7.9 & $-$ &&  11.7 & 6.9 &   35   &   5.0 &  1.6 &  2.3 &  3.3     &  $-$ &  11.6 & 0.8 \\
$\quad\cjv{\GT}1$                          &1.36 &0.18 &   58 &   43   & 1.3 & 6.6 & $-$ &&   6.9 & 5.6 &   14   &   3.0 &  1.3 &  1.3 &  2.0     &  $-$ &   6.8 & 0.6 \\
$\quad\olvlead{\LT}1$, $\olvsublead{\LT}1$ &1.42 &0.20 &   51 &   36   & 1.2 & 6.4 & $-$ &&   5.9 & 5.2 &   10.8 &   2.5 &  1.3 &  1.3 &  1.6     &  $-$ &   5.7 & 0.6 \\
$\quad{\mll, \dphill, \mTH}$               &2.53 &0.71 &   14 &    5.5 & 0.8 & 4.7 & $-$ &&   1.0 & 0.5 &    1.1 &   0.3 &  0.3 &  0.3 &  0.6     &  $-$ &   0.5 & 0.2 \\
\clineskip\clineskip
$\SFchan$ sample                           &0.99 &0.01 &26949 &27190   &31   &14   &10.1 && 594   &37   &23440   &1320   &230   &  8.6 &{\rm~137} &690   & 679   &16   \\
$\quad\Nbjet, \pTtot, \mtt\nq$             &1.03 &0.03 & 1344 & 1310   &13   & 8.0 & 4.0 && 229   &12.0 &  633   &  86   & 26   &  0.9 & 45       &187   &  76   & 1.5 \\
$\quad\mjj, \dyjj, \cjv, \olv$             &1.39 &0.28 &   26 &   19   & 0.4 & 2.9 & 0.0 &&   3.1 & 3.1 &    5.5 &   1.0 &  0.2 &  0.0 &  0.7     &  3.8 &   0.7 & 0.1 \\
$\quad{\mll, \dphill, \mTH}$               &1.63 &0.69 &    6 &    3.7 & 0.3 & 2.2 & 0.0 &&   0.4 & 0.2 &    0.6 &   0.2 &  0.2 &  0.0 &  0.1     &  1.5 &   0.3 & 0.1 \\
\dbline
\end{tabular*}
}
\end{table*}

The $\NjetGEtwo$ sample contains signal events produced by both the VBF and ggF
production mechanisms.  This section focuses on the former; the next section
focuses on the latter.

The sample is analyzed using a boosted decision tree multivariate
method~\cite{BDT} that considers VBF Higgs boson production as signal and the
rest of the processes as background, including ggF Higgs boson production.  A
cross-check analysis is performed using sequential selections on some of the variables that are used as inputs to
the BDT. Table~\ref{tab:sr_2jvbf} shows the sample composition after each of
the selection requirements in the cross-check analysis. For the $\WW$ and
$\ZDYtt$ backgrounds, the table separates contributions
from events with jets from QCD vertices and electroweak events with
VBS or VBF interactions (see Table~\ref{tab:mc}).

The VBF process is characterized by the kinematics of the pair of tag jets
(${j_1}$ and ${j_2}$) and the activity in the rapidity gap between them.
In general, this process results in two highly energetic forward jets with
$\dyjj{\GT}3$, where $\dyjj{\EQ}\ABS{\yleadjet{\MINUS}\ysubleadjet}$. The invariant mass of this tag-jet pair combines
$\dyjj$ with $\pTjet$ information since
$\mjj{\APPROX}\sqrt{\mbox{\smaller{$\pTleadjet{\CDOT}\pTsubleadjet$}}}e^{\dyjj/2}$ for large values of $\dyjj$.
Both $\dyjj$ and $\mjj$ are input variables to the BDT; for the cross-check
analysis, $\dyjj{\GT}3.6$ and $\mjj{\GT}600\GeV$ are required [see
Figs.~\ref{fig:2j}(a) and~\ref{fig:2j}(b)].

\begin{figure*}[bt!]
\includegraphics[width=1.00\textwidth]{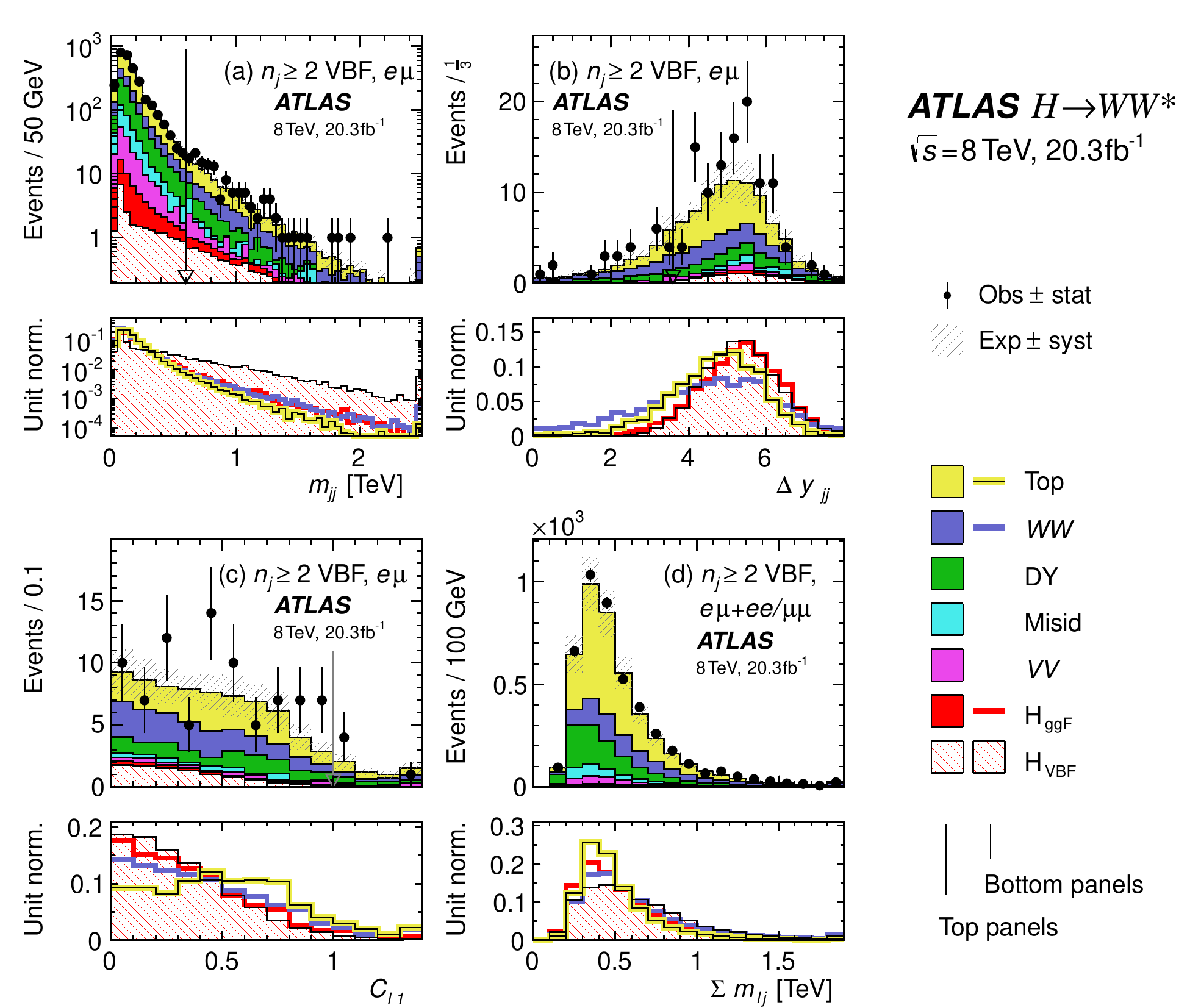}
\caption{
  Distributions of
  (a) $\mjj$,
  (b) $\dyjj$,
  (c) $\olvlead$, and
  (d) $\mlj$,
  for the $\NjetGEtwo$ VBF-enriched category.
  The plot in (a) is made after requiring all selections up to $\mjj$,
  (b) up to $\dyjj$ and (c) up to $\olvlead$  (see Table~\ref{tab:sr_2jvbf}).
  The signal is shown separately for the $\ggF$ and $\VBF$ production processes.
  The arrows mark the threshold of the selection requirements for the cross-check analysis in
  (a)--(c). There is no selection made on the variable in (d) since it is
  only used as an input to the training of the BDT.
  \HwwPlotSelectionDetail{See}.
}
\label{fig:2j}
\end{figure*}

The $\dyjj$ gap defines a ``central region,'' where a
relatively low level of hadronic activity is expected because the mediating
weak bosons do not exchange color.  The number of extra jets ($\Nextrajet$) in the
$\dyjj$ gap quantifies the activity.  Requiring the absence of such jets in
this region is known as a ``central-jet veto''~\cite{CJV} and it suppresses processes
where the jets are produced via QCD radiation.  The central-jet veto
uses jets with $\pT{\GT}20\GeV$, and this requirement is applied
in both the BDT and cross-check analyses. The selection can be expressed
in terms of jet centrality, defined as
\begin{equation}
  \cjv = \bigg|\, \etathirdjet - \frac{\sumetajj}{2} \,\bigg|\ \Big/\ \frac{\detajj}{2},
  \label{eqn:cenjet}
\end{equation}
\noindent
where $\etathirdjet$ is the pseudorapidity of an extra jet,
$\sumetajj{\EQ}\etaleadjet{\PLUS}\etasubleadjet$ and
$\detajj{\EQ}\ABS{\etaleadjet{\MINUS}\etasubleadjet}$.
The value of $\cjv$ increases from zero, when $\etathirdjet$ is
centered between the tag jets, to unity when $\etathirdjet$
is aligned in $\eta$ with either of the tag jets, and is greater than
unity when $\ABS{\etathirdjet}{\GT}\ABS{\etaleadjet}$ or $\ABS{\etathirdjet}{\GT}\ABS{\etasubleadjet}$.
The centrality of any extra jet in the event is required therefore to be
$\cjv{\GT}1$.

The Higgs boson decay products tend to be in the central rapidity region.  The centrality
of a given lepton, $\olv$, with respect to the tag jets is defined similarly to that
for extra jets in Eq.~(\ref{eqn:cenjet}).
A requirement of $\olv{\LT}1$ is applied to each lepton in the BDT and cross-check
analyses. The sum of lepton centralities
$\contolv{\EQ}\olvlead{\PLUS}\olvsublead$ is used as an input to the
BDT. The $\olvlead$ distribution is shown in Fig.~\ref{fig:2j}(c).

Top-quark pair production has a large cross section and the same final state
as VBF Higgs boson production, with the exception that its jets result from
$b$-quarks. A requirement of $\Nbjet{\EQ}0$ with $\pT{\GT}20\GeV$
is made in the BDT and cross-check analyses. This requirement is made on all jets in the event regardless of classification as tag jets.
Significant top-quark background still remains because of the limited $\eta$ coverage of the
tracker, the $\pT$ threshold applied to the $b$-jets, and the inefficiency of the $b$-jet
identification algorithm within the tracking region.
Further reductions are achieved through targeted kinematic selections and the BDT.

The pair production of top quarks occurs dominantly through gluon-gluon annihilation,
and is frequently accompanied by QCD radiation.
This radiation is used as a signature to further suppress top-quark backgrounds
using the summed vector $\vpT$ of the final-state objects,
$\vpTtot{\EQ}\vpTll{\PLUS}\vMPTj{\PLUS}\Sigma\,\vpTj$ where
the last term is a sum of the transverse momenta of all jets in the
event. Its magnitude is used as
input to the BDT and is required to be $\pTtot{\LT}15\GeV$ in the cross-check
analysis.

The sum of the four combinations of lepton-jet invariant mass,
$\mlj{\EQ}m_{\ell1,j1}{\PLUS}m_{\ell1,j2}{\PLUS}m_{\ell2,j1}{\PLUS}m_{\ell2,j2}$,
is also used as an input to the BDT. In the VBF topology, tag jets are
more forward whereas the leptons tend to be more central. This results
in differences in the shapes of the $\mlj$ distributions for the VBF signal and the background
processes, as can be seen in Fig.~\ref{fig:2j}(d). This variable is not
used in the cross-check analysis.

The other BDT input variables are those related to the $\HWWlvlv$
decay topology ($\mll$, $\dphill$, $\mTH$), which are also used in the $\NjetLEone$ categories.
The cross-check
analysis requires $\dphill{\LT}1.8$ and $\mll{\LT}50\GeV$.

Distributions from eight variables are input to the BDT:
$\contolv$, $\dyjj$, and $\mjj$ for VBF selection; $\pTtot$ and $\mlj$ for $\ttbar$
rejection; and $\dphill$, $\mll$, and $\mTH$ for their sensitivity to the $\HWWlvlv$ decay topology.
The BDT is trained
after the common preselection criteria (as listed in
Table~\ref{tab:selection}) and the $\Nbjet{\EQ}0$ requirement. This
event selection stage corresponds to the $\Nbjet{\EQ}0$ stage
presented for the cross-check analysis in Table~\ref{tab:sr_2jvbf}.
Additional criteria, common to the BDT and cross-check analyses, are
applied before the classification of the events based on the BDT output
(described below). They include requirements
on $\mtt$, $\cjv$ and $\olv$. The observed and expected event yields
after all these requirements are shown in Table~\ref{tab:sr_2jbdt}(a)
separately for the $\DFchan$ and $\SFchan$ samples. The dominant background processes include $\ttbar$
and $\ZDY$ production.  The normalization factors,
described in Sec.~\ref{sec:bkg}, are not applied to these backgrounds at this stage.

The BDT is trained using the MC samples after the above-mentioned selections.
The training starts with a single decision tree where an event is given a score of
$\pm\,1$ if it satisfies particular sets of decisions (${\PLUS}1$ leaf
contains signal-like events and ${\MINUS}1$ background-like ones).  A
thousand such trees are built and in each iteration
the weight of miscategorized events is relatively increased, or ``boosted.''
The final discriminant $\bdt$ for a given event is the weighted average of the binary scores
from the individual trees. The bin widths for the likelihood fit
are optimized for the expected significance while keeping
each bin sufficiently populated.  The chosen configuration is four bins with
boundaries at $[-1, -0.48, 0.3, 0.78, 1]$, and with corresponding bin numbers from $0$ to
$3$. The lowest bin contains the majority of background events and
has a very small signal-to-background ratio. It is therefore not used
in the likelihood fit.  The expected and observed
event yields after the classification in bins of $\bdt$ are shown in Table~\ref{tab:sr_2jbdt}(b).
Here the background yields are scaled by their corresponding normalization factors.

Appendix~\ref{sec:appendix_bdt} documents the BDT analysis in more
detail. It includes the distributions of all the input variables
(showing the comparison between the VBF signal and the
background processes) used in the training of the BDT as well as the data-to-MC
comparison after the BDT classification.
The distributions are in agreement and show that the
correlations are well understood.

\begin{table*}[t!]
\caption{
  Event selection for the $\NjetGEtwo$ VBF-enriched category in
  the $8\TeV$ BDT data analysis (see Table~\ref{tab:sr_0j} for presentation details).
  The event yields in (a) are shown after the preselection and the additional
  requirements applied before the BDT classification (see text).
  The event yields in (b) are given in bins in $\bdt$ after the classification,
  the normalization factors are applied to the yields (see
  Table~\ref{tab:nf}). In the specific case of (a), the normalization factors
  described in Sec.~\ref{sec:bkg} are not applied to the relevant backgrounds.
  The $\NggF$, $\NVBF$, and $\NVH$ expected yields are shown separately.
}
\label{tab:sr_2jbdt}
{
\small
  \centering
\begin{tabular*}{1\textwidth}{ l r@{$\PM$}l d{0}d{0}d{0}d{1}d{1} p{0.005\textwidth} d{0}d{0}d{0}d{0}d{0}d{0} d{1}d{1}d{0} d{2}d{1}d{1} }
\dbline
&\multicolumn{7}{c}{Summary}
&&\multicolumn{10}{c}{Composition of $\Nbkg$}
\\
\clineskip\cline{2-8}\cline{10-19}\clineskip
\multicolumn{1}{p{0.165\textwidth}}{Selection}
& \multicolumn{2}{p{0.050\textwidth}}{$\Nobs/\Nbkg\nq$}
& \multicolumn{1}{p{0.040\textwidth}}{$\Nobs\nq\no$}
& \multicolumn{1}{p{0.040\textwidth}}{$\Nbkg\np$}
& \multicolumn{3}{p{0.125\textwidth}}{~~~~~~$N_{\rm signal}$}
&
& \multicolumn{2}{l}{~~~~~$\NWW$}
& \multicolumn{2}{l}{~~~~~$\Ntop$}
& \multicolumn{2}{l}{~\,$\Nfakes$}
& \multicolumn{1}{p{0.048\textwidth}}{$~~~\NVV$}
& \multicolumn{3}{l}{~~~~~~~$\Ndrellyan$}
\\
\multicolumn{2}{l}{}
&
&
&
& \multicolumn{1}{l}{$\NggF\no$}
& \multicolumn{1}{l}{$\NVBF\no$}
& \multicolumn{1}{l}{$\NVH\no$}
&
& \multicolumn{1}{p{0.040\textwidth}}{$\NWWqcd$}
& \multicolumn{1}{p{0.035\textwidth}}{$\NWWew$}
& \multicolumn{1}{p{0.050\textwidth}}{~~~$\Nttbar\no$}
& \multicolumn{1}{p{0.035\textwidth}}{~$\Nt\nq$}
& \multicolumn{1}{p{0.035\textwidth}}{$\NWj\nq$}
& \multicolumn{1}{p{0.035\textwidth}}{$\Njj\nq$}
&
& \multicolumn{1}{p{0.045\textwidth}}{$\Nll\nq$}
& \multicolumn{1}{p{0.040\textwidth}}{$~\,\Ntautauqcd\nq$}
& \multicolumn{1}{p{0.050\textwidth}}{$~\,\Ntautauew\nq$}
\\
\sgline
\multicolumn{15}{l}{$\no$(a) Before the BDT classification}\\
\clineskip\clineskip
$\DFchan$ sample         &1.04 &0.04 &718   &689   &13   &15   &2.0 &&90   &11   &327   &42   &29    &23   &31    &2.2    &130   & 2   \\
$\SFchan$ sample         &1.18 &0.08 &469   &397   & 6.0 & 7.7 &0.9 && 37  & 3   &132   &17   & 5.2  &1.2  &10.1  &168    & 23   & 1   \\
\sgline
\multicolumn{19}{l}{$\no$(b) Bins in $\bdt$}\\
\clineskip\clineskip
$\DFchan$ sample   \\
$\quad$ Bin 0 (not used) &1.02 &0.04 &661   &650   & 8.8 & 3.0 &1.9 && 83  & 9   &313   &40   &26    &21   &28    &  2.2  &126   & 1   \\
$\quad$ Bin 1            &0.99 &0.16 & 37   & 37   & 3.0 & 4.2 &0.1 && 5.0 & 1.0 & 17   & 3.1 & 3.3  & 1.8 & 2.6  &  $-$  &  4.0 & 0.2 \\
$\quad$ Bin 2            &2.26 &0.63 & 14   &  6.2 & 1.2 & 4.2 &$-$ && 1.5 & 0.5 &  1.8 & 0.3 & 0.4  & 0.3 & 0.8  &  $-$  &  0.3 & 0.3 \\
$\quad$ Bin 3            &5.41 &2.32 &  6   &  1.1 & 0.4 & 3.1 &$-$ && 0.3 & 0.2 &  0.3 & 0.1 & $-$  & $-$ & 0.1  &  $-$  &  0.1 & 0.1 \\
\clineskip\clineskip
$\SFchan$ sample   \\
$\quad$ Bin 0 (not used) &1.91 &0.08 &396   &345   & 3.8 & 1.3 &0.8 && 33  & 2   &123   &16   & 4.1  &1.1  & 8.8  &137    & 20.5 & 0.5 \\
$\quad$ Bin 1            &0.82 &0.14 & 53   & 45   & 1.5 & 2.2 &0.1 && 3.0 & 0.5 & 10.4 & 1.8 & 0.8  &0.2  & 0.9  & 26    &  1.7 & 0.1 \\
$\quad$ Bin 2            &1.77 &0.49 & 14   &  7.9 & 0.6 & 2.5 &$-$ && 0.8 & 0.3 &  1.1 & 0.2 & 0.2  &$-$  & 0.3  &  4.4  &  0.3 & 0.1 \\
$\quad$ Bin 3            &6.52 &2.87 &  6   &  0.9 & 0.2 & 1.7 &$-$ && 0.1 & 0.2 &  0.2 & $-$ & $-$  &$-$  & $-$  &  0.7  &  $-$ & $-$ \\
\dbline
\end{tabular*}
}
\end{table*}

\subsection{\boldmath ggF-enriched $\NjetGEtwo$ category \label{sec:selection_2jggf}}

\begin{table*}[bt!]
\caption{
  Event selection for the $\NjetGEtwo$ ggF-enriched category in
  the $8\TeV$ data analysis (see Table~\ref{tab:sr_0j} for presentation details).
  The $\NggF$, $\NVBF$, and $\NVH$ expected yields are shown separately.
  The ``orthogonality'' requirements are given in the text.
}
\label{tab:sr_2jggf}
{
\small
  \centering
\begin{tabular*}{1\textwidth}{
  p{0.210\textwidth}
  r@{$\PM$}l
  d{2}d{2} cd{1}d{1}
  p{0.005\textwidth}
  rrrrr
}
\dbline
&\multicolumn{7}{c}{Summary}
&&\multicolumn{5}{c}{Composition of $\Nbkg$}
\\
\clineskip\cline{2-8}\cline{10-14}\clineskip
\multicolumn{1}{p{0.172\textwidth}}{Selection}
& \multicolumn{2}{p{0.090\textwidth}}{$\Nobs/\Nbkg$}
& \multicolumn{1}{p{0.040\textwidth}}{~$\Nobs\nq$}
& \multicolumn{1}{p{0.040\textwidth}}{~$\Nbkg$}
& \multicolumn{3}{p{0.095\textwidth}}{~~~~~~~~~~~~~$N_{\rm signal}$}
&
& \multicolumn{1}{p{0.060\textwidth}}{~~~~~$\NWW$}
& \multicolumn{1}{p{0.060\textwidth}}{~~~~$\Ntop$}
& \multicolumn{1}{p{0.060\textwidth}}{~~~~~$\Nfakes$}
& \multicolumn{1}{p{0.060\textwidth}}{~~~~~~$\NVV$}
& \multicolumn{1}{p{0.060\textwidth}}{~~~~~$\Ndy$}
\\
& \multicolumn{2}{l}{}
&
&
& \multicolumn{1}{p{0.050\textwidth}}{$\NggF$}
& \multicolumn{1}{p{0.050\textwidth}}{$\NVBF$}
& \multicolumn{1}{p{0.050\textwidth}}{$\NVH$}
&
&
&
&
&
&
\\
\sgline
$\emu$ category               & 0.99 & 0.00 & 56759 & 57180 & 76   & 29  & 24  && 1330 & 52020 & 959 & 324 & 2550 \\
\quad$\Nbjet{\EQ}0$           & 1.02 & 0.01 &  6777 &  6650 & 56   & 23  & 15  &&  964 &  3190 & 407 & 233 & 1850 \\
\quad$\mtt{\LT}\mZ{\MINUS}25$ & 1.06 & 0.02 &  3826 &  3620 & 49   & 19  & 12  &&  610 &  2120 & 248 & 152 &  485 \\
\quad VBF orthogonality  $\nq$& 1.05 & 0.02 &  3736 &  3550 & 44   & 9.0 & 12  &&  593 &  2090 & 241 & 148 &  477 \\
\quad VH orthogonality        & 1.04 & 0.02 &  3305 &  3170 & 40   & 8.6 & 7.4 &&  532 &  1870 & 212 & 132 &  423 \\
\quad$\mll{\LT}55$            & 1.09 & 0.03 &  1310 &  1200 & 35   & 7.5 & 5.0 &&  158 &   572 & 124 &  66 &  282 \\
\quad$\dphill{\LT}1.8$        & 1.06 & 0.03 &  1017 &   955 & 32   & 6.9 & 4.5 &&  140 &   523 &  99 &  60 &  133 \\
\quad$\mTHcut$                & 1.05 & 0.07 &   210 &   200 & 13.3 & 2.6 & 1.9 &&   35 &   131 &  16 &  15 &    3 \\
\dbline
\end{tabular*}
}
\end{table*}

The sample of $\NjetGEtwo$ events, which are neither in the
VBF-enriched category for the BDT analysis nor in the cross-check
analysis, are used to measure ggF production. In this category
only the $\emu$ final state is analyzed due to the relatively low
expected significance in the $\SFchan$ sample. Table~\ref{tab:sr_2jggf}
shows the signal and background yields after each selection requirement.

\begin{figure*}[tb!]
\includegraphics[width=0.350\textwidth]{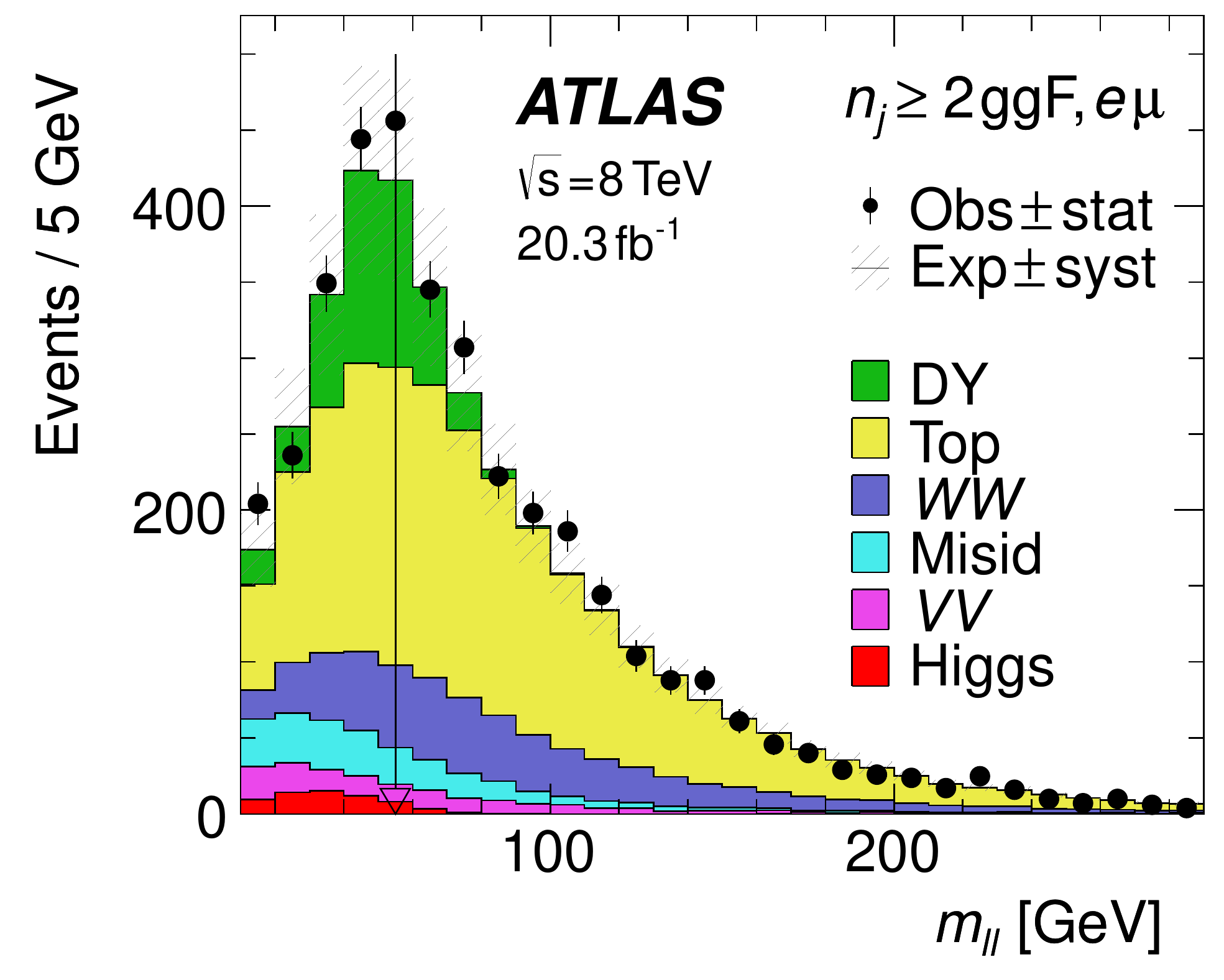} 
\caption{
  Distribution of dilepton invariant mass for the $\NjetGEtwo$ ggF-enriched category.
  The plot is made after requiring all selections up to $\mll$
  (see Table~\ref{tab:sr_2jggf}).
  \HwwPlotDetail{See}.
}
\label{fig:ggF2j}
\end{figure*}

\begin{figure*}[tb!]
\hspace{-3pt}\includegraphics[width=0.550\textwidth]{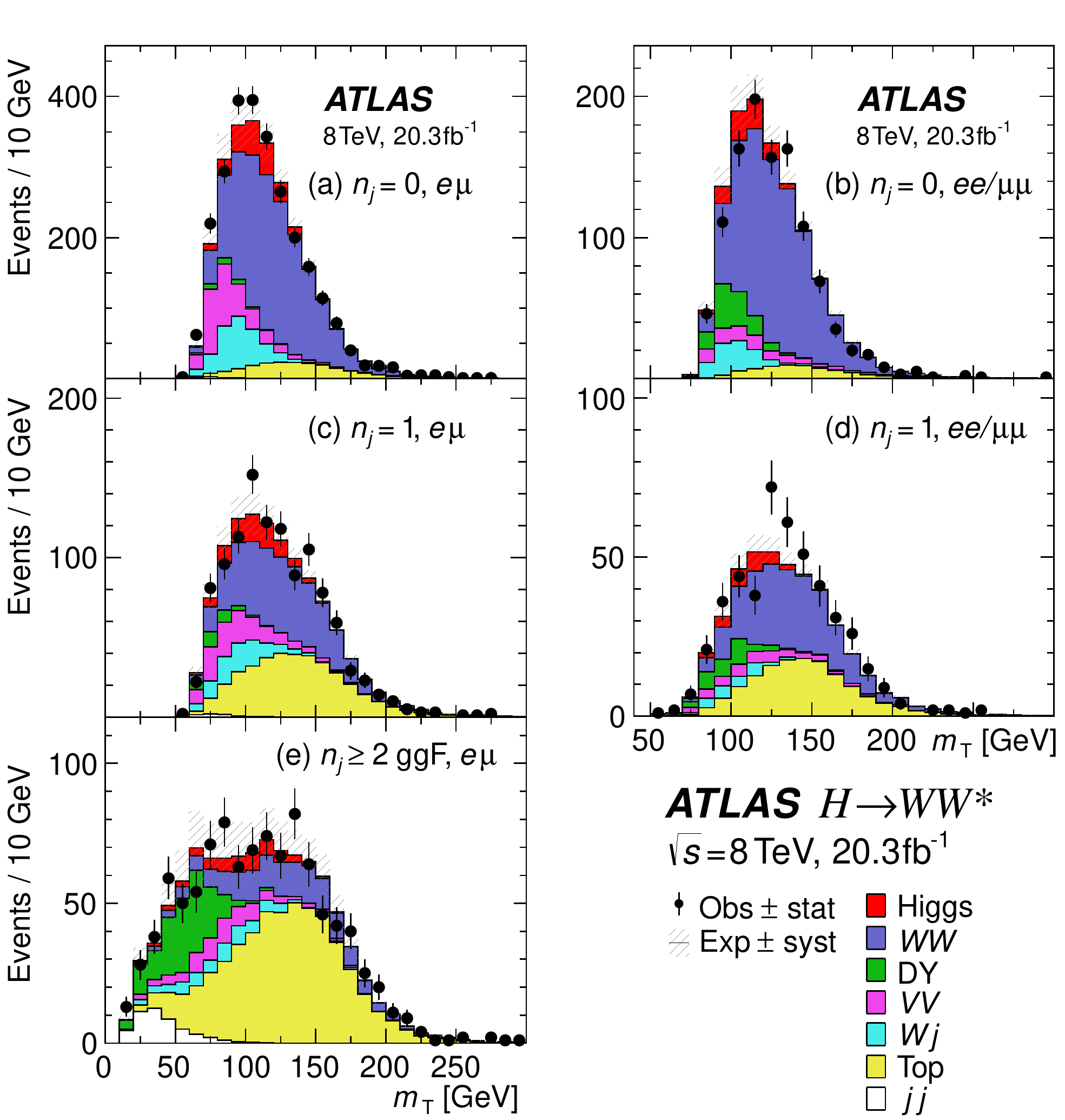} 
\caption{
  Distributions of the transverse mass $\mTH$ for the $\NjetLEone$ and $\NjetGEtwo$ ggF-enriched categories in the $8\TeV$ data analysis.
  The plots are made after requiring all selections up to $\mTH$ (see Tables~\ref{tab:sr_0j}, \ref{tab:sr_1j}, and \ref{tab:sr_2jggf}).
  \HwwPlotDetail{See}.
  The sum of the $\jj$ and $\Wj$ contributions corresponds to the label ``Misid.'' in Fig.~\ref{fig:MET}.
}
\label{fig:mT_finalstage8TeV}
\label{fig:mTH8}
\end{figure*}

The initial selection, $\Nbjet{\EQ}0$ and $\mtt{\LT}\mZ{\MINUS}25\GeV$, is
common to the other categories and reduces the top-quark and DY backgrounds.
The ggF-enriched sample is forced to be mutually exclusive to the VBF-enriched sample by inverting at least
one of the VBF-specific requirements:
$\cjv{\GT}1$, $\olv{\LT}1$, or $\bdt{\GT}-0.48$.
A similar inversion is done for the cross-check analysis:
$\dyjj{\GT}3.6$, $\mjj{\GT}600\GeV$, $\Nextrajet{\EQ}0$, or $\olv{\LT}1$.
Both sets of orthogonality requirements for the BDT and the cross-check are imposed for the $\NjetGEtwo$ ggF-enriched category.

The resulting sample contains events in a region sensitive to $\VH$ production where the associated
$W$ or $Z$ boson decays hadronically. This region is suppressed by rejecting
events in the region of $\detajj{\LE}1.2$ and $\ABS{\mjj{\MINUS}85}{\LT}15\GeV$.

Figure~\ref{fig:ggF2j} shows the $\mll$ distribution after
the VH orthogonality requirement.
The $\HWWlvlv$ topological selections, $\mll{\LT}55\GeV$ and $\dphill{\LT}1.8$,
further reduce the dominant top-quark background by $70\%$, resulting in a signal purity of $3.3\%$.

\subsection{\boldmath Modifications for $7\TeV$ data \label{sec:selection_7tev}}

The $7\TeV$ data analysis closely follows the selection
used in the $8\TeV$ analysis. The majority of the differences can be found in
the object definitions and identifications, as described in
Sec.~\ref{sec:atlas_detector}. The lower average pile-up allows
the loosening, or removal, of requirements on several pile-up sensitive variables from the selection.

The amount of DY background in the $\SFchan$ channel depends on the $\met$ resolution.
This background is reduced in a lower pile-up environment, allowing lower $\MET$ thresholds
in the $\SFchan$ samples for the $7\TeV$ data analysis. The $\MET$ requirement
is lowered to $35\GeV$, and the requirements on $\MPT$ are removed
entirely. The effect of the reduced $\MET$ thresholds
is partially compensated by an increased $\pTll$ requirement of $40\GeV$
in the $\NjetEQzero$ category and a $\pTllj{\GT}35\GeV$ requirement added to the
$\NjetEQone$ category.  The $\frecoil$ criteria are loosened to $0.2$ and $0.5$
in the $\NjetEQzero$ and $\NjetEQone$  categories, respectively.

In the $\NjetGEtwo$ category, only the VBF-enriched analysis is
considered; it follows an approach similar to the $8\TeV$ version. It
exploits the BDT multivariate method and it uses the same BDT classification and
output binning, as the $8\TeV$ data analysis. In the $\DFchan$ sample, a two-bin fit of the $\bdt$ is used (bins $2$ and $3$ are merged).
In the $\SFchan$ sample, a one-bin fit is used (bins $1$--$3$ are merged)
due to the smaller sample size.

The background estimation, signal modeling, final observed and expected
event yields, and the statistical analysis and results, are
presented in the next sections.

\subsection{\boldmath Summary \label{sec:selection_summary}}

This section described the event selection in the
$\Njet$ and lepton-flavor categories. Each of these categories is treated independently in the
statistical analysis using a fit procedure described in Sec.~\ref{sec:systematics}.
Inputs to the fit include the event yields and distributions at the
final stage of the event selection without the $\mTH$ requirement.

The total signal efficiency for $\HWWlvlv$ events
produced with $\ell{\EQ}e$ and $\mu$, including all signal categories
and production modes, is $10.2\%$ at $8\TeV$ for the ATLAS measured
mass value of $\HwwHiggsMass{ATLAS}\GeV$. The corresponding signal efficiency when
considering only the $\VBF$ production mode is $7.8\%$.

Figure~\ref{fig:mT_finalstage8TeV} shows the $\mTH$ distributions in
the $\NjetEQzero$, $\NjetEQone$ and $\NjetGEtwo$ ggF-enriched categories for the $8\TeV$ data.
The distributions for the $\NjetLEone$ categories are shown in
Fig.~\ref{fig:mT_finalstage7TeV} for the $7\TeV$ data. The final
$\bdt$ output distribution, for the VBF-enriched category, is shown in
Fig.~\ref{fig:bdt_finalstage78TeV} for the $7\TeV$ and $8\TeV$ data samples.

\begin{figure*}[tb!]
\includegraphics[width=0.75\textwidth]{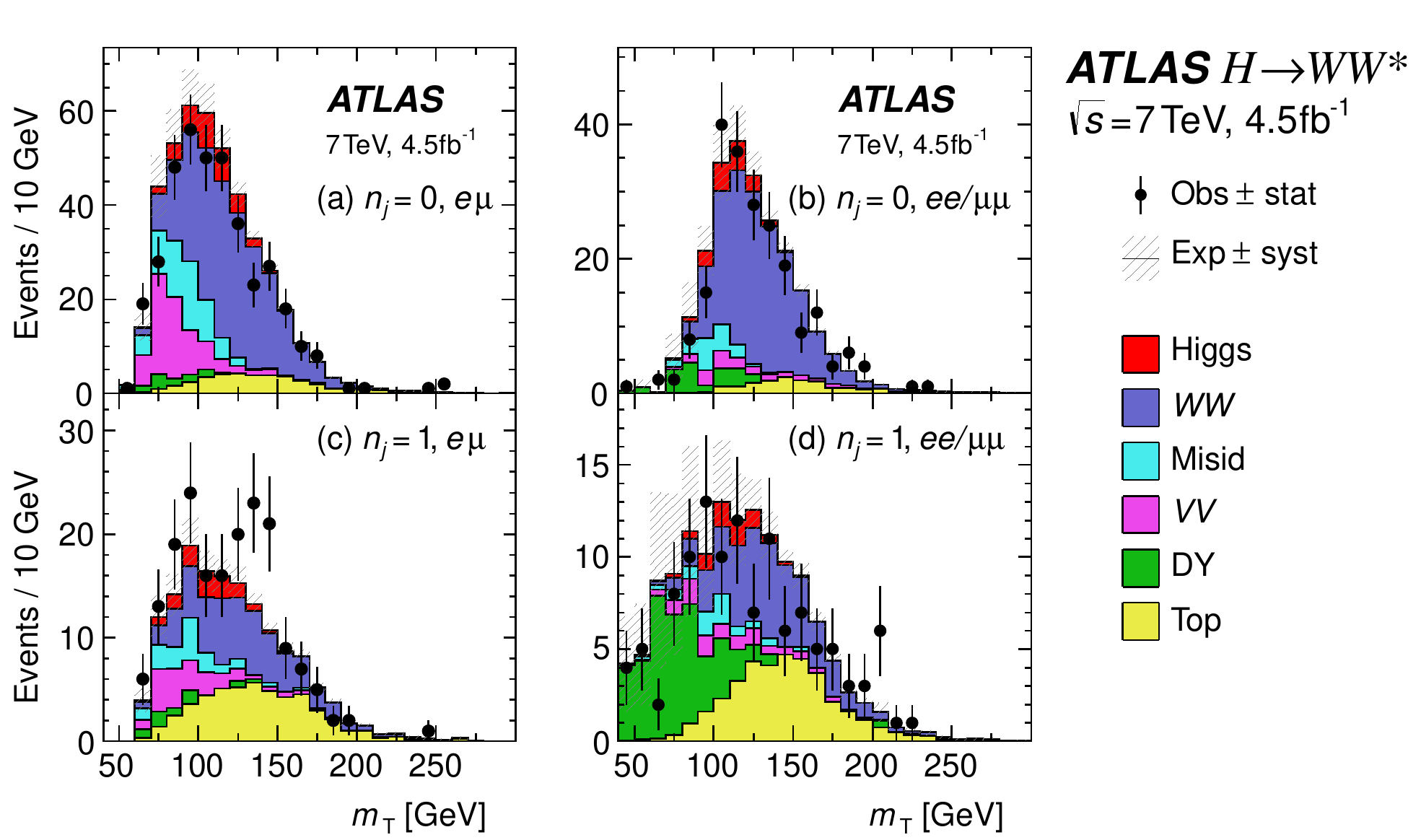}
\caption{
  Distributions of the transverse mass $\mTH$ for the $\NjetLEone$ categories in the $7\TeV$ data analysis.
  The plot is made after requiring all selections up to $\mTH$ (see Sec.~\ref{sec:selection_7tev}).
  \HwwPlotDetail{See}.
}
\label{fig:mT_finalstage7TeV}
\label{fig:mTH7}
\end{figure*}

\begin{figure*}[tb!]
\includegraphics[width=0.75\textwidth]{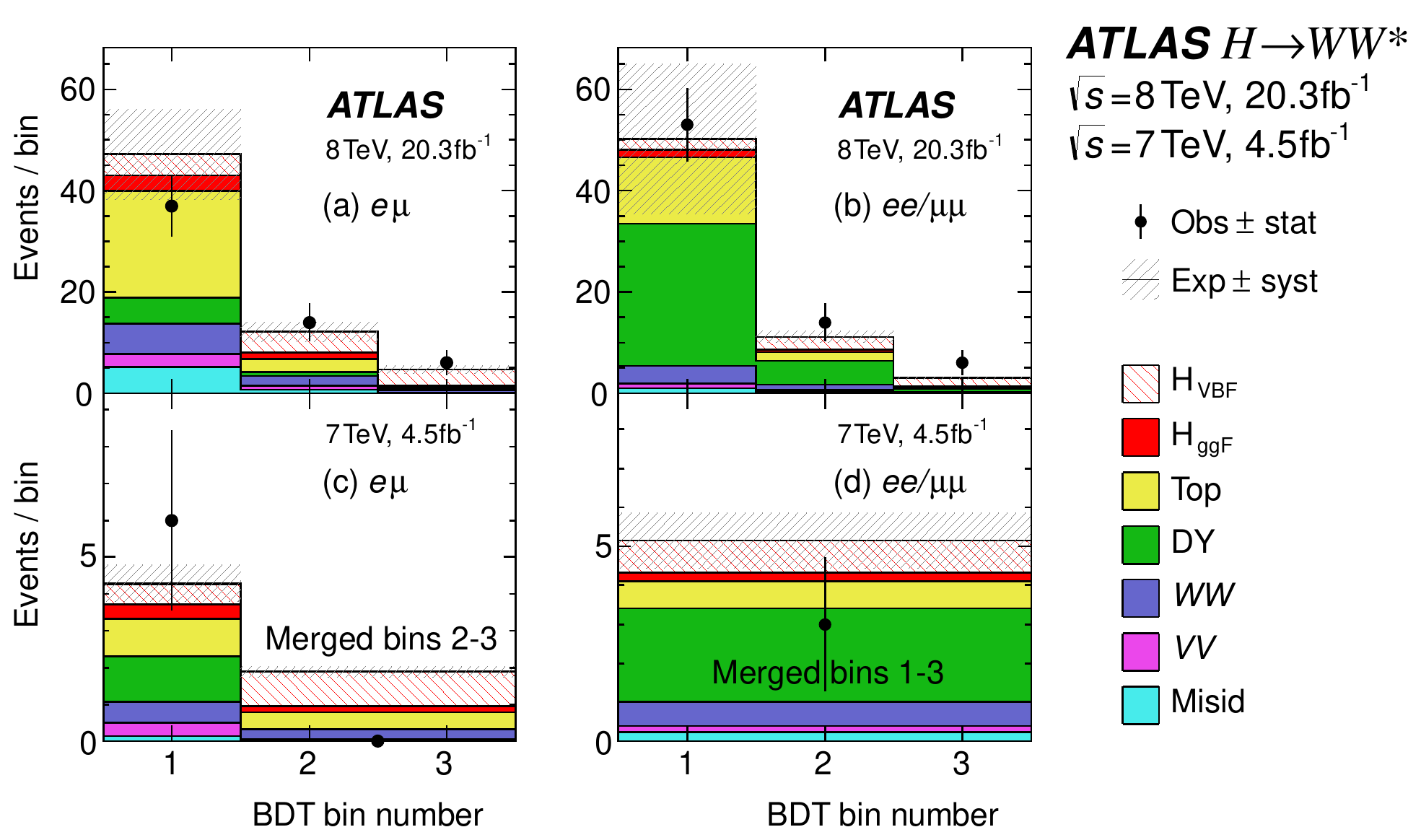}
\caption{
  Distributions of the BDT output in the $\NjetGEtwo$ VBF-enriched category
  in the $8$ and $7\TeV$ data analyses.
  The plot is made after requiring all the selections listed in Table~\ref{tab:selection} and after the BDT classification.
  \HwwPlotDetail{See}.
}
\label{fig:bdt_finalstage78TeV}
\end{figure*}

Figures~\ref{fig:pT} and~\ref{fig:mll} show the $\pTsublead$
and $\mll$ distributions at the end of the event selection in the
$\NjetLEone$ $\DFchan$ categories for the $8\TeV$ data analysis. The distributions are shown
for two categories of events based on the flavor of the lepton with the higher
$\pT$. This division is important for separating events based on the
relative contribution from the backgrounds from misidentified leptons
($\Wjets$ and multijets); see
Sec.~\ref{sec:bkg_misid} for details. The dependence of the
misidentified lepton and $\VV$ background distributions on $\pTsublead$
motivates the separation of the data sample into three bins of
$\pTsublead$. The variations in the background composition across the $\mll$ range motivate the division
into two bins of $\mll$. Figure~\ref{fig:pTmll_7TeV} shows the
corresponding distributions in the $\DFchan$ $\NjetLEone$ samples in the $7\TeV$
data analysis.

\begin{figure*}[tb!]
\includegraphics[width=0.75\textwidth]{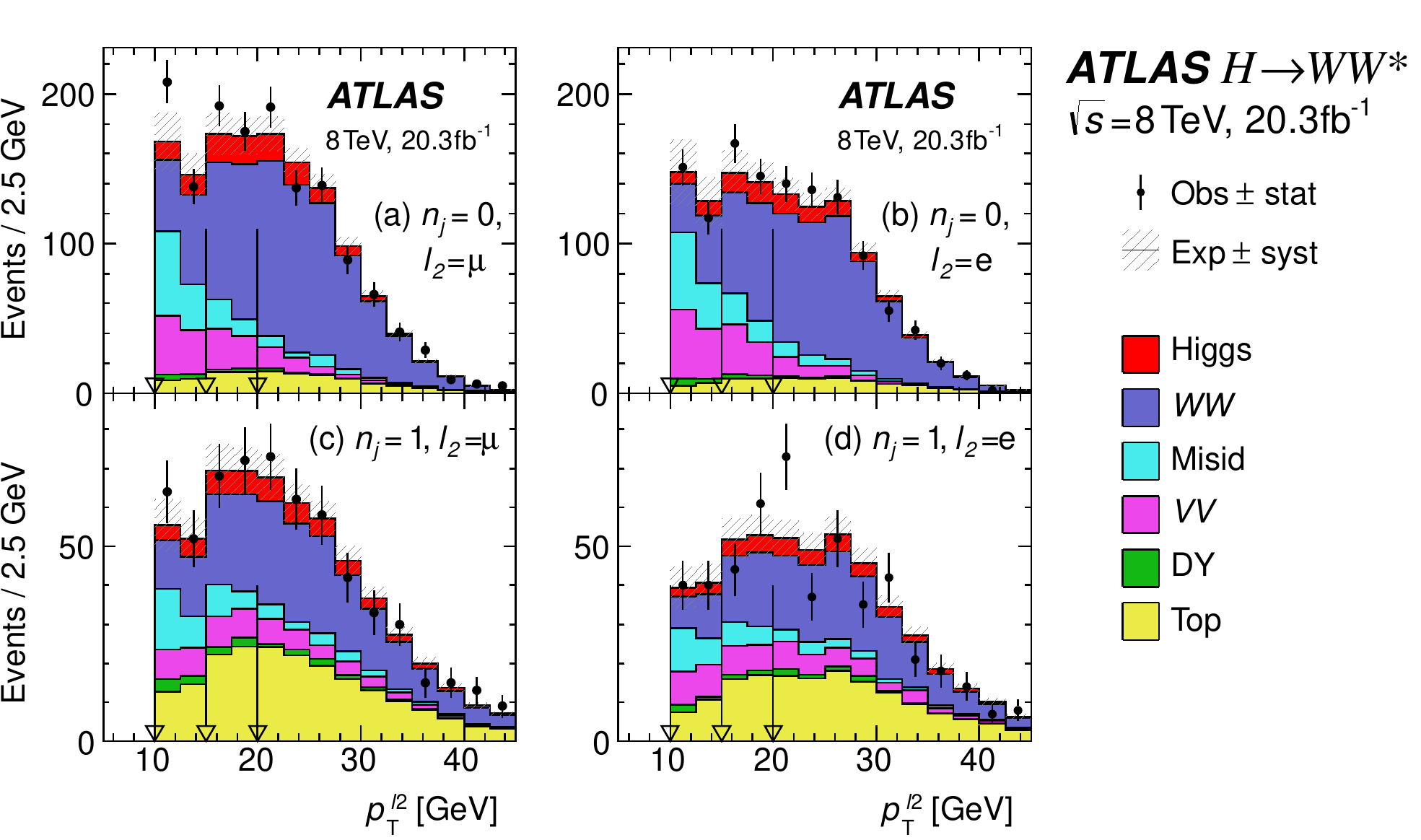}
\caption{
  Distributions of the subleading lepton $\pT$ for the $8\TeV$ data analysis in
  the $\DFchan$ sample used for the statistical analysis described in Sec.~\ref{sec:systematics}.
  The distributions are shown for two categories of events based on the flavor of the subleading lepton $\ell_2$.
  The plots are made after requiring all selections up to the $\mTH$
  requirement, as shown in Tables~\ref{tab:sr_0j} and \ref{tab:sr_1j}.
  The arrows indicate the bin boundaries;
  \HwwPlotDetail{see}.
}
\label{fig:pT}
\end{figure*}

\begin{figure*}[tb!]
\includegraphics[width=0.75\textwidth]{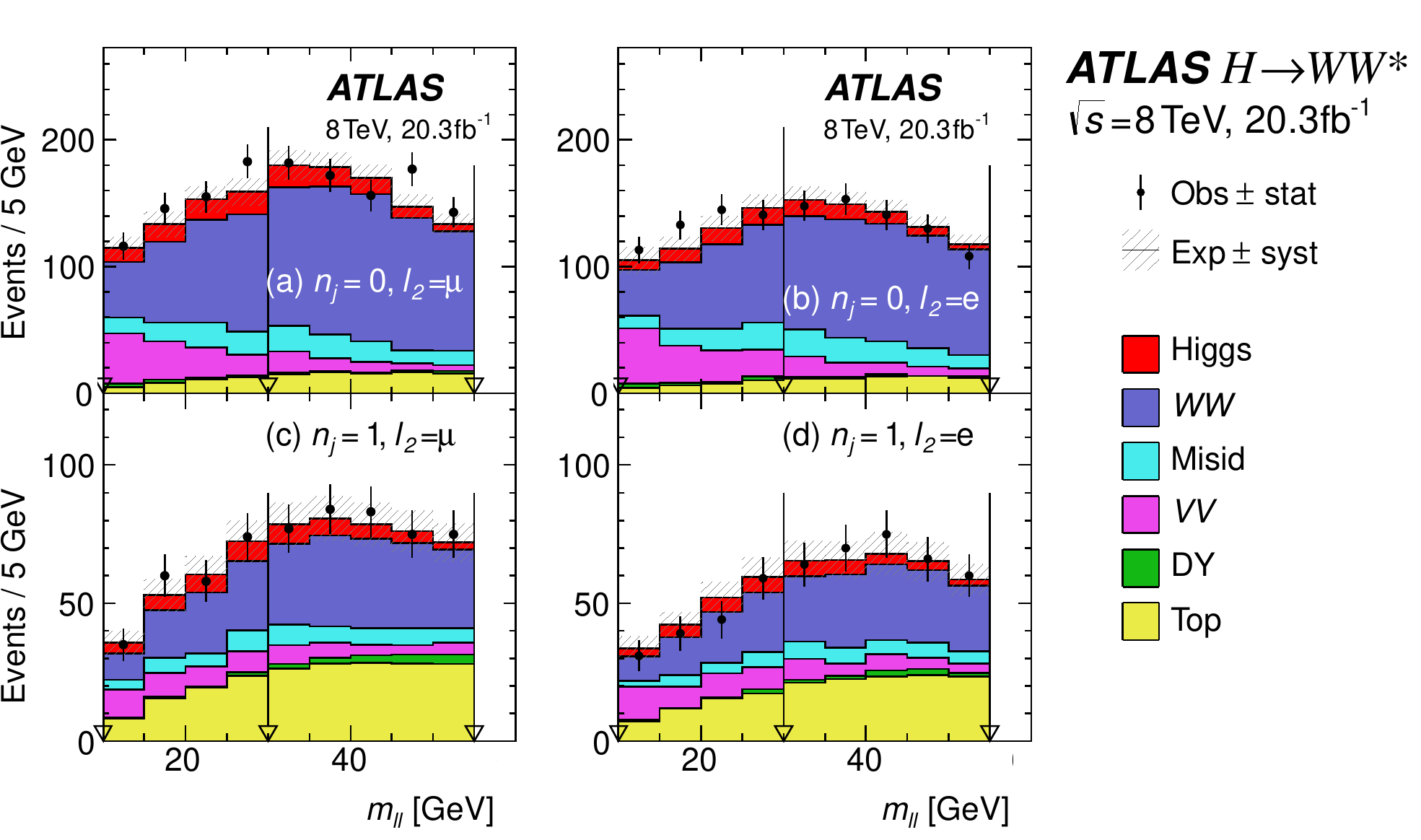}
\caption{
  Distributions of the dilepton invariant mass $\mll$ for the $8\TeV$ data analysis in
  the $\DFchan$ sample used for the statistical analysis described in Sec.~\ref{sec:systematics}.
  The distributions are shown for two categories of events based on the flavor of the subleading lepton $\ell_2$.
  The plot is made after requiring all selections up to the $\mTH$
  requirement, as shown in Tables~\ref{tab:sr_0j} and \ref{tab:sr_1j}.
  The arrows indicate the bin boundaries;
    \HwwPlotDetail{see}.
}
\label{fig:mll}
\end{figure*}

\begin{figure*}[tb!]
\includegraphics[width=0.75\textwidth]{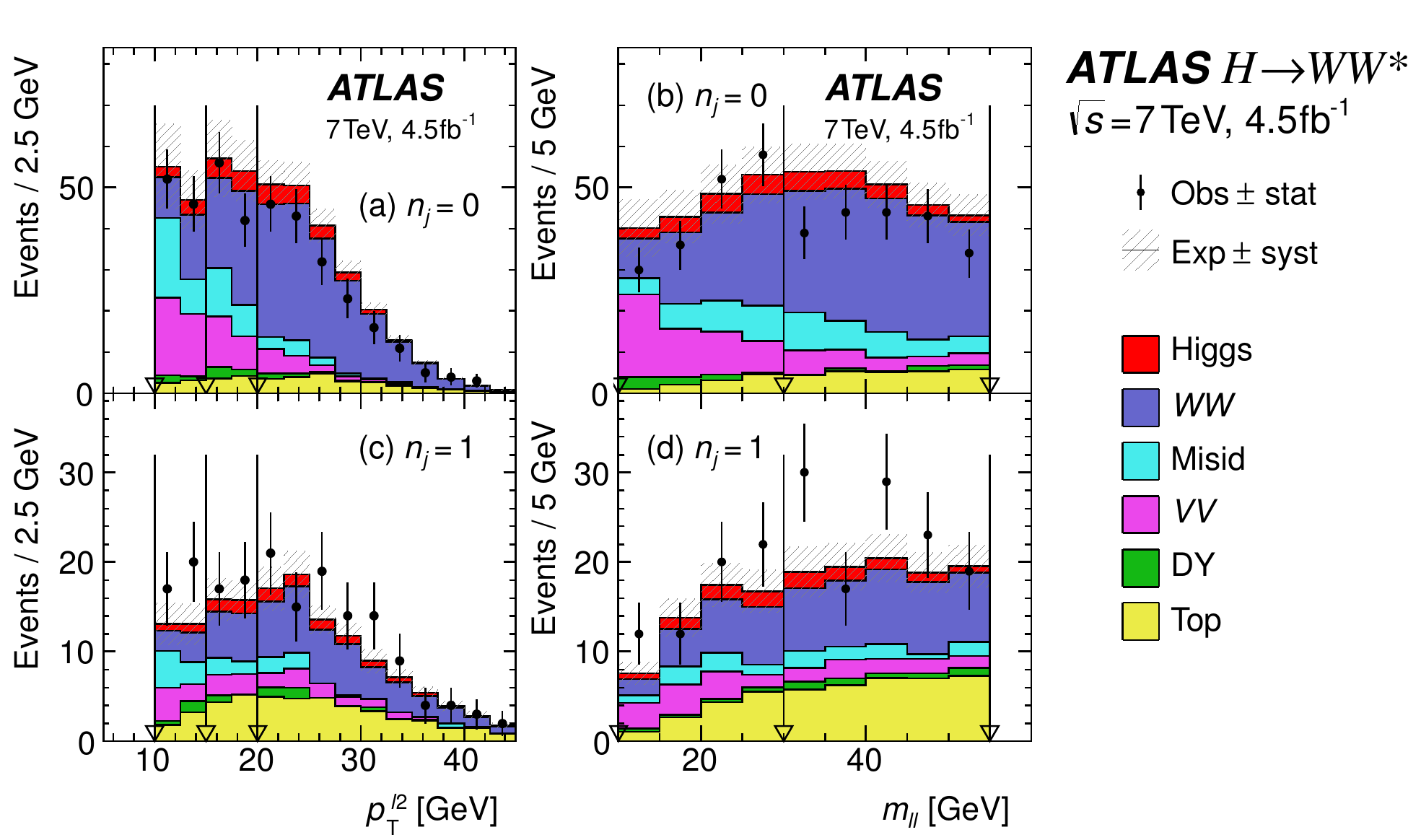}
\caption{
  Distributions of the subleading lepton $\pT$ and dilepton invariant mass for the $7\TeV$ data analysis in the $\DFchan$ sample.
  The plots are made after requiring all selections up to $\mTH$ (see Sec.~\ref{sec:selection_7tev}).
  The arrows indicate the bin boundaries;
  \HwwPlotDetail{see}.
}
\label{fig:pTmll_7TeV}
\end{figure*}
The event displays in Fig.~\ref{fig:display} show examples of the detector activity for
two signal candidates: one in the $\NjetEQzero$ $\DFchan$ category for the $7\TeV$ data analysis,
and one in the VBF-enriched $\NjetGEtwo$ $\DFchan$ category for the $8\TeV$ data analysis.
Both events have a small value of $\dphill$ as
is characteristic of the signal. The latter event shows two well-separated
jets that are characteristic of VBF production.

\begin{figure*}[p!]
{
  \raggedright
  \centering\includegraphics[width=0.85\textwidth]{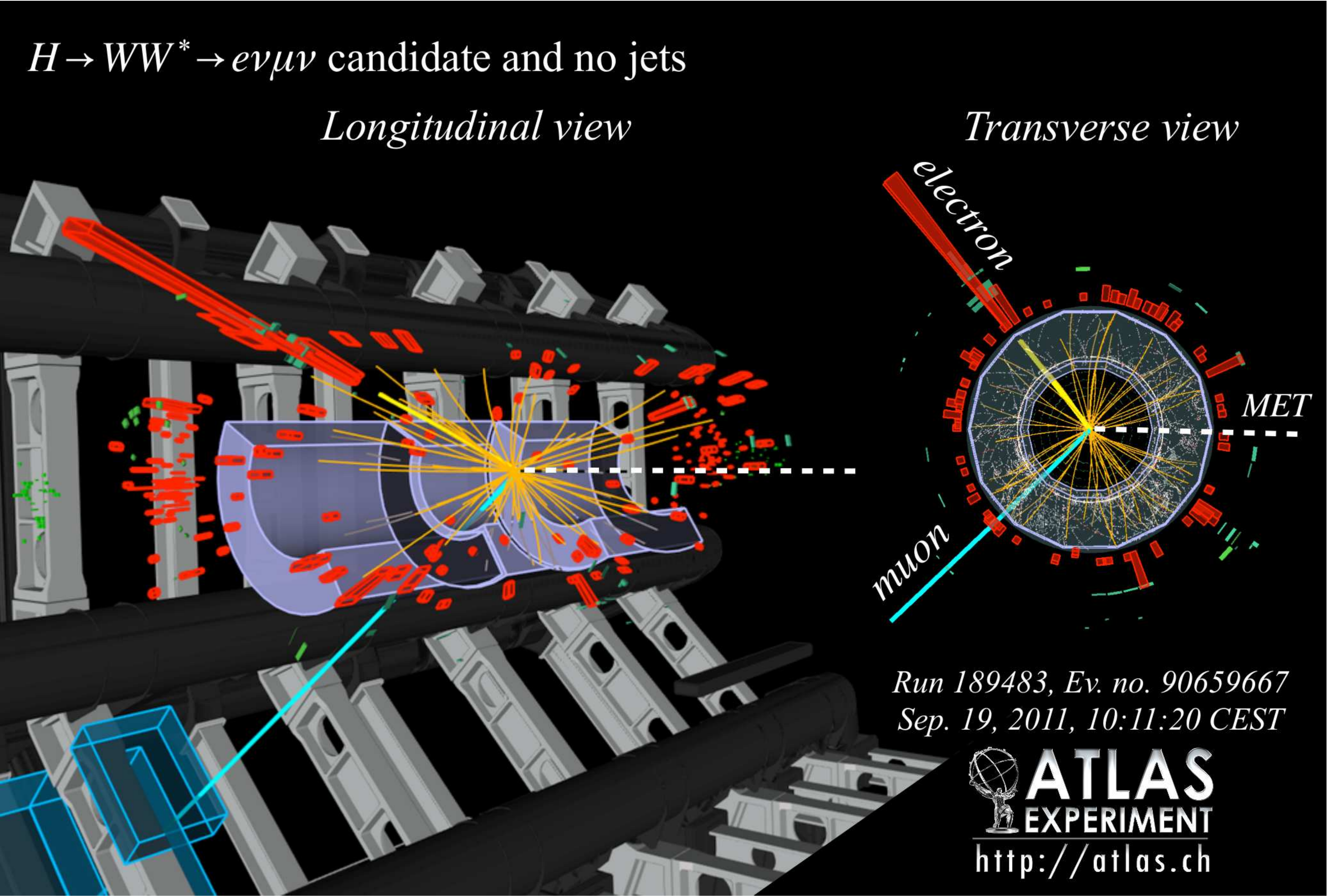}\\ \vspace{9pt}
  \centering\includegraphics[width=0.85\textwidth]{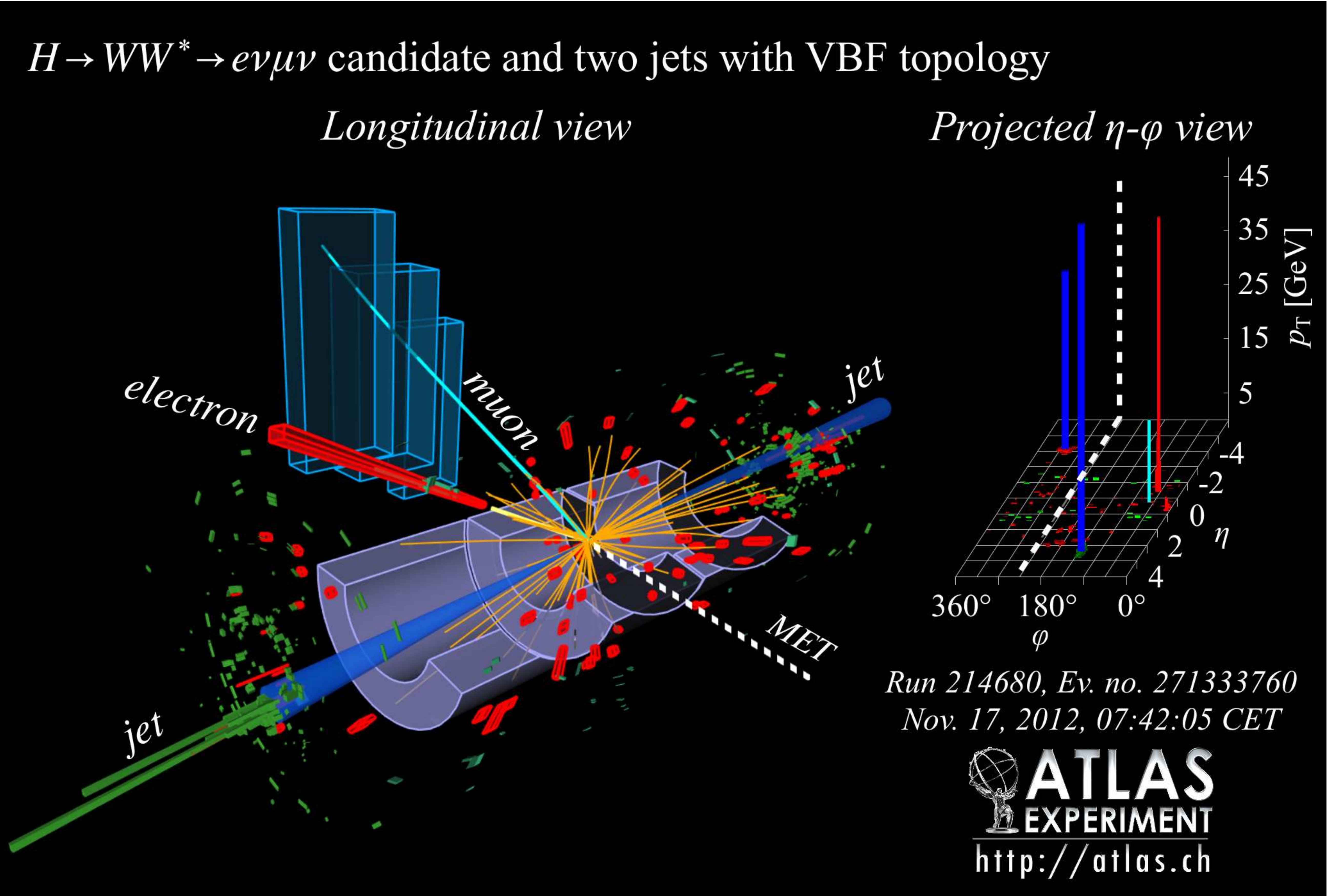}\\
}
\caption{
  Event displays of $\HWWevmv$ candidates in the $\NjetEQzero$ (top)
  and $\NjetGEtwo$ VBF-enriched (bottom) categories.
  The neutrinos are represented by $\met$ ($\METsc$, dotted line) that
  points away from the $\DFchan$ system.
  The properties of the first event are $\pT^{e}{\EQ}33\GeV$,
  $\pT^{\mu}{\EQ}24\GeV$, $\mll{\EQ}48\GeV$, $\dphill{\EQ}1.7$,
  $\MPTj{\EQ}37\GeV$, and $\mTH{\EQ}98\GeV$.
  The properties of the second event are $\pT^{e}{\EQ}51\GeV$,
  $\pT^{\mu}{\EQ}15\GeV$, $\mll{\EQ}21\GeV$, $\dphill{\EQ}0.1$,
  $\pTleadjet{\EQ}67\GeV$, $\pTsubleadjet{\EQ}41\GeV$,
  $\mjj{\EQ}1.4\TeV$, $\dyjj{\EQ}6.6$,
  $\MPTj{\EQ}59\GeV$, and $\mTH{\EQ}127\GeV$.
  Both events have a small value of $\dphill$, which is characteristic of the signal.
  The second event shows two well-separated jets that are characteristic of VBF production.
}
\label{fig:display}
\end{figure*}

\section{\boldmath Signal processes \label{sec:signal}}

The leading Higgs boson production processes are illustrated in
Fig.~\ref{fig:production}.  This section details the normalization
and simulation of the ggF and VBF production modes.  In both cases,
the production cross section has been calculated to NNLO in QCD and
next-to-leading order in the electroweak couplings.  Resummation
has been performed to NNLL for the ggF process.  For the decay, the
calculation of the branching fraction is computed using the $\HWW$
and $\HZZ$ partial widths from \PROPHECY4f~\cite{Bredenstein:2006rh}
and the width of all other decays from {\HDECAY}~\cite{Djouadi:1997yw}.
The $\HWW$ branching fraction is $22\%$ with a relative uncertainty of
$4.2\%$ for $m_H=125.36\GeV$~\cite{Heinemeyer:2013tqa}.  Interference
with direct $WW$ production~\cite{ggint} and uncertainties on VH
production~\cite{YellowReport} have a negligible impact on this
analysis.  Uncertainties on the ggF and VBF production processes
are described in the following subsections.

\subsection{\boldmath Gluon fusion \label{sec:signal_ggF}}

The measurement of Higgs boson production via gluon fusion, and the extraction of the
associated Higgs boson couplings, relies on detailed theoretical calculations and
Monte Carlo simulation.  Uncertainties on the perturbative calculations of the total
production cross section and of the cross sections exclusive in jet multiplicity are
among the leading uncertainties on the expected signal event yield and the extracted
couplings.  The {\POWHEG}~\cite{Bagnaschi:2011tu} generator matched to \PYTHIA8~is used
for event simulation and accurately models the exclusive jet multiplicities relevant
to this analysis.  The simulation is corrected to match higher-order calculations of
the Higgs boson $\pT$ distribution.

\begin{figure}[bt!]
\includegraphics[width=0.45\textwidth]{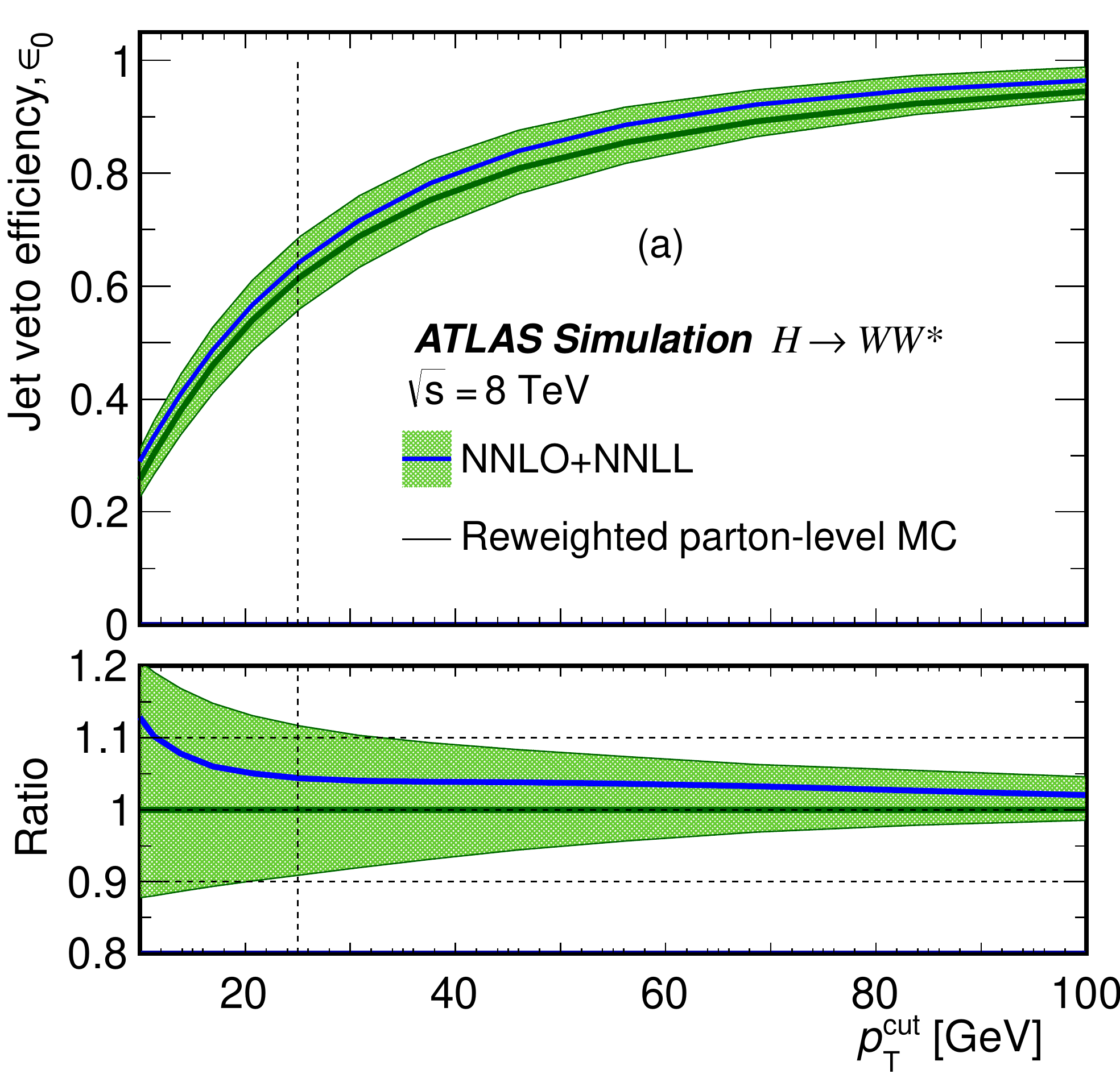}\\
\includegraphics[width=0.45\textwidth]{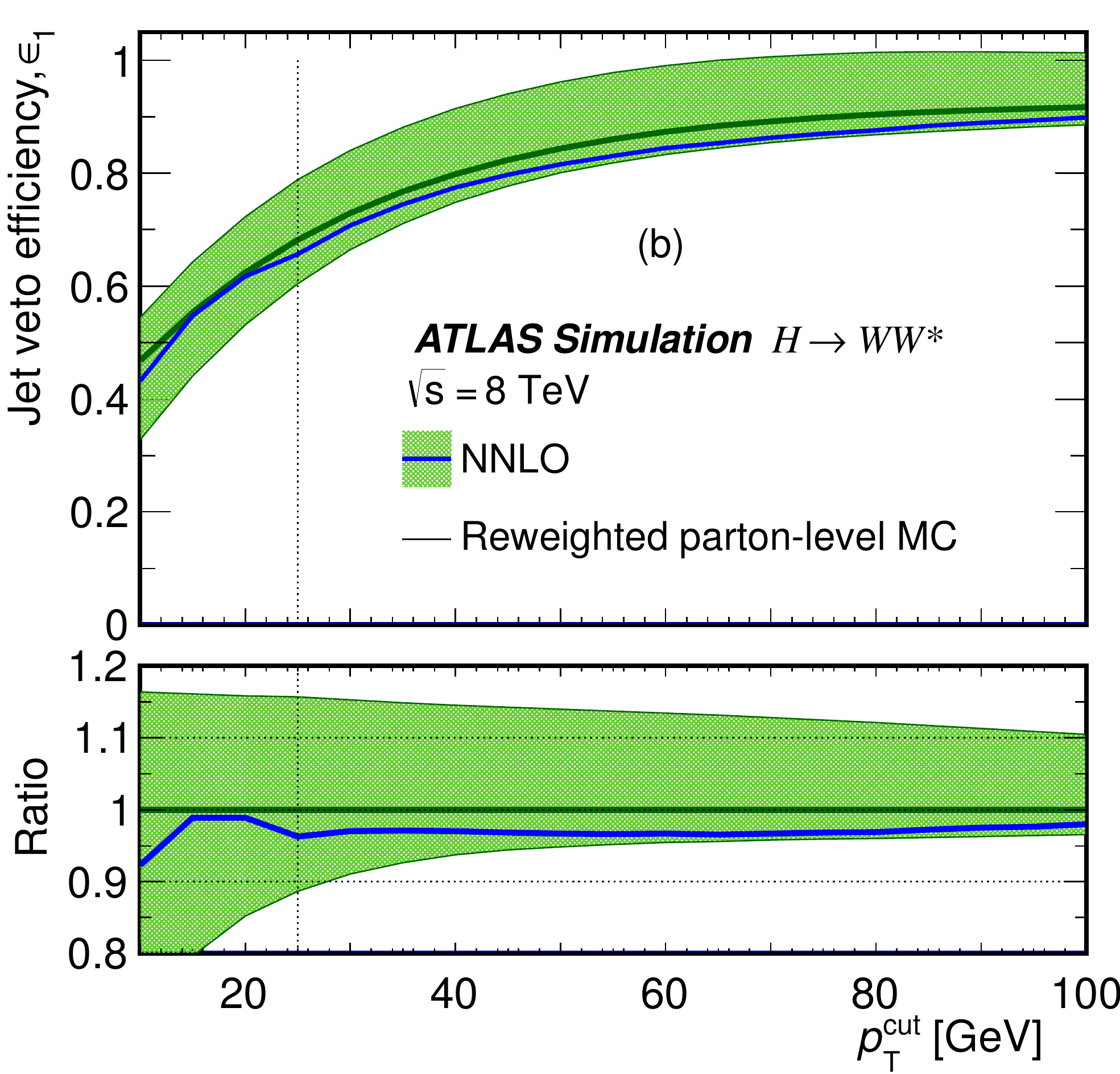}
\caption{
  Efficiencies of the veto of the (a) first jet and (b) second jet in
  inclusive ggF production of the Higgs boson, as a function of the veto-threshold $\pT$.
}
\label{fig:eps01}
\end{figure}

Production of a Higgs boson via gluon fusion proceeds dominantly through a top-quark
loop (the bottom-quark loop contributes $7\%$ to the cross section).  Higher-order QCD
corrections include radiation from the initial-state gluons and from the quark loop.
The total cross section is computed to NNLO~\cite{Anastasiou:2002yz} using the
$m_t{\TO}\infty$ approximation, where an effective point-like ggH coupling is
introduced.  Corrections for the finite top-quark mass have been computed to NLO
and found to be a few percent~\cite{Spira:1995rr}; this difference is applied as a
correction to the NNLO cross section.  Resummation of the soft QCD radiation has been
performed to NNLL~\cite{Catani:2003zt} in the $m_t{\TO}\infty$ approximation and
to the next-to-leading logarithms (NLL) for finite top- and bottom-quark masses.
Electroweak corrections to NLO~\cite{Aglietti:2004ki} are applied using the complete
factorization approximation~\cite{Actis:2008ug}.  Together, these calculations provide
the total inclusive cross section for the ggF process~\cite{deFlorian:2012yg}, which
is $19.15\pb$ for $\mH{\EQ}125.36\GeV$.  The uncertainty on the total cross section is
$10\%$, with approximately equal contributions from QCD scale variations ($7.5\%$) and
parton distribution functions ($7.2\%$).

The {\POWHEG} MC generator used to model ggF production~\cite{Bagnaschi:2011tu} is
based on an NLO calculation with finite quark masses and a running-width Breit-Wigner
distribution that includes electroweak corrections at next-to-leading order~\cite{deFlorian:2012yg}.  The
generator contains a scale for matching the resummation to the matrix-element
calculation, which is chosen to reproduce the NNLO+NLL calculation of the Higgs boson
$\pT$~\cite{Dittmaier:2012vm}.  To improve the modeling of this distribution, a
reweighting scheme is applied to reproduce the prediction of the NNLO+NNLL
dynamic-scale calculation given by the \HRES2.1 program~\cite{Grazzini:2013mca}.  The
scheme separately weights the $\pT$ spectra for events with ${\leq\,}1$ jet and events
with ${\geq\,}2$ jets, since the latter include jet(s) described purely by the
{\PYTHIA} shower model, which underestimates the rate of two balancing jets producing
low Higgs boson $\pT$.  Events with ${\geq\,}2$ jets are therefore reweighted to the
$\pT$ spectrum predicted by the NLO {\POWHEG} simulation of Higgs boson production in
association with two jets ($H{\PLUS}2$~jets)~\cite{Hamilton:2012np}.  The reweighting
procedure preserves agreement between the generated jet-multiplicity distribution and
the predictions of higher-order calculations.

The uncertainty on the jet multiplicity distribution is evaluated using the
jet-veto-efficiency (JVE) method~\cite{Dittmaier:2012vm,JVE} for the ggF categories and
the Stewart-Tackmann (ST) method~\cite{ST} for the VBF category.  The JVE method
factorizes the total cross section from the acceptances of the jet vetoes in the zero-jet
and one-jet channels, treating these components as uncorrelated.  Three calculations of
the jet-veto efficiency are defined based on ratios of cross sections with different
jet multiplicities and at different orders (for example,
$1{\MINUS}\sigma_{\Njet\geq 1}^{\mbox{\footnotesize\sc nlo}}/
\sigma_{\rm tot}^{\mbox{\footnotesize\sc nnlo}}$ for the veto efficiency of the
first jet).  The three calculations differ by next-to-next-to-next-to leading order terms in the inclusive
perturbative series, so their comparison provides an estimate of the perturbative
uncertainty on the jet veto.  A second estimate is obtained by individually varying
the factorization, renormalization, and resummation scales by factors of $2$ or
$1/2$, and by coherently varying the factorization and renormalization scales by
these factors.  These estimates are used to define an overall uncertainty, as
described below.

For the efficiency $\epsilon_0$ of the jet veto that defines the zero-jet channel, the
central value is evaluated at the highest available fixed order (NNLO), with NNLL
resummation.  The uncertainty is taken as the maximum effect of the scale variations
on the calculation, or the maximum deviation of the other calculations from this one.
The results using the {\JETVHETO} computation~\cite{Banfi:2012yh} are shown in
Fig.~\ref{fig:eps01}, along with the reweighted {\POWHEG}+{\PYTHIA8} prediction
evaluated without hadronization or the underlying event.  The results are consistent
to within a few percent for a jet $\pT$ threshold of $25\GeV$, and the relative
uncertainty at this threshold is $12\%$.

The efficiency of vetoing an additional jet, given the presence of a single jet, is
defined as $\epsilon_1$.  The NNLO $\Njet{\GE}1$ cross section needed for the
highest-order calculation of the jet-veto-efficiency method is not available, though
the other two calculations of the veto efficiency can be performed using the {\MCFM}
generator.  The highest-order calculation is in the range spanned by the other two calculations in
both the case of $\epsilon_0$ and in the case of $\epsilon_1$ evaluated using a
partial calculation of the NNLO $\Njet{\GE}1$ cross section~\cite{Boughezal:2013uia}.  The central value of
$\epsilon_1$ is thus estimated to be the average of the available calculations, with
the uncertainty given by the maximum difference with respect to these calculations or the scale-varied estimate of the average.  This
results in a relative uncertainty of $15\%$ on $\epsilon_1$, as shown in
Fig.~\ref{fig:eps01}.  The figure shows that the reweighted {\POWHEG}+{\PYTHIA8}
prediction for $\epsilon_1$ agrees with the calculation to within a few percent for
a jet $\pT$ threshold of $25\GeV$.

\begin{table}[b!]
\caption{
  Signal-yield uncertainties (in $\%$) due to the modeling of the gluon-fusion
  and vector-boson-fusion processes.  For the $\NjetEQzeroone$ categories the
  uncertainties are shown for events with same-flavor leptons; for events with
  different-flavor leptons the uncertainties are evaluated in bins of $\mll$ and
  $\pTsublead$.  For the $\NjetGEtwo$ VBF category the uncertainties are shown
  for the most sensitive bin of BDT output (bin $3$).
}
\label{tab:ggf_unc}
\begin{tabular*}{0.480\textwidth}{l d{1}d{1}d{1}d{1}}
\dbline
\multirow{2}{*}{Uncertainty source }
& \multicolumn{1}{p{0.065\textwidth}}{~~~~~$\NjetEQzero$}
& \multicolumn{1}{p{0.065\textwidth}}{~~~~~$\NjetEQone$}
& \multicolumn{1}{p{0.065\textwidth}}{~~~~~$\NjetGEtwo$}
& \multicolumn{1}{p{0.065\textwidth}}{~~~~~$\NjetGEtwo$} \\
&
&
& \multicolumn{1}{l}{~~~~ggF}
& \multicolumn{1}{l}{~~~~VBF} \\
\sgline
Gluon fusion \\
\quad Total cross section   & 10   & 10  & 10  & 7.2 \\
\quad Jet binning or veto   & 11   & 25  & 33  & 29 \\
\quad Acceptance & & & \\
\quad ~~Scale               & 1.4  & 1.9 & 3.6 & 48 \\
\quad ~~PDF                 & 3.2  & 2.8 & 2.2 & \multicolumn{1}{r}{-~~~~}  \\
\quad ~~Generator           & 2.5  & 1.4 & 4.5 & \multicolumn{1}{r}{-~~~~}  \\
\quad ~~UE/PS               & 6.4  & 2.1 & 1.7 & 15 \\
\clineskip\clineskip
Vector-boson fusion \\
\quad Total cross section   & 2.7   & 2.7  & 2.7  & 2.7 \\
\quad Acceptance & & & \\
\quad ~~Scale    & \multicolumn{1}{r}{-~~~} & \multicolumn{1}{r}{-~~~} & \multicolumn{1}{r}{-~~~} & 3.0  \\
\quad ~~PDF      & \multicolumn{1}{r}{-~~~} & \multicolumn{1}{r}{-~~~} & \multicolumn{1}{r}{-~~~} & 3.0 \\
\quad ~~Generator & \multicolumn{1}{r}{-~~~} & \multicolumn{1}{r}{-~~~} & \multicolumn{1}{r}{-~~~} & 4.2 \\
\quad ~~UE/PS    & \multicolumn{1}{r}{-~~~} & \multicolumn{1}{r}{-~~~} & \multicolumn{1}{r}{-~~~} & 14 \\
\dbline
\end{tabular*}
\end{table}

A prior ATLAS analysis in this decay channel~\cite{couplings} relied on the ST
procedure for all uncertainties associated with jet binning.  The JVE estimation
reduces uncertainties in the ggF categories by incorporating a resummation
calculation (in $\epsilon_0$) and the NLO calculation of $H{\PLUS}2$~jets (in $\epsilon_1$).
The uncertainties for the ST (JVE) procedure are $18\%$ ($15\%$), $43\%$ ($27\%$), and
$70\%$ ($34\%$) for the cross sections in the $\NjetEQzero$, $\NjetEQone$, and
$\NjetGEtwo$ ggF-enriched categories, respectively.  These uncertainties are reduced when
the categories are combined, and contribute a total of ${\APPROX}5\%$ to the
uncertainty on the measured ggF signal strength (see Table~\ref{tab:syst_mu}).

Additional uncertainties on the signal acceptance are considered in each signal
category.  The scale and PDF uncertainties are typically a few percent.  A generator
uncertainty is taken from a comparison between \POWHEG+\HERWIG~and \aMCATNLO+\HERWIG~\cite{Alwall:2014hca},
which differ in their implementation of the NLO matrix element and the matching of
the matrix element to the parton shower.  Uncertainties due to the underlying event
and parton shower models (UE/PS) are generally small, though in the $\NjetEQone$
category they are as large as $14\%$ in the signal regions where $\pTsublead{\LT}20\GeV$.
The UE/PS uncertainties are estimated by comparing predictions from \POWHEG+\HERWIG~and
\POWHEG+\PYTHIA8.

The evaluation of the ggF background to the $\NjetGEtwo$ VBF category includes an
uncertainty on the acceptance of the central-jet veto.  The uncertainty is evaluated
to be $29\%$ using the ST method, which treats the inclusive $H{\PLUS}2$-jet and $H{\PLUS}3$-jet
cross sections as uncorrelated.  Scale uncertainties are also evaluated in each
measurement range of the BDT output, and are $3$--$7\%$ in BDT bins $1$ and $2$, and $48\%$ in
BDT bin $3$.  Other uncertainties on ggF modeling are negligible in this category, except
those due to UE/PS, which are significant because the second jet in ggF $H{\PLUS}2$-jet
events is modeled by the parton shower in the \POWHEG+\PYTHIA8 sample.  A summary of
the uncertainties on the gluon-fusion and vector-boson-fusion processes is given in
Table~\ref{tab:ggf_unc}.  The table shows the uncertainties for same-flavor
leptons in the $\NjetLEone$ categories, since events with different-flavor leptons
are further subdivided according to $\mll$ and $\pTsublead$ (as described in
Sec.~\ref{sec:analysis}).

\subsection{\boldmath Vector-boson fusion \label{sec:signal_VBF}}

The VBF total cross section is obtained using an approximate QCD NNLO computation
provided by the {\VBFATNNLO} program~\cite{Bolzoni:2010xr}.  The calculation is
based on the structure-function approach~\cite{Han:1992hr} that considers the VBF
process as two deep-inelastic scattering processes connected to the colorless
vector-boson fusion producing the Higgs boson.  Leading-order contributions
violating this approximation are explicitly included in the computation; the
corresponding higher-order terms are negligible~\cite{YellowReport}.  Electroweak
corrections are evaluated at NLO with the {\HAWK} program~\cite{Ciccolini:2007ec}.
The calculation has a negligible QCD scale uncertainty and a $2.7\%$ uncertainty
due to PDF modeling.

The \POWHEG~\cite{Nason:2009ai} generator is used to simulate the VBF process (see Table~\ref{tab:mc}).
Uncertainties on the acceptance are evaluated for several sources: the impact of
the QCD scale on the jet veto and on the remaining acceptance; PDFs; generator
matching of the matrix element to the parton shower; and the underlying event and
parton shower.  Table~\ref{tab:ggf_unc} shows the VBF and ggF uncertainties in the
most sensitive bin of the BDT output (bin $3$).  The other bins have the same or
similar uncertainties for the VBF process, except for UE/PS, where the uncertainty
is $5.2\%$ (${<}\,1\%$) in bin $2$ (bin $1$).

\section{\boldmath Background processes \label{sec:bkg}}

The background contamination in the various signal regions (SR) comes from several physics
processes that were briefly discussed in Sec.~\ref{sec:analysis} and listed in
Table~\ref{tab:process}.  They are
\begin{itemize}
  \setlength{\itemsep}{0pt}
  \setlength{\parskip}{0pt}
  \setlength{\parsep}{0pt}
  \item[(i)] $\WW$: nonesonant $W$ pair production;
  \item[(ii)] top quarks (Top): pair production ($\ttbar$) and single-top
        production ($t$) both followed by the decay $t{\TO}Wb$;
  \item[(iii)] misidentified leptons (Misid): $W$ boson production in association with
        a jet that is misidentified as a lepton ($\Wj$) and dijet or multijet
        production with two misidentifications ($\jj$);
  \item[(iv)] other dibosons ($\VV$): $\Wg$, $\Wgs$, $\WZ$ and $\ZZ$; and
  \item[(v)] Drell-Yan (DY): $\ZDY$ decay to $e$ or $\mu$ pairs ($\SFchan$) and
        $\tau$ pairs ($\tautau$);
\end{itemize}%
the contamination of Higgs decays to non-$WW$ channels is small, but considered as signal.
A few background processes, such as $\Zg$ and $\WW$ produced by double
parton interactions, are not listed because their contributions are
negligible in the control and signal regions, but they are considered in the analysis
for completeness. Their normalizations and acceptances are taken from
Monte Carlo simulation.

For each background the event selection includes a targeted set of kinematic requirements (and sample selection)
to distinguish the background from the signal. The background estimate is made
with a control region (CR) that inverts some or all of these
requirements and in many cases enlarges the allowed range for certain
kinematic variables to increase the number of observed events in the
CR. For example, the relevant selections that suppress the $\WW$
background in the $\NjetEQzero$ SR are $\mll{\LT}55\GeV$ and $\dphill{\LT}1.8$.
The $\WW$ CR, in turn, is defined by requiring $55{\LT}\mll{\LT}110\GeV$ and
$\dphill{\LE}2.6$.

The most common use of a CR, like the $\WW$ example above, is to determine the
normalization factor $\fNorm$ defined by the ratio
of the observed to expected yields of $\WW$ candidates in the CR,
where the observed yield is obtained by subtracting the non-$\WW$
(including the Higgs signal) contributions from the data.
The estimate $B_\SR^{\est}$ of the expected background in the SR under consideration can be written as
\begin{equation}
\nq
 B_\SR^{\est}
  = B_\SR\,\CDOT\np\underbrace{N_\CR/B_\CR}_{\mbox{\scriptsize Normalization\,$\fNorm$}}\no
  = N_\CR\,\CDOT\np\underbrace{B_\SR/B_\CR}_{\no\mbox{\scriptsize Extrapolation\,$\fAlpha$}}\nq
\label{eqn:est_ww}
\end{equation}
where $\NCR$ and $\BCR$ are the observed yield and the MC estimate in the CR,
respectively, and $B_\SR$ is the MC estimate in the signal region.
The first equality defines the data-to-MC normalization factor in the CR,
$\fNorm$; the second equality
defines the extrapolation factor from the CR to the SR, $\fAlpha$, predicted
by the MC. With a sufficient number of events available in the CR, the large
theoretical uncertainties associated with estimating the background
directly from simulation are replaced by the combination of two
significantly smaller uncertainties, the statistical uncertainty on
$\NCR$ and the systematic uncertainty on $\fAlpha$.

When the SR is subdivided for reasons of increased signal sensitivity, as is
the case for the $\DFchan$ sample for $\NjetEQzero$, a corresponding $\fAlpha$ parameter is
computed for each of the subdivided regions.  The CR (hence the $\fNorm$ parameter),
however, is not subdivided for statistical reasons.

The uncertainties described in this section are inputs to the extraction of the
signal strength parameter using the likelihood fit, which is described in
Sec.~\ref{sec:systematics}.

\begin{table}[t!]
\caption{
  Background estimation methods summary.
  For each background process or process group, a set of three columns
  indicate whether data ($\bullet$) or MC ($\circ$) samples are used
  to normalize the SR yield ({\sc n}),
  determine the CR-to-SR extrapolation factor ({\sc e}), and
  obtain the SR distribution of the fit variable ({\sc v}).
  In general, the methods vary from one row to the next for a given background
  process; see Sec.~\ref{sec:bkg} for the details.
}\label{tab:cr}
{
\small
\centering
\begin{tabular*}{0.480\textwidth}{ l
  cccl
  cccl
  cccl
  cccl
  cccl
  cccl
}
\dbline
\multicolumn{1}{p{0.100\textwidth}}{ \multirow{2}{*}{Category }}
& \multicolumn{4}{p{0.050\textwidth}}{ \multirow{2}{*}{~\,$\WW$ }}
& \multicolumn{4}{p{0.050\textwidth}}{ \multirow{2}{*}{~\,Top }}
& \multicolumn{4}{p{0.050\textwidth}}{ \multirow{2}{*}{$\Fakes$ }}
& \multicolumn{4}{p{0.050\textwidth}}{ \multirow{2}{*}{~~$\VV$ }}
& \multicolumn{8}{p{0.120\textwidth}}{~~~Drell-Yan}
\\
&&&&
&&&&
&&&&
&&&&
& \multicolumn{4}{p{0.050\textwidth}}{ $\,\SFchan$ }
& \multicolumn{4}{p{0.040\textwidth}}{ ~~\,$\tautau$ }
\vspace{1mm}
\\
\cline{2-4}
\cline{6-8}
\cline{10-12}
\cline{14-16}
\cline{18-20}
\cline{22-24}
\clineskip
&  {\sc n} & {\sc e} & {\sc v}
&& {\sc n} & {\sc e} & {\sc v}
&& {\sc n} & {\sc e} & {\sc v}
&& {\sc n} & {\sc e} & {\sc v}
&& {\sc n} & {\sc e} & {\sc v}
&& {\sc n} & {\sc e} & {\sc v}
\\
\sgline
$\NjetEQzero$ \\
\quad$\DFchan$  &{\scriptsize $\bullet$} &{\scriptsize $\circ$}   &{\scriptsize $\circ$}   &
                &{\scriptsize $\bullet$} &{\scriptsize $\circ$}   &{\scriptsize $\circ$}   &
                &{\scriptsize $\bullet$} &{\scriptsize $\bullet$} &{\scriptsize $\bullet$} &
                &{\scriptsize $\bullet$} &{\scriptsize $\circ$}   &{\scriptsize $\circ$}   &
                &{\scriptsize $\circ$}   &{\scriptsize $\circ$}   &{\scriptsize $\circ$}   &
                &{\scriptsize $\bullet$} &{\scriptsize $\circ$}   &{\scriptsize $\circ$}   \\
\quad$\SFchan$  &{\scriptsize $\bullet$} &{\scriptsize $\circ$}   &{\scriptsize $\circ$}   &
                &{\scriptsize $\bullet$} &{\scriptsize $\circ$}   &{\scriptsize $\circ$}   &
                &{\scriptsize $\bullet$} &{\scriptsize $\bullet$} &{\scriptsize $\bullet$} &
                &{\scriptsize $\circ$}   &{\scriptsize $\circ$}   &{\scriptsize $\circ$}   &
                &{\scriptsize $\bullet$} &{\scriptsize $\bullet$} &{\scriptsize $\circ$}   &
                &{\scriptsize $\bullet$} &{\scriptsize $\circ$}   &{\scriptsize $\circ$}   \\
\clineskip\clineskip
$\NjetEQone$ \\
\quad$\DFchan$  &{\scriptsize $\bullet$} &{\scriptsize $\circ$}   &{\scriptsize $\circ$}   &
                &{\scriptsize $\bullet$} &{\scriptsize $\circ$}   &{\scriptsize $\circ$}   &
                &{\scriptsize $\bullet$} &{\scriptsize $\bullet$} &{\scriptsize $\bullet$} &
                &{\scriptsize $\bullet$} &{\scriptsize $\circ$}   &{\scriptsize $\circ$}   &
                &{\scriptsize $\circ$}   &{\scriptsize $\circ$}   &{\scriptsize $\circ$}   &
                &{\scriptsize $\bullet$} &{\scriptsize $\circ$}   &{\scriptsize $\circ$}   \\
\quad$\SFchan$  &{\scriptsize $\bullet$} &{\scriptsize $\circ$}   &{\scriptsize $\circ$}   &
                &{\scriptsize $\bullet$} &{\scriptsize $\circ$}   &{\scriptsize $\circ$}   &
                &{\scriptsize $\bullet$} &{\scriptsize $\bullet$} &{\scriptsize $\bullet$} &
                &{\scriptsize $\circ$}   &{\scriptsize $\circ$}   &{\scriptsize $\circ$}   &
                &{\scriptsize $\bullet$} &{\scriptsize $\bullet$} &{\scriptsize $\circ$}   &
                &{\scriptsize $\bullet$} &{\scriptsize $\circ$}   &{\scriptsize $\circ$}   \\
\clineskip\clineskip
\multicolumn{1}{l}{$\NjetGEtwo$ ggF} \\
\quad$\DFchan$  &{\scriptsize $\circ$}   &{\scriptsize $\circ$}   &{\scriptsize $\circ$}   &
                &{\scriptsize $\bullet$} &{\scriptsize $\circ$}   &{\scriptsize $\circ$}   &
                &{\scriptsize $\bullet$} &{\scriptsize $\bullet$} &{\scriptsize $\bullet$} &
                &{\scriptsize $\circ$}   &{\scriptsize $\circ$}   &{\scriptsize $\circ$}   &
                &{\scriptsize $\circ$}   &{\scriptsize $\circ$}   &{\scriptsize $\circ$}   &
                &{\scriptsize $\bullet$} &{\scriptsize $\circ$}   &{\scriptsize $\circ$}   \\
\clineskip\clineskip
\multicolumn{1}{l}{$\NjetGEtwo$ VBF} \\
\quad$\DFchan$  &{\scriptsize $\circ$}   &{\scriptsize $\circ$}   &{\scriptsize $\circ$}   &
                &{\scriptsize $\bullet$} &{\scriptsize $\circ$}   &{\scriptsize $\circ$}   &
                &{\scriptsize $\bullet$} &{\scriptsize $\bullet$} &{\scriptsize $\bullet$} &
                &{\scriptsize $\circ$}   &{\scriptsize $\circ$}   &{\scriptsize $\circ$}   &
                &{\scriptsize $\circ$}   &{\scriptsize $\circ$}   &{\scriptsize $\circ$}   &
                &{\scriptsize $\bullet$} &{\scriptsize $\circ$}   &{\scriptsize $\circ$}   \\
 \quad$\SFchan$ &{\scriptsize $\circ$}   &{\scriptsize $\circ$}   &{\scriptsize $\circ$}   &
                &{\scriptsize $\bullet$} &{\scriptsize $\circ$}   &{\scriptsize $\circ$}   &
                &{\scriptsize $\bullet$} &{\scriptsize $\bullet$} &{\scriptsize $\bullet$} &
                &{\scriptsize $\circ$}   &{\scriptsize $\circ$}   &{\scriptsize $\circ$}   &
                &{\scriptsize $\bullet$} &{\scriptsize $\bullet$} &{\scriptsize $\circ$}   &
                &{\scriptsize $\bullet$} &{\scriptsize $\circ$}   &{\scriptsize $\circ$}   \\
\dbline
\end{tabular*}
}
\end{table}

An extension of this method is used when it is possible to determine
the extrapolation factor $\fAlpha$ from data. As described in
Secs.~\ref{sec:bkg_misid} and~\ref{sec:bkg_dy}, this can be done for the misidentified
lepton backgrounds and in the high-statistics categories for the
$\ZDYll$ background.
For the former, the distribution of the discriminating variable of interest is also determined
from data.
For completeness, one should note that the smaller background sources are
estimated purely from simulation.

Table~\ref{tab:cr} summarizes, for all the relevant background
processes, whether data or MC
is used to determine the various aspects of the method.
In general, data-derived methods are preferred and MC simulation is used for
a few background processes that do not contribute significantly in the signal
region, that have a limited number of events in the control region, or both.
MC simulation is used (open circles) or a data sample is used (solid circles) for
each of the three aspects of a given method:
the normalization (N),
the extrapolation (E), and
the distribution of the discriminating variable of interest (V).
The plots in this section (Figs.~\ref{fig:cr_ww}--\ref{fig:cr_dy}) show
contributions that are normalized according to these methods.

This section focuses on the methodology for background predictions and
their associated theoretical uncertainties. The experimental
uncertainties also contribute to the total uncertainty on these
background predictions and are quoted here only for the backgrounds
from misidentified leptons, for which the total systematic
uncertainties are discussed in Sec.~\ref{sec:bkg_misid}.
Furthermore, although the section describes one background estimation
technique at a time, the estimates for most background contributions are
interrelated and are determined {\insitu} in the statistical part
of the analysis (see Sec.~\ref{sec:systematics}).

The section is organized as follows. Section~\ref{sec:bkg_ww}
describes the $\WW$ background in the various categories. This
background is the dominant one for the most sensitive $\NjetEQzero$
category. Section~\ref{sec:bkg_top} describes the background from top-quark
production, which is largest in the categories with one or more
high-$\pT$ jets. The data-derived estimate from misidentified
leptons is described in Sec.~\ref{sec:bkg_misid}.
The remaining backgrounds, $\VV$ and $\ZDY$, are
discussed in Secs.~\ref{sec:bkg_vv} and \ref{sec:bkg_dy},
respectively. The similarities and modifications for the background
estimation for the $7\TeV$ data analysis are described in Sec.~\ref{sec:bkg_7tev}.
Finally, Sec.~\ref{sec:bkg_summary} presents a summary of the
background predictions in preparation for the fit procedure described in Sec.~\ref{sec:systematics}.

\subsection{\boldmath $\WW$ dibosons \label{sec:bkg_ww}}

The nonresonant $\WW$ production process, with subsequent decay
$WW\rightarrow\lvlv$, is characterized by two well-separated charged
leptons.  By contrast, the
charged leptons in the $\HWWlvlv$ process tend to have a small opening
angle (see Fig.~\ref{fig:decay}).  The invariant mass of the charged
leptons, $\mll$, combines this angular information with the kinematic
information associated with the relatively low Higgs boson mass
($\mH{\LT}2\mW$), providing a powerful discriminant between the processes
(see Fig.~\ref{fig:0j}).  This variable is therefore used to define $\WW$
control regions in the $\NjetLEone$ categories, where the signal is selected
with the requirement $\mll{\LT}55\GeV$.  For the $\NjetGEtwo$ $\ggF$ and
$\VBF$ categories, the $\WW$ process is modeled with a merged multi-parton
{\SHERPA} sample and normalized to the NLO inclusive $\WW$ calculation from
{\MCFM}~\cite{mcfm6}, since the large top-quark backgrounds make a
control-region definition more challenging.

\subsubsection{\boldmath\textit{\textbf{$\mll$ extrapolation for $\NjetLEone$}}\label{sec:bkg_ww_01j}}

The $\NjetLEone$ analyses use a data-based normalization for the $WW$ background,
with control regions defined by a range in $\mll$ that does not overlap with the
signal regions.  The normalization is applied to the combined
$(\qq$~or~$qg){\TO}\WW$ and $gg{\TO}\WW$ background estimate, and theoretical
uncertainties on the extrapolation are evaluated.

To obtain control regions of sufficient purity, several requirements are applied.
In order to suppress the $\ZDY$ background, the CRs use $\emu$ events selected
after the $\pTll{\GT}30\GeV$ and $\mTlep{\GT}50\GeV$ requirements in the
$\NjetEQzero$ and $\NjetEQone$ categories, respectively.  The latter requirement
additionally suppresses background from multijet production.  A requirement of
$\pTsublead{\GT}15\GeV$ is applied to suppress the large $\Wjets$ background below
this threshold.  Additional $\ZDYtt$ reduction is achieved by requiring
$\dphill{\LT}2.6$ for $\NjetEQzero$, and $\ABS{\mtt{\MINUS}\mZ}{\GT}25\GeV$ for
$\NjetEQone$, where $\mtt$ is defined in Sec.~\ref{sec:selection_1j}.  The $\mll$
range is $55{\LT}\mll{\LT}110\GeV$ ($\mll{\GT}80\GeV$) for $\NjetEQzero$ ($1$), and
is chosen to maximize the signal significance.  Increasing the upper bound on $\mll$
for $\NjetEQzero$ increases the theoretical uncertainty on the $WW$ background
prediction.  The $\mTH$ distributions in the $WW$ control regions are shown in
Fig.~\ref{fig:cr_ww}.

\begin{figure}[tb!]
\includegraphics[width=0.45\textwidth]{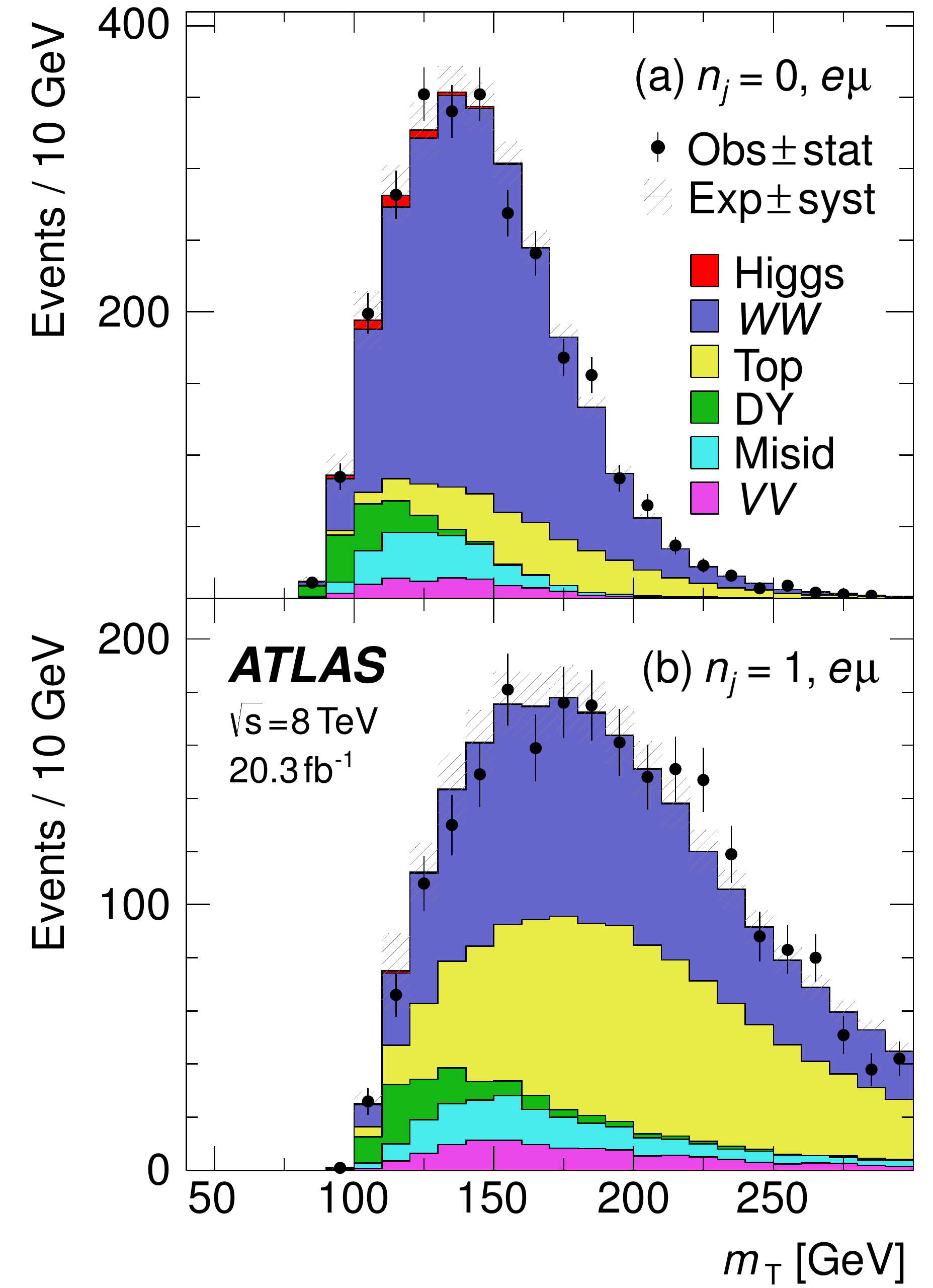}
\caption{
  $\WW$ control region distributions of transverse mass.
  The normalizations of all processes are as described in Sec.~\ref{sec:bkg}.
  \HwwPlotDetail{See}.
}
\label{fig:cr_ww}
\end{figure}

The $\WW$ estimate $B^{\est}_{\scWW,\,i}$ in each signal region $i$ is given by
Eq.~(\ref{eqn:est_ww}).  The control region is approximately $70\%$ ($45\%$)
pure in the $\NjetEQzero$ ($1$) category.  The contamination in the $\NjetEQone$
category is dominated by $\ttbar{\TO}WbWb$ events, where one jet is unidentified
and the other is misidentified as a light-quark jet.  The single-top contribution
is one-third the size of this background for $\NjetEQone$; for $\NjetEQzero$ this
ratio is about one-half.  All backgrounds are subtracted as part of the fit for
$\fNorm$ described in Sec.~\ref{sec:systematics_fit_lh}.

The CR-to-SR extrapolation factor has uncertainties due to the limited accuracy
of the MC prediction.  Uncertainties due to higher perturbative orders in QCD
not included in the MC simulation are estimated by varying the renormalization and
factorization scales independently by factors of $1/2$ and $2$, keeping the
ratio of scales in the range $1/2$ and $2$~\cite{Heinemeyer:2013tqa}.  An
uncertainty due to higher-order electroweak corrections is determined by reweighting
the MC simulation to the NLO electroweak calculation~\cite{Bierweiler:2013dja} and
taking the difference with respect to the nominal sample.  PDF uncertainties are
evaluated by taking the largest difference between the nominal CT10~\cite{Lai:2010vv}
PDF set and either the MSTW2008~\cite{Martin:2009iq} or the NNPDF2.3~\cite{Ball:2012cx}
PDF set, and adding in quadrature the uncertainty determined using the CT10 error
eigenvectors.  Additional uncertainties are evaluated using the same procedures as
for ggF production (Sec.~\ref{sec:signal_ggF}): uncertainties due to the modeling
of the underlying event, hadronization and parton shower are evaluated by comparing
predictions from \POWHEG+\PYTHIA6 and \POWHEG+\HERWIG; a generator uncertainty is
estimated with a comparison of \POWHEG+\HERWIG\ and \aMCATNLO+\HERWIG.  The detailed
uncertainties in each signal subregion are given in Table~\ref{tab:ww_alpha}.  The
corresponding uncertainties on the $\mTH$ distribution give a relative change
of up to $20\%$ between $90$ and $170\GeV$, depending on the signal region.

\begin{table}[b!]
\caption{
  $\WW$ theoretical uncertainties (in $\%$) on the
  extrapolation factor $\fAlpha$ for $\NjetLEone$.
  Total (Tot) is the sum in quadrature of the uncertainties due to
  the QCD factorization and renormalization scales (Scale), the PDFs,
  the matching between the hard-scatter matrix element to the UE/PS model (Gen),
  the missing electroweak corrections (EW), and the parton shower and
  underlying event (UE/PS).  The negative sign indicates anti-correlation
  with respect to the unsigned uncertainties for SR categories in the same
  column.  Energy-related values are given in $\!\GeV$.
}
\label{tab:ww_alpha}
{
\small
  \centering
\begin{tabular*}{0.480\textwidth}{
  p{0.145\textwidth}
  r
  r
  r
  r
  r
  r
  p{0.001\textwidth}
  r
}
\dbline
\multirow{1}{*}{SR category}
& \multicolumn{6}{c}{$\NjetEQzero$}
&& \multicolumn{1}{c}{${=\,}1$}
\\
\clineskip\cline{2-7}\cline{9-9}\clineskip
& \multicolumn{1}{c}{Scale}
& \multicolumn{1}{c}{$\!$PDF}
& \multicolumn{1}{c}{Gen}
& \multicolumn{1}{c}{EW}
& \multicolumn{1}{c}{$\!$UE/PS}
& \multicolumn{1}{c}{Tot}
&& \multicolumn{1}{l}{Tot$\no$}
\\
\sgline
\multicolumn{3}{l}{SR $\DFchan$, $10{\LT}\mll{\LT}30$} \\
\quad  $\pTsublead{\GT}20$                                & $0.7$~ & $0.6$~ & $3.1$~ & $-0.3$~ & $-1.9$~~ & $3.8$~ && $7.1$ \\
\quad  $15{\LT}\pTsublead{\LE}20$                         & $1.2$~ & $0.8$~ & $0.9$~ & $ 0.7$~ & $ 1.7$~~ & $2.6$~ && $3.9$ \\
\quad  $10{\LT}\pTsublead{\LE}15$                         & $0.7$~ & $1.0$~ & $0.4$~ & $ 1.2$~ & $ 2.2$~~ & $2.8$~ && $5.4$ \\
\clineskip\clineskip
\multicolumn{3}{l}{SR $\DFchan$, $30{\LT}\mll{\LT}55$} \\
\quad  $\pTsublead{\GT}20$                                & $0.8$~ & $0.7$~ & $3.9$~ & $-0.4$~ & $-2.4$~~ & $4.8$~ && $7.1$ \\
\quad  $15{\LT}\pTsublead{\LE}20$                         & $0.8$~ & $0.7$~ & $1.0$~ & $ 0.5$~ & $ 1.0$~~ & $2.0$~ && $4.5$ \\
\quad  $10{\LT}\pTsublead{\LE}15$                         & $0.7$~ & $0.8$~ & $0.5$~ & $ 0.8$~ & $ 1.5$~~ & $2.1$~ && $4.5$ \\
\clineskip\clineskip
\multicolumn{3}{l}{SR $\SFchan$, $12{\LT}\mll{\LT}55$} \\
\quad  $\pTsublead{\GT}10$                                & $0.8$~ & $1.1$~ & $2.4$~ & $ 0.1$~ & $-1.2$~~ & $2.9$~ && $5.1$ \\
\dbline
\end{tabular*}
}
\end{table}

The contribution from the $gg{\TO}\WW$ process is $5.8\%$ ($6.5\%$) of the total
$\WW$ background in the $\NjetEQzero$ ($1$) category in the signal region and
$4.5\%$ ($3.7\%$) in the control region.  Its impact on the extrapolation factor
is approximately given by the ratio of $gg{\TO}\WW$ to $q\bar{q}{\TO}\WW$ events
in the signal region, minus the corresponding ratio in the control region.  The
leading uncertainty on these ratios is the limited accuracy of the production
cross section of the gluon-initiated process, for which a full NLO calculation is
not available.  The uncertainty evaluated using renormalization and factorization
scale variations in the leading-order calculation is $26\%$ ($33\%$) in the
$\NjetEQzero$ ($1$) category~\cite{Melia:2012zg}.  An increase of the $gg{\TO}\WW$
cross section by a factor of $2$~\cite{higgscat} increases the measured $\sigmu$
value by less than $3\%$.

Boson pairs can be produced by double parton interactions (DPI) in $pp$ collisions.
The DPI contribution is very small---$0.4\%$ of $\WW$ production in the signal
regions---and is estimated using \PYTHIA8\ MC events normalized to the predicted
cross section (rather than the $\fNorm$ parameter from the $\WW$ CR).  The cross
section is computed using the NNLO $W^{\pm}$ production cross section and an effective
multi-parton interaction cross section, $\sigma_{\rm eff}{\EQ}15\mb$, measured by
ATLAS using $Wjj$ production~\cite{Aad:2013bjm}.  An uncertainty of $60\%$ is
assigned to the value of $\sigma_{\rm eff}$---and, correspondingly, to the DPI
yields---using an estimate of $\sigma_{\rm eff}{\APPROX}24\mb$ for $WW$
production~\cite{Blok:2013bpa}.  While these estimates rely on theoretical
assumptions, an increase of the DPI cross section by a factor of $10$ only increases
the measured $\sigmu$ by $1\%$.  Background from two $pp{\TO}W$ collisions in the same
bunch crossing is negligible.

In the $\NjetEQzero$ SR, the ratio of signal to $\WW$ background is about $1:5$,
magnifying the impact of background systematic uncertainties.  The definition
of the CR as a neighboring $\mll$ window reduces the uncertainty in the extrapolation
to low $\mll$.  To validate the assigned uncertainties, the CR normalization is
extrapolated to $\mll{\GT}110\GeV$ and compared to data.  The data are consistent
with the prediction at the level of $1.1$ standard deviations considering all
systematic uncertainties.

The normalization factors determined using predicted and observed event yields
are $\fNorm_{\scWW}^{0j}{\EQ}1.22 \pm{0.03}\,(\stat) \pm{0.10}\,(\syst)$ and
$\fNorm_{\scWW}^{1j}{\EQ}1.05{\PM}0.05\,(\stat){\PM}0.24\,(\syst)$, which are
consistent with the theoretical prediction at the level of approximately two
standard deviations.  Here the uncertainties on the predicted yields are included
though they do not enter into the analysis.  Other systematic uncertainties are
also suppressed in the full likelihood fit described in Sec.~\ref{sec:systematics_fit}.

\subsubsection{\boldmath\textit{\textbf{MC evaluation for $\NjetGEtwo$}}\label{sec:bkg_ww_2j}}

For the $\VBF$ and $\ggF$ $\NjetGEtwo$ analyses, the $\WW$ background
is estimated using {\SHERPA}.  The MC samples are generated as merged
multileg samples, split between the cases where final-state jets
result from QCD vertices or from electroweak vertices.
The interference between these diagrams is evaluated to be less than a
few percent using \MADGRAPH; this is included as an uncertainty on the
prediction.

For the processes with QCD vertices, uncertainties from higher orders
are computed by varying the renormalization and factorization scales
in \MADGRAPH\ and are found to be $27\%$ for the $\VBF$ category and
$19\%$ for the $\ggF$ category.  Differences between \SHERPA\ and
\MADGRAPH\ predictions after selection requirements are $8$--$14\%$
on the $\bdt$ distribution and $1$--$7\%$ on the $\mTH$ distribution,
and are taken as uncertainties.  The same procedures are used to
estimate uncertainties on processes with only electroweak vertices,
giving a normalization uncertainty of $10\%$ and an uncertainty on the
$\bdt$ ($\mTH$) distribution of $10$--$16\%$ ($5$--$17\%$).

\begin{figure}[b!]
\includegraphics[width=0.45\textwidth]{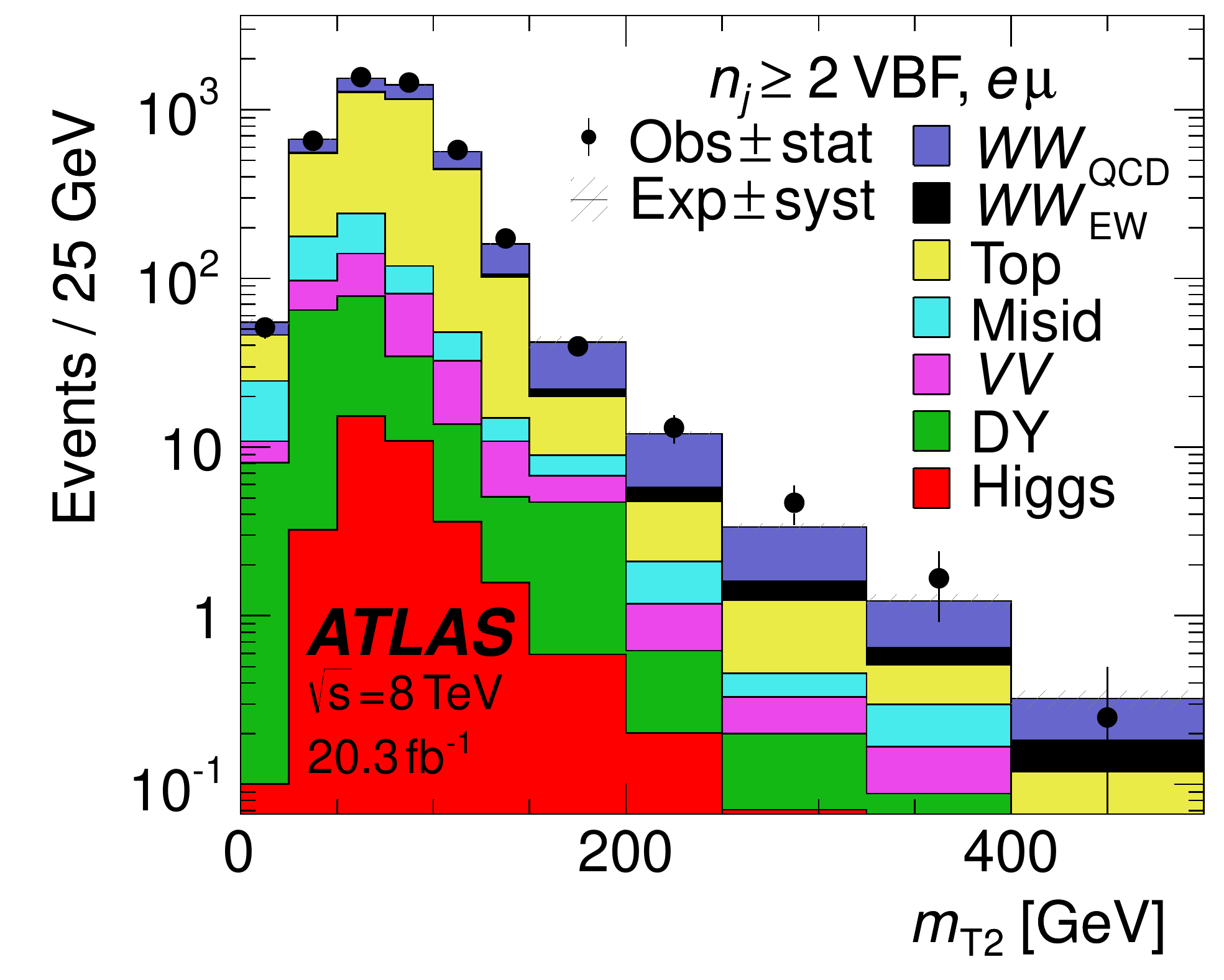}
\caption{
  $\WW$ validation region distribution of $\mTtwo$ in the $\NjetGEtwo$ VBF-enriched category.
  A requirement of $\mTtwo{\GT}160\GeV$ is used to define the validation region.
}
\label{fig:mT2}
\end{figure}

The MC prediction is validated using a kinematic selection that provides
a reasonably pure sample of $\WW{\PLUS}2$-jet events.  Events are selected
if they pass the preselection requirements on lepton $\pT$ and $\mll$,
have two jets, and $\NbjetEQzero$.  An additional requirement of
$\mTH{\GT}100\GeV$ is applied in order to enhance the $\WW$ contribution.
A final discriminant is the minimum of all possible calculations of
$\mTtwo$~\cite{Cheng:2008hk} that use the momenta of a lepton and a
neutrino, or the momenta of a lepton, a jet, and a neutrino.  The possible
momentum values of each neutrino, given $\MPTj$, are scanned in order to
calculate $\mTH$; this scan determines $\mTtwo$.  A requirement that the
minimum $\mTtwo$ be larger than $160\GeV$ provides a purity of $60\%$ for
$\WW{\PLUS}2$~jets (see Fig.~\ref{fig:mT2}).  The ratio of the observed
to the expected number of $\WW{\PLUS}2$-jet events in this region is
$1.15{\PM}0.19\,(\stat)$.

\subsection{\boldmath Top quarks \label{sec:bkg_top}}

At hadron colliders, top quarks are produced in pairs ($\ttbar$) or
in association with a $W$ boson ($Wt$) or quark(s) $q$ (single-$t$).
The leptonic decay of the $W$ bosons leads to a final state of two
leptons, \met\ and two $b$-jets (one $b$-jet) in $\ttbar$ ($Wt$)
production.  The single-$t$ production mode has only one $W$ boson in the final state
and the second, misidentified, lepton is produced by a jet.  The background from these events
is estimated together
with the $t\bar{t}$ and $Wt$ processes in spite of the different lepton production mechanism,
but the contribution from these processes to the top-quark background is small.  For example,
these events are $0.5\%$ of the top-quark background in the $\NjetEQzero$ category.
The top-quark background is estimated using the
normalization method, as described in Eq.~(\ref{eqn:est_ww}).
In the $\NjetEQzero$ category, the SR definition includes a jet veto but the CR has no jet requirements.
Because of this, the CR and the SR slightly overlap, but the $\NjetEQzero$ SR is only $3\%$ of
the CR and the expected total signal contamination is less than $1\%$, so the effect of the overlap on the results is negligible.
In the $\NjetEQone$ category, the SR definition requires $\NbjetEQzero$
but the CR has $\NbjetGEone$.
In the $\NjetEQtwo$ $\VBF$ category, the CR is defined requiring one
and only one $b$-tagged jet. Finally, in the $\NjetEQtwo$ $\ggF$ category,
to reduce the impact of $b$-tagging systematic uncertainties, the CR is
defined for $\NbjetEQzero$, and instead $\mll{\GT}80\GeV$ is applied
to remove overlap with the SR and minimize the signal contribution.

\subsubsection{\boldmath\textit{\textbf{Estimation of jet-veto efficiency for $\NjetEQzero$}}\label{sec:bkg_top_0j}}

For the $\NjetEQzero$ category, the CR  is defined after the
preselection $\met$  cut, using only the $\DFchan$ channel, with an additional requirement of
$\dphill{\LT}2.8$ to reduce
 the $\ZDY{\TO}\tautau$ background. The CR is
inclusive in the number of jets and has a purity of $74\%$ for top-quark events.
The extrapolation parameter $\fAlpha$ is the fraction of events
with zero reconstructed jets and is derived from the MC simulation.

The value of $\fAlpha$ is corrected using data in
a sample containing  at least one $b$-tagged jet. A parameter
$\fAlpha^{1b}$ is defined as the fraction of events with no additional
jets in this region. The
ratio $\big(\fAlpha^{1b}_{\data}/\fAlpha^{1b}_{\MC}\big)^2$
corrects systematic effects that have a similar impact on the $b$-tagged
  and inclusive regions, such as jet energy scale and resolution.
The square is applied to account for the presence of two jets in the
Born-level $\ttbar$ production. The prediction can be summarized as
\begin{equation}
B_{\top,0j}^{\est}
  =
  \NCR
  {\CDOT}\underbrace{\phantom{\big(}\!\!\BSR/\BCR}_{\displaystyle\fAlpha^{0j}_{\MC}}
  {\CDOT}\big(\underbrace{\fAlpha^{1b}_{\data}/\fAlpha^{1b}_{\MC}}_{\displaystyle~~\fbtag_{1b}^{\phantom{1b^{1b}}}}\big)^2
\label{eqn:est_top}
\end{equation}
where $\NCR$ is the observed yield in the CR and
$\BCR$ and $\BSR$ are the estimated yields from MC simulation in the CR and SR, respectively.

Theoretical uncertainties arise from the use of MC-simulated top-quark events in the computation of the
ratio $\fAlpha^{0j}_{\MC}/(\fAlpha^{1b}_{\MC})^2$. These uncertainties
include variations of the renormalization and factorization scales,
choice of PDFs, and the parton shower model. The procedure is sensitive to the
relative rates of $Wt$ and $\ttbar$ production, so an uncertainty is
included on this cross section ratio and on the interference between
these processes.
An additional theoretical uncertainty is evaluated on
the efficiency $\effrest$ of the additional selection after the $\NjetEQzero$ preselection, which is estimated purely from MC simulation.
Experimental uncertainties are also evaluated on the simulation-derived components
of the background estimate, with the main contributions coming from jet energy scale and resolution.
The uncertainties on $\fAlpha^{0j}_{\MC}/(\fAlpha^{1b}_{\MC})^2$ and on $\effrest$
are summarized in Table~\ref{tab:sys_alpha_top_1j}.
The resulting normalization factor is $\fNorm^{0j}_{\top}{\EQ}1.08{\PM}0.02\,({\rm stat})$, including the correction
factor $\big(\fAlpha^{1b}_{\data}/\fAlpha^{1b}_{\MC}\big)^2{\EQ}1.006$.
The total uncertainty on the background yield in the $\NjetEQzero$ signal region is $8\%$.

\subsubsection{\boldmath\textit{\textbf{Extrapolation from $\NbjetEQone$ for $\NjetEQone$}}\label{sec:bkg_top_1j}}

In the $\NjetEQone$ SR, top-quark production is the second leading background, after
nonresonant $\WW$ production.
Summing over all signal regions with no $\mTH$ requirement applied,
it is $36\%$ of the total expected background and the ratio of signal to top-quark background is approximately 0.2. It also
significantly contaminates the $\NjetEQone$ $\WW$ CR with a
 yield as large as that of nonresonant $\WW$ in this CR. Two parameters are
 defined for the extrapolation from the top CR, one to the SR
 ($\fAlpha_{\SR}$) and one to the $\WW$ CR ($\fAlpha_{\WW}$).

\begin{table}[b!]
\caption{
  Top-quark background uncertainties (in $\%$) for $\NjetLEone$.
  The uncertainties on the extrapolation procedure for $\NjetEQzero$
  are given in (a);
  the uncertainties on the extrapolation factor $\fAlpha_{\top}$ for $\NjetEQone$ are
  given in (b). The negative sign refers to the anti-correlation between
  the top-quark background predicted in the signal regions and in the
  $\WW$ CR. Only a relative sign between rows is meaningful; columns contain uncorrelated
  sources of uncertainty.
  Invariant masses are given in $\!\GeV$.
}
\label{tab:sys_alpha_top_1j}
{
\small
\centering
\begin{tabular*}{0.480\textwidth}{ llcccccc }
\dbline
\multicolumn{2}{l}{Uncertainty source}
& \multicolumn{2}{c}{$\fAlpha_{\MC}^{0j}/\big(\fAlpha^{1b}_{\MC}\big)^2$}
& \multicolumn{2}{l}{\quad\,$\effrest$}
& \multicolumn{1}{c}{Total}
\\
\sgline
\multicolumn{5}{l}{$\!$(a) $\NjetEQzero$} \\
\clineskip\clineskip
\multicolumn{2}{l}{Experimental                  }&\multicolumn{2}{c}{$4.4$}&\multicolumn{2}{l}{\quad~$1.2$}&\multicolumn{1}{c}{$4.6$}\\
\multicolumn{2}{l}{Non-top-quark subtraction$\nq$}&\multicolumn{2}{c}{  -  }&\multicolumn{2}{l}{\quad~~\,-} &\multicolumn{1}{c}{$2.7$}\\
\multicolumn{2}{l}{Theoretical                   }&\multicolumn{2}{c}{$3.9$}&\multicolumn{2}{l}{\quad~$4.5$}&\multicolumn{1}{c}{$4.9$}\\
\multicolumn{2}{l}{Statistical                   }&\multicolumn{2}{c}{$2.2$}&\multicolumn{2}{l}{\quad~$0.7$}&\multicolumn{1}{c}{$2.3$}\\
\multicolumn{2}{l}{Total                         }&\multicolumn{2}{c}{$6.8$}&\multicolumn{2}{l}{\quad~$4.7$}&\multicolumn{1}{c}{$7.6$}\\
\clineskip\clineskip
\dbline
\multicolumn{2}{l}{Regions}
& \multicolumn{1}{p{0.050\textwidth}}{~~Scale}
& \multicolumn{1}{p{0.055\textwidth}}{~~PDF}
& \multicolumn{1}{p{0.045\textwidth}}{~\,Gen}
& \multicolumn{1}{p{0.045\textwidth}}{$\!\!$UE/PS}
& \multicolumn{1}{p{0.040\textwidth}}{~\,Tot}
\\
\sgline
\multicolumn{8}{l}{$\!$(b) $\NjetEQone$.  See the caption of Table~\ref{tab:ww_alpha} for column headings.} \\
\clineskip\clineskip
\multicolumn{4}{l}{Signal region} \\
\quad $\DFchan$ &\multicolumn{1}{c}{($10{\LT}\mll{\LT}55$)}&$-1.1$     &$-0.12$ &$-2.4$ &$2.4$ &$3.6$ \\
\quad $\SFchan$ &\multicolumn{1}{c}{($12{\LT}\mll{\LT}55$)}&$-1.0$      &$-0.12$ &$-2.0$ &$3.0$ &$3.7$ \\
\clineskip\clineskip
\multicolumn{4}{l}{$\WW$ control region} \\
\quad $\DFchan$ &\multicolumn{1}{c}{($\mll{\GT}80$)       }&$\phantom{-}0.6$ &$\phantom{-}0.08$ &$\phantom{-}2.0$&$1.8$  &$2.8$ \\
\dbline
\end{tabular*}
}
\end{table}

The top CR is defined after the preselection in the $\DFchan$ channel
and requires the presence of exactly one jet, which must be
$b$-tagged.  There can be no additional $b$-tagged jet with $20{\LT}\pT{\LT}25\GeV$, following the SR requirement.
The requirement $\mTlep{\GT}50\GeV$ is also applied to reject $\jj$ background.
As in the $\WW$ case, only the $\emu$ events are used in order to suppress the $\ZDY$ contamination.
The $\mTH$ distribution in this control region is shown in Fig.~\ref{fig:cr_top_1j}(a).

\begin{figure}[t!]
\includegraphics[width=0.45\textwidth]{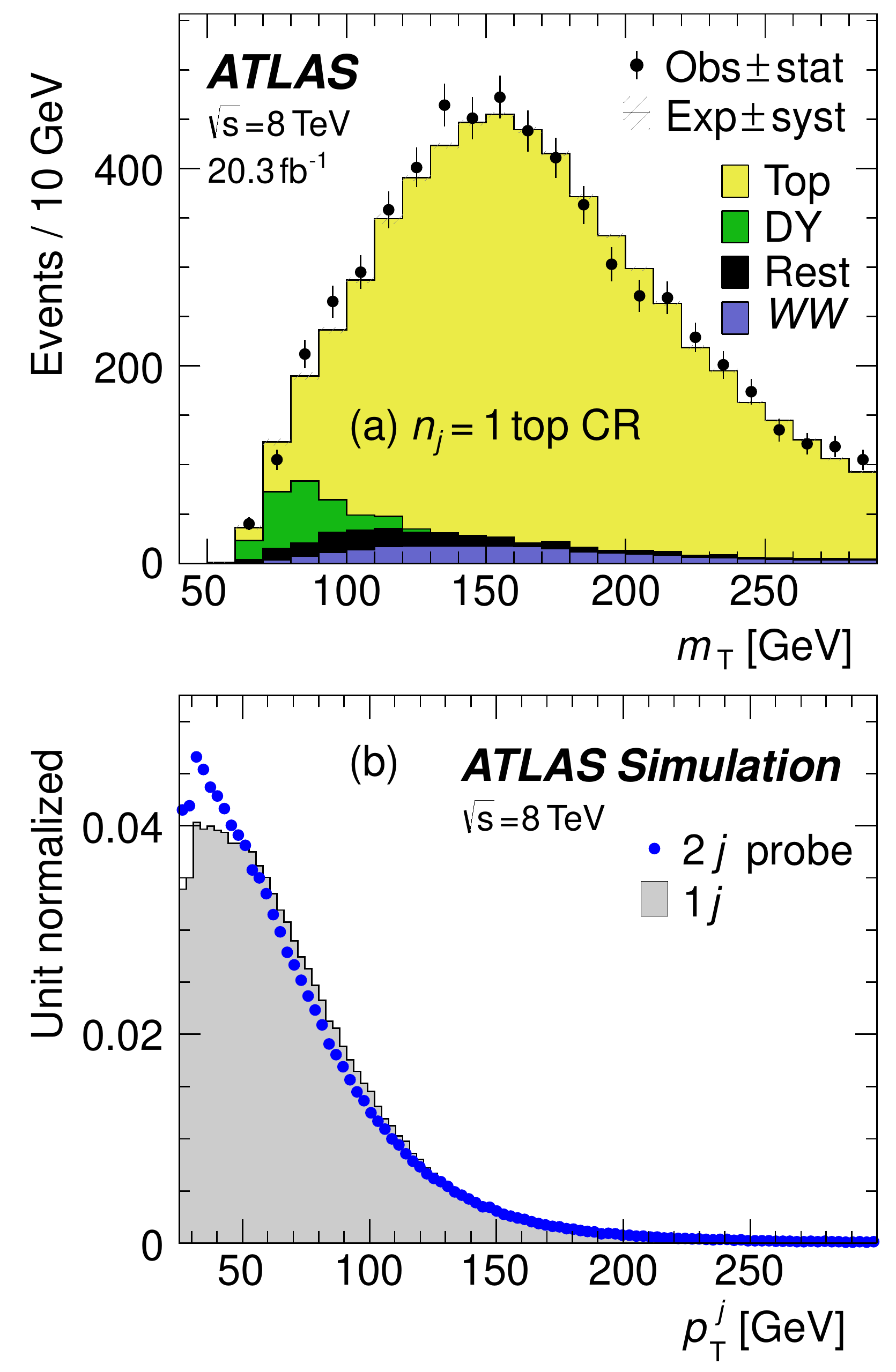}
\caption{
  Top-quark control region (CR) distributions of (a) transverse mass and (b) jet $\pT$.
   The $\mTH$ plot in (a) scales the top-quark contributions with the normalization factor $\fNorm_{\top}$.
   The $\pTj$ plot in (b) compares the jet $\pT$ distribution in top-quark MC--both the $\ttbar$ and the $Wt$
   processes--in $\NjetEQtwo$ ($2j$ probe) events to $\NjetEQone$ ($1j$) events.  For each $\NjetEQtwo$ event,
   one of the two jets is chosen randomly and the $\pT$ of that jet enters the distribution if the other jet is tagged.
   \HwwPlotDetail{See}.
}
\label{fig:cr_top_1j}
\end{figure}

The CR requires at least one $b$-jet, but the SR requires zero.
In the case of a simple extrapolation using the ratio of the predicted
yields in the signal and control regions, the impact of the $b$-tagging efficiency
uncertainty on the measurement is substantial. A systematic uncertainty of $5\%$ on
the $b$-tagging efficiency would induce an uncertainty of about $20\%$ on the
estimated yield in the SR. In order to reduce this effect, the
$b$-tagging efficiency $\effbest$ is estimated from data.  The efficiency $\effbtwoj$
is the probability to tag an individual jet, measured
in a sample selected similarly to the SR but containing exactly two jets, at
least one of which is $b$-tagged.  It can be measured in data and MC simulation, because
a high-purity top sample can be selected.
Most of the events in this sample are $t\bar{t}$ events with reconstructed jets from
$b$-quarks, although there is some contamination from light-quark jets from
initial state radiation when a $b$-quark does not produce a reconstructed
jet. Similarly, $\effbonej$ is the efficiency to tag a jet in a sample with one jet,
in events passing the signal region selection.

The efficiency measurement $\effbtwojdata$ is extrapolated from the
$\NjetEQtwo$ sample to the $\NjetEQone$ samples using
$\fcor{\EQ}\effbonej/\effbtwoj$, which is evaluated using MC simulation.
Jets in the $\NjetEQtwo$ and jets in the $\NjetEQone$ samples have similar kinematic features;
one example, the jet $\pT$, is illustrated in Fig.~\ref{fig:cr_top_1j}(b).  In this figure, the $\NjetEQtwo$
distribution contains the $\pT$ of one of the two jets, chosen at random, provided that the other jet is
tagged, so that the distribution contains the same set of jets as is used in the extrapolation to $\NjetEQone$.
Residual disagreements between the distributions are
reflected in the deviation of $\fcor$ from unity, which is small.
The value of $\fcor$ is $1.079{\PM}0.002\,({\rm stat})$
with an experimental uncertainty of $1.4\%$ and
a theoretical uncertainty of $0.8\%$.
The experimental uncertainty is almost entirely due to uncertainties on the $b$-tagging efficiency.
The theoretical uncertainty is due to the PDF model, renormalization and factorization scales, matching of the
matrix element to the parton shower, top-quark cross sections,
and interference between top-quark single and pair production.

The estimated $b$-tagging efficiency in the $\NjetEQone$ data is
$\effbest{\EQ}\fcor{\CDOT}\effbtwojdata$ and the top-quark background estimate in the SR is then:
\begin{equation}
B_{\top,1j}^{\est}
  =
  \NCR
  \cdot
  \underbrace{\bigg(\frac{1-\effbest}{\effbest}\bigg)}_{\displaystyle\fAlpha^{1j}_{\data}}
\label{eqn:est_jbee}
\end{equation}
The theoretical systematic uncertainties are
summarized in Table \ref{tab:sys_alpha_top_1j}.
The normalization factor for this background is $\fNorm^{1j}_{\top}{\EQ}1.06{\PM}0.03\,({\rm stat})$, and the total
uncertainty on the estimated background in the $\NjetEQone$ signal region is $5\%$.

\subsubsection{\boldmath\textit{\textbf{Extrapolation from $\NbjetEQone$ for VBF-enriched $\NjetGEtwo$}}\label{sec:bkg_top_vbf}}

The $\NjetGEtwo$ categories have a large contribution from top-quark background events even after
selection requirements, such as the $b$-jet veto, that are applied to reduce them,
because of the two $b$ quarks in $\ttbar$ events.
The majority of the residual top-quark events have a light-quark jet from initial-state radiation and
a $b$-quark jet that is not identified by the $b$-tagging algorithm. The CR requires
exactly one $b$-tagged jet to mimic this topology, so that at first order the CR-to-SR
extrapolation factor ($\fAlpha$) is the ratio of $b$-jet efficiency to $b$-jet inefficiency.
The CR includes events from $\DFchan$ and $\SFchan$ final states because the $\ZDY$ contamination
is reduced by the jet selection.

The $\bdt$ discriminant is based on variables, such as $\mjj$, that depend on the jet kinematics,
so the acceptance for top-quark events in each $\bdt$ bin is strongly dependent on the Monte Carlo generator and modeling.
Motivated by the large variation of top-quark event kinematics as a function of the $\bdt$ bin, the
top-quark background is normalized independently in each bin, which reduces the modeling uncertainties.
Figure \ref{fig:mjj_topCR} shows the $\mjj$ and $\bdt$ distributions
in the top CR used for the VBF category. The two bins with the highest $\bdt$ score are merged to
improve the statistical uncertainty on the estimated background.
The uncertainties on the extrapolation from the single bin in the CR to
the two bins in the SR are separately evaluated.

\begin{figure}[b!]
\includegraphics[width=0.40\textwidth]{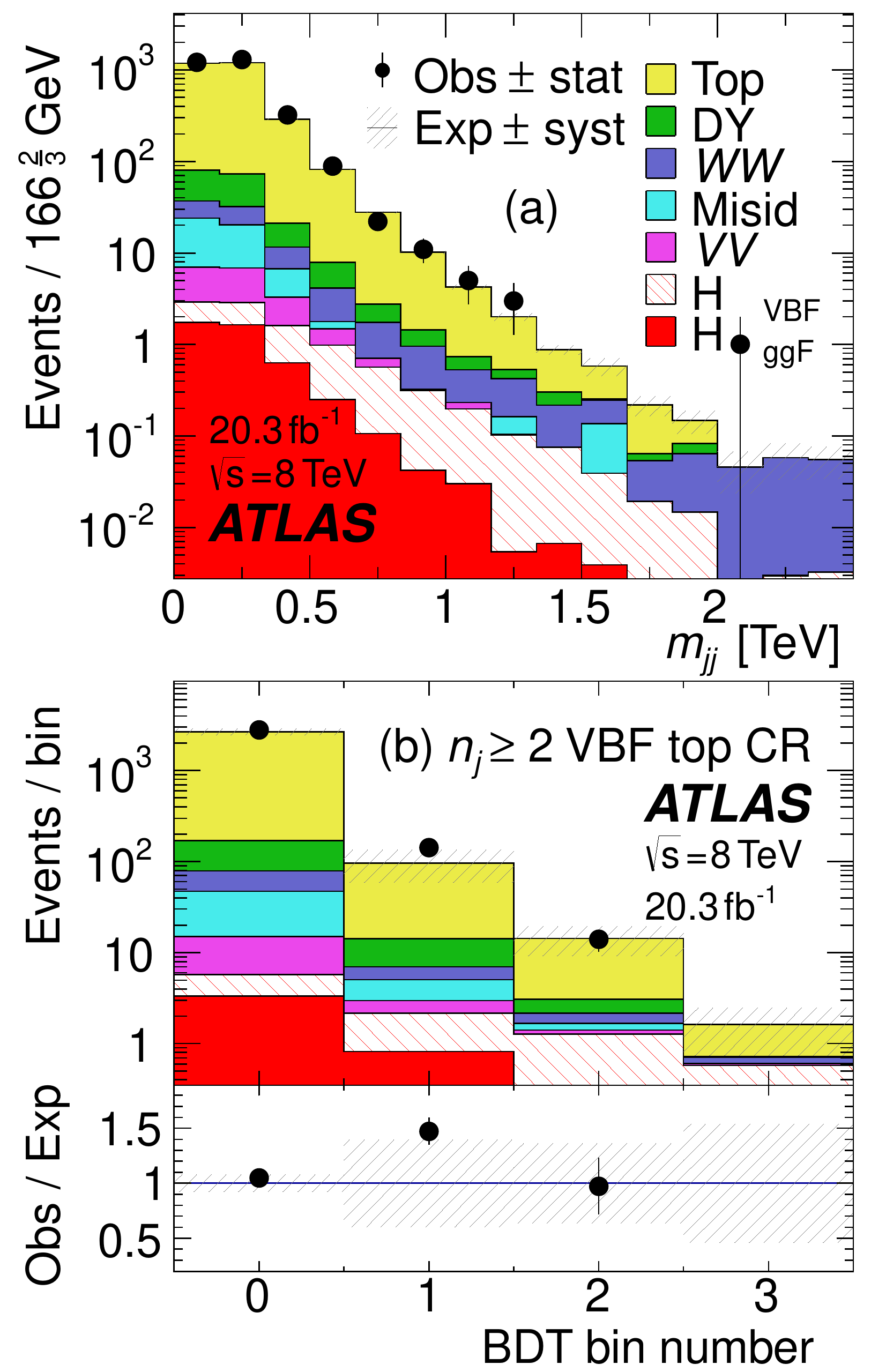}
\caption{
  Top-quark control region (CR) distributions in the VBF-enriched $\NjetGEtwo$ category for:
  (a) $\mjj$ and (b) BDT output.
  For the plot in (b) the shaded band in the ratio shows the uncertainty on the normalization of each bin.
  No events are observed in bin 3.
  \HwwPlotDetail{See}.
}
\label{fig:mjj_topCR}
\label{fig:BDT_topCR}
\end{figure}

Table~\ref{tab:top_unc_vbf} shows the normalization factors $\fNorm_i$ and their uncertainties for
each $\bdt$ bin, as well as the theoretical uncertainties on the extrapolation factors $\fAlpha_j$ to
the corresponding SR bins.

\begin{table}[t!]
\caption{
  Top-quark background uncertainties (in $\%$) for $\NjetGEtwo$ VBF on
  the extrapolation factor $\fAlpha$ and normalization factor $\fNorm$.
  The contributions are given in bins of $\bdt$.
  The systematic uncertainty on $\fNorm$ does not affect the measurement,
  but is shown to illustrate the compatibility of the normalization factor with unity.
  The values of $\fNorm$ are also shown; bins 2 and 3 use a common value of $\fNorm$.
  Bin $0$ is unused, but noted for completeness.
}
\label{tab:top_unc_vbf}
{\small
  \centering
\begin{tabular*}{0.480\textwidth}{
    p{0.140\textwidth}
    ccccc
}
\dbline
$\bdt$ bins
& \multicolumn{1}{c}{$\Delta\fAlpha/\fAlpha$}
& \multicolumn{1}{c}{$\Delta\fNorm$ statistical}
& \multicolumn{1}{c}{$\Delta\fNorm$ systematic}
& \multicolumn{1}{c}{$\fNorm$}
\\
\sgline
SR bin 0 (unused) & 0.04 & 0.02 & 0.05 & 1.09 \\
SR bin 1          & 0.10 & 0.15 & 0.55 & 1.58 \\
SR bin 2          & 0.12 & 0.31 & 0.36 & 0.95 \\
SR bin 3          & 0.21 & 0.31 & 0.36 & 0.95 \\
\dbline
\end{tabular*}
}
\end{table}

The uncertainties on $\fAlpha$ were evaluated with the same procedure used for the $\WW$ background
(see Sec.~\ref{sec:bkg_ww_01j}). The only significant source is a modeling uncertainty evaluated by
taking the maximum spread of predictions from {\POWHEG}+{\HERWIG}, {\ALPGEN}+{\HERWIG} and {\MCATNLO}+{\HERWIG}~\cite{Frixione:2002ik}.
The generators are distinguished by the merging of LO matrix-element evaluations of up to three jets produced in
association with $\ttbar$ ({\ALPGEN}+{\HERWIG}) or by differences in procedures for matching a NLO matrix-element
calculation to the parton shower ({\MCATNLO}+{\HERWIG} and {\POWHEG}+{\HERWIG}). The systematic uncertainty is
dominated by the {\ALPGEN}+{\HERWIG}--{\MCATNLO}+{\HERWIG} difference, and for this reason the theoretical
uncertainties shown in Table~\ref{tab:top_unc_vbf} are fully correlated between bins.

\subsubsection{\boldmath\textit{\textbf{Extrapolation in $\mll$ for ggF-enriched $\NjetGEtwo$}}\label{sec:bkg_top_2j}}

In the more inclusive phase space of the $\ggF$-enriched $\NjetGEtwo$ category, the $\ttbar$ background remains dominant
after the $\NbjetEQzero$ requirement, as is the case for the $\VBF$-enriched category. The CR is defined with $\mll{\GT}80\GeV$
to distinguish it from the signal region (see Fig.~\ref{fig:ggF2j}) and reduce signal contamination. The CR is approximately
$70\%$ pure in top-quark events,
and a normalization factor of $\fNorm{\EQ}1.05{\PM}0.03\,(\stat)$ is obtained. The uncertainties on the extrapolation factor
$\fAlpha$ to the SR are $3.2\%$ from the comparison of {\MCATNLO}+{\HERWIG}, {\ALPGEN}+{\HERWIG}, and {\POWHEG}+{\PYTHIA};
$1.2\%$ for the parton shower and underlying-event uncertainties from the comparison of {\POWHEG}+{\PYTHIA6} and {\POWHEG}+{\HERWIG};
$1\%$ from the missing higher-order contribution, evaluated by varying the renormalization and factorization scales;
$0.3\%$ from the PDF envelope evaluated as described in Sec.~\ref{sec:bkg_ww_01j}; and $0.7\%$ from the experimental uncertainties.
The effect of the same set of variations on the predicted $\mTH$ distribution in the signal region was also
checked.  The variations from the nominal distribution are small, at most $4\%$ in the tails, but they are included as a shape
systematic in the fit procedure.

\subsection{\boldmath Misidentified leptons \label{sec:bkg_misid}}

Collisions producing $W$~bosons in association with one or more
jets---referred to here as $W$+jets---may enter the signal sample when a jet is
misidentified as a prompt lepton.
In this background, there is a prompt lepton and a transverse
momentum imbalance from the leptonic decay of the $W$~boson.
Background can also arise from multijet production when two jets
are misidentified as prompt leptons and a
transverse momentum imbalance is reconstructed.

\subsubsection{\boldmath\textit{\textbf{$\Wjets$}}\label{sec:bkg_misid_wj}}

The $\Wjets$ background contribution is estimated using a control sample
of events where one of the two lepton candidates satisfies the identification
and isolation criteria for the signal sample, and the other lepton
fails to meet these criteria but satisfies less restrictive criteria
(these lepton candidates are denoted ``anti-identified'').
Events in this sample are otherwise required to satisfy all of the
signal selection requirements.
The dominant component of this sample ($85\%$ to $90\%$)
is due to $\Wjets$ events in which
a jet produces an object reconstructed as a lepton.
This object may be either a nonprompt lepton from the decay of a
hadron containing a heavy quark, or else
a particle (or particles) from a jet reconstructed as a lepton candidate.

\begin{figure*}[!tb]
\includegraphics[width=0.75\textwidth]{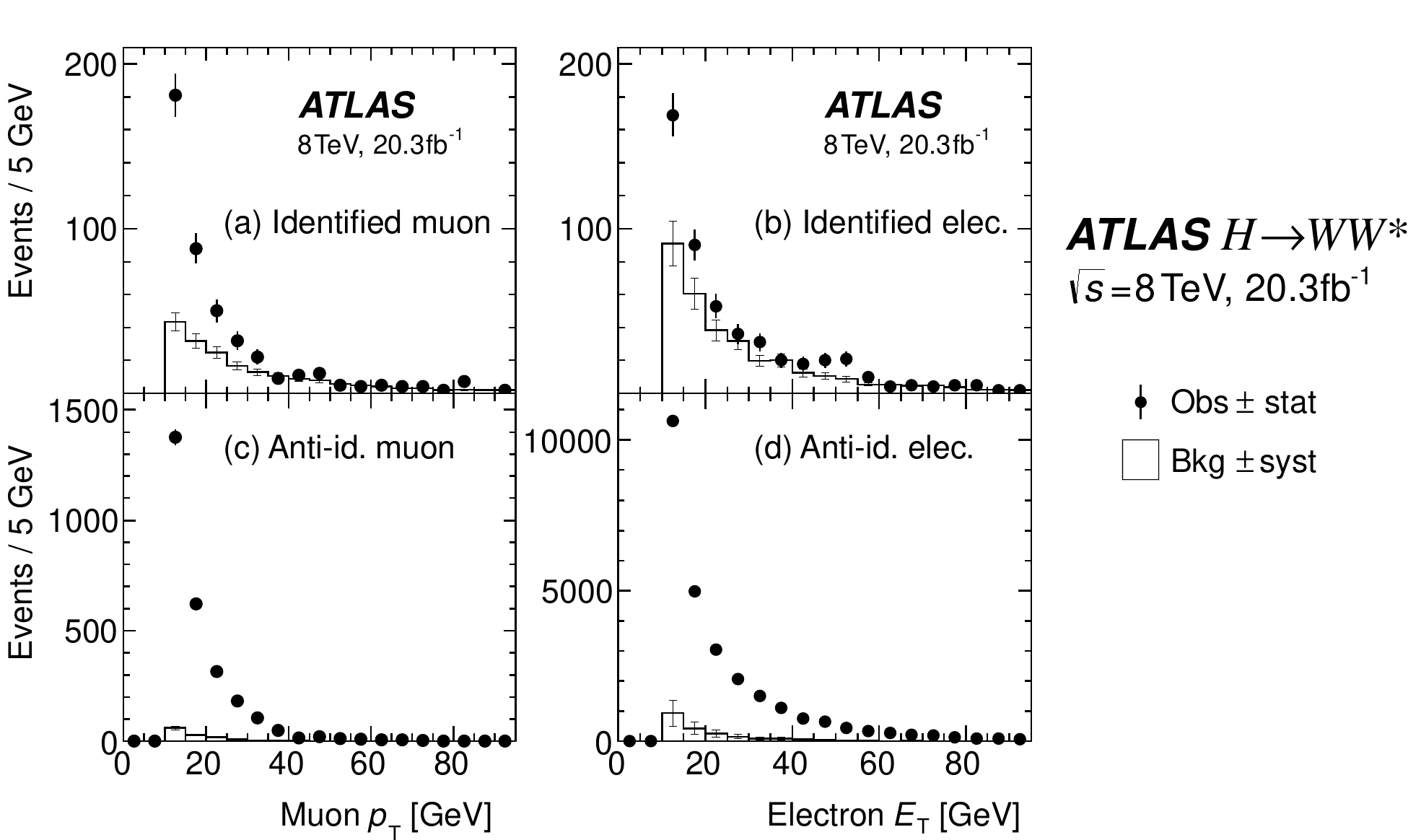}
\caption{
  Misidentified lepton sample distributions of $\pT$
  in the $\Zjets$ control sample:
  (a) identified muon,
  (b) identified electron,
  (c) anti-identified muon, and
  (d) anti-identified electron.
  The symbols represent the data (Obs);
  the histograms are the background MC estimates (Bkg) of the sum of electroweak processes
  other than the associated production of a $Z$~boson and jets.
}
\label{fig:fake_distributions}
\end{figure*}

The $\Wjets$ contamination
in the signal region is obtained by scaling the number
of events in the data control sample by an extrapolation factor.
This extrapolation factor is measured in a data sample of jets produced in
association with $Z$~bosons reconstructed in either the
$e^+e^-$ or $\mu^+\mu^-$ final state
(referred to as the $\Zjets$ control sample below).
The factor is the ratio of the number of identified lepton
candidates satisfying all lepton selection criteria to the number of
anti-identified leptons measured in bins of anti-identified lepton
$\pT$ and $\myeta$.
Anti-identified leptons must explicitly not satisfy the signal selection
criteria (so that leptons counted in the numerator of this ratio exclude
the anti-identified leptons counted in the denominator of this ratio)
and the signal requirements for isolation and track impact parameters
are either relaxed or removed.
In addition, for anti-identified electrons
the identification criteria specifically targeting conversions are
removed and the anti-identified electron is explicitly required to
fail the ``medium'' electron identification
requirement specified in Ref.~\cite{ElectronEff2012}.

Figure~\ref{fig:fake_distributions} shows the $\pT$ distributions of
identified muons          [Fig.~\ref{fig:fake_distributions}(a)],
identified electrons      [Fig.~\ref{fig:fake_distributions}(b)],
anti-identified muons     [Fig.~\ref{fig:fake_distributions}(c)], and
anti-identified electrons [Fig.~\ref{fig:fake_distributions}(d)] in
the $\Zjets$ control sample.
The extrapolation factor in a given $\pT$ bin is the number of
identified leptons divided by the number of anti-identified leptons
in that particular bin.
Each number is corrected for the presence of processes not due to
$Z$+jets.
The $\Zjets$ sample is contaminated by other production processes that
produce additional prompt leptons (\eg, $\WZ{\TO}\ell\nu\ell\ell$)
or nonprompt leptons not originating from jets
(\eg, $\ZDY$ and $\Zg$) that
create a bias in the extrapolation factor.
Kinematic criteria suppress about $80\%$ of
the contribution from these other processes in the $\Zjets$ sample.
The remaining total contribution of these other processes after applying
these kinematic criteria is shown
in the histograms in Fig.~\ref{fig:fake_distributions}.
The uncertainty shown in these histograms is the $10\%$ systematic
uncertainty assigned to the contribution from these other processes,
mainly due to cross section uncertainties.
This remaining contribution from other processes is estimated
using Monte Carlo simulation and removed from the event yields before
calculating the extrapolation factor.

The composition of the associated jets---namely the fractions of jets due
to the production of heavy-flavor quarks, light-flavor quarks and
gluons---in the $\Zjets$ sample and the $\Wjets$ sample may be different.
Any difference would lead to a systematic error in the estimate of the
$\Wjets$ background due to applying the extrapolation factor determined with
the $\Zjets$ sample to the $\Wjets$ control sample,
so Monte Carlo simulation is used to determine a correction factor
that is applied to the extrapolation factors determined with the
$\Zjets$ data sample.
A comparison of the extrapolation factors determined with the
$\Zjets$ sample and the $\Wjets$ sample is made for three
Monte Carlo simulations:
{\ALPGEN}+{\PYTHIA6}, {\ALPGEN}+{\HERWIG} and {\POWHEG}+{\PYTHIA8}.
For each combination of matrix-element and parton-shower simulations,
a ratio of the extrapolation factors for $\Wjets$ versus $\Zjets$ is calculated.
These three ratios are used to determine a correction factor and an
uncertainty that is applied to the extrapolation factors determined
with the $\Zjets$ data sample:
this correction factor is $0.99{\PM}0.20$
for anti-identified electrons and
$1.00{\PM}0.22$
for anti-identified muons.

The total uncertainties on the corrected extrapolation factors are
summarized in Table~\ref{tab:wj_ff}.
In addition to the systematic uncertainty on the correction factor
due to the sample composition, the other important uncertainties on the
$\Zjets$ extrapolation factor are due to the limited
number of jets that meet the lepton selection criteria in the
$\Zjets$ control sample and the uncertainties on the contributions
from other physics processes in
the identified and anti-identified lepton samples.
The total systematic uncertainty on the corrected
extrapolation factors varies as a
function of the $\pT$ of the anti-identified lepton;
this variation is from $29\%$ to $61\%$ for anti-identified electrons
and $25\%$ to $46\%$ for anti-identified muons.
The systematic uncertainty on the corrected extrapolation factor dominates
the systematic uncertainty on the $\Wjets$ background.

\begin{table}[!b]
\caption{
  Uncertainties (in $\%$) on the extrapolation factor $\fAlpha_{\fakes}$
  for the determination of the $\Wjets$ background.
  Total is the quadrature sum of the uncertainties due to the correction
  factor determined with MC simulation (Corr.\ factor), the number of jets misidentified as
  leptons in the
  $\Zjets$ control sample (Stat)\ and the subtraction of other processes (Other bkg).
  As described in the text, Corr.\ factor is classified as theoretical and the rest as experimental.
  OC (SC) refers to the uncertainties in the opposite-charge (same-charge) $\Wjets$ CR.
}
\label{tab:wj_ff}
{\small
\begin{tabular*}{0.480\textwidth}{ l cc p{0.005\textwidth} c c cc }
\dbline
\multicolumn{1}{p{0.120\textwidth}}{\multirow{2}{*}{SR $\pT$ range}}
& \multicolumn{2}{p{0.090\textwidth}}{~~~Total}
&
& \multicolumn{2}{p{0.090\textwidth}}{$\!$Corr.\,factor}
& \multicolumn{1}{p{0.045\textwidth}}{\multirow{2}{*}{~Stat$\nq$}}
& \multicolumn{1}{p{0.080\textwidth}}{\multirow{2}{*}{Other\,bkg}}
\\
& OC
& SC
&
& OC
& SC
&
&
\\
\sgline
Electrons \\
\quad $10$--$15\GeV$  & 29 & 32 && 20 & 25 & 18 & 11           \\
\quad $15$--$20\GeV$  & 44 & 46 && 20 & 25 & 34 & 19           \\
\quad $20$--$25\GeV$  & 61 & 63 && 20 & 25 & 52 & 25           \\
\quad ${\ge\,}25\GeV$ & 43 & 45 && 20 & 25 & 30 & 23           \\
\clineskip\clineskip
Muons \\
\quad $10$--$15\GeV$  & 25 & 37 && 22 & 35 & 10 & \phantom{0}3 \\
\quad $15$--$20\GeV$  & 37 & 46 && 22 & 35 & 18 & \phantom{0}5 \\
\quad $20$--$25\GeV$  & 37 & 46 && 22 & 35 & 29 & \phantom{0}9 \\
\quad ${\ge\,}25\GeV$ & 46 & 53 && 22 & 35 & 34 & 21           \\
\dbline
\end{tabular*}
}
\end{table}

The uncertainties on the signal strength $\sigmu$ are classified into
experimental, theoretical, and other components,
as described in Sec.~\ref{sec:results} and  Table~\ref{tab:syst_mu}.
The uncertainty on $\sigmu$ due to the correction factor applied to the extrapolation
factor is classified as theoretical because the uncertainty on the correction factor
is derived from a comparison of predictions from different combinations of
Monte Carlo generators and parton shower algorithms.
The uncertainty on $\sigmu$ due to the other uncertainties on the extrapolation factor
($Z$+jet control sample statistics and the subtraction of other processes from this
control sample)  is classified as experimental.

Figure~\ref{fig:extrapolation_factors} shows the extrapolation factor
measured in the $\Zjets$ data compared to the predicted extrapolation
factor determined using Monte Carlo simulated samples
({\ALPGEN}+{\PYTHIA6}) of $\Zjets$ and $\Wjets$ for
anti-identified muons [Fig.~\ref{fig:extrapolation_factors}(a)] and
anti-identified electrons [Fig.~\ref{fig:extrapolation_factors}(b)].
The values of the extrapolation factors are related to the
specific criteria used to select
the anti-identified leptons and, as a result, the extrapolation factor
for anti-identified muons is about one order of magnitude larger than the
extrapolation factor for anti-identified electrons.
This larger extrapolation factor does not indicate a
larger probability for a jet to be misidentified as a muon compared
to an electron.
In fact, misidentified electrons contribute a larger portion of the
$\Wjets$ background in the signal region.

\begin{figure}[!bt]
\includegraphics[width=0.40\textwidth]{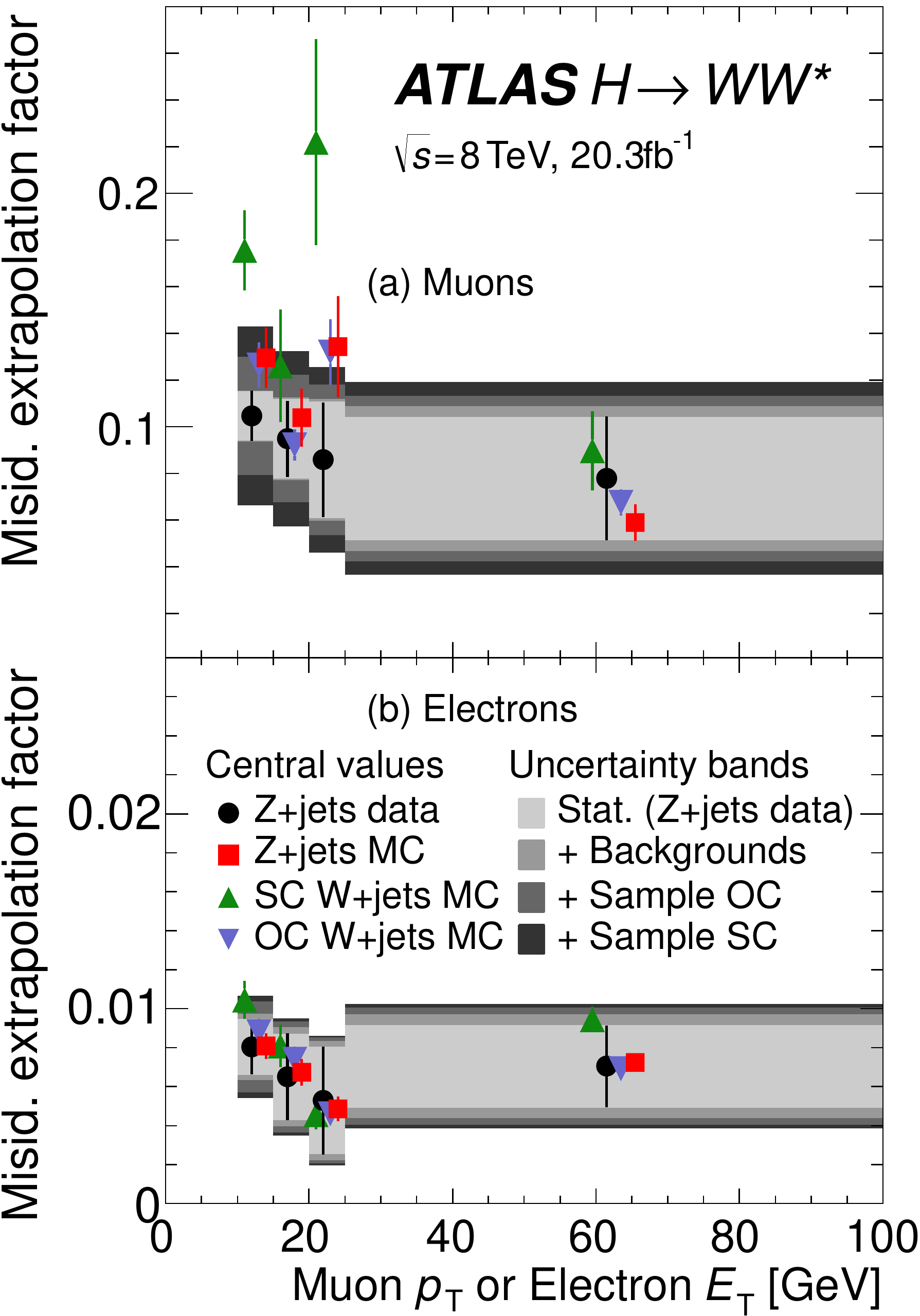}
\caption{
  Misidentified lepton extrapolation factors, $\fAlpha_{\fakes}$,
  for anti-identified (a) muons and (b) electrons before applying the
  correction factor described in the text.
  The symbols represent the central values of
  the $\Zjets$ data and
  the three {\ALPGEN}+{\PYTHIA6} MC samples:
  $\Zjets$, opposite-charge (OC) $\Wjets$, and same-charge (SC) $\Wjets$.
  The bands represent the uncertainties:
  Stat~refers to
  the statistical component, which is dominated by the number of jets
  identified as leptons in $\Zjets$ data;
  Background is due to the subtraction
  of other electroweak processes present in $\Zjets$ data;
  and Sample is due to the
  variation of the $\fAlpha_{\fakes}$ ratios
  in $\Zjets$ to OC $\Wjets$ or to SC $\Wjets$ in the three MC samples.
  The symbols are offset from each other for presentation.
}
\label{fig:extrapolation_factors}
\end{figure}

The $\Wjets$ background in the signal region is determined using a
control sample in which the lepton and the anti-identified lepton
are required to have opposite charge.
A prediction of the $\Wjets$ background is also used for
a data control sample consisting of events that satisfy all of the
Higgs~boson signal requirements {\it except} that the two lepton candidates
are required to have the same charge.
This same-charge control region is described in Sec.~\ref{sec:bkg_vv}.

The $\Wjets$ process is not expected to produce equal numbers of same-charge
and opposite-charge candidates.
In particular, associated production processes such as $Wc$, where the
second lepton comes from the semileptonic decay of a charmed hadron,
produce predominantly opposite-charge candidates.
Therefore,
a separate extrapolation factor is applied to the same-charge $W$+jets
control sample.

The procedure used to determine the same-charge extrapolation factor
from the $\Zjets$ data is the same as the one used for the signal region.
Because of the difference in jet composition of the same-charge $W$+jets
control sample, a different correction factor is derived from MC
simulation to correct the extrapolation factor determined with the $Z$+jets
data sample for application to the same-charge $\Wjets$ sample.
Figure~\ref{fig:extrapolation_factors} compares
the extrapolation factors in same-charge $\Wjets$ with the ones in $Z$+jets.
The correction factor is $1.25{\PM}0.31$
for anti-identified electrons and
$1.40{\PM}0.49$
for anti-identified muons;
as with the opposite-charge correction factors, these factors and their
systematic uncertainty are determined by comparing the factors
determined with the three different samples of MC simulations
mentioned previously in the text
({\ALPGEN}+{\PYTHIA6}, {\ALPGEN}+{\HERWIG} and {\POWHEG}+{\PYTHIA8}).
The total uncertainties on the corrected extrapolation factors used
to estimate the $\Wjets$ background in the same-charge control region
are shown in Table~\ref{tab:wj_ff}.
The correlation between the systematic uncertainties on the opposite-charge and same-charge correction factors
reflects the composition of the jets producing objects misidentified as leptons.
These jets have a component that is charge-symmetric with respect to
the charge of the $W$~boson as well as a component unique to opposite-charge $\Wjets$ processes.
Based on the relative rates of same- and opposite-charge $\Wjets$ events,
$60\%$ of the opposite-charge correction factor uncertainty is correlated with $100\%$
of the corresponding same-charge uncertainty.

\subsubsection{\boldmath\textit{\textbf{Multijets}}\label{sec:bkg_misid_jj}}

The background in the signal region due to multijets is determined using
a control sample that has two anti-identified lepton candidates, but
otherwise satisfies all of the signal region selection requirements.
A separate extrapolation factor---using a multijet sample---is measured for
the multijet background and applied twice to this control sample.
The sample used to determine the extrapolation factor is expected to have
a similar sample composition (in terms of heavy-flavor jets, light-quark
jets and gluon jets) to the control sample.
Since the presence of a misidentified lepton in a multijet sample
influences the sample composition---for example by increasing the fraction of
heavy-flavor processes in the multijet sample---corrections to the extrapolation factor
are made that take into account this correlation.
The event-by-event corrections vary between $1.0$ and $4.5$
depending on the lepton flavor
and $\pT$  of both misidentified leptons in the event;
the electron extrapolation factor corrections are larger than
the muon extrapolation factor corrections.

\begin{table}[b!]
\caption{
  $\Wjets$ and multijets estimated yields in the $\DFchan$ category.
  For $\NjetEQzeroone$, yields for both the opposite-charge (OC) and same-charge (SC) leptons are given.
  The yields are given before the $\mTH$ fit for the ggF-enriched categories
  and after the VBF-selection for the VBF-enriched categories.
  The uncertainties are from a combination of statistical and systematic sources.
}\label{tab:wj_evts}
{
\small
\begin{tabular*}{0.480\textwidth}{
  p{0.110\textwidth} r@{$\PM$}l
  p{0.015\textwidth} r@{$\PM$}l
  p{0.005\textwidth} r@{$\PM$}l
  p{0.015\textwidth} r@{$\PM$}l
}
\dbline
\multirow{2}{*}{Category}
                  & \multicolumn{5}{c}{$\Wjets$ yield $\NWj$}
                 && \multicolumn{5}{c}{Multijets yield $\Njj$} \\
                  & \multicolumn{2}{c}{OC}
                  &
                  & \multicolumn{2}{c}{SC}
                  &
                  & \multicolumn{2}{c}{OC}
                  &
                  & \multicolumn{2}{c}{SC}
                  \\
\sgline
$\NjetEQzero$     & 278 & 71   && 174 & 54                  && 9.2   & 4.2  && 5.5 & 2.5             \\
$\NjetEQone$      & 88  & 22   && 62  & 18                  && 6.1   & 2.7  && 3.0 & 1.3             \\
$\NjetGEtwo$ ggF  & 50  & 22   && \multicolumn{2}{c}{~~\,-} && 49    & 22   && \multicolumn{2}{c}{-} \\
$\NjetGEtwo$ VBF  & 3.7 & 1.2  && \multicolumn{2}{c}{~~\,-} && 2.1   & 0.8  && \multicolumn{2}{c}{-} \\
\dbline
\end{tabular*}
}
\end{table}

\subsubsection{\boldmath\textit{\textbf{Summary}}\label{sec:bkg_misid_summary}}

Table~\ref{tab:wj_evts} lists
the estimated event counts for the multijet and $\Wjets$ backgrounds
in the $\DFchan$ channel for the various jet multiplicities.
The values are given before the $\mTH$ fit for the ggF-enriched categories and
after the VBF-selection for the VBF-enriched categories.
The uncertainties are the combination of the statistical and systematic
uncertainties and are predominantly systematic.
The dominant systematic uncertainty is from the extrapolation factors.
In the case of the $\Wjets$ background, these uncertainties are summarized
in Table~\ref{tab:wj_ff}; in the case of the multijet background, the
largest contribution is the uncertainty introduced by the correlations
between extrapolation factors in an event with two misidentified leptons.

For the $\NjetEQzero$ and $\NjetEQone$ categories, the expected backgrounds
are provided for both the opposite-charge signal region and the same-charge
control region (described in Sec.~\ref{sec:bkg_vv}),
and the multijet background is expected to be less than
$10\%$ of the $\Wjets$ background in these two categories.
For higher jet multiplicities, the multijet background is expected to be
comparable to the $\Wjets$ background because there is no selection criterion
applied to $\mTlep$.
In this case, however, the multijet background has a very different
$\mTH$ distribution than the Higgs~boson signal, so it is not necessary
to suppress this background to the same extent as
in the lower jet multiplicity categories.

\subsection{\boldmath Other dibosons \label{sec:bkg_vv}}

There are backgrounds that originate from the production of two
vector bosons other than $\WW$.
These include $\Wg$, $\Wgs$, $\WZ$ and $\ZZ$
production and are referred to here as $\VV$.
The $\VV$ processes add up to about $10\%$ of the total estimated background
in the $\NjetLEone$ channels and are of the same magnitude as the signal.
The dominant sources of these backgrounds are the production of
$\Wg$ and $\Wgs/\WZ$, where this latter background is a combination of
the associated production of a $W$~boson with a nonresonant $\ZDY$
or an on-shell $Z$~boson.

\begin{figure*}[tb!]
\includegraphics[width=0.75\textwidth]{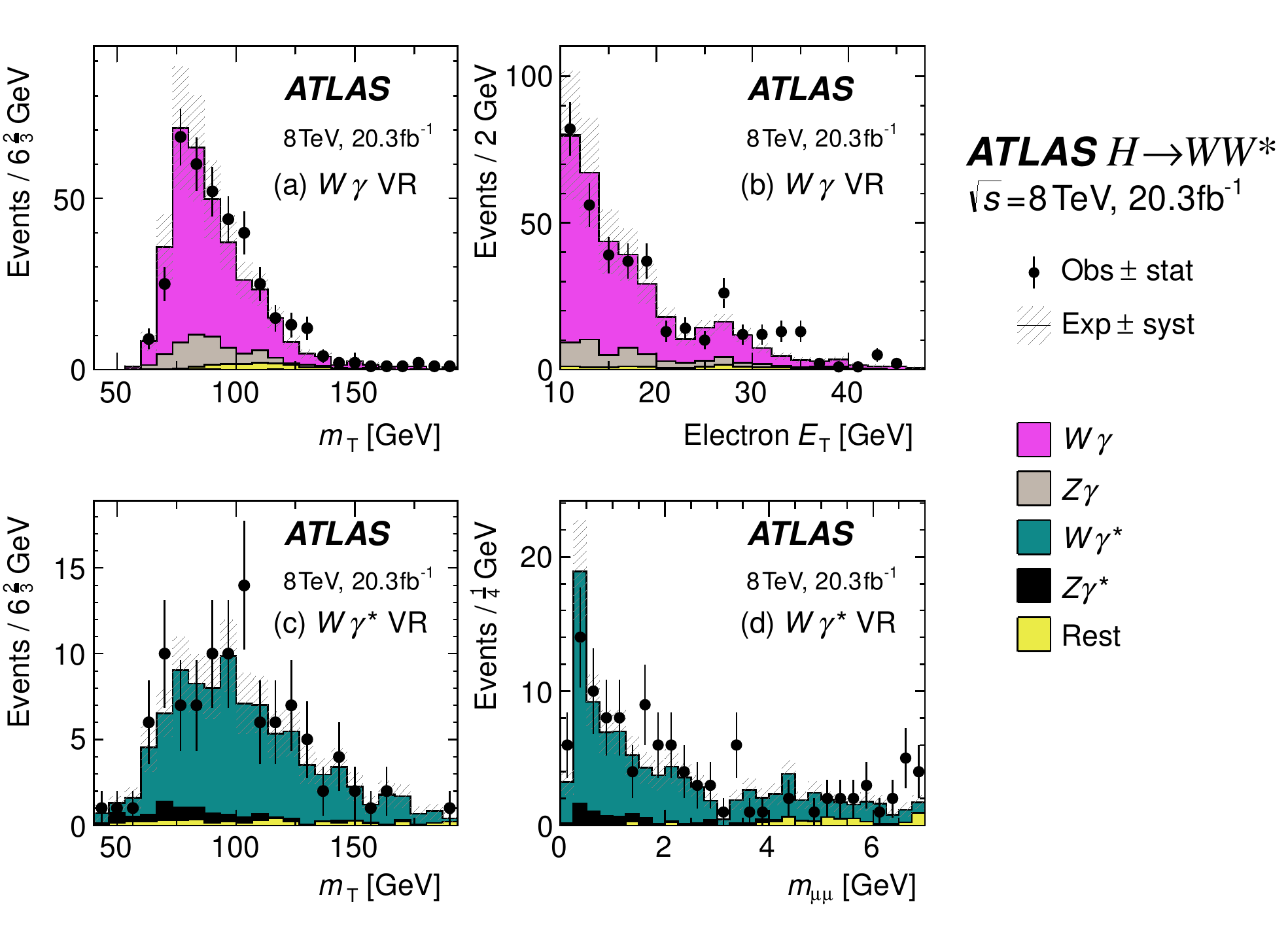}
\caption{
  $\Wg$ and $\Wgs$ validation region (VR) distributions:
  (a) $\Wg$ transverse mass,
  (b) $\Wg$ electron $\eT$,
  (c) $\Wgs$ transverse mass using the leading two leptons, and
  (d) $\Wgs$ dimuon invariant mass.
  The $\Wg$ ($\Wgs$) plots use the data in the $\NjetEQzero$ (all $\Njet$) category.
  ``Rest'' consists of contributions not listed in the legend.
  All processes are normalized to their theoretical cross sections.
  \HwwPlotDetail{See}.
}
\label{fig:vr_wgamma}
\label{fig:vr_wgstar}
\end{figure*}

The normalization of the $\VV$ background processes
in the $\emu$ channel is determined from the data
using a same-charge control region, which is described below.
The distribution of these various contributing processes in the different
signal bins is determined using MC simulation.
In the $\SFchan$ channels,
both the normalization and the distributions
of the $\VV$ processes are estimated
with MC simulation.
The details of these simulations are provided in
Sec.~\ref{sec:atlas_samples_mc}.

Several specialized data sample selections are used to validate the simulation
of the rate and the shape of distributions of various kinematic quantities
of the $\Wg$ and $\Wgs$ processes and the
simulation of the efficiency for rejecting electrons from photon conversions.

The $\Wg$ background enters the signal region when the $W$~boson decays
leptonically and the photon converts into an $e^{+}e^{-}$ pair in the
detector material.
If the pair is very asymmetric in $\pT$, then it is possible that
only the electron or positron satisfies the electron selection criteria,
resulting in a Higgs~boson signal candidate.
This background has a prompt electron or muon and missing transverse momentum
from the $W$~boson decay and a nonprompt electron or positron.
The prompt lepton and the conversion product
are equally likely to have opposite electric charge (required in the
signal selection) and the same electric charge,
since the identification is not charge dependent.

A sample of nonprompt electrons from photon conversions can be selected
by reversing two of the electron signal selection requirements:
the electron track
should be part of a reconstructed photon conversion vertex candidate
and the track
should have no associated hit on the innermost layer of the pixel detector.
Using these two reversed criteria, a sample of $\emu$ events
that otherwise satisfy all of the kinematic requirements imposed on
Higgs~boson signal candidates is selected;
in the $\NjetEQzero$ category ($\NjetEQone$ category),
$83\%$ ($87\%$) of this sample originates from $\Wg$ production.
This sample is restricted to events selected online with a muon trigger
to avoid biases in the electron selection introduced by the online electron
trigger requirements.
Figures~\ref{fig:vr_wgamma}(a) and \ref{fig:vr_wgamma}(b) show the $\mTH$ distribution
and the $\pT$ distribution of the electron of the $\NjetEQzero$
category of this $\Wg$ validation sample
compared to expectations from the MC simulation.
Verifying that the simulation correctly models the efficiency of
detecting photon conversions is important to ensure that the $\Wg$
background normalization and distributions are accurately modeled.
To evaluate the modeling of photon conversions,
a $\Zmumug$ validation sample consisting of either $\Zg$ or
$Z$~boson production with final-state radiation is selected.
The $Z$~boson is reconstructed in the $\mu^+\mu^-$ decay channel,
and an electron (or positron) satisfying all the electron selection criteria
except the two reversed criteria specified above is selected.
The $\mu^+\mu^-e^\pm$ invariant mass is required to be within 15~\GeV\ of
$\mZ$ to reduce contributions from the associated production of a $Z$~boson
and hadronic jets.
The resulting data sample is more than $99\%$ pure in the $\Zmumug$ process.
A comparison between this data sample and a $\Zmumug$ MC simulation
indicates some potential mismodeling of the rejection of nonprompt electrons
in the simulation.
Hence a $\pT$-dependent systematic uncertainty ranging from $25\%$ for
$10{\LT}\pT{\LT}15\GeV$ to $5\%$ for $\pT{\GT}20\GeV$ is assigned to the
efficiency for nonprompt electrons from photon conversions to satisfy the
rejection criteria.

The $\Wgs$ background originates from the associated production of a $W$~boson
that decays leptonically and a virtual photon $\gstar$ that produces
an $e^+e^-$ or $\mu^+\mu^-$ pair in which only one lepton of the pair
satisfies the lepton selection criteria.
This background is most relevant in the $\NjetEQzero$ signal category, where it
contributes a few percent of the total background and is equivalent to
about $25\%$ of the expected Higgs~boson signal.

The modeling of the $\Wgs$ background is studied with a specific selection
aimed at isolating a sample of $\Wgs\rightarrow e\nu\mu\mu$ candidates.
Events with an electron and a pair of opposite-charge muons are selected with
$\M_{\mu\mu}{\LT}7\GeV$, $\MPTj{\GT}20\GeV$
and both muons must satisfy $\Delta\myphi(e, \mu){\LT}2.8$.
Muon pairs consistent with originating from the decay of a $\Jpsi$ meson
are rejected.
The electron and the highest $\pT$ muon are required to satisfy the signal region
lepton selection criteria and $\pT$~thresholds; however,
the subleading-muon $\pT$ threshold is reduced to $3\GeV$.
The isolation criteria for the higher-$\pT$ muon are modified to
take into account the presence of the lower-$\pT$ muon.
The \SHERPA\ $\Wgs$ simulation sample with $\M_{\gstar}{\LT}7\GeV$ is compared
to the data selected with the above criteria;
the distributions of the $\mTH$ calculated using
the electron and the higher-$\pT$ muon and the invariant mass of the two
muons $\M_{\mu\mu}$ are shown in Figs.~\ref{fig:vr_wgstar}(c) and \ref{fig:vr_wgstar}(d).

The $\WZ$ and $\ZZ$ backgrounds are modeled with MC simulation.
No special samples are selected to validate the simulation of these processes.
The $\ZZ$ background arises primarily when one $Z$~boson decays to
$e^+e^-$ and the other to $\mu^+\mu^-$ and an electron and a muon are not detected.
This background is very small, amounting to less than $3\%$ of the $\VV$
background.
Background can also arise from $\Zgs$ and $\Zg$ production if
the $Z$~boson decays to $\ell^+\ell^-$ and one of the leptons is not
identified and the photon results in a second lepton.
These backgrounds are also very small, and the $\Zgs$ background is neglected.

\begin{figure*}[tb!]
\includegraphics[width=0.75\textwidth]{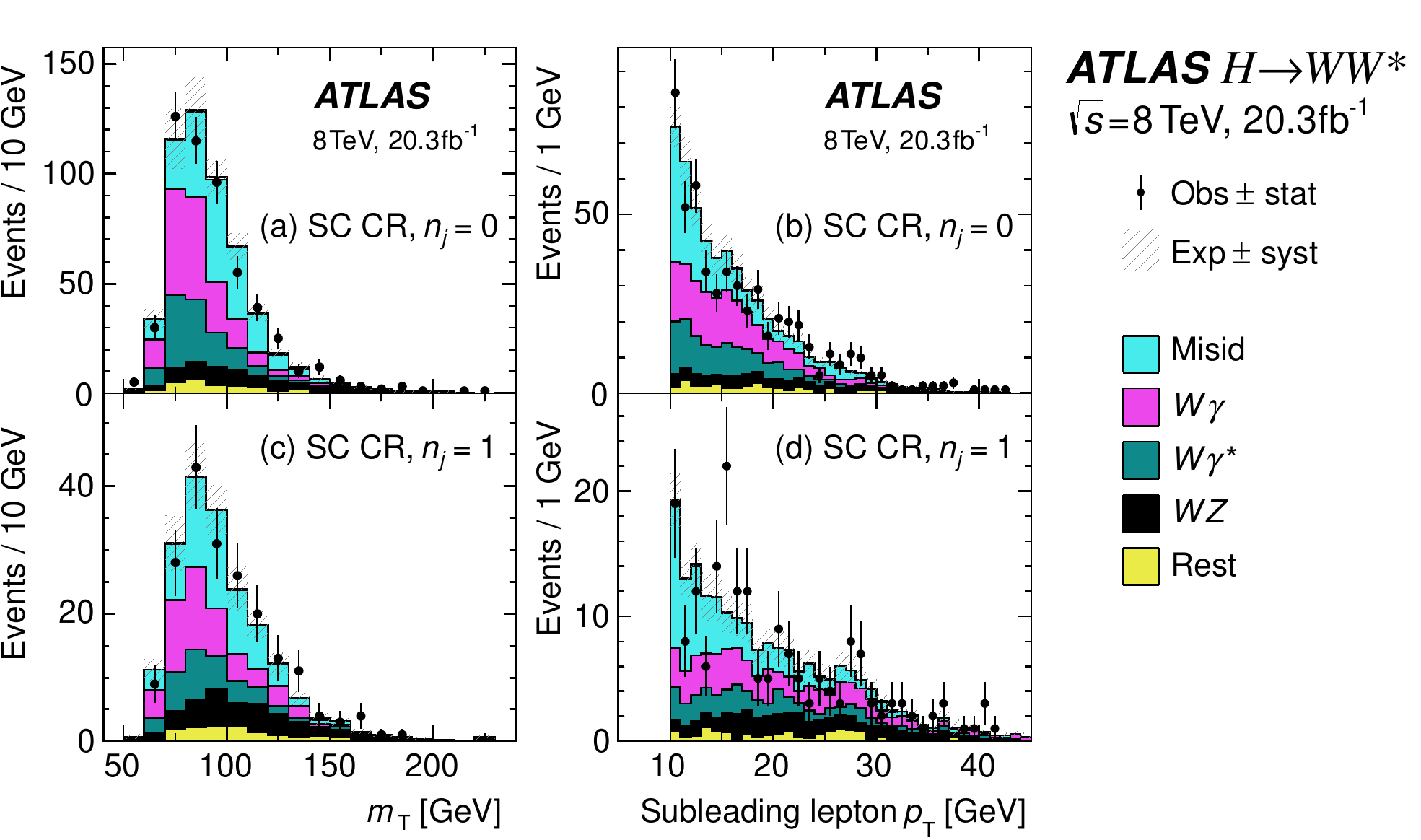}
\caption{
  Same-charge control region (CR) distributions:
  (a) transverse mass         in the $\NjetEQzero$ category,
  (b) subleading lepton $\pT$ in the $\NjetEQzero$ category,
  (c) transverse mass         in the $\NjetEQone$ category, and
  (d) subleading lepton $\pT$ in the $\NjetEQone$ category.
  ``Rest'' consists of contributions not listed in the legend.
  \HwwPlotDetail{See}.
}
\label{fig:ss_cr}
\end{figure*}

The $\VV$ backgrounds arising from $\Wg$, $\Wgs$, and $\WZ$ are equally
likely to result in a second lepton that has the same charge or opposite charge
compared to the lepton from the $W$~boson decay.
For this reason, a selection of $e\mu$ events that is identical to the
Higgs~boson candidate selection except that it requires the two leptons to have
the same charge is used to define a same-charge control region.
The same-charge control region is dominated by $\VV$ processes.
The other process that contributes significantly to the same-charge sample
are the $\Wjets$ process and---to a much lesser extent---the multijet process.
The same-charge data sample can be used to normalize the $\VV$ processes
once the contribution from the $\Wjets$ process is taken into account,
using the method described in Sec.~\ref{sec:bkg_misid}.

Figure~\ref{fig:ss_cr} shows the distributions of
the transverse mass [\ref{fig:ss_cr}(a) and \ref{fig:ss_cr}(c)] and
the subleading lepton $\pT$ [\ref{fig:ss_cr}(b) and \ref{fig:ss_cr}(d)]
for the same-charge data compared with the MC
simulations after normalizing the sum of these MC predictions
to the same-charge data.
A single normalization factor is applied simultaneously
to all four MC simulations of the $\VV$ backgrounds (shown
separately in the figures).
These normalization factors are
$\fNorm_{0j}{\EQ}0.92{\PM}0.07\,{(\stat)}$  and $\fNorm_{1j}{\EQ}0.96{\PM}0.12\,{(\stat)}$
for the $\DFchan$ channels in the $\NjetLEone$ categories.
The $\VV$ processes comprise about $60\%$ of the total in both
the zero-jet and one-jet same-charge data samples,
with $30\%$ coming from the $\Wjets$ process.

Theoretical uncertainties on the $\VV$ backgrounds are dominated by the
scale uncertainty on the prediction for each jet bin.
For the $\Wg$ process, a relative uncertainty of $6\%$ on the total cross section
is correlated across jet categories, and the uncorrelated jet-bin uncertainties
are $9\%$, $53\%$, and $100\%$ in the
$\NjetEQzero$, $\NjetEQone$, and $\NjetGEtwo$
categories, respectively.
For the $\Wgs$ process, the corresponding
uncertainties are $7\%$ (total cross section),
$7\%$ ($\NjetEQzero$), $30\%$ ($\NjetEQone$), and $26\%$ ($\NjetGEtwo$).
No uncertainty is applied for the extrapolation of these backgrounds from
the same-charge control region to the opposite-charge signal region,
since it was verified in the simulation that these processes contribute
equal numbers of opposite-charge and same-charge events.

\subsection{\boldmath Drell-Yan \label{sec:bkg_dy}}

The DY processes produce two oppositely charged leptons
and some events are reconstructed with significant missing transverse momentum. This is mostly due to
neutrinos produced in the $Z$ boson decay in the case of the
$\ZDYtt$ background to the $\DFchan$ channels. In contrast, in the case of the $\ZDYll$
background to the $\SFchan$ channels, it is mostly due to detector resolution that is degraded at high pile-up
and to neutrinos produced in $b$-hadron or $c$-hadron decays (from jets
produced in association with the $Z$ boson). Preselection requirements,
such as the one on $\MPTj$, reduce the bulk of this background, as shown
in Fig.~\ref{fig:MET}, but the residual background is significant in all
categories, especially in the $\SFchan$ samples. The estimation of the
$\ZDYtt$ background for the $\DFchan$ samples is done using a control region, which is defined in a very similar way across
all $\Njet$ categories, as described below. Since a significant
contribution to the $\ZDYll$ background to the $\SFchan$ categories
arises from mismeasurements of the missing transverse momentum, more
complex data-derived approaches are used to estimate this background, as described below.

Mismodeling of $\pTZDY$, reconstructed as $\pTll$, was observed
in the $\ZDY$-enriched region in the $\NjetEQzero$ category.
The \ALPGEN${\PLUS}$\HERWIG\ MC generator does not adequately model the parton
shower of soft jets that balance $\pTll$ when there are
no selected jets in the event.
A correction, based on the weights derived from a data-to-MC
comparison in the $Z$ peak, is therefore applied
to MC events in the $\NjetEQzero$ category, for all leptonic final states from Drell-Yan production.

\subsubsection{\boldmath\textit{\textbf{$\ZDYtt$}}\label{sec:bkg_dy_tautau}}

The $\ZDYtt$ background prediction is normalized
to the data using control regions.
The contribution from this background process is negligible in the $\SFchan$
channel, and in order to remove the potentially large $\ZDYll$ contamination,
the CR is defined using the $\DFchan$ samples in all categories except
the $\NjetGEtwo$ VBF-enriched one.

\begin{figure}[bt!]
\includegraphics[width=0.40\textwidth]{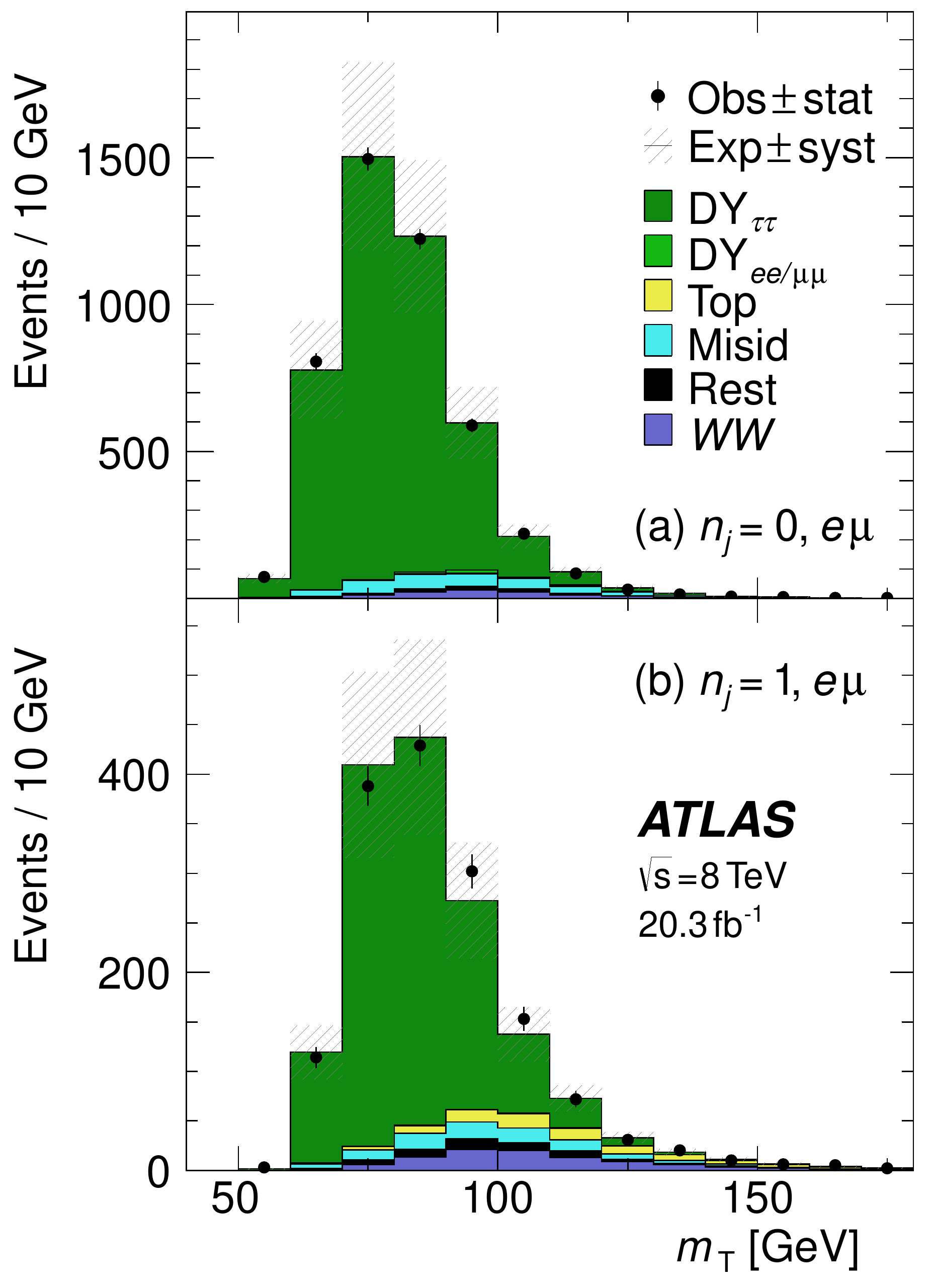}
\caption{
  $\ZDYtt$ control region distributions of transverse mass.
  \HwwPlotDetail{See}.
}
\label{fig:cr_dy}
\end{figure}
The control region in the $\NjetEQzero$ category is defined by the requirements $\mll{\LT}80\GeV$
and $\dphill{\GT}2.8$, which select a $91\%$-pure region and
result in a normalization factor $\fNorm_{0j}{\EQ}1.00{\PM}0.02$\,(stat).
In the $\NjetEQone$ category, the invariant mass of the $\tautau$
system, calculated with the collinear mass approximation, and defined in
Sec.~\ref{sec:selection_1j}, can be used since the
dilepton system is boosted. An $80\%$-pure region is
selected with $\mll{\LT}80\GeV$ and
$\mtt{\GT}(\mZ{\MINUS}25\GeV)$. The latter requirement ensures that there is no overlap with the signal region selection.
The resulting normalization factor is $\fNorm_{1j}{\EQ}1.05{\PM}0.04$\,(stat).
The $\NjetGEtwo$ ggF-enriched category uses a CR
selection of $\mll{\LT}70\GeV$ and
$\dphill{\GT}2.8$ providing $74\%$ purity and a normalization factor
$\fNorm_{2j}{\EQ}1.00{\PM}0.09$\,(stat). Figure~\ref{fig:cr_dy} shows the
$\mTH$ distributions in the control regions in the
$\NjetEQzero$ and $\NjetEQone$ categories. High purity and good
data/MC agreement is observed.

In order to increase the available statistics in the $\Ztt$ control region in the
$\NjetGEtwo$ VBF-enriched category, $\SFchan$ events are also considered. The
contribution from $\ZDYll$ decays is still negligible. The control
region is defined by the invariant mass requirements: $\mll{\LT}80\GeV$
($75\GeV$ in $\SFchan$) and $\ABS{\mtt-\mZ}{\LT}25\GeV$. The resulting normalization
factor is derived after summing all three bins in $\bdt$ and yields
$\fNorm{\EQ}0.9{\PM}0.3$\,(stat).

Three sources of uncertainty are considered on the
extrapolation of the $\ZDYtt$ background from the control region: QCD scale variations, PDFs and
generator modeling. The latter are evaluated based on a comparison of
\ALPGEN${\PLUS}$\HERWIG\ and \ALPGEN${\PLUS}$\PYTHIA\ generators. An
additional uncertainty on the $\pTZDY$ reweighting procedure is
applied in the $\NjetEQzero$ category. It is estimated by comparing the different effects
of reweighting with the nominal weights and with an alternative
set of weights derived with a $\MPTj{\GT}20\GeV$ requirement applied
in the $Z$-peak region. This requirement follows the event selection
criteria used in the $\DFchan$ samples where the $\ZDYtt$ background
contribution is more important. Table~\ref{tab:sys_alpha_ztt} shows
the uncertainties on the extrapolation factor $\fAlpha$ to the signal regions and the
$WW$ control regions in the $\NjetLEone$ and $\NjetGEtwo$ ggF-enriched categories.

\begin{table}[t!]
\caption{
  $\Ztt$ uncertainties (in $\%$) on the extrapolation factor
  $\fAlpha$, for the $\NjetLEone$ and $\NjetGEtwo$ ggF-enriched
  categories. Scale, PDF and generator modeling (Gen) uncertainties
  are reported. For the $\NjetEQzero$ category,
  addtional uncertainty due to $\pTZDY$ reweighting is shown. The negative sign indicates anti-correlation with
  respect to the unsigned uncertainties in the same column.
}
\label{tab:sys_alpha_ztt}
{\small
  \centering
\begin{tabular*}{0.480\textwidth}{
  p{0.165\textwidth}
  rc rc
}
\dbline
Regions
& \multicolumn{1}{p{0.050\textwidth}}{~~Scale}
& \multicolumn{1}{p{0.065\textwidth}}{~~PDF}
& \multicolumn{1}{p{0.075\textwidth}}{~~~~Gen}
& \multicolumn{1}{p{0.075\textwidth}}{~~~$\pTZDY$}
\\
\sgline
\multicolumn{2}{l}{Signal regions} \\
\quad $\NjetEQzero$                         & $-1.6\quad$ &$1.4$ & $5.7\quad$~~& $19$\\
\quad $\NjetEQone$                          &  $4.7\quad$ &$1.8$ &$-2.0\quad$~~& \multicolumn{1}{c}{-}\\
\quad $\NjetGEtwo$ ggF                      &$-10.3\quad$ &$1.1$ &$10.4\quad$~~& \multicolumn{1}{c}{-}\\
\clineskip\clineskip
\multicolumn{3}{l}{$\WW$ control regions} \\
\quad $\NjetEQzero$                         & $-5.5\quad$ &$1.0$ &$-8.0\quad$~~& $16$\\
\quad $\NjetEQone$                          & $-7.2\quad$ &$2.1$ & $3.2\quad$~~& \multicolumn{1}{c}{-}\\
\dbline
\end{tabular*}
}
\end{table}

\subsubsection{\boldmath\textit{\textbf{$\ZDYll$ in $\NjetLEone$}}\label{sec:bkg_dy_ll_ggf}}

The $\frecoil$ variable (see Sec.~\ref{sec:selection}) shows a clear shape difference between DY
and all processes with neutrinos in the final state, including
signal and $\ZDYtt$, which are collectively referred to as ``non-DY''. A method based on a measurement of the selection
efficiency of a cut on $\frecoil$ from data, and an estimate of the
remaining DY contribution after such a cut, is used in the $\SFchan$
category. A sample of events is divided into two bins based on whether they pass or
fail the $\frecoil$ requirement, and the former defines the signal
region. The efficiency of this cut, $\fEff{\EQ}\Npass/({\Npass{\PLUS}\Nfail})$, measured separately in data for
the DY and non-DY processes, is used together with the fraction of the observed
events passing the $\frecoil$ requirement to estimate
the final DY background. It is analytically equivalent to
inverting the matrix:
\begin{equation}
  \Bigg[\!\!
  \begin{array}{c}
    \Npass
    \vspace{1.5mm}
    \\
    \Npass{\PLUS}\Nfail
  \end{array}\!\!\Bigg]
  {\EQ}
  \Bigg[\!\!
  \begin{array}{cc}
    1 & 1
    \vspace{1.5mm}
    \\
    1/\fEff_{\scDY} &
    1/\fEff_{\scnonDY}
  \end{array}
  \!\!\Bigg]
  \CDOT
  \Bigg[\!\!
  \begin{array}{c}
    B_{\scDY}
    \vspace{1.5mm}
    \\
    B_{\scnonDY}
  \end{array}\!\!\Bigg],
  \label{eqn:est_dy}
\end{equation}
and solving for $B_{\scDY}$, which gives the fully data-derived estimate of the DY yield
in the $\SFchan$ signal region. The $\mTH$ distribution for this background
is taken from the Monte Carlo prediction, and the $\mTH$ shape
uncertainties due to the $\pTZDY$ reweighting are found to be negligible.

The non-DY selection efficiency $\fEff_{\scnonDY}$ is evaluated using
the $\DFchan$ sample, which is almost entirely composed of non-DY
events. Since this efficiency is applied to the
non-DY events in the final $\SFchan$ signal region, the event selection is modified to match the $\SFchan$ signal
region selection criteria. This efficiency is used for the signal and for all non-DY backgrounds.
The DY selection efficiency $\fEff_{\scDY}$ is evaluated using the $\SFchan$ sample satisfying
the $\ABS{\mll{\MINUS}\mZ}{\LT}15\GeV$ requirement, which selects the
$Z$-peak region. An additional non-DY efficiency
$\fEff_{\scnonDY}'$ is introduced to account for the non-negligible
non-DY contribution in the $Z$-peak, and is used in the evaluation of
$\fEff_{\scDY}$. It is calculated using the same
$\mll$ region but in $\DFchan$ events. Numerical values for these
$\frecoil$ selection efficiencies are shown in Table~\ref{tab:dy_pacman}(a).

\begin{table}[t!]
\caption{
  The $\frecoil$ summary for the $\ZDYll$ background in the $\NjetLEone$ categories.
  The efficiency for Drell-Yan and non-DY processes are given in (a);
  the associated systematic uncertainties (in $\%$) are given in (b).
  For each group in (b), the subtotal is given first.
  The last row gives the total uncertainty on the estimated $B_{\scDY}$ yield in the SR.
}
\label{tab:dy_pacman}
{\small
  \centering
\begin{tabular*}{0.480\textwidth}{ll rr}
\dbline
\multicolumn{2}{p{0.260\textwidth}}{Efficiency type}
& \multicolumn{1}{p{0.070\textwidth}}{~~~~~~$\NjetEQzero$}
& \multicolumn{1}{p{0.080\textwidth}}{~~~~~~~~~~$\NjetEQone$}
\\
\sgline
\multicolumn{4}{l}{$\!$(a) $\frecoil$ selection efficiencies (in $\%$)} \\
\clineskip\clineskip
$\fEff_{\scnonDY}$,  & efficiency for non-DY events & \multicolumn{1}{r}{$69{\PM}1$} & \multicolumn{1}{r}{$64{\PM}2$} \\
$\fEff_{\scDY}$,     & efficiency for DY events     & \multicolumn{1}{r}{$14{\PM}5$} & \multicolumn{1}{r}{$13{\PM}4$} \\
$\fEff_{\scnonDY}'$, & efficiency for non-DY when   & \multicolumn{1}{r}{$68{\PM}2$} & \multicolumn{1}{r}{$66{\PM}3$} \\
                     & determining the prev.\ row  \\
\clineskip\clineskip
\dbline
  \multicolumn{2}{p{0.260\textwidth}}{Source}
& \multicolumn{1}{p{0.070\textwidth}}{~~~~~~$\NjetEQzero$}
& \multicolumn{1}{p{0.080\textwidth}}{~~~~~~~~~~$\NjetEQone$}
\\
\sgline
\multicolumn{4}{l}{$\!$(b) Systematic uncertainties (in $\%$) on the above efficiencies} \\
\clineskip\clineskip
\multicolumn{2}{l}{Uncertainty on $\fEff_{\scnonDY}$                     }& $1.9$  & $3.2$ \\
\multicolumn{2}{l}{\quad from statistical                                }& $1.8$  & $3.0$ \\
\multicolumn{2}{l}{\quad from using $\DFchan$ CR to extrapolate to $\nqq$}& $0.8$  & $1.2$ \\
\multicolumn{2}{l}{\quad the SR ($\SFchan$ category)                     }\\
\clineskip\clineskip
\multicolumn{2}{l}{Uncertainty on $\fEff_{\scDY}$                        }& $38\Z$ & $32\Z$ \\
\multicolumn{2}{l}{\quad from statistical                                }& $9.4$  & $16\Z$ \\
\multicolumn{2}{l}{\quad from using $Z$-peak to extrapolate to $\nqq$    }& $32\Z$ & $16\Z$ \\
\multicolumn{2}{l}{\quad the SR ($12{\LT}\mll{\LT}55\GeV$)               }\\
\clineskip\clineskip
\multicolumn{2}{l}{Uncertainty on $\fEff_{\scnonDY}'$                    }& $3.1$  & $4.5$ \\
\multicolumn{2}{l}{\quad from statistical                                }& $1.9$  & $3.9$ \\
\multicolumn{2}{l}{\quad from using $\DFchan$ CR to extrapolate to $\nqq$}& $2.5$  & $2.4$ \\
\multicolumn{2}{l}{\quad the SR ($\SFchan$ category)                     }\\
\clineskip\clineskip
\multicolumn{2}{l}{Total uncertainty on yield estimate $B_{\scDY}\nqq$   }& $49\Z$ & $45\Z$ \\
\dbline
\end{tabular*}
}
\end{table}

For the non-DY $\frecoil$ selection efficiencies $\fEff_{\scnonDY}$
and $\fEff_{\scnonDY}'$, the systematic uncertainties are based on
the $\DFchan$-to-$\SFchan$ extrapolation. They are evaluated with MC simulations by
taking the full difference between the selection efficiencies for $\DFchan$ and $\SFchan$
events in the $Z$-peak and SR. Obtained uncertainties are validated with
alternative MC samples and with data, and are added in quadrature
to the statistical uncertainties on the efficiencies. The difference
in the $\frecoil$ selection efficiencies for the signal
and the other non-DY processes is taken as an additional
uncertainty on the signal, and is $9\%$ for the $\NjetEQzero$ category and
$7\%$ for the $\NjetEQone$ one. Systematic uncertainties on the
efficiencies related to the sample composition of the non-DY
background processes were found to be negligible.

The systematic uncertainties on $\fEff_{\scDY}$ are based on the
extrapolation from the $Z$ peak to the SR and are evaluated with MC simulation by
comparing the $\frecoil$ selection efficiencies in these two
regions. This procedure is checked with several generators, and the
largest difference in the selection efficiency is taken as the systematic uncertainty on the efficiency.
It is later added in quadrature to the statistical uncertainty. The procedure is also validated with
the data. Table~\ref{tab:dy_pacman}(b) summarizes all the uncertainties. The largest uncertainties are on $\fEff_{\scDY}$ but
since the non-DY component dominates in the signal region, the uncertainties on its $\frecoil$ efficiency are the dominant
contribution to the total uncertainty on the estimated $B_{\scDY}$ yield.

\begin{table*}[t!]
\caption{
  Control region event yields for $8\TeV$ data.
  All of the background processes are normalized with the corresponding
  $\fNorm$ given in Table~\ref{tab:nf} or with the data-derived methods
  as described in the text; each row shows the composition of one CR.
  The $\Nsig$ column includes the contributions from all signal production processes.
  For the VBF-enriched $\NjetGEtwo$, the values for the bins in $\bdt$ are given.
  The entries that correspond to the target process for the CR are given in
  bold; this quantity corresponds to $N_{\rm bold}$ considered in the last
  column for the purity of the sample (in $\%$).
  The uncertainties on $\Nbkg$ are due to sample size.
}
\label{tab:cr_summary}
{\small
  \centering
\begin{tabular*}{1\textwidth}{
  l
  d{1} r@{$\PM$}l d{1}
  p{0.005\textwidth}
  d{1}d{1}
  d{1}d{1}d{1}d{1}
  p{0.005\textwidth}
  ll
}
\dbline
&\multicolumn{4}{c}{Summary}
&&\multicolumn{6}{c}{Composition of $\Nbkg$}
&
& \multicolumn{2}{c}{Purity}
\\
\clineskip\cline{2-5}\cline{7-12}\cline{14-15}\clineskip
\multicolumn{1}{p{0.235\textwidth}}{Control regions}
& \multicolumn{1}{p{0.060\textwidth}}{~~$\Nobs\nq$}
& \multicolumn{2}{p{0.075\textwidth}}{~~~$\Nbkg$}
& \multicolumn{1}{p{0.060\textwidth}}{~~~$\Nsig$}
&
& \multicolumn{1}{p{0.060\textwidth}}{~$\NWW$}
& \multicolumn{1}{p{0.060\textwidth}}{~~$\Ntop$}
& \multicolumn{1}{p{0.060\textwidth}}{$\Nfakes$}
& \multicolumn{1}{p{0.050\textwidth}}{$\NVV$}
& \multicolumn{2}{c}{~~$\Ndy$}
&
& \multicolumn{2}{c}{$N_{\rm bold}/\Nbkg$}
\\
\multicolumn{4}{l}{}
&
&
&
&
&
&
& \multicolumn{1}{p{0.060\textwidth}}{~~~$\Nll$}
& \multicolumn{1}{p{0.060\textwidth}}{~~~$\Ntautau$}
&
& \multicolumn{2}{c}{($\%$)}
\\
\sgline
\multicolumn{4}{l}{$\NjetEQzero$} \\
\quad CR for $\WW$                    & 2713&$2680$ &$ 9  $& 28&&\MBFr{1950\zz}&  335          &184 &  97         &8.7&  106         &&\multicolumn{2}{c}{73} \\
\quad CR for top quarks               &76013&$75730$&$50  $&618&&     8120     &\MBFr{56210\zz}&2730&  1330       &138&  7200        &&\multicolumn{2}{c}{74} \\
\quad CR for $\VV$                    &  533&$ 531$ &$ 8  $&2.2&&     2.5      &  1.1          &180 &\MBFr{327\zz}& 19&  2.7         &&\multicolumn{2}{c}{62} \\
\quad CR for $\ZDYtt$                 & 4557&$4530$ &$30  $& 23&&     117      & 16.5          &239 &  33         & 28&\MBFr{4100\zz}&&\multicolumn{2}{c}{91} \\
\clineskip
\multicolumn{4}{l}{$\NjetEQone$} \\
\quad CR for $\WW$                    & 2647&$2640$ &$12  $&4.3&&\MBFr{1148\zz}& 1114          &165 & 127         & 17&   81         &&\multicolumn{2}{c}{43} \\
\quad CR for top quarks               & 6722&$6680$ &$12  $& 17&&     244      &\MBFr{6070\zz} &102 &  50         &  6&  204         &&\multicolumn{2}{c}{91} \\
\quad CR for $\VV$                    &  194&$ 192$ &$ 4  $&1.9&&       1      &  3.1          & 65 &\MBFr{117\zz}&4.7&  0.8         &&\multicolumn{2}{c}{61} \\
\quad CR for $\ZDYtt$                 & 1540&$1520$ &$14  $& 18&&     100      &   75          & 84 &  27         &  7&\MBFr{1220\zz}&&\multicolumn{2}{c}{80} \\
\clineskip
\multicolumn{4}{l}{$\NjetGEtwo$ ggF}\\
\quad CR for top quarks               & 2664&$2660$ &$10  $&4.9&&     561      &\MBFr{1821\zz} &129 & 101         & 10&   44         &&\multicolumn{2}{c}{68} \\
\quad CR for $\ZDYtt$                 &  266&$ 263$ &$ 6  $&2.6&&      13      &   34          & 18 & 4.1         &0.1&\MBFr{194\zz} &&\multicolumn{2}{c}{74} \\
\clineskip
\multicolumn{4}{l}{$\NjetGEtwo$ VBF}\\
\quad CR for top quarks, bin 1        &  143&$ 142$ &$ 2  $&2.1&&     1.9      &\MBFr{130\Z}   &2.1 & 0.8         &6.3&  1.1         &&\multicolumn{2}{c}{92} \\
\quad CR for top quarks, bin 2--3     &   14&$14.3$ &$ 0.5$&1.8&&     0.6      &\MBFr{11.6}    &0.2 & 0.2         &0.9&  0.2         &&\multicolumn{2}{c}{81} \\
\quad CR for $\ZDYtt$                 &   24&$20.7$ &$0.9 $&2.4&&     0.9      & 1.2           &0.6 & 0.2         &0.8&\MBFr{17\zz}  &&\multicolumn{2}{c}{82} \\
\dbline
\end{tabular*}
}
\end{table*}

\subsubsection{\boldmath\textit{\textbf{$\ZDYll$ in VBF-enriched $\NjetGEtwo$}}\label{sec:bkg_dy_ll_vbf}}

The $\ZDYll$ background in the VBF-enriched channel is estimated using
an ``{\sc abcd}'' method.
The BDT shape for this process is taken from a high-purity data
sample with low $\mll$ and low $\MPTj$ (region {\sc b}). It is then
normalized with a $\MPTj$ cut efficiency, derived from the data using
the $Z$-peak region separated into low- and high-$\MPTj$
regions ({\sc c} and {\sc d}, respectively). It yields
$0.43{\PM}0.03$. The final estimate in the signal region
({\sc a}) is corrected with a nonclosure factor derived from the MC,
representing the differences in $\MPTj$ cut efficiencies between the
low-$\mll$ and $Z$-peak regions. It yields $0.83{\PM}0.22$. Bins 2 and 3 of
$\bdt$ are normalized using a common factor due to the low number of events in
the highest $\bdt$ bin in region {\sc b}. The normalization
factors, applied to the $\ZDYll$ background in the $\SFchan$ channel
in the signal region, are $\fNorm_{\rm bin1}{\EQ}1.01{\PM}0.15$\,(stat) and $\fNorm_{\rm bin 2+3}{\EQ}0.89{\PM}0.28$\,(stat).

The uncertainty on the nonclosure factor is $17\%$ (taken as its deviation from
unity), and is fully correlated across all $\bdt$ bins.
Uncertainties are included on the $\bdt$
shape due to QCD scale variations, PDFs, and the parton shower model, and are $11\%$ in the bin with the highest
$\bdt$ score.
No dependence of the BDT response on $\MPTj$ is observed in MC, and an uncertainty
is assigned based on the assumption that they are uncorrelated ($4\%$, $10\%$, and $60\%$ in the bins with increasing $\bdt$
score).

\subsection{\boldmath Modifications for $7\TeV$ data \label{sec:bkg_7tev}}

The background estimation techniques in the $\NjetLEone$ channels for $7\TeV$ data closely follow the ones applied to $8\TeV$ data.
The definitions of the control regions of $\WW$, top-quark, and $\ZDYtt$ are the same. The $\ZDYll$ background is estimated with the same method based on the $\frecoil$ selection \
efficiencies. The $\frecoil$ requirements are loosened (see Sec.~\ref{sec:selection_7tev}).
The calculation of the extrapolation factor in the $W+$jets estimate uses a multijet sample instead of a $Z$+jets
sample, which has a limited number of events.
The $\VV$ backgrounds are estimated using Monte Carlo predictions because of the small number of events in the same-charge region.
In the $\NjetGEtwo$ VBF-enriched category, the background estimation techniques are the same as in the $8\TeV$ analysis.
The normalization factors from the control regions are given in Table~\ref{tab:nf} in the next section along with the values for the $8\TeV$ analysis.

The theoretical uncertainties on the extrapolation factors used in the $\WW$, top-quark, and $\ZDYtt$ background estimation methods
are assumed to be the same as in the $8\TeV$ analysis. Uncertainties due to experimental sources are unique to the $7\TeV$ analysis and are
taken into account in the likelihood fit. The uncertainties on the $\frecoil$ selection efficiencies used in the $\ZDYll$ background
estimation were evaluated following the same technique as in the $8\TeV$ analysis.
The dominant uncertainty on the extrapolation factor in the $\Wjets$ estimate is due to the uncertainties
on the differences in the compositions of the jets in the multijet and
$\Wjets$ sample and is $29\%$ ($36\%$) for muons (electrons).

\subsection{\boldmath Summary \label{sec:bkg_summary}}

This section described the control regions used to
estimate, from data, the main backgrounds to the various categories in the analysis. An overview of the observed and
expected event yields in these control regions is provided in Table~\ref{tab:cr_summary}
for the $8\TeV$ data. This shows the breakdown of each control
region into its targeted physics process (in bold) and its purity,
together with the other contributing physics processes. The $\WW$ CR
in the $\NjetEQone$ category is relatively low in $\WW$ purity
but the normalization for the large contamination by $\Ntop$ is determined by
the relatively pure CR for top quarks.

The normalization factors $\fNorm$ derived from these control regions are
summarized in Table~\ref{tab:nf}, for both the $7$ and $8\TeV$ data samples. Only the statistical uncertainties
are quoted and in most of the cases the normalization factors agree
with unity within the statistical uncertainties. In two cases where
a large disagreement is observed, the systematic uncertainties on $\fNorm$ are
evaluated. One of them is the $\WW$ background in the $\NjetEQzero$
category, where adding the systematic uncertainties reduces the
disagreement to about two standard deviations:
$\fNorm{\EQ}1.22{\PM}0.03\,(\stat){\PM}0.10\,(\syst)$.
The systematic component includes the experimental uncertainties and
additionally the theoretical uncertainties on the cross section and acceptance, and the uncertainty
on the luminosity determination. Including the systematic
uncertainties on the normalization factor for the top-quark background in
the first bin in the $\NjetGEtwo$ VBF-enriched category reduces the
significance of the deviation of the normalization factor with unity:
$\fNorm{\EQ}1.58{\PM}0.15\,(\stat){\PM}0.55\,(\syst)$. In
this case, the uncertainty on MC generator modeling is also included.
The systematic uncertainties quoted here do not have an
impact on the analysis since the background estimation in the signal
region is based on the extrapolation factors and their associated
uncertainties, as quoted in the previous subsections. In addition, the sample statistics of the control region,
the MC sample statistics and the uncertainties on the background
subtraction all affect the estimation of the backgrounds normalized to
data.
\vfill

\begin{table*}[t!]
\caption{
  Control region normalization factors $\fNorm$.
  The $\fNorm$ values scale the corresponding estimated yields in the signal region;
  those that use MC-based normalization are marked with a dash.
  For the VBF-enriched $\NjetGEtwo$ category, the values in bins of $\bdt$ are
  given for top quarks; a combined value is given for $\Ztt$.
  The uncertainties are due to the sample size of the corresponding control regions.
}
\label{tab:nf}
{\small
  \centering
\begin{tabular*}{1\textwidth}{l
  r@{${\PM}$}l
  r@{${\PM}$}l
  r@{${\PM}$}l
  r@{${\PM}$}l
}
\dbline
\multicolumn{1}{p{0.250\textwidth}}{Category}
& \multicolumn{2}{p{0.170\textwidth}}{~~~~$\WW$}
& \multicolumn{2}{p{0.170\textwidth}}{Top quarks}
& \multicolumn{2}{p{0.170\textwidth}}{~~~~$\VV$}
& \multicolumn{2}{p{0.170\textwidth}}{$\ZDYtt$}
\\
\sgline
\multicolumn{2}{l}{$8\TeV$ sample} \\
$\quad\NjetEQzero$                   &$1.22$ &$0.03$     &$1.08$ &$0.02$ &$0.92$ &$0.07$     &$1.00$ &$0.02$                     \\
$\quad\NjetEQone$                    &$1.05$ &$0.05$     &$1.06$ &$0.03$ &$0.96$ &$0.12$     &$1.05$ &$0.04$                     \\
$\quad\NjetGEtwo$, ggF               &\MCOL{2}{\qquad\,-}&$1.05$ &$0.03$ &\MCOL{2}{\qquad\,-}& $1.00$ &$0.09$                    \\
$\quad\NjetGEtwo$, VBF bin $1$       &\MCOL{2}{\qquad\,-}&$1.58$ &$0.15$ &\MCOL{2}{\qquad\,-}&\MCOL{2}{\MROW{2}{$0.90{\PM}0.30$}}\\
$\quad\NjetGEtwo$, VBF bins $2$--$3$ &\MCOL{2}{\qquad\,-}&$0.95$ &$0.31$ &\MCOL{2}{\qquad\,-}&\MCOL{2}{}                         \\
\clineskip\clineskip
\multicolumn{2}{l}{$7\TeV$ sample} \\
$\quad\NjetEQzero$                   &$1.09$ &$0.08$     &$1.12$ &$0.06$ &\MCOL{2}{\qquad\,-}&$0.89$ &$0.04$                     \\
$\quad\NjetEQone$                    &$0.98$ &$0.12$     &$0.99$ &$0.04$ &\MCOL{2}{\qquad\,-}&$1.10$ &$0.09$                     \\
$\quad\NjetGEtwo$, VBF bins $1$--$3$ &\MCOL{2}{\qquad\,-}&$0.82$ &$0.29$ &\MCOL{2}{\qquad\,-}&$1.52$ &$0.91$                     \\
\dbline
\end{tabular*}
}
\end{table*}

\section{\boldmath Fit procedure and uncertainties \label{sec:systematics}}

The signal yields and cross sections are obtained from a
statistical analysis of the data samples described in Sec.~\ref{sec:selection}.
A likelihood function---defined to simultaneously model, or ``fit'' the yields
of the various subsamples---is maximized.

\begin{table*}[tb!]
\caption{
  Fit region definitions for the Poisson terms in the likelihood,
  Eq.~(\ref{eqn:likelihood}), not including the terms used for MC statistics.
  The signal region categories $i$ are given in (a).
  The definitions for bins $b$ are given by listing the bin edges, except for $\mTH$ and $\bdt$, and
  are given in the text and noted as the fit variables on the right-most column.
  The background control regions are given in (b),
  which correspond to the ones indicated as using data in Table~\ref{tab:cr}.
  The profiled CRs are
  marked by $\bullet$ and the others are marked by $\circ$.
  ``Sample'' notes the lepton flavor composition of the
  CR that is used for all the SR regions for a given $\Njet$ category:
  ``$\DFchan$'' means that a $\DFchan$ CR sample is used for all SR regions;
  the $\Wj$ and $\jj$ CRs use the same lepton-flavor samples in the SR
  (same), \ie, ``$\DFchan$'' CR for ``$\DFchan$'' SR and ``$\SFchan$'' CR for ``$\SFchan$'' SR;
  the DY, $\SFchan$ sample is used only for the $\SFchan$ SR;
  the two rows in $\NjetGEtwo$ VBF use a CR that combines the two samples (both);
  see text for details.
  Energy-related quantities are in $\!\GeV$.
}
\label{tab:fitregions}
{\small
  \centering
\begin{tabular}{c p{0.005\textwidth} c}
\begin{tabular*}{0.465\textwidth}{
  p{0.085\textwidth}
  l
  l
  l
  c
  c
}
\\
\multicolumn{4}{l}{$\no$(a) Signal region categories} \\
\dbline
\multicolumn{4}{c}{SR category $i$}
&
&\multicolumn{1}{l}{\multirow{2}{*}{Fit var.}}
\\
\clineskip
\cline{1-4}
\clineskip
$\Njet$, flavor
&${\otimes\,}\mll$
&${\otimes\,}\pTsublead$
&${\otimes\,}\ell_2$
&
&
\\
\sgline
$\NjetEQzero$ \\
\quad $\DFchan$     &${\otimes\,}[10,30,55]$ &${\otimes\,}[10,15,20,\infty]$ &${\otimes\,}[e,\mu]$ &&$\mTH$ \\
\quad $\SFchan$     &${\otimes\,}[12,55]$    &${\otimes\,}[10,\infty]$       &                     &&$\mTH$ \\
\sgline
$\NjetEQone$ \\
\quad $\DFchan$     &${\otimes\,}[10,30,55]$ &${\otimes\,}[10,15,20,\infty]$ &${\otimes\,}[e,\mu]$ &&$\mTH$ \\
\quad $\SFchan$     &${\otimes\,}[12,55]$    &${\otimes\,}[10,\infty]$       &                     &&$\mTH$ \\
\sgline
$\NjetGEtwo$ ggF \\
\quad $\DFchan$     &${\otimes\,}[10,55]$    &${\otimes\,}[10,\infty]$       &                     &&$\mTH$ \\
\sgline
\multicolumn{2}{l}{$\NjetGEtwo$ VBF} \\
\quad $\DFchan$     &${\otimes\,}[10,50]$    &${\otimes\,}[10,\infty]$       &                     &&$\bdt$ \\
\quad $\SFchan$     &${\otimes\,}[12,50]$    &${\otimes\,}[10,\infty]$       &                     &&$\bdt$ \\
\dbline
\phantom{X}\\
\phantom{X}\\
\phantom{X}\\
\phantom{X}\\
\phantom{X}\\
\phantom{X}\\
\clineskip
\clineskip
\clineskip
\clineskip
\clineskip
\clineskip
\clineskip
\clineskip
\clineskip
\end{tabular*}
&
&
\begin{tabular*}{0.505\textwidth}{
  p{0.070\textwidth}
  c
  p{0.070\textwidth}
  l
}
\\
\multicolumn{4}{l}{$\no$(b) Control regions that are profiled ($\bullet$) and nonprofiled ($\circ$)}\\
\dbline
\multicolumn{1}{l}{CR}
&\multicolumn{1}{l}{Profiled?}
&\multicolumn{1}{l}{Sample}
&\multicolumn{1}{l}{Notable differences vs.\ SR} \\
\sgline
$\NjetEQzero$ \\
\quad $\WW$                               &~~$\bullet$ $\np\nq$ &$\DFchan$ &$55{<}\mll{<}110$, $\dphill{<}2.6$, $\pTsublead{>}15$ \\
\quad Top                                 &~~$\circ$   $\np\nq$ &$\DFchan$ &$\NjetEQzero$ after presel., $\dphill{\LT}2.8$\\
\quad $\Wj$                               &~~$\circ$   $\np\nq$ & same     &one anti-identified $\ell$ \\
\quad $\jj$                               &~~$\circ$   $\np\nq$ & same     &two anti-identified $\ell$ \\
\quad $\VV$                               &~~$\bullet$ $\np\nq$ &$\DFchan$ &same-charge $\ell$ \,(only used in $\DFchan$) \\
\quad DY,\,$\SFchan$$\nq\nq$              &~~$\bullet$ $\np\nq$ &$\SFchan$ &$\frecoil{\GT}0.1$ (only used in $\SFchan$) \\
\quad DY,\,$\tautau$$\nq$                 &~~$\bullet$ $\np\nq$ &$\DFchan$ &$\mll{\LT}80$, $\dphill{\GT}2.8$ \\
\sgline
$\NjetEQone$ \\
\quad $\WW$                               &~~$\bullet$ $\np\nq$ &$\DFchan$ &$\mll{>}80$, $|\mtt{-}\mZ|{>}25$, $\pTsublead{>}15$ \\
\quad Top                                 &~~$\bullet$ $\np\nq$ &$\DFchan$ &$\NbjetEQone$ \\
\quad $\Wj$                               &~~$\circ$   $\np\nq$ & same     &one anti-identified $\ell$ \\
\quad $\jj$                               &~~$\circ$   $\np\nq$ & same     &two anti-identified $\ell$ \\
\quad $\VV$                               &~~$\bullet$ $\np\nq$ &$\DFchan$ &same-charge $\ell$ \,(only used in $\DFchan$) \\
\quad DY,\,$\SFchan$$\nq\nq$              &~~$\bullet$ $\np\nq$ &$\SFchan$ &$\frecoil{\GT}0.1$ (only used in $\SFchan$)\\
\quad DY,\,$\tautau$$\nq$                 &~~$\bullet$ $\np\nq$ &$\DFchan$ &$\mll{\LT}80$, $\mtt{\GT}\mZ{\MINUS}25$ \\
\sgline
\multicolumn{2}{l}{$\NjetGEtwo$ ggF} \\
\quad Top                                 &~~$\bullet$ $\np\nq$ &$\DFchan$ &$\mll{\GT}80$ \\
\quad $\Wj$                               &~~$\circ$   $\np\nq$ & same     &one anti-identified $\ell$ \\
\quad $\jj$                               &~~$\circ$   $\np\nq$ & same     &two anti-identified $\ell$ \\
\quad DY,\,$\tautau$$\nq$                 &~~$\bullet$ $\np\nq$ &$\DFchan$ &$\mll{\LT}70$, $\dphill{\GT}2.8$ \\
\sgline
\multicolumn{2}{l}{$\NjetGEtwo$ VBF} \\
\quad Top                                 &~~$\bullet$ $\np\nq$ & both     &$\NbjetEQone$ \\
\quad $\Wj$                               &~~$\circ$   $\np\nq$ & same     &one anti-identified $\ell$ \\
\quad $\jj$                               &~~$\circ$   $\np\nq$ & same     &two anti-identified $\ell$ \\
\quad DY,\,$\SFchan$$\nq\nq$              &~~$\circ$   $\np\nq$ &$\SFchan$ &$\MET{<}45$\,(only\,used\,in\,$\SFchan$) \\
\quad DY,\,$\tautau$$\nq$                 &~~$\circ$   $\np\nq$ & both     &$\mll{\LT}80$, $\ABS{\mtt-\mZ}{\LT}25$ \\
\dbline
\end{tabular*}
\end{tabular}
}
\end{table*}

The signal strength parameter $\sigmu$, defined in Sec.~\ref{sec:introduction}, is the ratio of the
measured signal yield to the expected SM value.
Its expected value ($\sigmu_\exp$) is unity by definition.
A measurement of zero corresponds to no
signal in the data.  The observed value $\sigmu_\obs$, reported in
Sec.~\ref{sec:results}, is one of the central results of this \paper.

In this section, the fit regions are described in
Sec.~\ref{sec:systematics_region} followed by the details of the
likelihood function and the test statistic in
Sec.~\ref{sec:systematics_fit}.  Section~\ref{sec:systematics_sources}
summarizes the various sources of uncertainty that affect the results.  A
check of the results is given in Sec.~\ref{sec:systematics_pulls}.

\subsection{\boldmath Fit regions \label{sec:systematics_region}}

The fit is performed over data samples defined by fit regions listed in
Table~\ref{tab:fitregions}, which consist of
\begin{itemize}
  \setlength{\itemsep}{0pt}
  \setlength{\parskip}{0pt}
  \setlength{\parsep}{0pt}
  \item[(i)] signal region categories [Table~\ref{tab:fitregions}(a)] and
  \item[(ii)] profiled control regions
        (rows in Table~\ref{tab:fitregions}(b) marked by solid circles).
\end{itemize}
The nonprofiled control regions (rows in Table~\ref{tab:fitregions}(b) marked
by open circles) do not have explicit terms in the likelihood, but are listed in the
table for completeness.

The profiled CRs determine the normalization of the corresponding backgrounds through a Poisson
term in the likelihood, which, apart from the Drell-Yan $\tautau$ CR, use
the $\DFchan$ sample.  The nonprofiled CRs do not have a Poisson term and enter the
fit in other ways.  The details are described in the next section.

\begin{figure*}[b!]
\hspace{0pt}\includegraphics{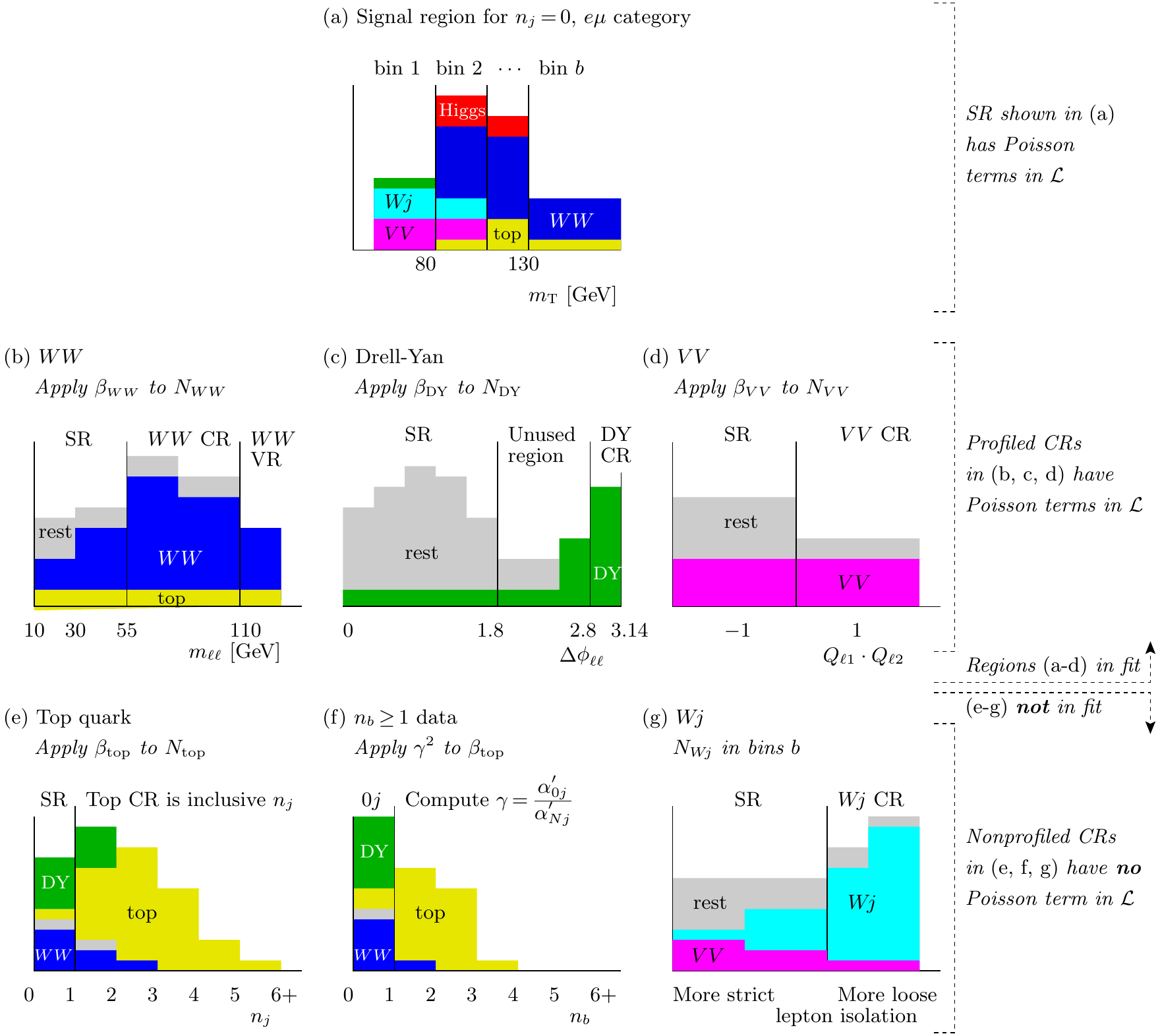}
\caption{
  Simplified illustration of the fit regions for $\NjetEQzero$, $\emu$ category.
  The figure in (a) is the variable-binned $\mTH$ distribution in the signal
  region for a particular range of $\mll$ and $\pTsublead$ specified in
  Table~\ref{tab:fitregions}; the $\mTH$ bins are labeled $b{\EQ}1, 2, \ldots$;
  the histograms are stacked for the five principal background
  processes---$\WW$, $\top$, Misid.\ (mostly $\Wj$), $\VV$,  DY (unlabeled)---and the Higgs signal process.
  The figures in (b, c, d) represent the distributions that define the various
  profiled control regions used in the fit with a corresponding Poisson term in
  the likelihood $\likelihood$.  Those in (e, f, g) represent the nonprofiled
  control regions that do not have a Poisson term in $\likelihood$, but
  determine parameters that modify the background yield predictions.
  A validation region (VR) is also defined in (b); see text.
}
\label{fig:fitregions}
\end{figure*}

The SR categories $i$ and fit distribution bins $b$ that contribute to the
likelihood were briefly motivated in Sec.~\ref{sec:analysis}.

The $\DFchan$ samples in the $\NjetLEone$, the most signal-sensitive of all channels,
are each divided into $12$ kinematic regions ($12{\EQ}2{\CDOT}3{\CDOT}2$):
two regions in $\mll$,
three regions in $\pTsublead$, and
two regions for the subleading lepton flavors.
In contrast, the less sensitive $\SFchan$ samples for the
$\NjetLEone$ categories use one range of $\mll$ and $\pTsublead$.

The $\mTH$ distribution is used to fit all of the ggF-enriched categories.
Its distribution for the signal process has an upper kinematic edge at $\mH$, but, in practice,
$\mTH$ can exceed $\mH$ because of detector resolution.  There is also a
kinematic suppression below a value of $\mTH$ that increases with increasing
values of $\mll$ and $\pTsublead$ due to the kinematic requirements in each of
the $\NjetLEone$ categories.

The $\mTH$ distribution for the $\NjetEQzero$ category in the $\DFchan$ ($\SFchan$) samples uses a
variable binning scheme that is optimized for each of the $12$ (one) kinematic regions.
In the kinematically favored range of the $\DFchan$ and $\SFchan$ samples,
there are $8$ bins that are approximately $6\GeV$ wide between a range
of {\sc x} to {\sc y}, where {\sc x} is approximately $80\GeV$ and {\sc y}
is approximately $130\GeV$.  A single bin at low $\mTH$, from $0$ to
{\sc x}, has a few events in each category; another bin at high $\mTH$---from
{\sc y} to $\infty$---is populated dominantly by $\WW$ and top-quark events,
constraining these backgrounds in the fit.

The $\mTH$ distribution for the $\NjetEQone$ category follows the above scheme
with six bins.  The bins are approximately $10\GeV$ wide in the same range as
for $\NjetEQzero$.

The $\mTH$ distribution of the $\emu$ events in the ggF-enriched $\NjetGEtwo$ uses four bins
specified by the bin boundaries $[0, 50, 80, 130, \infty]\GeV$.

The $\bdt$ distribution is used to fit the VBF-enriched $\NjetGEtwo$ samples.  The signal
purity increases with increasing value of $\bdt$, so the bin widths decrease
accordingly.  The bin boundaries are $[-1, -0.48, 0.3, 0.78, 1]$ and define
four bins that are labeled $0$ through $3$.  Only bins $1$, $2$, and $3$
are used in the fit.  The selection-based cross-check analysis uses two bins in $\mjj$
($[600, 1000, \infty]\GeV$) and four bins in the $\mTH$ distribution as defined
above for the ggF-enriched $\NjetGEtwo$.

In general, the bin boundaries are chosen to maximize the expected signal significance
while stabilizing the statistical fluctuations associated with the subtraction of the backgrounds.
For the $\mTH$ fits, this is accomplished by maintaining an approximately constant
signal yield in each of the bins.  The exact values of the $\mTH$ bins are given in
Appendix~\ref{sec:appendix_bins} in Table~\ref{tab:bins}.

The interplay of the various fit regions is illustrated for one kinematic region of the $\NjetEQzero$
in Fig.~\ref{fig:fitregions}.
The shape of the $\mTH$ distribution is used in the fit to discriminate between
the signal and the background as shown in the top row for the SR.
Three profiled CRs determine
the normalization factors ($\fNorm_k$) of the respective background contributions \insitu.
The variable and selections used to separate the SR from the CRs regions are given in the second row:
for $\WW$ the $\mll$ variable divides the SR and CR, but also the validation region (VR) used to test the $\WW$ extrapolation (see Sec.~\ref{sec:bkg_ww});
for DY the $\dphill$ variable divides the SR and CR with a region separating the two;
for $\VV$ the discrete same/opposite charge variable is used.
The last row shows the backgrounds whose normalizations are not
profiled in the fit, but are computed prior to the fit.

The treatment of a given region as profiled or nonprofiled CR depends from the
complexity related to its implementation in the fit, the impact of the
estimated background in the analysis, and  the level of contamination of the
other process in the relative CR. Subdominant backgrounds and those whose
estimation is not largely affected by the postfit yield of the other
backgrounds, like $\Wj$ and multijet backgrounds, are not profiled.

\subsection{\boldmath Likelihood, exclusion, and significance \label{sec:systematics_fit}}

The statistical analysis involves the use of the likelihood
$\likelihood(\sigmu, \nuipars\SPACE{|}\MBF{N})$, which is a function of the
signal strength parameter $\sigmu$ and a set of nuisance parameters
$\nuipars{\EQ}\{\nuipar_a, \nuipar_b, \ldots\}$ given a set of the numbers of
events $\MBF{N}{\EQ}\{N_A, N_B, \ldots\}$.  Allowed ranges of $\sigmu$ are found using
the distribution of a test statistic $\qmu$.

\begin{widetext}
\begin{equation}
\no\!\likelihood{\EQ}\!
\underbrace{
  \raisebox{-3pt}{\ensuremath{\phantom{\Bigg[}\!\!}}
  \np\mathop{\prod_{i, b}
             ^{\mbox{\tiny\ref{tab:fitregions}a}}
            }^{\mbox{\scriptsize Table}}
  \!{f}\Big(
    N_{ib}
    \SPACE{\Big|}
    \sigmu{\,\cdot\,}S_{ib}{\,\cdot}\no
    \raisebox{2pt}{\scriptsize\ensuremath
                    {\mathop{\prod\limits
                             _{\raisebox{-4.0pt}{\scriptsize $r$}}
                             ^{\raisebox{1.8pt}{\tiny Sec.\,\ref{sec:signal}}}\nq\nq\!
                            }^{\raisebox{4.5pt}{\scriptsize Syst\,in}}
                    }
                  }
    \no\!\responsefn_{\!br}\big(\nuipar_{r}\big){+}
    \raisebox{2pt}{\scriptsize\ensuremath
                    {\mathop{\sum\limits
                             _{\raisebox{-4.0pt}{\scriptsize $k$}}
                             ^{\raisebox{1.8pt}{\tiny\ref{tab:process}}}\np\nq
                            }^{\raisebox{4.5pt}{\scriptsize Table}}
                    }
                  }
    \!\fNorm_k{\cdot}{B}_{kib}{\,\cdot}\np\!
    \raisebox{2pt}{\scriptsize\ensuremath
                     {\mathop{\prod\limits
                              _{\raisebox{-4.0pt}{\scriptsize $s$}}
                              ^{\raisebox{1.8pt}{\tiny Sec.\,\ref{sec:systematics_sources}}}\nq\nq\!
                             }^{\np\,\raisebox{4.5pt}{\scriptsize Syst\,in}}
                     }
                  }
    \np\responsefn_{\!bs}\big(\nuipar_{s}\big)
    \Big)
    \raisebox{-3pt}{\ensuremath{\phantom{\Bigg]}\!\!}}
}_{\footnotesize\mbox{Poisson for SR with signal strength $\sigmu$; predictions $S$, $B$}}
{\!\cdot\!}
\underbrace{
  \raisebox{-3pt}{\ensuremath{\phantom{\Bigg[}\!\!}}\no\!\!
  \mathop{\prod
          _{\raisebox{0pt}{\scriptsize $l$}}
          ^{\raisebox{0pt}{\tiny\ref{tab:fitregions}b}\,\bullet}
         }^{\raisebox{0pt}{\scriptsize Table~}}\!\!
  f\Big(N_l\SPACE{\Big|}
  \raisebox{2pt}{\scriptsize\ensuremath
                   {\mathop{\scriptsize\sum\limits
                            _{\raisebox{-4.0pt}{\scriptsize $k$}}
                            ^{\raisebox{1.8pt}{\tiny\ref{tab:process}}}\np\nq
                           }^{\raisebox{4.5pt}{\scriptsize Table}}
                   }
                }
  \fNorm_k{\cdot}B_{kl}\Big)
  \raisebox{-3pt}{\ensuremath{\phantom{\Bigg]}\!\!}}
}_{\footnotesize\mbox{Poisson for profiled CRs}}
{\!\cdot\!}
\underbrace{
  \raisebox{-3pt}{\ensuremath{\phantom{\Bigg[}\!\!}}\np\!
  \mathop{\prod
          _{\raisebox{0pt}{\scriptsize $t$}}
          ^{\raisebox{0pt}{\tiny\{$r,\,s$\}}}\no
         }^{\raisebox{0pt}{\scriptsize Syst\,in}}\no
  g\big(\vartheta_t{\big|}\nuipar_t\big)
  \raisebox{-3pt}{\ensuremath{\phantom{\Bigg]}\!\!}}
}_{\footnotesize\mbox{Gauss.\,for\,syst}}
{\!\cdot\!}
\underbrace{
  \raisebox{-3pt}{\ensuremath{\phantom{\Bigg[}\!\!}}\np
  \mathop{\prod
          _{\raisebox{0pt}{\scriptsize $k$}}
          ^{\raisebox{0pt}{\tiny\ref{tab:process}}}
         }^{\raisebox{0pt}{\scriptsize Table}}\no
  f\big(\fBsignif_k\SPACE{\big|}\fBvarsignif_k{\cdot}\nuipar_k\big)
  \raisebox{-3pt}{\ensuremath{\phantom{\Bigg]}\!\!\!\!}}
}_{\footnotesize\mbox{Poiss.\,for\,MC\,stats}}
\label{eqn:likelihood}
\end{equation}
\end{widetext}

\subsubsection{\boldmath\textit{\textbf{Likelihood function}}\label{sec:systematics_fit_lh}}

The likelihood function $\likelihood$ [Eq.~(\ref{eqn:likelihood}) below] is the
product of four groups of probability distribution functions:
\begin{itemize}
  \setlength{\itemsep}{0pt}
  \setlength{\parskip}{0pt}
  \setlength{\parsep}{0pt}
  \item[(i)] Poisson function $f(N_{ib}\SPACE{|}\ldots)$ used to model the event yield in each
        bin $b$ of the variable fit to extract the signal yield for each category $i$;

  \item[(ii)] Poisson function $f(N_l\SPACE{|}\Sigma_k\,\fNorm_{k}\,B_{kl})$ used to model the
        event yield in each control region $l$ with the total background yield summed
        over processes $k$ ($B_{kl}$);

  \item[(iii)] Gaussian functions $g(\Nuipar_t\SPACE{|}\nuipar_t)$ used to model the systematic
        uncertainties $t$; and

  \item[(iv)] Poisson functions $f(\fBvarsignif_k\SPACE{|}\ldots)$ used to account for the MC
        statistics $k$.
\end{itemize}

The statistical uncertainties are considered explicitly in the
first, second, and fourth terms.  The first and second terms treat the random
error associated with the predicted value, \ie, for a background yield
estimate $B$ the $\sqrt{B}$ error associated with it.  The fourth term treats
the sampling error associated with the finite sample size used for the
prediction, \eg, the $\sqrt{N_{\scMC}}$ ``MC statistical errors'' when MC is used.  All of the
terms are described below and summarized in Eq.~(\ref{eqn:likelihood}).

The first term of $\likelihood$ is a
Poisson function $f$ for the probability of observing $N$ events given
$\lambda$ expected events, $f(N\SPACE{|}\lambda){\EQ}e^{-\lambda}\lambda^N/N!$.
The expected value $\lambda$ is the sum of event yields from signal ($S$) and
the sum of the background contributions ($\Sigma_k\,B_k$) in a given signal
region, \ie, $\lambda{\EQ}\sigmu{\CDOT}S{\PLUS}\Sigma_k\,B_k$.  The
parameter of interest, $\sigmu$, multiplies $S$; each background yield in the
sum is evaluated as described in Sec.~\ref{sec:bkg}.
In our notation, the yields are scaled by the response functions
$\responsefn$ that parametrize the impact of the systematic
uncertainties $\nuipar$.  The $\responsefn$ and $\nuipar$ are described in
more detail below when discussing the third term of $\likelihood$.

The second term constrains the background yields with Poisson components that
describe the profiled control regions.  Each term is of the form
$f(N_l\SPACE{|}\lambda_l)$ for a given CR labeled by $l$, where $N_l$ is the
number of observed events in $l$, \ie,
$\lambda_l{\EQ}\Sigma_k\,\fNorm_k{\CDOT}B_{kl}$ is the predicted yield in $l$,
$\fNorm_k$ is the normalization factor of background $k$, and $B_{kl}$ is the
MC or data-derived estimate of background $k$ in $l$.
The $\fNorm_k$ parameters are the same as those that appear in the first
Poisson component above.

The third term constrains the systematic uncertainties with Gaussian
terms.  Each term is of the form
$g(\Nuipar\SPACE{|}\nuipar){\EQ}e^{-(\Nuipar-\nuipar)^2/2}/\sqrt{2\pi}$,
where $\Nuipar$ represents the central value of the measurement and $\nuipar$ the associated
nuisance parameter for a given systematic uncertainty.
The effect of the systematic uncertainty on the yields
is through an exponential response function
$\responsefn(\nuipar){\EQ}(1{\PLUS}\epsilon)^{\nuipar}$ for normalization
uncertainties that have no variations among bins $b$ of the fit variable,
where $\epsilon$ is the value of the uncertainty in question.
In this case, $\responsefn$ follows a log-normal distribution \cite{LHCstat}.
In this notation, $\epsilon{\EQ}3\%$ is written if the uncertainty that
corresponds to one standard deviation affects the associated yield by
${\pm\,}3\%$ and corresponds to $\nuipar{\EQ}{\pm\,}1$, respectively.

For the cases where the systematic uncertainty affects a given
distribution differently in each bin $b$, a different linear response function
is used in each bin; this function is written as $\responsefn_b(\nuipar){\EQ}1{\PLUS}\epsilon_b{\CDOT}\nuipar$.
In this case, $\responsefn_b$ is normally distributed
around unity with width $\epsilon_b$, and is truncated by the
$\responsefn_b{\GT}0$ restriction to avoid unphysical values.
Both types of response function impact the predicted $S$ and
$B_k$ in the first Poisson component.

The fourth term treats the sample error due to the finite sample size~\cite{Barlow:1993dm},
\eg, the sum of the number of generated MC events for all background processes, $B{\EQ}\Sigma_k\,B_k$.
The quantity $B$ is constrained with a Poisson term
$f(\fBsignif\SPACE{|}\lambda)$, where $\fBsignif$ represents the central
value of the background estimate and $\lambda{\EQ}\fBvarsignif{\CDOT}\nuipar$.
The $\fBvarsignif{\EQ}(B/\delta)^2$ defines the quantity with the statistical uncertainty of $B$ as $\delta$.
For instance,
if a background yield estimate $B$ uses $\NMC$ MC events that correspond to
a data sample with effective luminosity $L_\MC$, then for a data-to-MC
luminosity ratio $r{\EQ}L_{\rm data}/L_{\MC}$ the background estimate is
$B{\EQ}r{\CDOT}\NMC$, and the uncertainty (parameter) in question is
$\delta{\EQ}r{\CDOT}\sqrt{\NMC}$ ($\fBvarsignif{\EQ}\NMC$).
In this example,
the Poisson function is evaluated at $\NMC$ given $\lambda{\EQ}\nuipar{\CDOT}\NMC$.
Similar to the case for the third term, a linear response
function $\responsefn(\nuipar){\EQ}\nuipar$ impacts the predicted $S$ and $B_k$
in the first Poisson component.

In summary, the likelihood is the product of the four above-mentioned
terms and can be written schematically as done in Eq.~(\ref{eqn:likelihood}),
where the $\responsefn_{\!br}$ and $\responsefn_{\!bs}$
are implicitly products over all three types of response functions---normalization, shape of the distribution, and finite MC sample size---whose
parameters are constrained by the second, third, and fourth terms, respectively.
In the case of
finite MC sample size, $\nuipar$ is unique to each bin, which is not
shown in Eq.~(\ref{eqn:likelihood}).
The statistical treatments of two quantities---the $\ZDYll$ estimate in $\NjetLEone$ and the top-quark
estimate in $\NjetEQone$---are constrained with additional multiplicative terms
in the likelihood (see Appendix~\ref{sec:appendix_stat}).

To determine the observed value of the signal strength, $\sigmu_\obs$, the likelihood is maximized with
respect to its arguments, $\sigmu$ and $\nuipars$, and evaluated at
$\Nuipar{\EQ}0$ and $\fBsignif{\EQ}\fBvarsignif$.

\subsubsection{\boldmath\textit{\textbf{Test statistic}}\label{sec:systematics_fit_cls}}

The profiled likelihood-ratio test statistic~\cite{asimov} is used to test the background-only
or background-and-signal hypotheses. It is defined as
\begin{equation}
  q(\sigmu) = -2 \ln
  \frac{\likelihood(\sigmu, \nuipars)}{\Lmax}
  {\bigg|}_{\footnotesize\nuipars{\EQ}\hatnuiparsmu},
\label{eqn:qmu}
\end{equation}
and it is also written as $\qmu$; the argument of the logarithm is written as $\Lambda$ in later plots.
The denominator of Eq.~(\ref{eqn:qmu}) is unconditionally maximized over all possible values of $\sigmu$ and $\nuipars$, while the numerator is maximized over $\nuipars$ for a con\
ditional value of $\sigmu$.
The latter takes the values $\hatnuiparsmu$, which are $\nuipars$ values that maximize $\likelihood$ for a
given value of $\sigmu$. When the denominator is maximized, $\sigmu$ takes the value of $\hatsigmu$.

The $\pzero$ value is computed for the test statistic $\qzero$, Eq.~(\ref{eqn:qmu}), evaluated at $\sigmu{\EQ}0$,
and is defined to be the probability to obtain a value of $\qzero$ larger than
the observed value under the background-only hypothesis. There are no boundaries on $\hatsigmu$, although $\qzero$ is defined to be negative if $\hatsigmu{\LE}0$.
All $\pzero$ values are computed using the asymptotic approximation that $-2 \ln \Lambda(\sigmu)$
follows a $\chi^2$ distribution~\cite{asimov}.

Local significance is defined as the one-sided tail of a Gaussian distribution,
$Z0{\EQ}\sqrt{2}\,\textrm{erf}^{-1}(1{\MINUS}2\,p_0)$.

A modified frequentist method known as $\CLs$~\cite{CLs_2002} is used to compute
the one-sided $95\%$ confidence level ({C.L.}) exclusion regions.

\subsubsection{\boldmath\textit{\textbf{Combined fit}}\label{sec:systematics_fit_combo}}

The combined results for the $7$ and $8\TeV$ data samples account for the
correlations between the analyses due to common systematic uncertainties.

The correlation of all respective nuisance parameters is assumed to be $100\%$
except for those that are statistical in origin or have a different source for
the two data sets.  Uncorrelated systematics include the statistical component
of the jet energy scale calibration and the luminosity uncertainty.
All theoretical uncertainties are treated as correlated.

\subsection{\boldmath Systematic uncertainties \label{sec:systematics_sources}}

Uncertainties enter the fit as nuisance parameters in the likelihood
function [Eq.~(\ref{eqn:likelihood})].  Uncertainties (both theoretical
and experimental) specific to individual processes are described in
Secs.~\ref{sec:signal} and~\ref{sec:bkg}; experimental uncertainties
common to signal and background processes are described in this
subsection. The impact on the yields and distributions from both sources of
uncertainty is also discussed.

\subsubsection{\boldmath\textit{\textbf{Sources of uncertainty}}\label{sec:systematics_exp}}

The dominant sources of experimental uncertainty on
the signal and background yields are the jet
energy scale and resolution, and the $b$-tagging efficiency. Other sources of uncertainty are the
lepton resolutions and identification and trigger efficiencies, missing transverse momentum
measurement, and the luminosity calculation. The uncertainty on the integrated luminosity in the $8\TeV$ data analysis
is $2.8\%$. It is derived following the same methodology as in Ref.~\cite{luminosity}, from a preliminary
calibration of the luminosity scale derived from beam-separation
scans. The corresponding uncertainty in the $7\TeV$ data analysis
is $1.8\%$.

The jet energy scale is determined from a combination of test beam, simulation,
and \insitu\ measurements~\cite{Aad:2014bia}. Its uncertainty is
split into several independent categories:
modeling and statistical uncertainties on the extrapolation
of the jet calibration from the central region ($\myeta$ intercalibration), high-$\pT$ jet behavior,
MC nonclosure uncertainties, uncertainties on the calorimeter
response and calibration of the jets originating from light quarks or gluons,
the $b$-jet energy scale uncertainties, uncertainties due to modeling of in-time and
out-of-time pile-up,
and uncertainties on \insitu\ jet energy corrections. All of these categories
are further subdivided by the physical source of the uncertainty.
For jets used in this analysis ($\pT{\GT}25\GeV$ and
$\ABS{\myeta}{\LE}4.5$), the jet energy scale uncertainty ranges from
$1\%$ to $7\%$, depending on $\pT$ and $\myeta$.
The jet energy resolution varies from $5\%$ to $20\%$ as a function of the jet $\pT$ and $\myeta$.
The relative uncertainty on the resolution, as determined from \insitu\ measurements, ranges from
$2\%$ to $40\%$, with the largest value of the resolution and relative uncertainty occurring at
the $\pT$ threshold of the jet selection.

The method used to evaluate the $b$-jet tagging efficiency uses a sample dominated by dileptonic decays of top-quark pairs.
This method is based on a likelihood fit to the data, which
combines the per-event jet-flavor information and the expected momentum correlation between the
jets to allow the $b$-jet tagging efficiency to be measured to
high precision~\cite{ATLAS-CONF-2014-046}. To further improve the precision,
this method is combined with a second calibration method, which is
based on samples containing muons reconstructed in the vicinity of the
jet. The uncertainties related to $b$-jet identification are
decomposed into six uncorrelated components using an
eigenvector method~\cite{ATLAS-CONF-2014-004}. The number of components is
equal to the numbers of $\pT$ bins used in the calibration, and the
uncertainties range from ${\LT}1\%$ to $7.8\%$. The uncertainties
on the misidentification rate for light-quark jets depend on $\pT$ and $\eta$,
and have a range of $9$--$19\%$.
The uncertainties on $c$-jets reconstructed as
$b$-jets range between $6\%$ and $14\%$ depending on $\pT$ only.

The reconstruction, identification, isolation, and trigger efficiencies for electrons and muons, as well
as their momentum scales and resolutions, are estimated
using $Z{\TO}ee, \mu\mu$, $\Jpsi{\TO}ee, \mu\mu$, and $W{\TO}e\nu, \mu\nu$ decays~\cite{MCPpaper2014,ElectronEff2012}.
The uncertainties on the lepton and trigger
efficiencies are smaller than $1\%$ except for the uncertainty on the electron identification effficiency,
which varies between $0.2\%$ and $2.7\%$ depending on $\pT$ and $\eta$, and the uncertainties on the isolation
efficiencies, which are the largest for $\pT{\LT}15\GeV$ and yield $1.6\%$ and $2.7\%$ for electrons and muons, respectively.

The changes in jet energy and lepton momenta due to varying them by their systematic
uncertainties are propagated to $\MET$; the changes in the
high-$\pT$ object momenta and in $\MET$ are,
therefore, fully correlated~\cite{met-perf}.
Additional contributions to the $\MET$ uncertainty arise from the modeling of low-energy particle measurements
(soft terms). In the calorimeter, these particles are measured as calibrated clusters of cells that are above a noise threshold but not associated
with reconstructed physics objects.
The longitudinal and perpendicular (with respect to the hard
component of the missing transverse momentum) components of the soft terms are fit with Gaussian functions in data and
MC DY samples in order to assess the associated uncertainties.  The
uncertainties are parametrized as a function of the magnitude of the summed
$\vpT$ of the high-$\pT$ objects, and they are evaluated in bins of the average number of interactions per bunch crossing.
Differences of the mean and width of the soft term
components between data and MC result in variations on the mean of the
longitudinal component of about $0.2\GeV$, while the uncertainty on the
resolution of the longitudinal and perpendicular components is $2\%$ on average.

Jet energy and lepton momentum scale uncertainties are also propagated to the $\MPTj$
calculation. The systematic uncertainties related to the track-based
soft term are based on the balance between tracks not
associated with charged leptons and jets and the total transverse
momentum of the hard objects in the event.
These uncertainties are calculated by comparing the properties of $\MPTj$ in $Z{\TO}ee, \mu\mu$
events in real and simulated data, as a function of the magnitude of
the summed $\vpT$ of the hard $\pT$ objects in the event. The
variations on the mean of the longitudinal component
are in the range $0.3$--$1.4\GeV$ and the uncertainties on the resolution on
the longitudinal and perpendicular components are in the range
$1.5$--$3.3\GeV$, where the lower and upper bounds correspond to the
range of the sum of the hard $\pT$ objects below $5\GeV$ and above
$50\GeV$, respectively.

\subsubsection{\boldmath\textit{\textbf{Impact on yields and distributions}}\label{sec:systematics_tables}}

\begin{table*}[t!]
\caption{
  Sources of systematic uncertainty (in $\%$) on the predicted signal yield ($\Nsig)$ and the
  cumulative background yields ($\Nbkg$).
  Entries marked with a dash (-) indicate that the corresponding uncertainties either do not apply or are less than
  $0.1\%$.
  The values are postfit and given for the $8\TeV$ analysis.
}
\label{tab:syst_yield}
{\small
  \centering
\begin{tabular*}{0.960\textwidth}{p{0.420\textwidth} cccc}
\dbline
&\multicolumn{1}{p{0.110\textwidth}}{~~~~~$\NjetEQzero$}
&\multicolumn{1}{p{0.110\textwidth}}{~~~~~$\NjetEQone$}
&\multicolumn{1}{p{0.110\textwidth}}{$\NjetGEtwo$ ggF}
&\multicolumn{1}{p{0.110\textwidth}}{$\NjetGEtwo$ VBF}
\\
\sgline
\multicolumn{4}{l}{$\!$(a) Uncertainties on $\Nsig$ (in $\%$)}\\
\clineskip\clineskip
ggF $H$, jet veto for $\NjetEQzero$, $\epsilon_0\nqq$           &  8.1  &   14  &   12  & -     \\
ggF $H$, jet veto for $\NjetEQone$, $\epsilon_1\nqq$            &   -   &   12  &   15  & -     \\
ggF $H$, $\NjetGEtwo$ cross section                             &   -   &   -   &   -   & 6.9   \\
ggF $H$, $\NjetGEthree$ cross section                           &   -   &   -   &   -   & 3.1   \\
ggF $H$, total cross section                                    &  10  &  9.1  &  7.9  & 2.0    \\
ggF $H$ acceptance model                                        &  4.8  &  4.5  &  4.2  & 4.0   \\
VBF $H$, total cross section                                    &   -   &  0.4  &  0.8  & 2.9   \\
VBF $H$ acceptance model                                        &   -   &  0.3  &  0.6  & 5.5   \\
$\HWW$ branch.\ fraction$\nqq$                                  &  4.3  &  4.3  &  4.3  & 4.3   \\
Integrated luminosity                                           &  2.8  &  2.8  &  2.8  & 2.8   \\
Jet energy scale \& reso.$\nqq$                                 &  5.1  &  2.3  &  7.1  & 5.4   \\
$\MPTj$ scale \& resolution                                     &  0.6  &  1.4  &  0.1  & 1.2   \\
$\frecoil$ efficiency                                           &  2.5  &  2.1  &  -    & -     \\
Trigger efficiency                                              &  0.8  &  0.7  &  -    & 0.4   \\
Electron identification, isolation, reconstruction eff.         &  1.4  &  1.6  &  1.2  & 1.0   \\
Muon identification, isolation, reconstruction eff.             &  1.1  &  1.6  &  0.8  & 0.9   \\
Pile-up model                                                   &  1.2  &  0.8  &  0.8  & 1.7   \\
\clineskip
\sgline
\multicolumn{4}{l}{$\!$(b) Uncertainties on $\Nbkg$ (in $\%$)}\\
\clineskip\clineskip
$WW$ theoretical model                                          & 1.4  &  1.6  &  0.7  &  3.0   \\
Top theoretical model                                           &  -   &  1.2  &  1.7  &  3.0   \\
$VV$ theoretical model                                          &  -   &  0.4  &  1.1  &  0.5   \\
$\ZDYtt$ estimate                                               & 0.6  &  0.3  &  1.6  &  1.6   \\
$\ZDYll$ est.\ in VBF                                           &  -   &   -   &   -   &  4.8   \\
$\Wj$ estimate                                                  & 1.0  &  0.8  &  1.6  &  1.3   \\
$\jj$ estimate                                                  & 0.1  &  0.1  &  1.8  &  0.9   \\
Integrated luminosity                                           &  -   &   -   &  0.1  &  0.4   \\
Jet energy scale \& reso.$\nqq$                                 & 0.4  &  0.7  &  0.9  &  2.7   \\
$\MPTj$ scale \& resolution                                     & 0.1  &  0.3  &  0.5  &  1.6   \\
$b$-tagging efficiency                                          &  -   &  0.2  &  0.4  &  2.0   \\
Light- and $c$-jet mistag                                       &  -   &  0.2  &  0.4  &  2.0   \\
$\frecoil$ efficiency                                           & 0.5  &  0.5  &   -   &   -    \\
Trigger efficiency                                              & 0.3  &  0.3  &  0.1  &   -    \\
Electron identification, isolation, reconstruction eff.         & 0.3  &  0.3  &  0.2  &  0.3   \\
Muon identification, isolation, reconstruction eff.             & 0.2  &  0.2  &  0.3  &  0.2   \\
Pile-up model                                                   & 0.4  &  0.5  &  0.2  &  0.8   \\
\dbline
\end{tabular*}
}
\end{table*}

In the likelihood fit, the experimental uncertainties are varied
in a correlated way across all backgrounds and all signal and control
regions, so that uncertainties on the extrapolation factors $\fAlpha$
described in Sec.~\ref{sec:bkg} are correctly propagated.
If the normalization uncertainties are less than
$0.1\%$ they are excluded from the fit. If the shape uncertainties
(discussed below) are less than $1\%$ in all bins, they are excluded
as well. Removing such small uncertainties increases the performance
and stability of the fit.

In the fit to the $\mTH$ distribution to extract the signal yield,
the predicted $\mTH$ shape from simulation is used for all of the
backgrounds except $\Wjets$ and multijets.
The impact of experimental uncertainties on the
$\mTH$ shapes for the individual backgrounds and signal are evaluated, and no
significant impact is observed for the majority of the experimental uncertainties.
Those experimental uncertainties that do produce statistically significant variations
of the shape have no appreciable effect on the final results, because
the uncertainty on the $\mTH$ shape of the total background is dominated by the
uncertainties on the normalizations of the individual backgrounds. The
theoretical uncertainties on the $\WW$ and $\Wgs$ $\mTH$ shape are
considered in the $\NjetLEone$ categories, as discussed in
Secs.~\ref{sec:bkg_ww_01j} and~\ref{sec:bkg_vv}. In the $\NjetGEtwo$
ggF-enriched category, only the theoretical uncertainties on the top-quark
$\mTH$ shape are included (see Sec.~\ref{sec:bkg_top_2j}).

\begin{table}[t!]
\caption{
  Composition of the postfit uncertainties (in $\%$) on the total signal ($\Nsig$),
  total background ($\Nbkg$), and individual background yields in the signal
  regions.
  The total uncertainty (Total) is decomposed into three
  components: statistical (Stat), experimental (Expt) and
  theoretical (Theo).  Entries marked with
  a dash (-) indicate that the corresponding uncertainties either do not
  apply or are less than $1\%$.
  The values are given for the $8\TeV$ analysis.
}
\label{tab:syst_bkg}
\centering
{\small
  \centering
\begin{tabular*}{0.480\textwidth}{
    p{0.120\textwidth}
    d{1}
    d{1}
    d{1}
    d{1}
}
\dbline
\multicolumn{1}{l}{Sample} &
\multicolumn{1}{p{0.070\textwidth}}{~~~~Total} &
\multicolumn{1}{p{0.070\textwidth}}{~~~~Stat$\nq$} &
\multicolumn{1}{p{0.070\textwidth}}{~~~~~~~~Expt} &
\multicolumn{1}{p{0.070\textwidth}}{~~~~~~~~Theo} \\
&
\multicolumn{1}{l}{~~~~error}  &
\multicolumn{1}{l}{~~~~error}  &
\multicolumn{1}{l}{~~~~syst\,err$\np$}  &
\multicolumn{1}{l}{~~~~syst\,err$\np$} \\
\sgline
$\NjetEQzero$ \\
\quad $\Nsig$                   & 16  & \multicolumn{1}{r}{-~~~} & 6.7 & 15  \\
\quad $\Nbkg$                   & 2.5 & 1.5                      & 1.2 & 1.7 \\
\qquad $\NWW$                   & 4.2 & 2.4                      & 2.3 & 2.6 \\
\qquad $\Ntop$                  & 7.4 & 2.3 & 4.2 & 5.6 \\
\qquad $\Nfakes$                & 17  & \multicolumn{1}{r}{-~~~} & 9.9 & 14  \\
\qquad $\NVV$                   & 9.9 & 4.8                      & 4.6 & 7.4 \\
\qquad $\Ntautau$\,(DY)$\nqq$   & 34  & 1.7                      & 33  & 7.2 \\
\qquad $\Nll$\,(DY)$\nqq$       & 30  & 14                       & 26  & 5.5 \\
\clineskip\clineskip
$\NjetEQone$ \\
\quad $\Nsig$                   & 22  & \multicolumn{1}{r}{-~~~} & 5.3 & 22  \\
\quad $\Nbkg$                   & 3 & 1.7                      & 1.4 & 2.1 \\
\qquad $\NWW$                   & 7.7 & 5.5                      & 2.7 & 4.6 \\
\qquad $\Ntop$                  & 5   & 3.4                      & 2.9 & 2.3 \\
\qquad $\Nfakes$                & 18  & \multicolumn{1}{r}{-~~~} & 11  & 14  \\
\qquad $\NVV$                   & 14  & 8.9                      & 6.1 & 8.5 \\
\qquad $\Ntautau$\,(DY)$\nqq$   & 27  & 3.3                      & 26  & 6.3 \\
\qquad $\Nll$\,(DY)$\nqq$       & 39  & 27                       & 26  & 7.4 \\
\clineskip\clineskip
\multicolumn{3}{l}{$\NjetGEtwo$ ggF-enriched} \\
\quad $\Nsig$                   & 23  & \multicolumn{1}{r}{-~~~} & 8.6 & 22  \\
\quad $\Nbkg$                   & 4.2 & 1.5                      & 2.2 & 3.2 \\
\qquad $\NWW$                   & 20  & \multicolumn{1}{r}{-~~~} & 8.7 & 18  \\
\qquad $\Ntop$                  & 7.9 & 2.6                      & 3.4 & 6.7 \\
\qquad $\Nfakes$                & 29  & \multicolumn{1}{r}{-~~~} & 16  & 24  \\
\qquad $\NVV$                   & 32  & \multicolumn{1}{r}{-~~~} & 9.6 & 31  \\
\qquad $\Ntautau$\,(DY)$\nqq$   & 18  & 8                        & 13  & 10  \\
\qquad $\Nll$\,(DY)$\nqq$       & 15  &  \multicolumn{1}{r}{-~~~}  & 14  & 4 \\
\clineskip\clineskip
\multicolumn{3}{l}{$\NjetGEtwo$ VBF-enriched} \\
\quad $\Nsig$                   & 13  & \multicolumn{1}{r}{-~~~} & 6.8 & 12  \\
\quad $\Nbkg$                   & 9.2 & 4.7                      & 6.4 & 4.5 \\
\qquad $\NWW$                   & 32  & \multicolumn{1}{r}{-~~~} & 14  & 28  \\
\qquad $\Ntop$                  & 15  & 9.6                      & 7.6 & 8.5 \\
\qquad $\Nfakes$                & 22  & \multicolumn{1}{r}{-~~~} & 12  & 19  \\
\qquad $\NVV$                   & 20  & \multicolumn{1}{r}{-~~~} & 12  & 15  \\
\qquad $\Ntautau$\,(DY)$\nqq$   & 40  & 25 & 31  & 2.9 \\
\qquad $\Nll$\,(DY)$\nqq$       & 19  & 11   & 15  & \multicolumn{1}{r}{-~~~}\\
\dbline
\end{tabular*}
}
\end{table}

The $\bdt$ output distribution is fit in the $\NjetGEtwo$ VBF-enriched
category, and as with the $\mTH$ distribution its shape is taken
from the MC simulation, except for the $\Wjets$ and multijet background
processes. The theoretical uncertainties on the top-quark $\bdt$ shape are
included in the analysis, as described in Sec.~\ref{sec:bkg_top_2j}.

Table~\ref{tab:syst_yield}(a) shows the relative uncertainties on the
combined predicted signal yield, summed over all the lepton-flavor
channels, for each $\Njet$ category for the $8\TeV$
analysis. They represent the final postfit uncertainties on the estimated yields.
The first two entries show the perturbative uncertainties on the ggF
jet-bin acceptances in the exclusive $\NjetEQzeroone$ categories.
The following entries are specific to the
QCD scale uncertainties on the inclusive $\NjetGEtwo$ and $\NjetGEthree$
cross sections, and on the total cross section and the acceptance. The
latter includes the uncertainties due to the PDF variations, UE/PS and
generator modeling, as described in Table~\ref{tab:ggf_unc}. The
uncertainties on the VBF production process are also shown but
are of less importance. The dominant uncertainties on the signal
yields are theoretical. The uncertainties on the $\frecoil$ selection
efficiency (relevant to the $\ZDYll$ estimate in the $\NjetLEone$ categories) are applied only in the $\SFchan$ channels.

Table~\ref{tab:syst_yield}(b) shows the leading uncertainties on the
cumulative background yields for each $\Njet$ category. The first three entries are theoretical and apply to the $\WW$, top-quark, and $\VV$
processes (see Sec.~\ref{sec:bkg}). The remaining uncertainties arise from the modeling of specific backgrounds and
from experimental uncertainties.

Table~\ref{tab:syst_bkg} summarizes
the above postfit uncertainties on the total signal and backgrounds
yields. The uncertainties shown are divided into three categories: statistical, experimental and
theoretical. The statistical uncertainties are only relevant in the
cases where the background estimates rely on the data. For example,
the entry under $\NWW$ in $\NjetEQzero$ represents the uncertainty on
the sample statistics in the $\WW$ control region. The uncertainties on
$\Ntop$ in the $\NjetLEone$ categories also include the uncertainties on the
corrections applied to the normalization factors. The uncertainties
from the number of events in the control samples used to derive the
$\Wjets$ and multijet extrapolation factors are listed under the
experimental category, as discussed in Sec.~\ref{sec:bkg_misid}.
Uncertainties on the total $\Wjets$ estimate are reduced compared
to the values quoted in Table~\ref{tab:wj_ff}, because they are
largely uncorrelated between lepton $\pT$ bins (statistical
uncertainties on the $\Zjets$ data sample) and between the lepton
flavors (systematic uncertainties on the OC correction factor).
The uncertainty due to the limited sample of background MC events for all the considered
processes is included in the experimental component.

Background contamination in the control regions causes
anti-correlations between different background processes, resulting in an
uncertainty on the total background smaller than the sum in quadrature of the individual process uncertainties.
This effect is called ``cross talk'' and is most
prominent between the $WW$ and top-quark backgrounds in the $\NjetEQone$ category. The
uncertainties on the background estimates, as described in
Sec.~\ref{sec:bkg}, cannot be directly compared to the ones presented
in Table~\ref{tab:syst_bkg}. The latter uncertainties are postfit and
are subject to subtle effects, such as the cross talk mentioned above, and
also pulls and data-constraints (defined below) on the various nuisance parameters.

\subsection{\boldmath Checks of fit results \label{sec:systematics_pulls}}

\begin{table*}[btp!]
\caption{Impact on the signal strength $\hatsigmu$ from the prefit and postfit variations of the nuisance parameter uncertainties, $\Delta_\nuipar$.
  The $+$ ($-$) column header indicates the positive (negative)
  variation of $\Delta_{\nuipar}$ and
  the resulting change in $\hatsigmu$ is noted in the entry (the sign
  represents the direction of the change).
  The right-hand side shows the pull of $\nuipar$ and the data-constraint of $\Delta_\nuipar$.
  The pulls are given in units of standard deviations ($\sigma$) and $\Delta_\nuipar$ of ${\pm}1$ means no data-constraint.
  The rows are ordered by the size of a change in $\hatsigmu$ due to varying $\theta$ by the postfit uncertainty $\Delta_\nuipar$.
}
\label{tab:syst_pulls}
{\small
  \centering
\begin{tabular*}{1\textwidth}{l p{0.005\textwidth} l}
\dbline
\begin{tabular*}{0.800\textwidth}{
  l llll l
}
\clineskip
\clineskip
& \multicolumn{5}{c}{Impact on $\hatsigmu$}
\\
\clineskip
\cline{2-6}
\clineskip
\clineskip
\multicolumn{1}{p{0.400\textwidth}}{Systematic source}
& \multicolumn{2}{p{0.100\textwidth}}{~~Prefit~$\Delta_{\hatsigmu}$}
& \multicolumn{2}{p{0.100\textwidth}}{~~Postfit~$\Delta_{\hatsigmu}$}
& \multicolumn{1}{p{0.150\textwidth}}{Plot of postfit ${\PM}\Delta_{\hatsigmu}\nqq$}
\\
& \,~~~$+$ & \,~~~$-$
& \,~~~$+$ & \,~~~$-$
\\
\sgline
\balkenscale{135}{-10}
ggF $H$, PDF variations on cross section                                &$-0.06$ &$+0.06$ &$-0.06$ &$+0.06$ &{\myb\Balkenx{0}{6.2}{5.8}{0}{0}} \\
ggF $H$, QCD scale on total cross section                               &$-0.05$ &$+0.06$ &$-0.05$ &$+0.06$ &{\myb\Balkenx{0}{5.4}{5.9}{0}{0}} \\
$\WW$, generator modeling                                               &$-0.07$ &$+0.06$ &$-0.05$ &$+0.05$ &{\myb\Balkenx{0}{4.6}{4.6}{0}{0}} \\
Top quarks, generator modeling on $\fAlpha_{\rm top}$ in ggF cat.\      &$+0.03$ &$-0.03$ &$+0.03$ &$-0.03$ &{\myb\Balkenx{0}{2.9}{2.8}{0}{0}} \\
Misid.\ of $\mu$, OC uncorrelated corr.\ factor $\fAlpha_\fakes$, 2012  &$-0.03$ &$+0.03$ &$-0.03$ &$+0.03$ &{\myb\Balkenx{0}{2.8}{2.8}{0}{0}} \\
Integrated luminosity, 2012                                             &$-0.03$ &$+0.03$ &$-0.03$ &$+0.03$ &{\myb\Balkenx{0}{2.6}{2.8}{0}{0}} \\
Misid.\ of $e$, OC uncorrelated corr.\ factor $\fAlpha_\fakes$, 2012    &$-0.03$ &$+0.03$ &$-0.02$ &$+0.03$ &{\myb\Balkenx{0}{2.3}{2.5}{0}{0}} \\
ggF $H$, PDF variations on acceptance                                   &$-0.02$ &$+0.02$ &$-0.02$ &$+0.02$ &{\myb\Balkenx{0}{2.2}{2.2}{0}{0}} \\
Jet energy scale, $\myeta$ intercalibration                             &$-0.02$ &$+0.02$ &$-0.02$ &$+0.02$ &{\myb\Balkenx{0}{2.0}{2.2}{0}{0}} \\
VBF $H$, UE/PS                                                          &$-0.02$ &$+0.02$ &$-0.02$ &$+0.02$ &{\myb\Balkenx{0}{1.8}{2.2}{0}{0}} \\
ggF $H$, QCD scale on $\epsilon_1$                                      &$-0.01$ &$+0.03$ &$-0.01$ &$+0.03$ &{\myb\Balkenx{0}{1.3}{2.7}{0}{0}} \\
Muon isolation efficiency                                               &$-0.02$ &$+0.02$ &$-0.02$ &$+0.02$ &{\myb\Balkenx{0}{1.8}{2.0}{0}{0}} \\
$\VV$, QCD scale on acceptance                                          &$-0.02$ &$+0.02$ &$-0.02$ &$+0.02$ &{\myb\Balkenx{0}{1.9}{2.0}{0}{0}} \\
ggF $H$, UE/PS                                                          &~~~~-   &$-0.02$ &~~~~-   &$-0.02$ &{\myb\Balkenx{0}{0.0}{1.9}{0}{0}} \\
ggF $H$, QCD scale on acceptance                                        &$-0.02$ &$+0.02$ &$-0.02$ &$+0.02$ &{\myb\Balkenx{0}{1.6}{1.7}{0}{0}} \\
Light jets, tagging efficiency                                          &$+0.02$ &$-0.02$ &$+0.02$ &$-0.02$ &{\myb\Balkenx{0}{1.6}{1.6}{0}{0}} \\
ggF $H$, generator modeling on acceptance                               &$+0.01$ &$-0.02$ &$+0.01$ &$-0.02$ &{\myb\Balkenx{0}{1.4}{1.7}{0}{0}} \\
ggF $H$, QCD scale on $\NjetGEtwo$ cross section                        &$-0.01$ &$+0.02$ &$-0.01$ &$+0.02$ &{\myb\Balkenx{0}{1.4}{1.8}{0}{0}} \\
Top quarks, generator modeling on $\fAlpha_{\rm top}$ in VBF cat.\      &$-0.01$ &$+0.02$ &$-0.01$ &$+0.02$ &{\myb\Balkenx{0}{1.4}{1.6}{0}{0}} \\
Electron isolation efficiency                                           &$-0.02$ &$+0.02$ &$-0.02$ &$+0.02$ &{\myb\Balkenx{0}{1.5}{1.5}{0}{0}} \vspace{1.0mm}\\
                                                                        &        &        &        &        &
\renewcommand{\bcfontstyle}{\bfseries}%
\renewcommand{\bcfontstyle}{}%
\hspace{-13.4pt}%
\begin{bchart}[step=0.05,min=-0.1,max=0.1,width=0.150\textwidth,scale=1.0]\end{bchart}%
\hspace{-5.0mm}%
\end{tabular*}
&
&
\begin{tabular*}{0.180\textwidth}{ll}
\multicolumn{2}{c}{Impact on $\hatnuipar$}
\\
\clineskip
\cline{1-2}
\clineskip
  \multicolumn{1}{p{0.080\textwidth}}{Pull,}
& \multicolumn{1}{p{0.080\textwidth}}{Constr.,}
\\
$\hat{\nuipar}$ ($\sigma$)
& $\Delta_{\nuipar}$
\\
\sgline
          $-0.06$ & ${\pm\,}1$ \\
          $-0.05$ & ${\pm\,}1$ \\
$\phantom{+}0$ & ${\pm\,}0.7$  \\
          $-0.40$ & ${\pm\,}0.9$ \\
$\phantom{+}0.48$ & ${\pm\,}0.8$\\
$\phantom{+}0.08$ & ${\pm\,}1$ \\
          $-0.06$ & ${\pm\,}0.9$  \\
          $-0.03$ & ${\pm\,}1$ \\
$\phantom{+}0.45$ & ${\pm\,}0.95$ \\
$\phantom{+}0.26$ & ${\pm\,}1$ \\
          $-0.10$ & ${\pm\,}0.95$ \\
$\phantom{+}0.13$ & ${\pm\,}1$ \\
$\phantom{+}0.09$ & ${\pm\,}1$  \\
$\phantom{+}0$    & ${\pm\,}0.9$ \\
$\phantom{+}0$    & ${\pm\,}1$ \\
$\phantom{+}0.21$ & ${\pm\,}1$  \\
$\phantom{+}0.10$ & ${\pm\,}1$ \\
          $-0.04$ & ${\pm\,}1$ \\
          $-0.16$ & ${\pm\,}1$  \\
          $-0.14$ & ${\pm\,}1$ \vspace{1.0mm}\\
\vspace{2.0mm}
\end{tabular*}
\vspace{-1.0mm}
\\
\dbline
\end{tabular*}
}
\end{table*}

The fit simultaneously extracts the signal strength $\sigmu$ and the set of
auxiliary parameters $\nuipars$. This process adjusts the initial prefit
estimation of every parameter $\nuipar$ as well as its uncertainty,
$\Delta_\nuipar$. The fit model is designed to
avoid any significant constraints on the input uncertainties
to minimize the assumptions on the correlations
between the phase spaces in which they are measured and applied.
This is achieved by having mostly single-bin control regions.
Of central importance is the prefit and postfit comparison of how
the variation of a given systematic source translates to an uncertainty on $\sigmu$.

The impact of a single nuisance parameter $\nuipar$ is assessed by considering its
effect on the signal strength, \ie,
\begin{equation}
\Delta_{\hatsigmu,\pm} = \hatsigmu(\hatnuipar\pm\Delta_{\nuipar}) - \hatsigmu(\hatnuipar),
  \label{eqn:impact_mu}
\end{equation}
where $\hatsigmu$ is the postfit value of the signal strength.  In the
following, quantities with a hat represent postfit parameter values
or their uncertainties.

The values $\hatsigmu(\hatnuipar\pm\Delta_{\nuipar})$ are the result of a fit with one
$\nuipar$ varied by $\pm\Delta_{\nuipar}$ around the postfit value for $\nuipar$,
namely $\hatnuipar$. All other $\nuipar$ are floating in these
fits. In the prefit scenario, the $\Delta_{\nuipar}$ are taken as their
prefit values of $\pm$ 1, as $\nuipar$ is constrained by a unit
Gaussian. The postfit scenario is similar, but with $\hatnuipar$ varied
by its postfit uncertainty of $\Delta_{\hatnuipar}$. This uncertainty is found by a scan about the maximum so that
the likelihood ratio takes the values
$-2\ln(\likelihood(\hatnuipar{\PM}\Delta_{\hatnuipar})/\likelihood(\hatnuipar)){\EQ}1$. The
corresponding impact on $\hatsigmu$ is $\Delta_{\hatsigmu}$.

When $\Delta_{\nuipar}$ is less than the prefit value, $\nuipar$
is said to be data constrained. In this case the systematic
uncertainty is reduced below its input value given the information
from the data. This can result from
the additional information that the data part of the likelihood
injects. As can be seen from Table~\ref{tab:syst_pulls}, only a few of
the uncertainties are data constrained, and only one of them is data constrained
by more than $20\%$. That is the $WW$ generator modeling that includes
the $\mTH$ shape uncertainties correlated with the uncertainties on the
extrapolation factor $\fAlpha_{\WW}$. The data-constraint in this case comes from the
high-$\mTH$ tail of the signal region, which contains a large fraction of $WW$ events.

The postfit values for $\nuipar$ modify the rates of signal and background processes,
and the data-constraints affect the corresponding uncertainties. The results of
these shifts are summarized in Table~\ref{tab:syst_pulls} for a set of $20$
nuisance parameters ordered by the magnitude of $\Delta_{\hatsigmu}$ (Higgs signal hypothesis
is taken at $\mH{\EQ}125\GeV$).
The highest-ranked nuisance parameter is the uncertainty on the total ggF cross
section due to the PDF variations. It changes $\hatsigmu$ by
${-0.06}/{+0.06}$ when varied up and down by $\Delta_\nuipar$,
respectively. It is followed by the uncertainty on the total ggF cross section due to QCD scale variations and
$\WW$ generator modeling uncertainty.
Other uncertainties that have a significant impact on $\hatsigmu$ include
the effects of generator modeling on $\fAlpha_{\rm top}$, the
systematic uncertainties on $\fAlpha_\fakes$ originating from a correction for
oppositely charged electrons and muons, the luminosity determination for $8\TeV$ data, and
various theoretical uncertainties on the ggF and VBF signal production processes.
In total there are 253 nuisance parameters which are divided into three main categories:
experimental uncertainties (137 parameters), theoretical uncertainties (72 parameters)
and normalisation uncertainties (44 parameters). They are further divided into more categories as shown in Table~\ref{tab:syst_mu}.

\section{\boldmath Yields and distributions \label{sec:yields}}

The previous section described the different parameters of
the simultaneous fit to the various signal categories defined in the
preceding sections. In particular, the signal and background rates and shapes are allowed to vary
in order to fit the data in both the signal and control regions, within
their associated uncertainties.

\begin{table*}[t!]
\caption{
  Signal region yields with uncertainties.
  The tables give the ggF- and VBF-enriched postfit yields for each
  $\Njet$ category, separated for the $8$ and $7\TeV$ data analyses.
  The $N_{\rm signal}$ columns show the expected signal yields from the ggF
  and VBF production modes, with values scaled to the observed combined signal strength (see Sec.~\ref{sec:results_mu}).
  For each group separated by a horizontal line, the first line gives
  the combined values for the different subchannels or BDT bins.
  The yields and the
  uncertainties take into account the pulls and data-constraints of the
  nuisance parameters, and the correlations between the channels and
  the background categories.
  The quoted uncertainties include
  the theoretical and experimental systematic sources and those due to sample statistics.
  Values less than $0.1$ ($0.01$) events are written as $0.0$ (-).
}
\label{tab:sr_summary}
{\small
  \centering
\begin{tabular*}{1\textwidth}{
  lr
  r@{${\PM}$}l
  r@{${\PM}$}l
  r@{${\PM}$}l
  p{0.005\textwidth}
  p{0.001\textwidth}
  r@{${\PM}$}l
  r@{${\PM}$}l
  r@{${\PM}$}l
  r@{${\PM}$}l
  r@{${\PM}$}l
  r@{${\PM}$}l
  r@{${\PM}$}l
  l
}
\dbline
& \multicolumn{7}{c}{Summary}
&& \multicolumn{14}{c}{Composition of $\Nbkg$}
\\
\clineskip\cline{2-8}\cline{10-24}\clineskip
\multicolumn{1}{p{0.095\textwidth}}{Channel}
& \multicolumn{1}{p{0.055\textwidth}}{~$\Nobs$}
& \multicolumn{2}{p{0.074\textwidth}}{~~~$\Nbkg$}
& \multicolumn{4}{c}{$N_{\rm signal}$}
&
&
& \multicolumn{2}{p{0.064\textwidth}}{~$\NWW$}
& \multicolumn{4}{l}{~~~~~~~~~~$\Ntop$}
& \multicolumn{4}{l}{~~~~~~~~$\Nfakes$}
& \multicolumn{2}{p{0.060\textwidth}}{~~$\NVV$}
& \multicolumn{2}{p{0.060\textwidth}}{~~~~$\Ndy$}
\\
& \multicolumn{2}{c}{}
& \multicolumn{2}{p{0.074\textwidth}}{~~~~~~~~~~~~~$\NggF$}
& \multicolumn{2}{p{0.074\textwidth}}{~~~~~~~$\NVBF$}
&
&
&
& \multicolumn{2}{c}{}
& \multicolumn{2}{p{0.074\textwidth}}{~~~~~~~$\Nt$}
& \multicolumn{2}{p{0.074\textwidth}}{~$\Nttbar$}
& \multicolumn{2}{p{0.064\textwidth}}{~~~~$\NWj$}
& \multicolumn{2}{p{0.064\textwidth}}{~~~$\Njj$}
& \multicolumn{2}{c}{}
& \multicolumn{2}{c}{}
&
\\
\sgline
\multicolumn{10}{l}{$\!$(a) $8\TeV$ data sample}\\
\clineskip\clineskip
\multicolumn{1}{l}{$\NjetEQzero$}    &   3750\Z&$\no$3430&90 &300 &50 &8\Z &4  &&&$\np$2250&95 &$\no$112 &9  &$\no$195 &15 &$\no$360&60 &$\no$16\Z&5   &420 &40  &78      &21 \\
\quad $\DFchan$, $\ell_2{\EQ}\mu$    &   1430\Z&$\no$1280&40 &129 &20 &3.0 &2.1&&&$\np$830 &34 &$\no$41  &3  &$\no$73  &6  &$\no$149&29 &$\no$10.1&3.6 &167 &21  &14      &2.4\\
\quad $\DFchan$, $\ell_2{\EQ}e$      &   1212\Z&$\no$1106&35 & 97 &15 &2.5 &0.6&&&$\np$686 &29 &$\no$33  &3  &$\no$57  &5  &$\no$128&31 &$\no$3.8 &1.5 &184 &23  &14      &2.4\\
\quad$\SFchan$                       &   1108\Z&$\no$1040&40 & 77 &15 &2.4 &1.7&&&$\np$740 &40 &$\no$39  &3  &$\no$65  &5  &$\no$ 82&16 &$\no$2\Z &0.5 &68  & 7  &50      &21 \\
\clineskip\clineskip
\multicolumn{1}{l}{$\NjetEQone$}     &   1596\Z&$\no$1470&40 &102 &26 &17\z&5  &&&$\np$630 &50 &$\no$150 &10 &$\no$385 &20 &$\no$108&20 &$\no$8.2 &3.0 &143 &20  &51\z    &13 \\
\quad $\DFchan$, $\ell_2{\EQ}\mu$    &    621\Z&$\no$569 &19 & 45 &11 &7.4 &2  &&&$\np$241 &20 &$\no$58  &4  &$\no$147 &7  &$\no$51 &11 &$\no$5.7 &2.0 & 53 &10  &13.8    &3.3\\
\quad $\DFchan$, $\ell_2{\EQ}e$      &    508\Z&$\no$475 &18 & 35 &9  &6.1 &1.4&&&$\np$202 &17 &$\no$45  &3  &$\no$119 &6  &$\no$37 &9  &$\no$2.3 &0.9 & 60 &10  &9.3     &2.5\\
\quad$\SFchan$                       &    467\Z&$\no$427 &21 & 22 &6  &3.6 &1.8&&&$\np$184 &15 &$\no$46  &4  &$\no$119 &10 &$\no$19 &4  &$\no$0.2 &0.1 & 31 & 4  &28\z    &12 \\
\clineskip\clineskip
\multicolumn{1}{l}{$\NjetGEtwo$,\,ggF\,$\DFchan\nq$}
                                     &   1017\Z&$\no$960 &40 & 37 &11 &13\z&1.4&&&$\np$138 &28 &$\no$56  &5  &$\no$480 &40 &$\no$54 &25 &$\no$62  &22  &56  &18  &117     &21 \\
\clineskip\clineskip
\multicolumn{1}{l}{$\NjetGEtwo$, VBF}&    130\Z&$\no$99\z&9  &7.7 &2.6&21\z&3  &&&$\np$11\z&3.5&$\no$5.5 &0.7&$\no$29\z&5  &$\no$4.7&1.4&$\no$2.8 &1.0 &4.4 &0.9 &38\z    &7  \\
\quad$\DFchan$ bin~1                 &     37\Z&$\no$36\z&4  &3.3 &1.2&4.9 &0.5&&&$\np$5.0 &1.5&$\no$3.0 &0.6&$\no$15.6&2.6&$\no$3.2&1.0&$\no$2.3 &0.8 &2.3 &0.7 &3.6     &1.5\\
\quad$\DFchan$    bin~2              &     14\Z&$\no$6.5 &1.3&1.4 &0.5&4.9 &0.5&&&$\np$1.7 &0.7&$\no$0.3 &0.4&$\no$2.0 &1.0&$\no$0.4&0.1&$\no$0.3 &0.1 &0.7 &0.2 &0.6     &0.2\\
\quad$\DFchan$    bin~3              &      6\Z&$\no$1.2 &0.3&0.4 &0.3&3.8 &0.7&&&$\np$0.3 &0.1&$\no$0.1 &0.0&$\no$0.3 &0.1&\mcolz      &\mcolz        &0.1 &0.0 &0.2     &0.1\\
\quad$\SFchan$~bin\,1   $\no$        &$\np$53\Z&$\no$46\z&6  &1.7 &0.6&2.6 &0.3&&&$\np$3.1 &1.0&$\no$1.7 &0.3&$\no$10.1&1.6&$\no$0.9&0.2&$\no$0.2 &0.1 &1.0 &0.3 &28\z    &5  \\
\quad$\SFchan$~\,bin\,2$\no$         &$\np$14\Z&$\no$8.4 &1.8&0.7 &0.3&3.0 &0.4&&&$\np$0.9 &0.3&$\no$0.3 &0.2&$\no$1.2 &0.5&$\no$0.2&0.1&\mcolz        &0.3 &0.1 &5.2     &1.7\\
\quad$\SFchan$~\,bin\,3$\no$         &$\np$ 6\Z&$\no$1.1 &0.4&0.2 &0.2&2.1 &0.4&&&$\np$0.1 &0.1&$\no$0.1 &0.0&$\no$0.2 &0.1&\mcolz      &\mcolz        &\mcolz   &0.5     &0.3\\
\clineskip\clineskip
\sgline
\multicolumn{10}{l}{$\!$(b) $7\TeV$ data sample}\\
\clineskip\clineskip
\multicolumn{1}{l}{$\NjetEQzero$}    &   594\Z &$\no$575 &24 &49  &8  &1.4 &0.2&&&$\np$339 &24 &$\no$20.5&2.1&$\no$38  &4  &$\no$74 &15 &$\no$1.3 &0.6 &79  & 10 &23\z    &6  \\
\quad $\DFchan$, $\ell_2{\EQ}\mu$    &   185\Z &$\no$186 &8  &19  &3  &0.5 &0.0&&&$\np$116 &8  &$\no$7\Z &1  &$\no$14  &2  &$\no$19 &5  &\mcolz        &24  & 3  &4.8     &1  \\
\quad $\DFchan$, $\ell_2{\EQ}e$      &   195\Z &$\no$193 &12 &15  &2.4&0.5 &0.0&&&$\np$95  &7  &$\no$5.3 &0.5&$\no$10  &1  &$\no$37 &9  &$\no$1.1 &0.5 &41  & 6  &4.1     &0.9\\
\quad$\SFchan$                       &   214\Z &$\no$196 &11 &16  &3.1&0.5 &0.1&&&$\np$128 &10 &$\no$8\Z &1  &$\no$14  &2  &$\no$18 &4  &$\no$0.2 &0.1 &14  & 2  &14\z    &5  \\
\clineskip\clineskip
\multicolumn{1}{l}{$\NjetEQone$}     &   304\Z &$\no$276 &15 &16\z&4  &3.2 &0.3&&&$\np$103 &15 &$\no$22  &2  &$\no$58  &6  &$\no$20 &4  &$\no$3.2 &1.6 &32  & 8  &38\z    &6  \\
\quad $\DFchan$, $\ell_2{\EQ}\mu$    &    93\Z &$\no$75  &4  &5.7 &1.6&1.2 &0.1&&&$\np$33  &5  &$\no$7   &1  &$\no$18  &2  &$\no$5  &1  &\mcolz        &9   & 2  &2.7     &0.4\\
\quad $\DFchan$, $\ell_2{\EQ}e$      &    91\Z &$\no$76  &5  &4.5 &1.2&0.9 &0.1&&&$\np$28  &4  &$\no$6   &1  &$\no$16  &2  &$\no$10 &2  &$\no$0.7 &0.3 &14  & 4  &2.3     &0.7\\
\quad$\SFchan$                       &   120\Z &$\no$125 &9  &5.3 &1.6&1.2 &0.2&&&$\np$43  &6  &$\no$9   &1  &$\no$24  &3  &$\no$5  &1  &$\no$2.5 &1.4 & 9  & 2  &33\z    &6  \\
\clineskip\clineskip
\multicolumn{1}{l}{$\NjetGEtwo$, VBF}&     9\Z &$\no$7.8 &1.8&0.9 &0.3&2.7 &0.3&&&$\np$1.2 &0.4&$\no$0.3 &0.1&$\no$1.6 &0.8&$\no$0.4&0.1&$\no$0.1 &0.0 &0.5 & 0.2&3.4     &1.5\\
\quad$\DFchan$ bin~1                 &     6\Z &$\no$3.0 &0.9&0.4 &0.2&0.6 &0.1&&&$\np$0.5 &0.2&$\no$0.2 &0.1&$\no$0.9 &0.5&$\no$0.1&0.0&$\no$0.1 &0.0 &0.3 & 0.1&$\no$0.8&0.6\\
\quad$\DFchan$   bin~2--3            &     0\Z &$\no$0.7 &0.2&0.2 &0.1&1.1 &0.1&&&$\np$0.2 &0.1&\mcolz       &$\no$0.3 &0.2&\mcolz      &\mcolz        &\mcolz   &\mcolz      \\
\quad$\SFchan$~bins\,1--3$\nq$       &$\nq$3\Z &$\no$4.1 &1.3&0.3 &0.1&1.0 &0.1&&&$\np$0.5 &0.2&$\no$0.1 &0.0&$\no$0.4 &0.3&$\no$0.3&0.1&\mcolz        &0.2 & 0.1&$\no$2.5&1.1\\
\dbline
\end{tabular*}
}
\end{table*}

In the figures and tables presented in this section,
background processes are individually normalized to their postfit rates,
which account for changes in the normalization
factors ($\fNorm$) and for pulls of the nuisance parameters ($\theta$).
The varying background composition as a function of $\mTH$ (or $\bdt$
in the $\NjetGEtwo$ VBF-enriched category) induces a shape uncertainty
on the total estimated background. As described in
Sec.~\ref{sec:systematics_sources}, additional specific shape uncertainties are
included in the fit procedure and are accounted for in the results
presented in Sec.~\ref{sec:results}. No specific $\mTH$ shape uncertainties
are applied to the figures since their contribution to the total
systematic uncertainty band was found to be negligible.
The Higgs boson signal rate is normalized to the observed signal strength reported in Sec.~\ref{sec:results}.

\begin{figure*}[t!]
\includegraphics[width=0.75\textwidth]{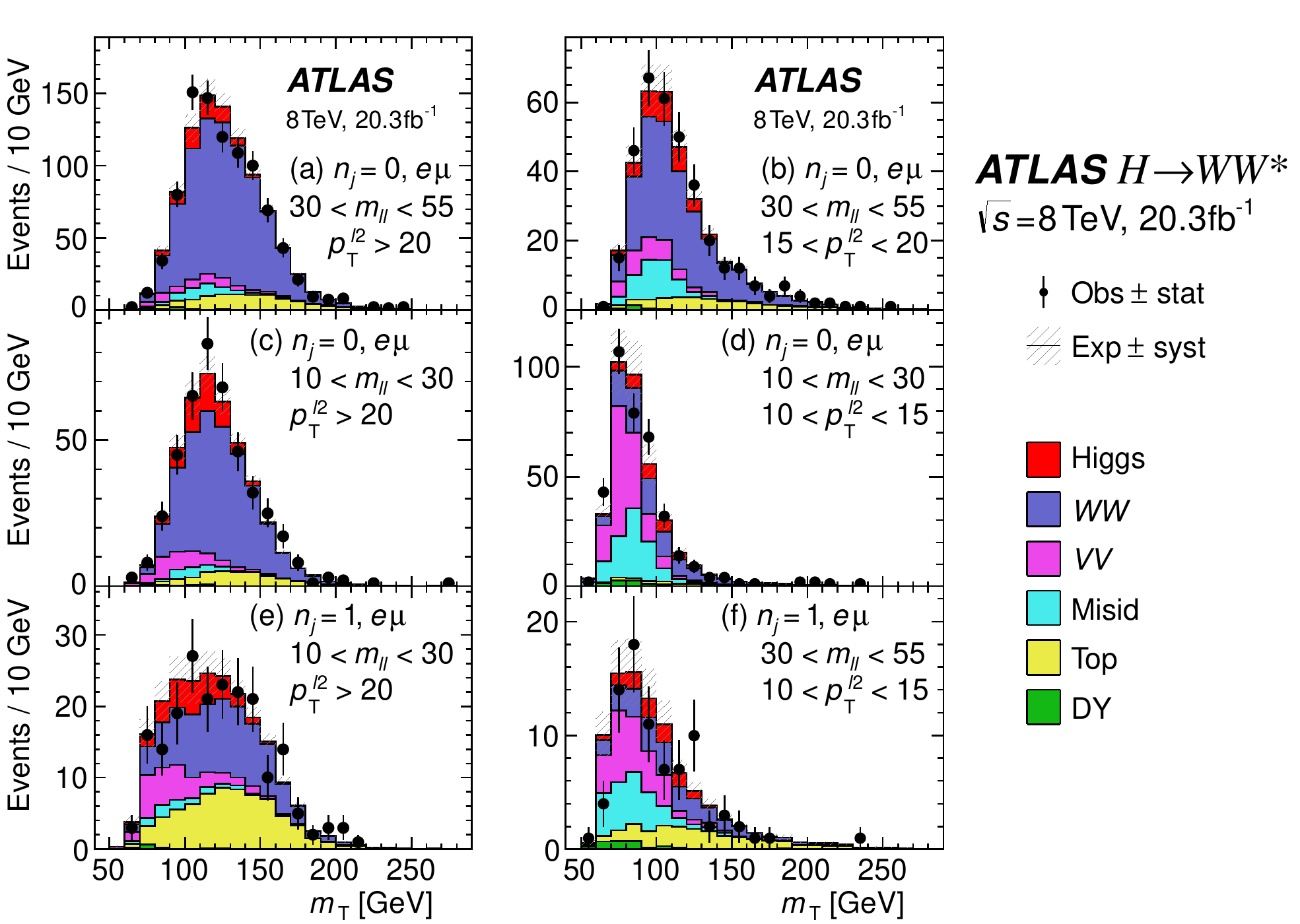}
\caption{
  Postfit transverse mass distributions in the $\DFchan$ $\NjetLEone$
  categories in the $8\TeV$ data analysis, for specific $\mll$ and
  $\pTsublead$ ranges. The plots are made after applying all the selection requirements (see Tables~\ref{tab:sr_0j} and \ref{tab:sr_1j}).
  The signal processes are scaled with the observed signal
  strength $\sigmu$ from the fit to all the regions and the background
  normalizations include the postfit $\fNorm$ values and effects from the pulls of the nuisance parameters.
  \HwwPlotDetail{See}.
}
\label{fig:mT_fitDF}
\label{fig:mT_fitDF_0j}
\label{fig:mT_fitDF_1j}
\end{figure*}

This section is organized as follows. The event yields are presented in
Sec.~\ref{sec:yields_tables} for each signal category including the statistical and systematic
uncertainties. The relevant distributions in
the various signal regions are shown in Sec.~\ref{sec:yields_distributions}. Section~\ref{sec:yields_differences} summarizes the
differences in the event and object selection, the signal treatment and
the background estimates with respect to the previously published analysis~\cite{couplings}.

\subsection{\boldmath Event yields \label{sec:yields_tables}}

Table~\ref{tab:sr_summary} shows the postfit
yields for all of the fitted categories in the $8\TeV$
[Table~\ref{tab:sr_summary}(a)] and $7\TeV$ [Table~\ref{tab:sr_summary}(b)]
data analyses. The signal yields are scaled with the observed signal
strength derived from the simultaneous combined fit to all of the
categories. All of the background processes are normalized to the
postfit $\fNorm$ values (where applicable) and additionally their rates take
into account the pulls of the nuisance parameters. The
observed and expected yields are shown, for each $\Njet$ category,
separately for the $\DFchan$ and $\SFchan$ channels. The sum of the
expected and observed yields is also reported. The uncertainties include
both the statistical and systematic components.

As described in the previous section, the changes in the normalization factors and the
pulls of the nuisance parameters can affect the expected rates of the
signal and  background processes. The differences between the prefit
(tables in Sec.~\ref{sec:selection}) and postfit (Table~\ref{tab:sr_summary}) expected rates
for each background process are compared to the total uncertainty on that expected background,
yielding a significance of the change.
In the analysis of the $\NjetLEone$ category of the $8\TeV$ data, most of the
changes are well below one standard deviation.
In the $\DFchan$ $\NjetEQzero$ sample, the expected multijet
background is increased by $1.3$ standard deviations (equivalent to a $30\%$
increase in the expected multijet background prediction which corresponds to $2\%$ of the signal prediction) due to the
positive pulls of the three nuisance parameters assigned to the
uncertainties on the extrapolation factor.
A negative pull of the nuisance parameter associated with the uncertainties on
the DY $\frecoil$ selection efficiency changes the
$\ZDYll$ yield in the $\SFchan$ $\NjetEQzero$ sample by $1.6$ standard deviations (equivalent to
a $40\%$ decrease in DY in this category which corresponds to $25\%$ of the signal prediction).

\subsection{\boldmath Distributions \label{sec:yields_distributions}}

The transverse mass formed from the dilepton and missing transverse momenta ($\mTH$) is
used as the final discriminant in the extraction of the signal strength
in the $\NjetLEone$ and $\NjetGEtwo$ ggF-enriched categories.
The likelihood fit exploits the differences in $\mTH$ shapes between
the signal and background processes.

Several of the $\mTH$ distributions for the $\DFchan$ sample
(corresponding to different choices of the $\mll$ and $\pTsublead$
bins) in the $\NjetLEone$ categories are shown in Fig.~\ref{fig:mT_fitDF_0j}.
The background composition, signal contribution, and the separation in the
$\mTH$ distributions between signal and background are different for each region.
In general, as shown in Figs.~\ref{fig:mT_fitDF_0j}(a)-(c),
the $\WW$ process dominates the background contributions in regions with $\NjetEQzero$;
the difference between these distributions is due to the varying signal contribution and background $\mTH$ shape.
In contrast, Fig.~\ref{fig:mT_fitDF_0j}(d) shows that
$VV$ and $\Wjets$ processes are dominant backgrounds in the $10{\LT}\mll{\LT}30\GeV$ and $10{\LT}\pTsublead{\LT}15\GeV$ region.
For most of the distributions shown in Fig.~\ref{fig:mT_fitDF_0j}, agreement between data and MC is improved qualitatively when including
the expected signal from a Standard Model Higgs boson with $\mH{\EQ}125\GeV$.

The $\mTH$ distributions for the $\SFchan$ samples in the $\NjetLEone$ categories are shown in Fig.~\ref{fig:mT_fitSF}.
In contrast to the $\DFchan$ distributions, the residual DY background is present in these samples at low values of $\mTH$.

For the ggF-enriched $\NjetGEtwo$ category, Fig.~\ref{fig:mT_fit2j} shows the $\mTH$ distribution.
In contrast to the $\NjetLEone$ distributions, the dominant backgrounds arise from top-quark and $\ZDYtt$ production
(shown together with the negligible contribution from $\ZDYll$).

\begin{figure}[t!]
\includegraphics[width=0.40\textwidth]{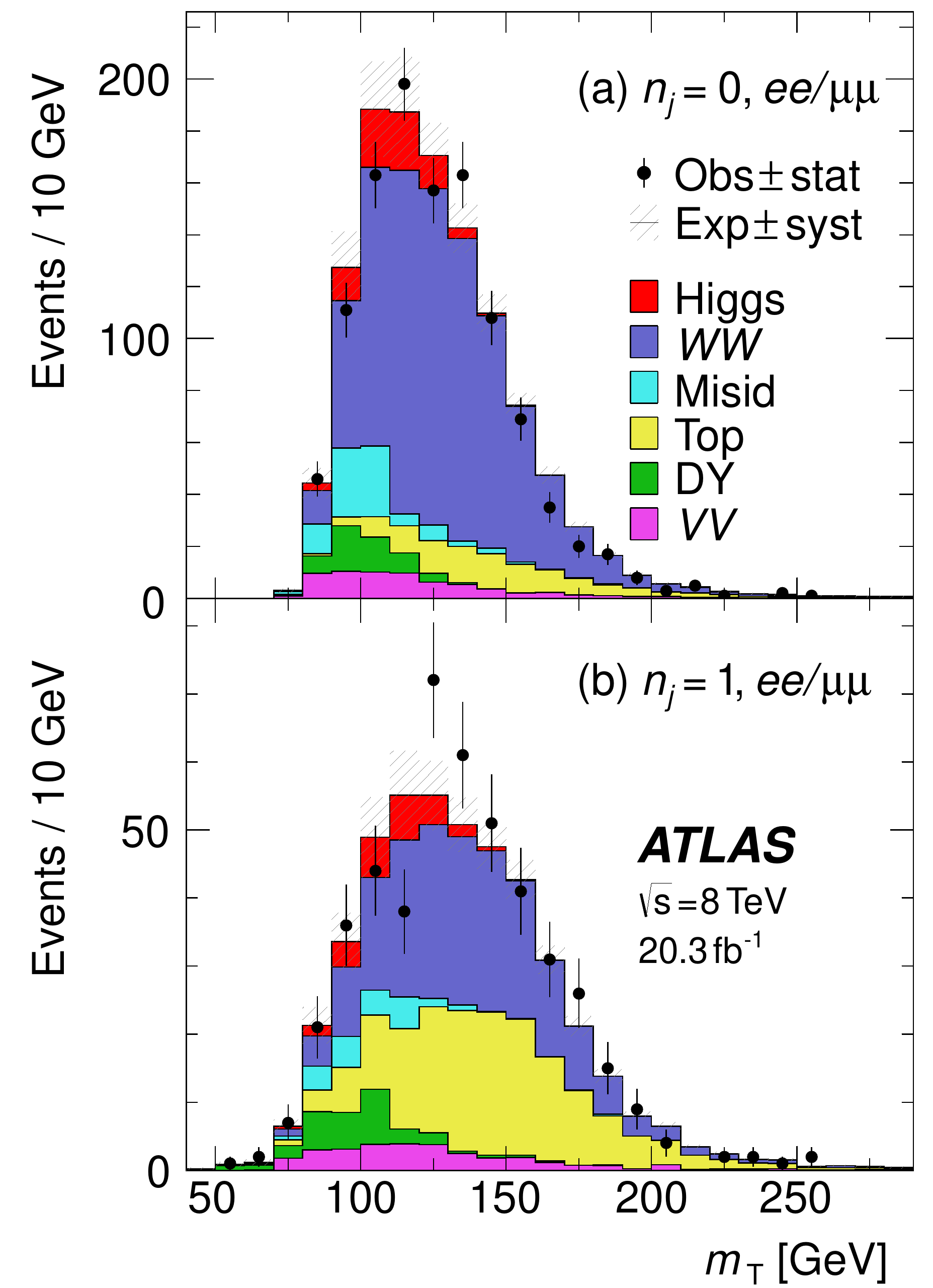}
\caption{
  Postfit transverse mass distributions in the $\NjetLEone$, $\SFchan$ categories in the $8\TeV$ analysis.
  \HwwPlotFitDetailShort{See}.
}
\label{fig:mT_fitSF}
\end{figure}

For the VBF-enriched $\NjetGEtwo$ category,
a selection-based analysis,
which uses the $\mTH$ distribution as the discriminant,
is used as a cross-check of the BDT result.
In this case, $\mTH$ is divided into three bins (with boundaries at $80$ and $130\GeV$) and an
additional division in $\mjj$ at $1\TeV$ is used in the $\DFchan$ channel to
profit from the difference in shapes between signal and background processes.
Figure~\ref{fig:bdtcb_fit}(a) shows the $\mTH$ distribution
before the division into the high- and low-$\mjj$ regions.
Figure~\ref{fig:bdtcb_fit}(b) shows the scatter plot of $\mjj$ versus $\mTH$.
The areas with the highest signal-to-background ratio are characterized by low $\mTH$ and high $\mjj$.

Figures~\ref{fig:mT_fit2jvbf}(a) and~\ref{fig:mT_fit2jvbf}(c) show
the $\bdt$ outputs in the $\DFchan$ and $\SFchan$ samples, respectively.
In terms of VBF signal production, the third BDT bin provides the highest purity,
with a signal-to-background ratio of approximately $2$. The $\mTH$ variable is
an input to the BDT and its distributions after the
BDT classification are shown in Figs.~\ref{fig:mT_fit2jvbf}(b) and~\ref{fig:mT_fit2jvbf}(d),
combining all three BDT bins, for the $\DFchan$ and $\SFchan$ samples, respectively.

\begin{figure}[t!]
\includegraphics[width=0.40\textwidth]{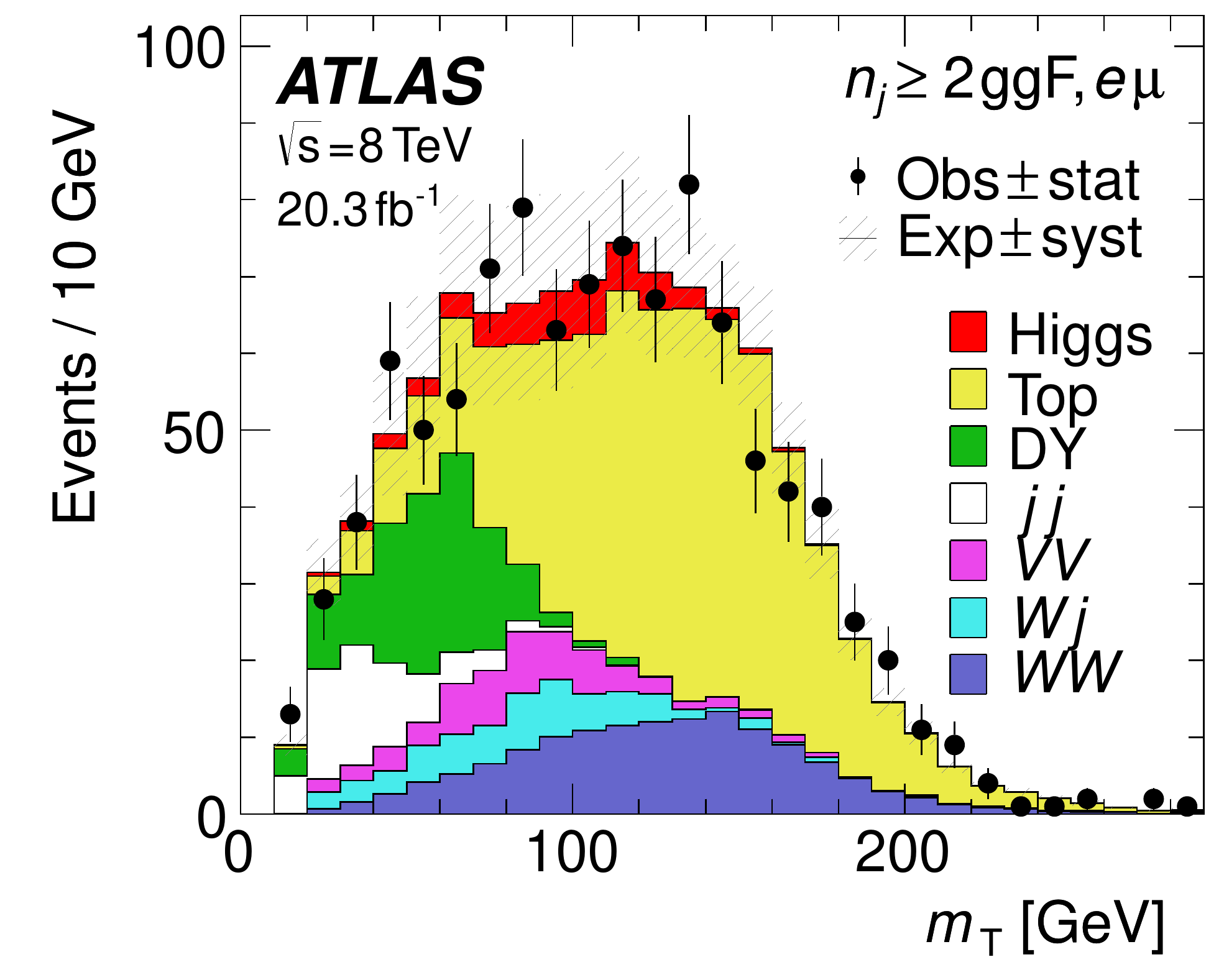}\\
\caption{
  Postfit transverse mass distribution in the $\NjetGEtwo$ ggF-enriched category in the $8\TeV$ analysis.
  \HwwPlotFitDetailShort{See}.
}
\label{fig:mT_fit2j}
\label{fig:mT_fitDF_2jggf}
\label{fig:mT_fit2jggf}
\end{figure}

\begin{figure}[b!]
\includegraphics[width=0.40\textwidth]{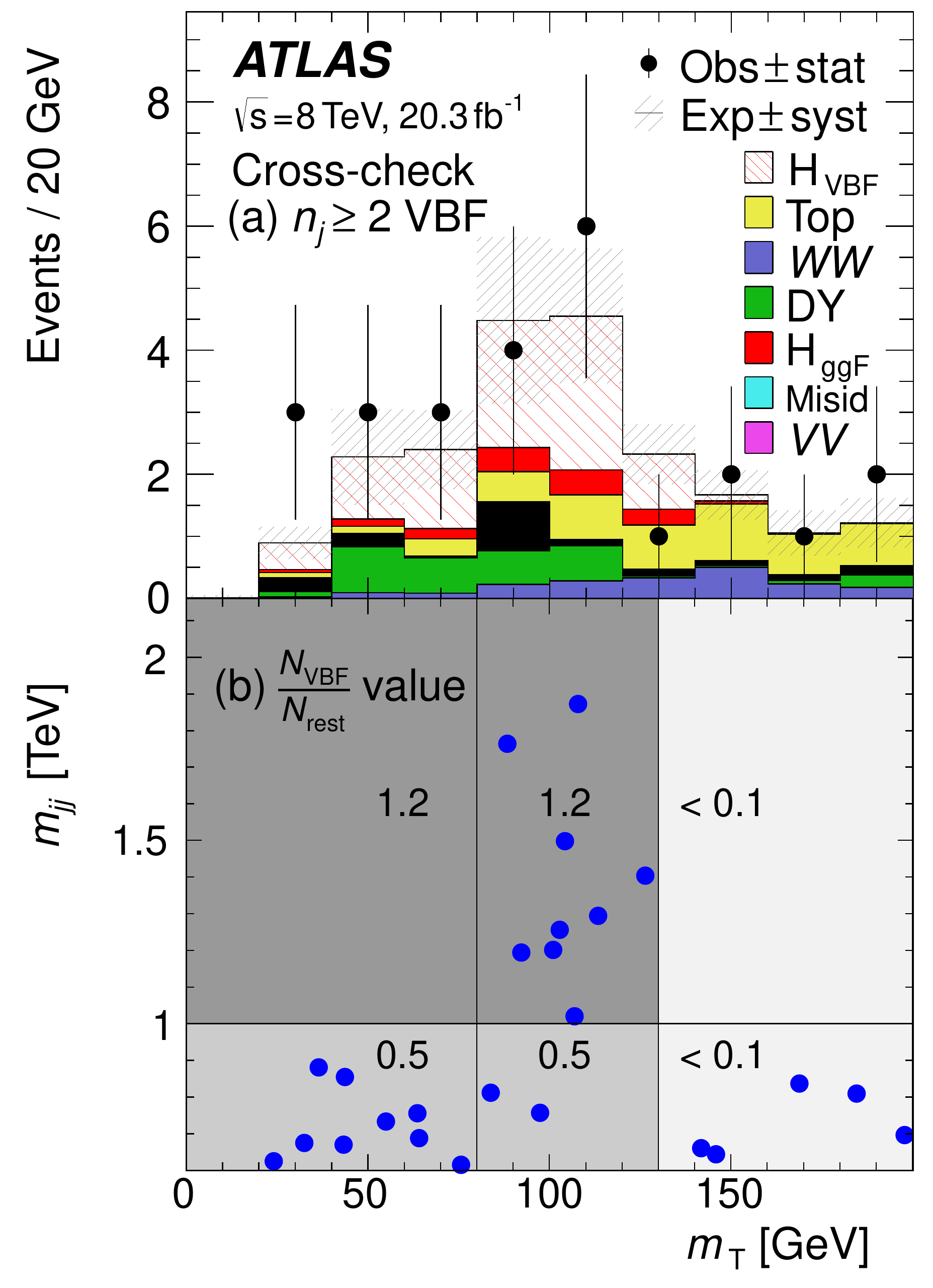}
\caption{
  Postfit distributions in the cross-check analysis in the $\DFchan{\PLUS}\SFchan$ $\NjetGEtwo$ VBF-enriched category in the $8\TeV$ data analysis:
  (a) $\mTH$ and
  (b) $\mjj$ versus $\mTH$ scatter plot for data.
  For each bin in (b), the ratio $\NVBF/N_\mathrm{rest}$ is stated in the plot, where $N_\mathrm{rest}$ includes all processes other than the VBF signal.
  \HwwPlotFitDetailShort{See}.
}
\label{fig:bdtcb_fit}
\end{figure}

\begin{figure*}[t!]
\includegraphics[width=0.75\textwidth]{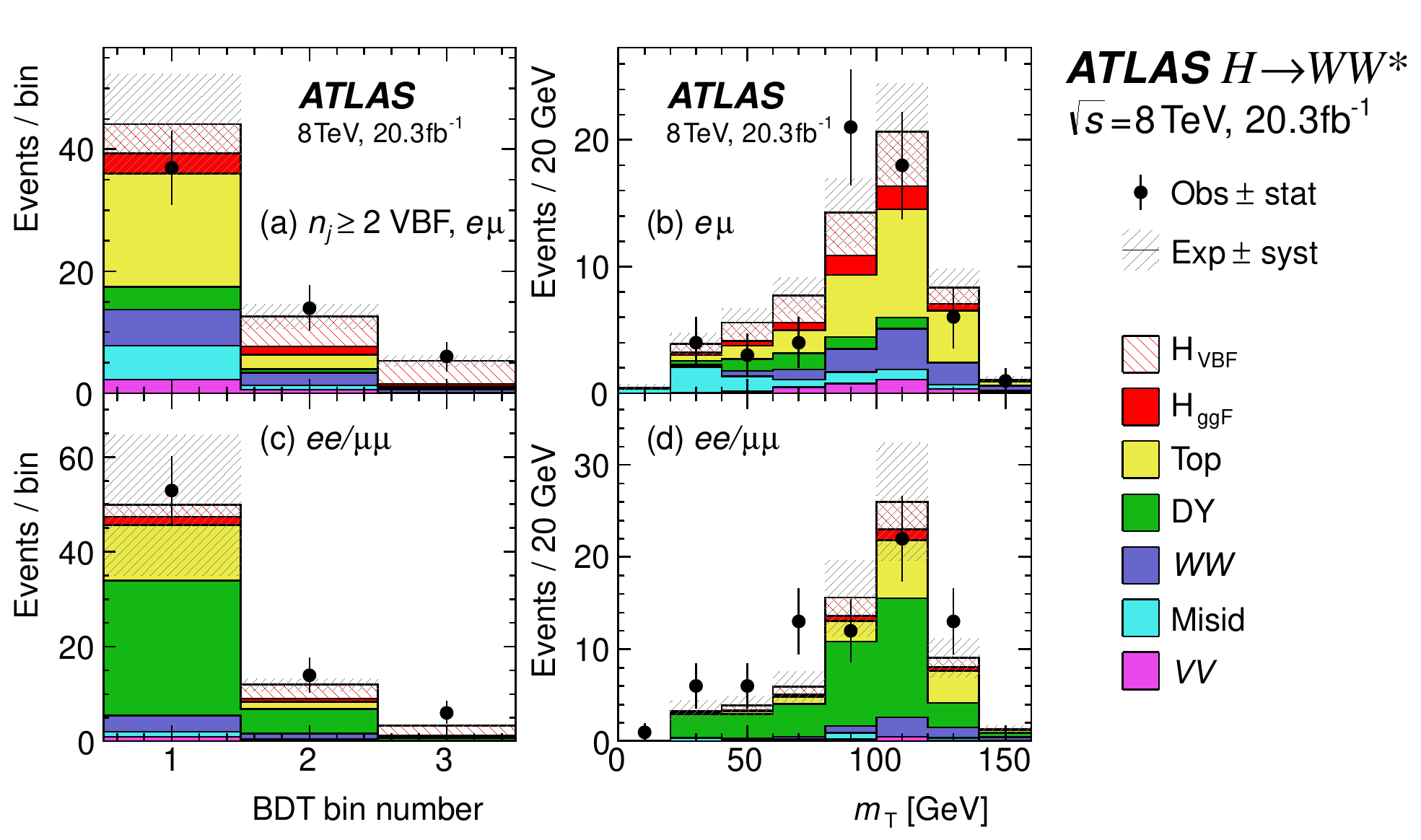}
\caption{
  Postfit BDT and transverse mass distributions in the $\NjetGEtwo$ VBF-enriched category in the $8\TeV$ data analysis:
  (a) BDT output in $\DFchan$,
  (b) $\mTH$ in $\DFchan$,
  (c) BDT output in $\SFchan$, and
  (d) $\mTH$ in $\SFchan$.
  For (b) and (d), the three BDT bins are combined.
  \HwwPlotFitDetailShort{See}.
}
\label{fig:mT_fit2jvbf}
\end{figure*}

\begin{figure*}[bt!]
\includegraphics[width=0.75\textwidth]{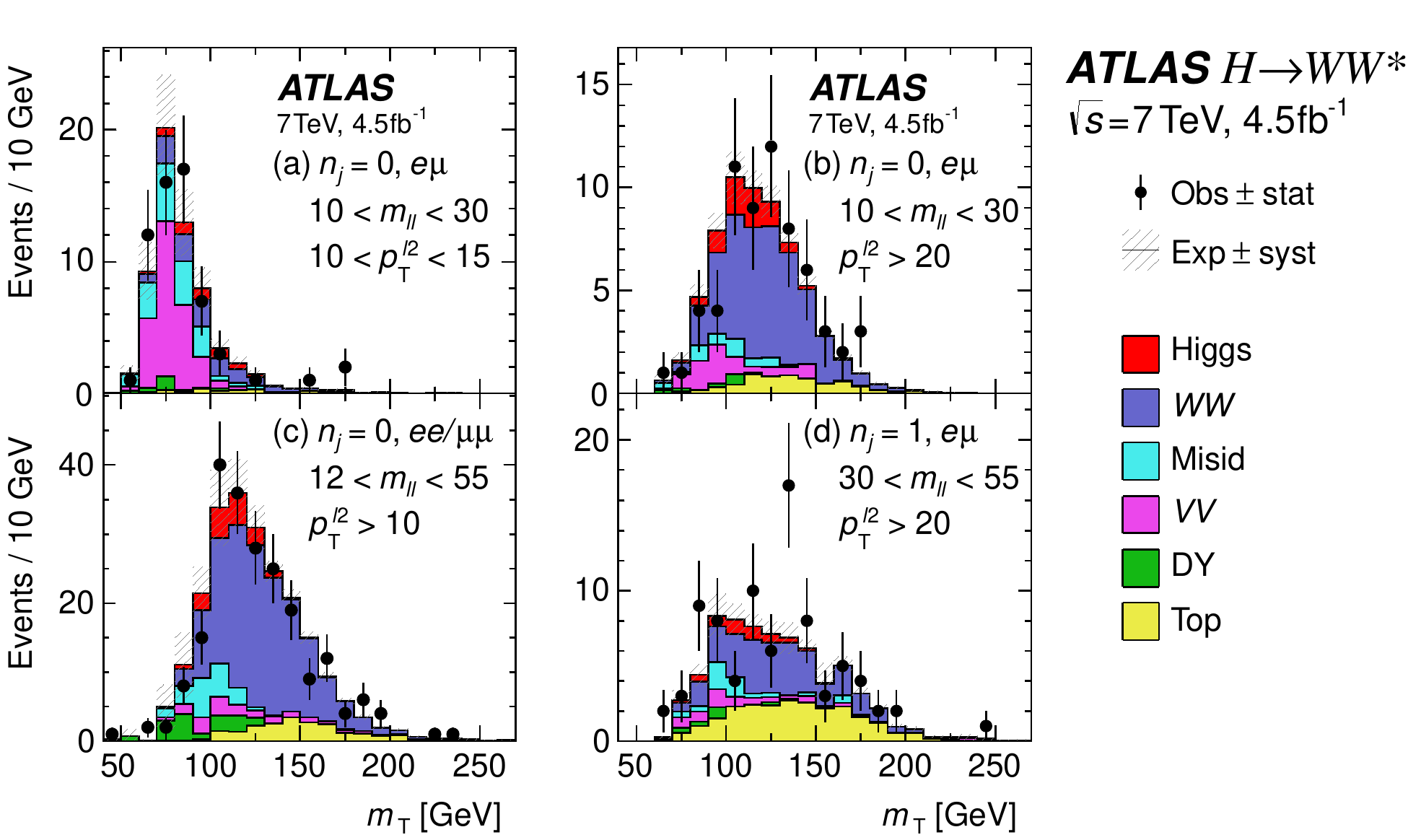}
\caption{
  Postfit transverse mass distributions in the $\NjetLEone$
  categories in the $7\TeV$ data analysis, for specific $\mll$ and
  $\pTsublead$ ranges. The plots are made after applying all the
  selection requirements (see Sec.~\ref{sec:selection_7tev}).
  \HwwPlotFitDetailShort{See}.
}
\label{fig:7TeV_fit}
\end{figure*}

Figure~\ref{fig:7TeV_fit} shows the $\mTH$ distributions in the
$7\TeV$ analysis in the various signal regions in the $\NjetLEone$
categories. Characteristics similar to those in the $8\TeV$ analysis are observed,
but with fewer events.

Finally, Fig.~\ref{fig:mTbkgsubtr}(a) shows the combined $\mTH$ distribution, summed
over the lepton-flavor samples and the $\NjetLEone$ categories for the $7$ and
$8\TeV$ data analyses.
To illustrate the significance of the excess of events observed in data with
respect to the total background, the systematic uncertainty on the signal is omitted.
The uncertainty band accounts for the correlations between the signal regions, including
between the $7$ and $8\TeV$ data, and for the varying size of the uncertainties
as a function of $\mTH$.
Figure~\ref{fig:mTbkgsubtr}(b) shows the residuals of the data with respect to
the total estimated background compared to the expected $\mTH$ distribution of an SM Higgs
boson with $\mH{\EQ}125\GeV$ scaled by the observed combined signal strength (see Sec.~\ref{sec:results}).
The level of agreement observed in Fig.~\ref{fig:mTbkgsubtr}(b) between the
background-subtracted data and the expected Higgs boson signal strengthens the
interpretation of the observed excess as a signal from Higgs boson decay.

\begin{figure}[tb!]
\includegraphics[width=0.45\textwidth]{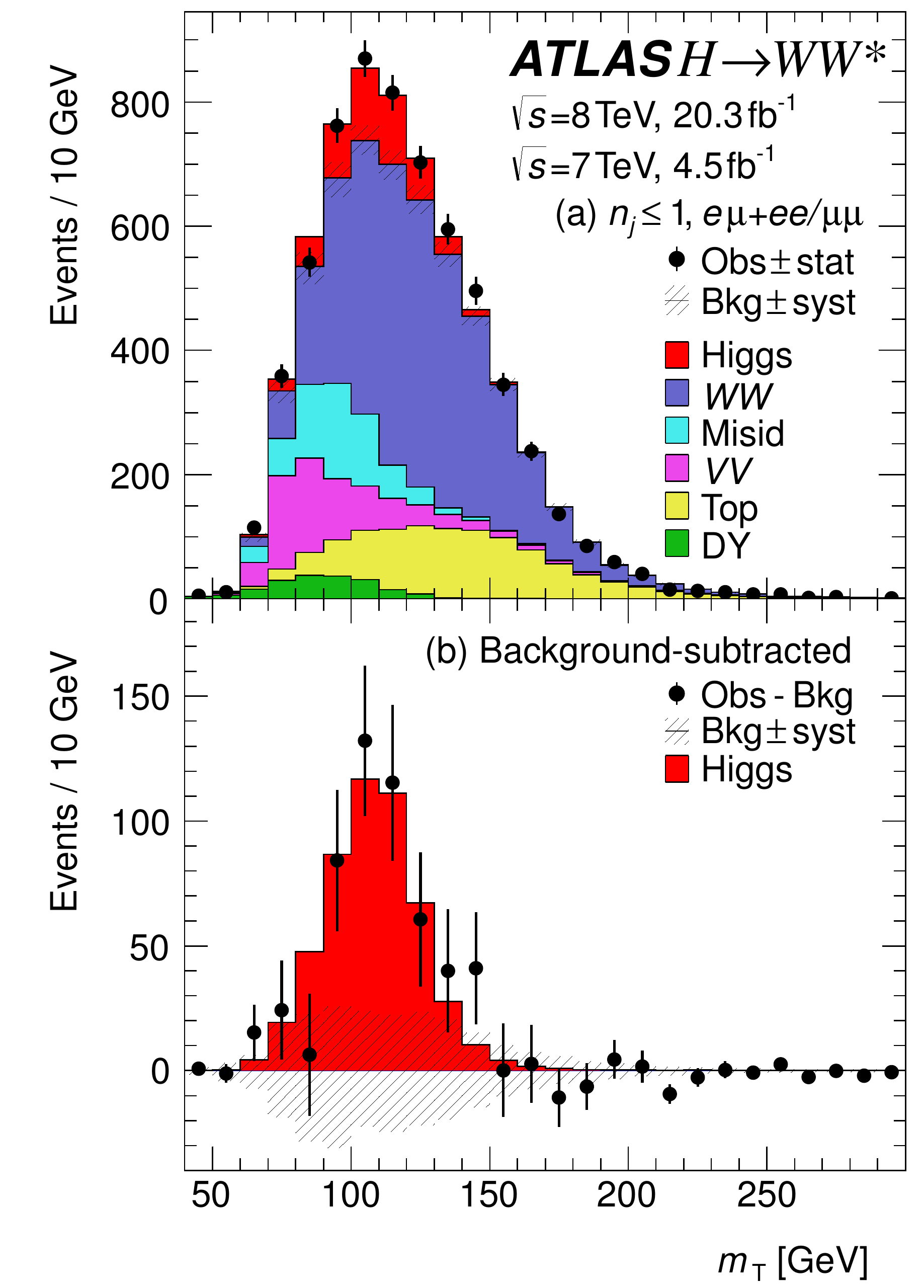}
\caption{
  Postfit combined transverse mass distributions for $\NjetLEone$ and for
  all lepton-flavor samples in the $7$ and $8\TeV$ data analyses.
  The plot in (b) shows the residuals of the data with respect to
  the estimated background compared to the expected distribution for an SM Higgs boson with $\mH{\EQ}125\GeV$;
  the error bars on the data are statistical ($\sqrt{\Nobs}$).
  The uncertainty on the background (shown as the shaded band around $0$) is at most about $25$ events per $\mTH$ bin and partially correlated between bins.
  Background processes are scaled by postfit normalization factors
  and the signal processes by the observed signal strength
  $\sigmu$ from the likelihood fit to all regions. Their
  normalizations also include effects from the pulls of the
  nuisance parameters.
}
\label{fig:mTbkgsubtr}
\end{figure}

\subsection{\boldmath Differences with respect to previous results \label{sec:yields_differences}}

The analysis presented in this \paper\ has better sensitivity than the previous ATLAS analysis~\cite{couplings}. The most important
changes---described in detail below---include improvements
in the object identification, the signal acceptance, the background
estimation and modeling, and the fit procedure.

Electron identification is based on a likelihood technique~\cite{ElectronEff2012} that
improves background rejection. An improved definition of missing transverse momentum, $\MPTj$ based on tracks, is
introduced in the analysis since it is robust against pile-up and provides improved resolution with respect to
the true value of $\met$.

Signal acceptance is increased by $75\%$ ($50\%$) in the
$\NjetEQzero$ (1) category. This is achieved by lowering the $\pTsublead$
threshold to $10\GeV$. Dilepton triggers are included in addition to single lepton
triggers, which allows reduction of the $\pTlead$ threshold to
$22\GeV$.  The signal kinematic region in the $\NjetLEone$ categories
is extended from $50$ to $55\GeV$. The total signal efficiency, including all signal categories
and production modes, at $8\TeV$ and for a Higgs boson mass of $125.36\GeV$ increased from $5.3\%$ to $10.2\%$.

The methods used to estimate nearly all of the background contributions in the signal region are
improved.  These improvements lead to a better understanding of the normalizations
and thus the systematic uncertainties. The rejection of the top-quark background is improved by applying a veto
on $b$-jets with $\pT{\GT}20\GeV$, which is below the nominal $25\GeV$ threshold in the analysis.
A new method of estimating the jet $b$-tagging efficiency
is used. It results in the cancellation of
the $b$-tagging uncertainties between the top-quark control region and
signal regions in the $\NjetEQone$ categories.
The $\ZDYtt$ background process is
normalized to the data in a dedicated high-statistics control
region in the $\NjetLEone$ and $\NjetGEtwo$ ggF-enriched categories.
The $\VV$ backgrounds are normalized
to the data using a new control region, based on a sample with two same-charge leptons.
Introducing this new control region results in the cancellation of
most of the theoretical uncertainties on the $\VV$
backgrounds. The multijet background is now explicitly estimated
with an extrapolation factor method using a sample with two anti-identified
leptons. Its contribution is negligible in the $\NjetLEone$ category, but it
is at the same level as $\Wjets$ background in the $\NjetGEtwo$  ggF-enriched
category. A large number of improvements are applied to the
estimation of the $\Wjets$ background, one of them being an estimation
of the extrapolation factor using $Z+$jets instead of dijet data
events.

Signal yield uncertainties are smaller than in the previous analysis. The uncertainties on the jet multiplicity
distribution in the ggF signal sample, previously estimated with the
Stewart-Tackmann technique~\cite{ST}, are now estimated with the jet-veto-efficiency method~\cite{JVE}. This method yields more precise estimates of the
signal rates in the exclusive jet bins in which the analysis is performed.

The $\NjetGEtwo$ sample is divided into VBF- and ggF-enriched categories.
The BDT technique, rather than a selection-based approach, is used for the VBF category. This improves the sensitivity
of the expected VBF results by $60\%$ relative to the
previously published analysis. The ggF-enriched category is a new subcategory
that targets ggF signal production in this sample.

In summary, the analysis presented in this \paper\ brings a gain of $50\%$
in the expected significance relative to the previous published
analysis~\cite{couplings}.

\section{\boldmath Results and interpretations \label{sec:results}}

Combining the 2011 and 2012 data in all categories,
a clear excess of signal over the background is seen in Fig.~\ref{fig:mTbkgsubtr}.
The profile likelihood fit described in Sec.~\ref{sec:systematics_fit} is used
to search for a signal and characterize the production rate in the $\ggF$ and $\VBF$ modes.
Observation of the inclusive Higgs boson signal, and evidence for the $\VBF$ production
mode, are established first.
Following that, the excess in data is characterized using the SM Higgs boson as the signal
hypothesis, up to linear rescalings of the production cross sections and decay modes.
Results include the inclusive signal strength as well
as those for the individual $\ggF$ and $\VBF$ modes.  This information is also
interpreted as a measurement of the vector-boson and fermion couplings of the Higgs boson,
under the assumptions outlined in Ref.~\cite{Heinemeyer:2013tqa}.  Because this is the first observation in the $\WWlvlv$
channel using ATLAS data, the exclusion sensitivity and observed exclusion limits as
a function of $\mH$ are also
presented to illustrate the improvements with respect to the version of this
analysis used in the 2012 discovery \cite{discovery}.
Finally, cross-section measurements, both inclusive and in specific fiducial regions, are presented.
All results in this section are quoted for a Higgs boson mass corresponding to the central
value of the ATLAS measurement in the $ZZ\to 4\ell$ and $\gamma\gamma$ decay modes,
$\mH{\EQ}\HwwHiggsMass{ATLAS}{\PM}\HwwHiggsMassError{ATLAS}\GeV$~\cite{atlasMassPaper}.

\subsection{\boldmath Observation of the $\HWW$ decay mode \label{sec:results_obs}}

The test statistic $\qmu$, defined in Sec.~\ref{sec:systematics_fit}, is used
to quantify the significance of the excess observed in Sec.~\ref{sec:yields}.
The probability that the background can fluctuate to produce an excess at least
as large as the one observed in data is called $\pzero$ and is computed using $\qmu$ with $\sigmu{\EQ}0$.  It depends on the
mass hypothesis $\mH$ through the
distribution used to extract the signal ($\mTH$ or $\bdt$).
The observed and expected $\pzero$ are shown as a function of $\mH$ in Fig.~\ref{fig:p0}.
The observed curve presents a broad minimum centered around $\mH\APPROX{130}\GeV$, in contrast with the higher
$\pzero$ values observed for lower and higher values of $\mH$.
The shapes of the observed and expected curves are in good agreement.

\begin{figure}[t!]
\includegraphics[width=0.45\textwidth]{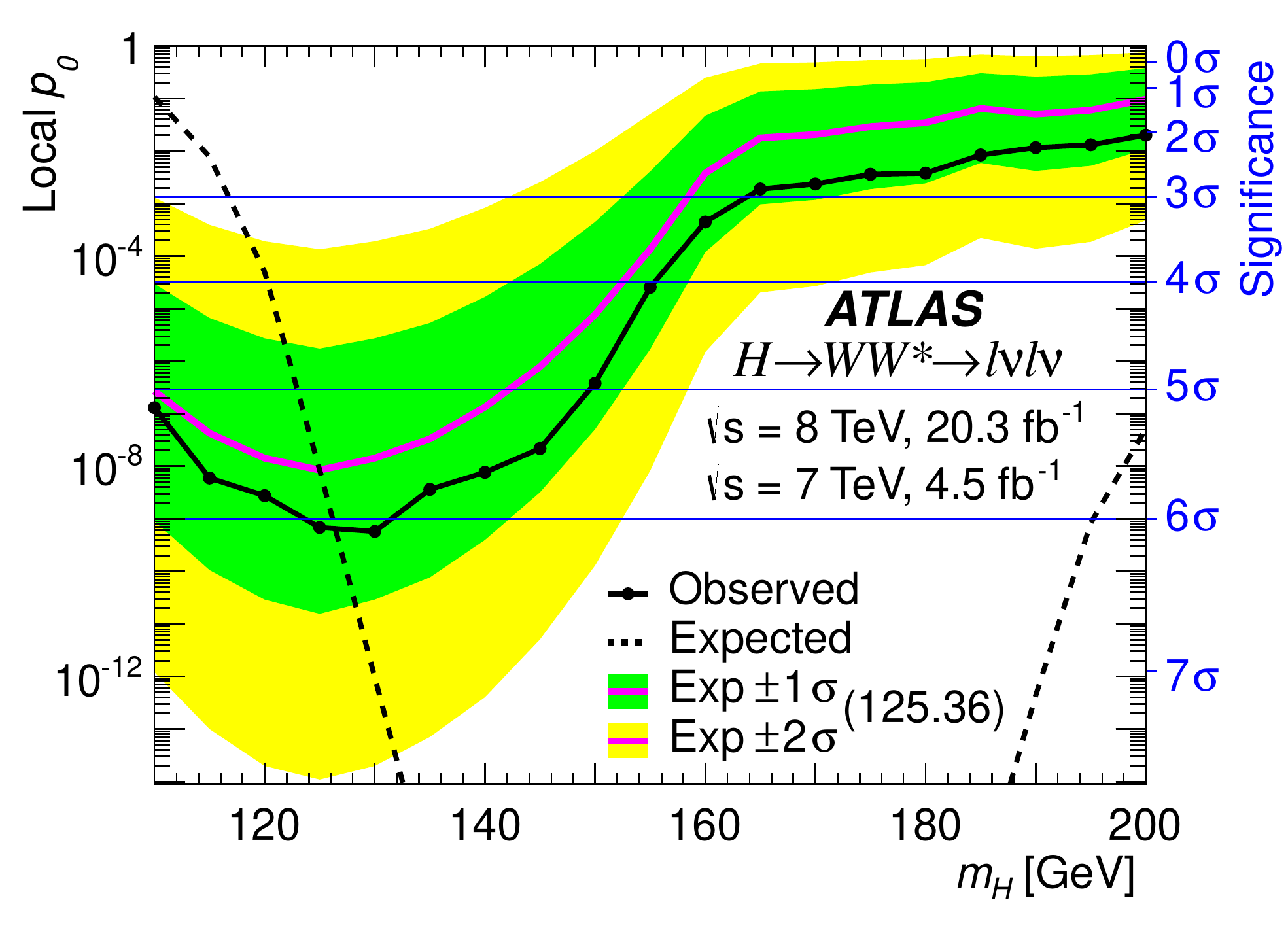}
\caption{
  Local $\pzero$ as a function of $\mH$.
  The observed values are shown as a solid line with points where $\pzero$ is evaluated.
  The dashed line shows the expected values given the presence of a signal at each $x$-axis value.
  The expected values for $\mH{\EQ}125.36\GeV$ are given as a solid line without points;
  the inner (outer) band shaded darker (lighter) represents the one (two) standard deviation uncertainty.
}
\label{fig:p0}
\end{figure}

\begin{figure}[b!]
\includegraphics[width=0.45\textwidth]{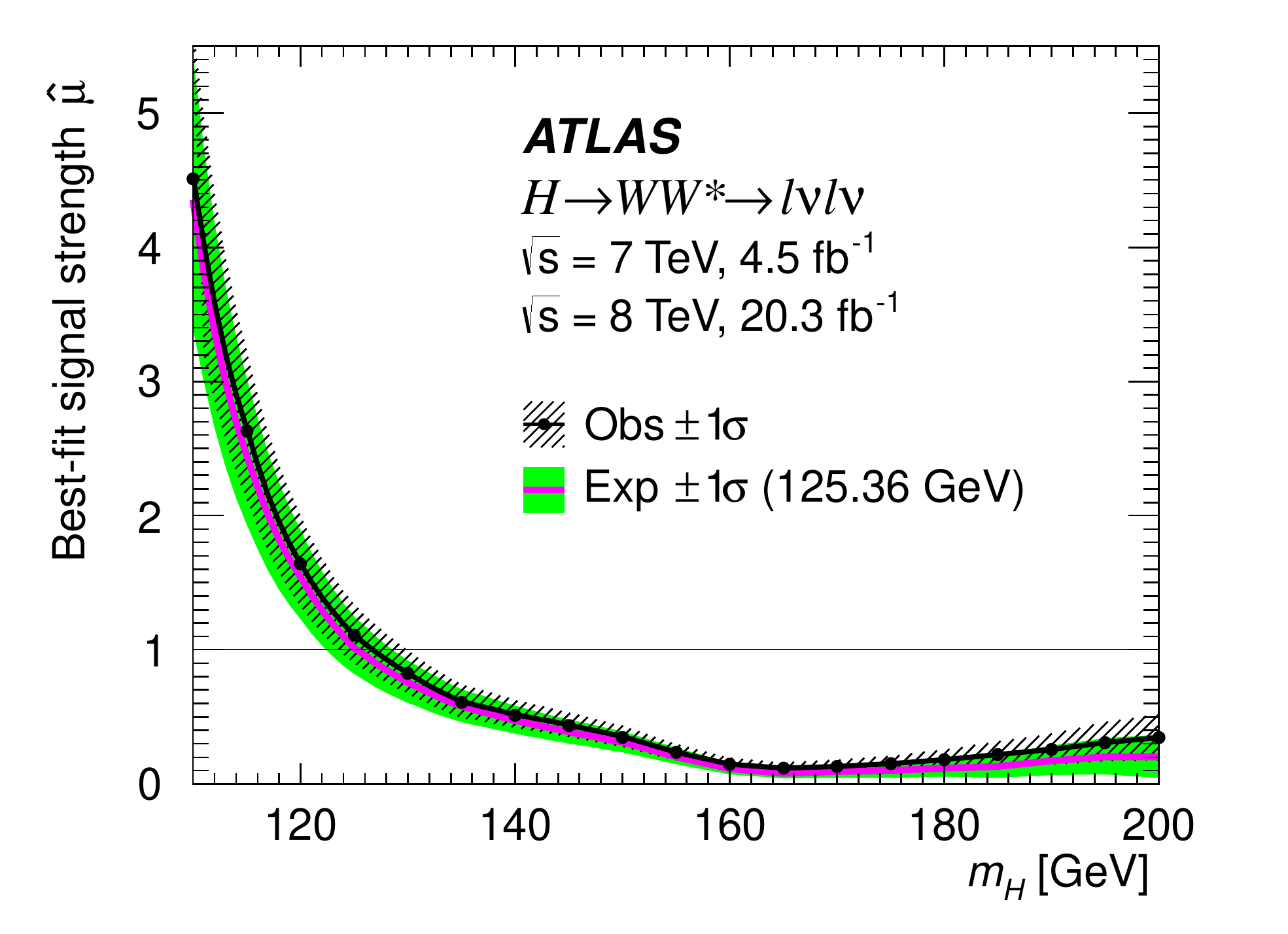}
\caption{Best-fit signal strength $\hat{\sigmu}$ as a function of $\mH$.
  The observed values are shown as a solid line with points where $\hat{\sigmu}$ is evaluated.
  The expected values for $\mH{\EQ}125.36\GeV$ are shown as a solid line without points.
  The dashed and shaded (solid) bands represent the one standard deviation uncertainties for the observed (expected)
  values.
}
\label{fig:mu}
\end{figure}

The probability $\pzero$ can equivalently be expressed in terms of the number of standard deviations,
referred to as the local significance ($Z_0$ defined in Sec.~\ref{sec:systematics_fit_cls}).
The value of $\pzero$ as a function of $\mH$ is found by scanning $\mH$ in $5\GeV$ intervals.
The minimum $\pzero$ value is found at $\mH{\EQ}\HwwHiggsMass{minp0}\GeV$ and corresponds to
a local significance of $\HwwSignif{obs}{minp0}$ standard deviations.
The same observed significance within the quoted precision is found for $\mH{\EQ}\HwwHiggsMass{ATLAS}\GeV$.
This result establishes a discovery-level signal in the $\HWWlvlv$ channel alone.
The expected significance for a SM Higgs boson at the same mass is $\HwwSignif{exp}{all}$ standard deviations.

In order to assess the compatibility with the SM expectation for a
Higgs boson of mass $\mH$, the observed best-fit $\hatsigmu$ value as a function of
$\mH$ is shown in Fig.~\ref{fig:mu}.
The observed $\hatsigmu$ is close to zero for $\mH{\GT}160\GeV$ and
crosses unity at $\mH{\APPROX}125\GeV$.  The increase
of $\sigmu$ for small values of $\mH$ is expected in the presence of a
signal with mass $\mH{\EQ}\HwwHiggsMass{ATLAS}\GeV$, as is also shown in Fig.~\ref{fig:mu}.
The dependence of $\hatsigmu$ on the value of $\mH$ arises mostly from
the dependence of the $\WWs$ branching fraction on $\mH$.

The assumption that the total yield is predicted by the SM is relaxed to evaluate the two-dimensional
likelihood contours of ($\mH$, $\sigmu$), shown in Fig.~\ref{fig:banana}.
The value ($\sigmu{\EQ}1$, $\mH{\EQ}\HwwHiggsMass{ATLAS}\GeV$) lies well within
the $68\%$ C.L.\ contour, showing that the signal observed is compatible with
those in the high-resolution channels.

\begin{figure}[t!]
\includegraphics[width=0.45\textwidth]{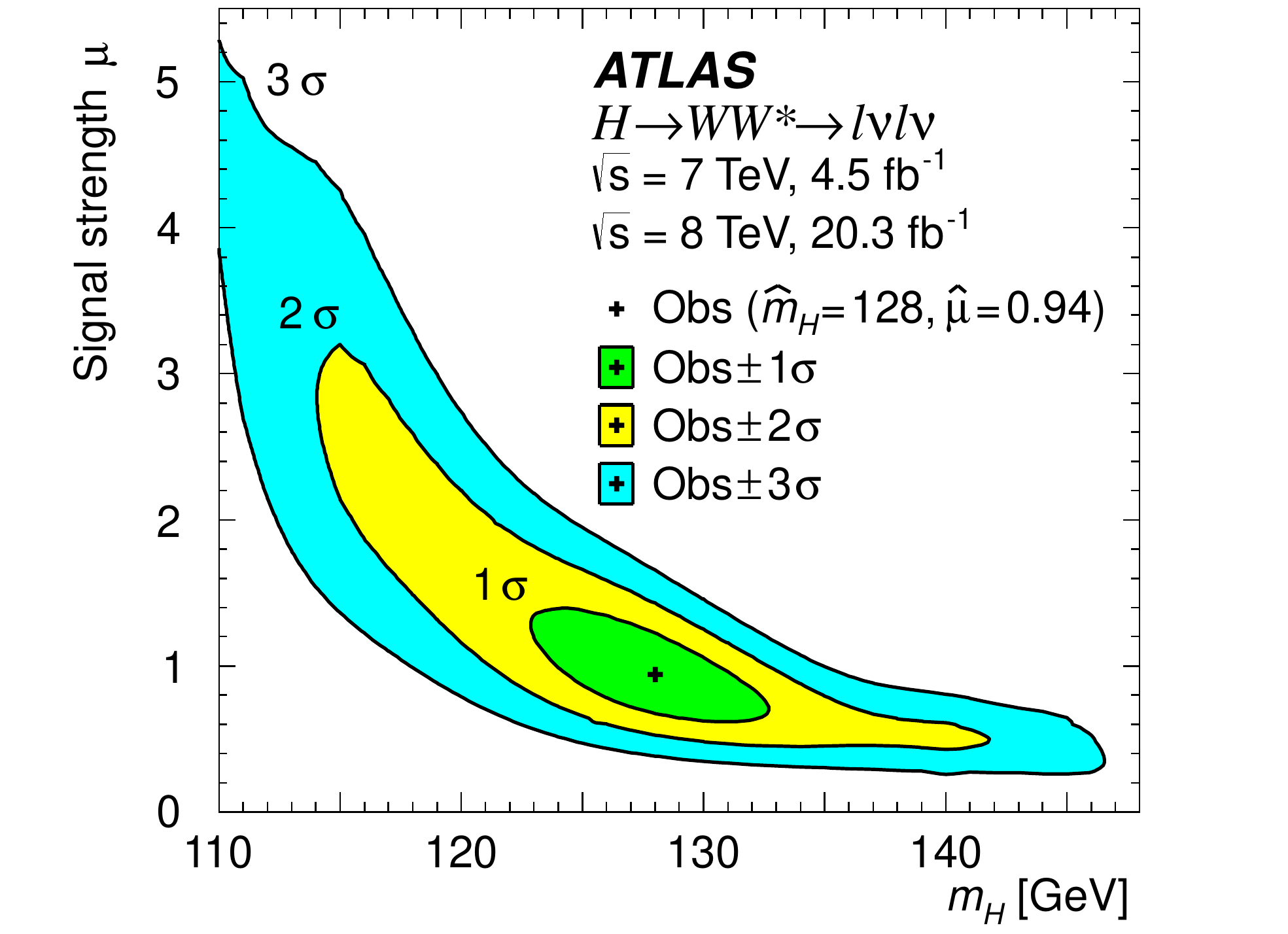}
\caption{
  Observed signal strength $\sigmu$ as a function of $\mH$ as evaluated by the
  likelihood fit.
  The shaded areas represent the one, two, and three standard deviation contours with
  respect to the best fit values $\hat{m}_H$ and $\hatsigmu$.
}
\label{fig:banana}
\end{figure}

\subsection{\boldmath Evidence for $\VBF$ production \label{sec:results_vbf}}

The $\NjetGEtwo$ $\VBF$-enriched signal region was optimized for its specific
sensitivity to the $\VBF$ production process, as described in particular in Sec.~\ref{sec:selection}.
Nevertheless, as can be seen in Table~\ref{tab:sr_summary}, the $\ggF$ contribution to this
signal region is large, approximately~$30\%$, so it has to be profiled by the global fit together
with the extraction of the significance of the signal strength of the $\VBF$ production process.

The global likelihood can be evaluated as a function of the ratio $\sigmu_{\scVX}/\sigmu_{\scggF}$,
with both signal strengths varied independently. The result is illustrated in
Fig.~\ref{fig:likelihood_vs_ratio}, which has a best-fit value for the
ratio of
\begin{equation}
  \frac{\sigmu_{\scVX}}{\sigmu_{\scggF}}
  = 1.26\,^{+0.61}_{-0.45}\,(\rm{stat})\,^{+0.50}_{-0.26}\,(\rm{syst})
  = 1.26\,^{+0.79}_{-0.53}
  .
\end{equation}

\begin{figure}[t!]
  \includegraphics[width=0.45\textwidth]{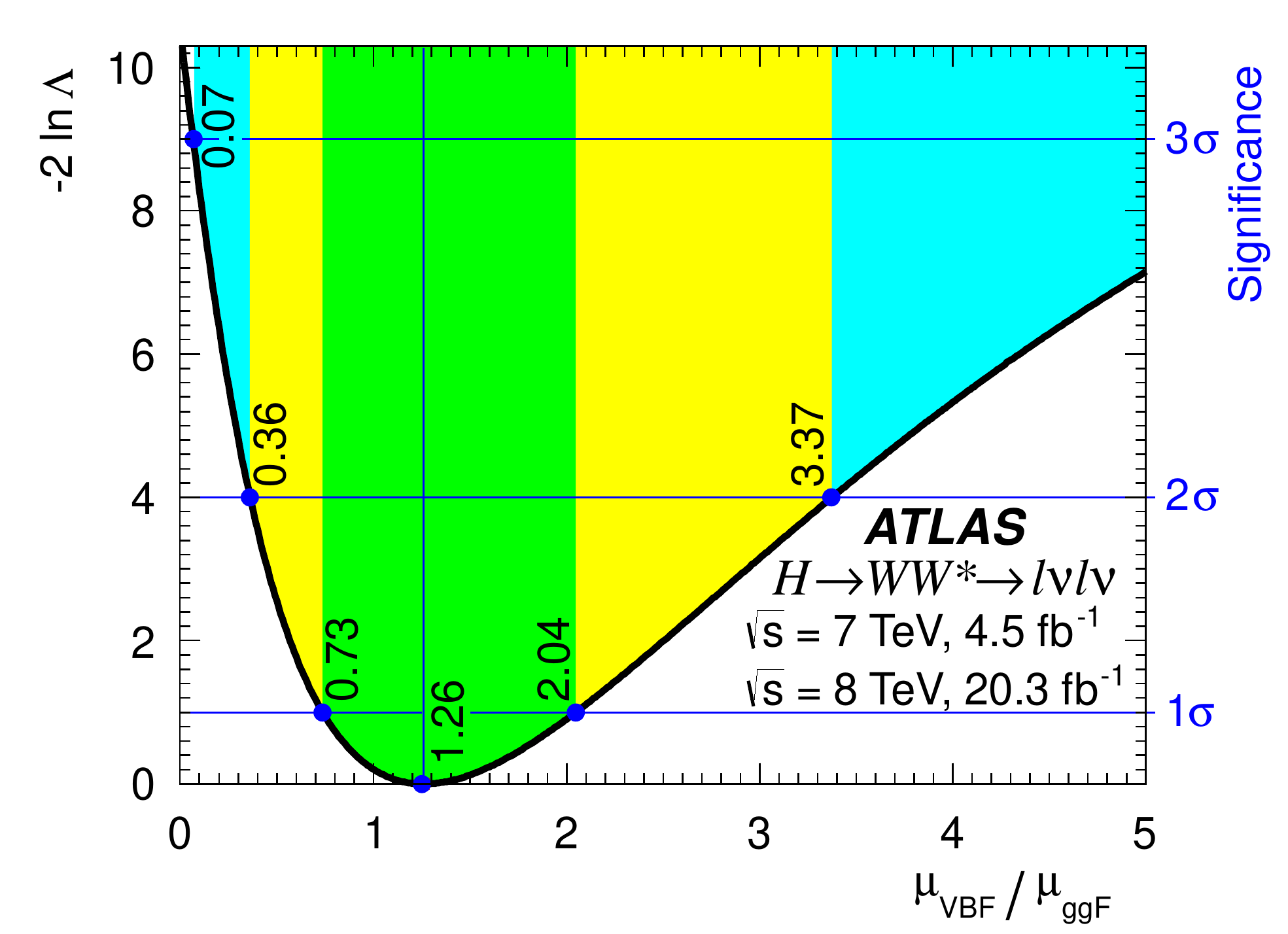}
\caption{Likelihood scan as a function of $\sigmu_{\scVX}/\sigmu_{\scggF}$ for $\mH{\EQ}125.36\GeV$.
  The value of the likelihood at $\sigmu_{\scVX}/\sigmu_{\scggF}{\EQ}0$ gives the
  significance of the $\VBF$ signal at $3.2$ standard deviations.
  The inner (middle) [outer] band shaded darker (lighter) [darker] represents the one (two) [three]
  standard deviation uncertainty around the central value represented by the vertical line.
}
\label{fig:likelihood_vs_ratio}
\end{figure}

The value of the likelihood at $\sigmu_{\scVX}/\sigmu_{\scggF}{\EQ}0$ can be interpreted as the
observed significance of the $\VBF$ production process for $\mH{\EQ}\HwwHiggsMass{ATLAS}\GeV$,
and corresponds to $3.2$ standard deviations; the expected significance
is $2.7$ standard deviations. This establishes the evidence for the $\VBF$ production mode in the $\HWWlvlv$ final state.
The significance derived from testing the ratio $\sigmu_{\scVX}/\sigmu_{\scggF}{\EQ}0$ is equivalent to
the significance of testing $\sigmu_{\scVX}{\EQ}0$, though testing the ratio is conceptually
advantageous since the branching fraction cancels in this parameter, while it is
implicit in $\sigmu_{\scVX}$.

This result was verified with the cross-check analysis described in Sec.~\ref{sec:selection_2jvbf},
in which the multivariate discriminant
is replaced with a series of event selection requirements motivated by the VBF topology.
The expected and observed significances at $\mH{\EQ}\HwwHiggsMass{ATLAS}\GeV$ are
$2.1$ and $3.0$ standard deviations, respectively.
The compatibility of the $8\TeV$ results from the cross-check and $\bdt$ analyses was checked with
pseudo-experiments, considering the statistical uncertainties only and fixing $\sigmu_{\scggF}$ to $1.0$.
With those caveats, the probability that the difference in $Z_0$ values is
larger than the one observed is $79\%$, reflecting good agreement.

\subsection{\boldmath Signal strength $\sigmu$ \label{sec:results_mu}}

The parameter $\sigmu$ is used to characterize the inclusive Higgs boson signal strength
as well as subsets of the signal regions or individual production modes.
First, the $\ggF$ and $\VBF$ processes can be distinguished by using
the normalization parameter $\sigmu_{\scggF}$ for the signal predicted for the $\ggF$
signal process, and $\sigmu_{\scVBF}$ for the signal predicted for the $\VBF$ signal process.
This can be done for a fit to any set of the signal regions in the various categories.
In addition, to check that the measured value is consistent among categories,
different subsets of the signal regions can be fit.
For example, the $\NjetEQzero$ and $\NjetEQone$ categories can be compared, or the $\DFchan$
and $\SFchan$ categories.  To derive these results, only the signal regions are
separated; the control region definitions do not change.  In particular, the control regions
defined using only $\DFchan$ events are used, even when only $\SFchan$ signal regions are considered.

The combined Higgs signal
strength $\sigmu$, including $7$ and $8\TeV$ data and all signal region categories, is
\begin{widetext}
\begin{equation}
\begin{array}{llllll}
 \sigmu
 &= 1.09
 &^{+0.16}_{-0.15}\,(\textrm{stat})
 &^{+0.08}_{-0.07}\,\Big(\!\begin{tabular}{c}{\rm\footnotesize expt}\\ \noalign{\vskip -0.10truecm}{\rm\footnotesize syst}\end{tabular}\!\Big)
 &^{+0.15}_{-0.12}\,\Big(\!\begin{tabular}{c}{\rm\footnotesize theo}\\ \noalign{\vskip -0.10truecm}{\rm\footnotesize syst}\end{tabular}\!\Big)
 &{\PM}0.03\,\Big(\!\begin{tabular}{c}{\rm\footnotesize lumi}\\ \noalign{\vskip -0.10truecm}{\rm\footnotesize syst}\end{tabular}\!\Big)
 \\
 \clineskip
 &= 1.09 &^{+0.16}_{-0.15}\,\textrm{(stat)} &^{+0.17}_{-0.14}\,\textrm{(syst)}
 \\
 \clineskip
 \clineskip
 &= 1.09 &^{+0.23}_{-0.21}.
\end{array}
\label{eqn:mu}
\end{equation}
\end{widetext}

The uncertainties are divided according to their source.
The statistical uncertainty accounts for the number of observed
events in the signal regions and profiled control regions.  The statistical
uncertainties from Monte Carlo simulated samples, from nonprofiled
control regions, and from the extrapolation factors used in the $\Wjets$ background estimate
are all included in the experimental uncertainties here and for all results in this section.
The theoretical uncertainty includes uncertainties on the signal acceptance and cross
section as well as theoretical uncertainties on the background
extrapolation factors and normalizations.
The expected value of $\sigmu$ is
$1\,^{+0.16}_{-0.15}\,\textrm{(stat)}\,^{+0.17}_{-0.13}\,\textrm{(syst)}$.

\begin{figure}[t!]
\includegraphics[width=0.40\textwidth]{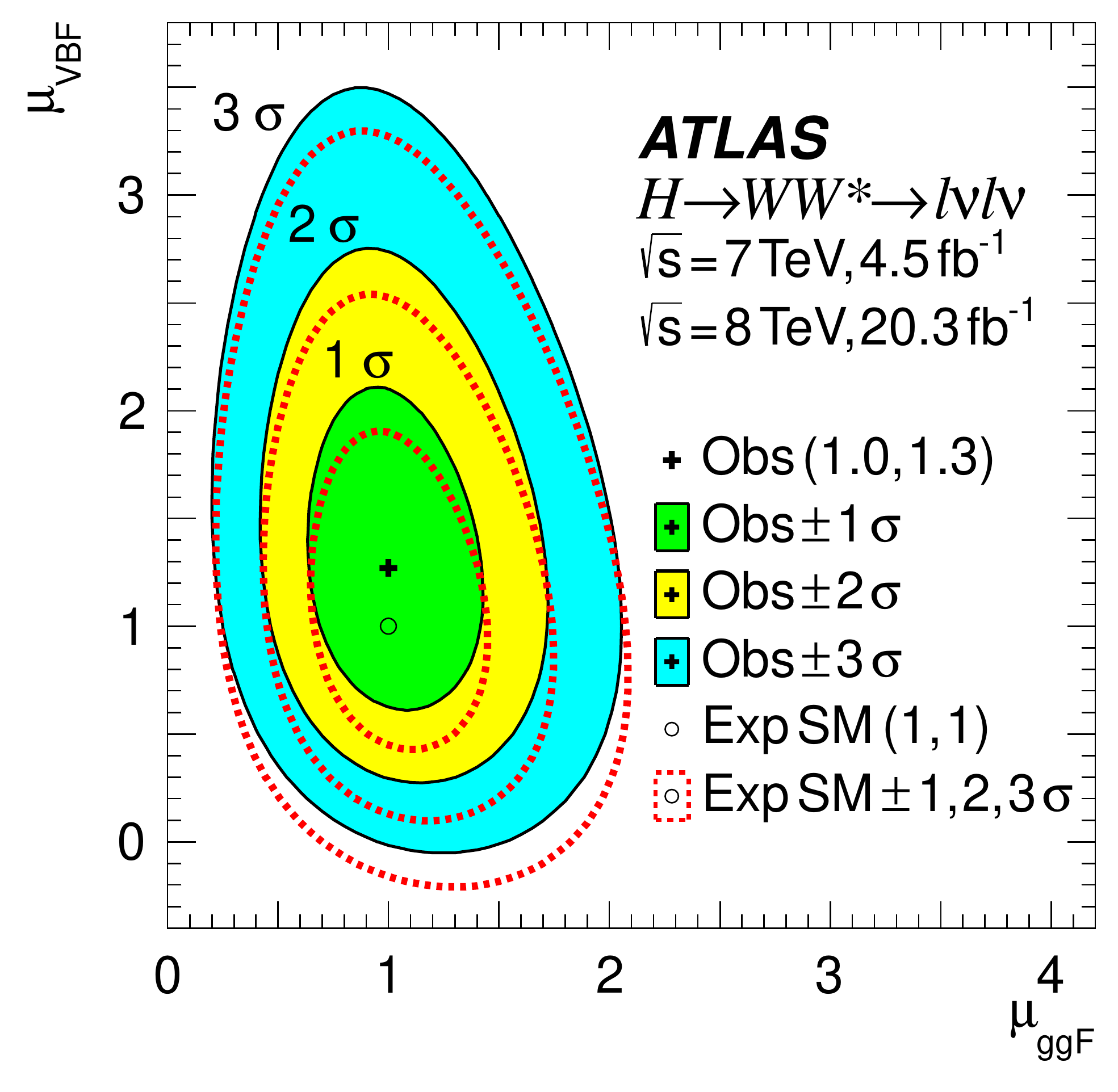}
\caption{
  Likelihood scan as a function of $\sigmu_{\scggF}$ and $\sigmu_{\scVX}$.
  The best-fit observed (expected SM) value is represented by the cross symbol (open circle)
  and its one, two, and three standard deviation contours are shown by solid lines surrounding the filled areas (dotted lines).
  The $x$- and $y$-axis scales are the same to visually highlight the relative sensitivity.
}
\label{fig:muVX_vs_ggF}
\end{figure}

In order to check the compatibility with the SM predictions of the
$\ggF$ and $\VBF$ production processes, $\sigmu_{\scggF}$ and $\sigmu_{\scVX}$
can be simultaneously determined through a fit to all categories because of the
different sensitivity to these processes in the various categories.
In this fit, the $\VH$ contribution is included although there is no dedicated category for it,
and the SM value for the ratio $\sigma_{\scVBF}/\sigma_{\scVH}$ is assumed.
Technically, the signal strength $\sigmu_{\scVBFVH}$ is measured, but because the contribution
from $\VH$ is negligible, the notation $\sigmu_{\scVBF}$ is used.
The corresponding two-dimensional likelihood contours as a function of
$\sigmu_{\scggF}$ and $\sigmu_{\scVX}$ are shown in Fig.~\ref{fig:muVX_vs_ggF}.
Using the same treatment, the separate signal strengths can be measured.  The results
are
\begin{equation}
\begin{array}{llllll}
\no\sigmu_{\scggF}\no&=1.02\no&{\PM}0.19           \np&^{+0.22}_{-0.18}\np&=1.02&^{+0.29}_{-0.26} \\ \clineskip
\no\sigmu_{\scVX} \no&=1.27\no&\,\,^{+0.44}_{-0.40}\np&^{+0.30}_{-0.21}\np&=1.27&^{+0.53}_{-0.45}.\\ \clineskip
                     &        & ~\textrm{(stat)}      &\textrm{(syst)}\np
\end{array}
\end{equation}

\begin{table*}[tb!]
\caption{
  Summary of uncertainties on the signal strength $\sigmu$.
  The table gives the relative uncertainties for inclusive Higgs production (left),
  $\ggF$ production (middle), and $\VX$ production (right).
  For each group separated by a horizontal line, the first line gives the combined result.
  The ``profiled signal region'' indicates the contribution of the uncertainty on the $\ggF$ signal yield
  to the $\sigmu_\scVX$ measurement and vice versa.
  The ``misid.\ factor'' is the systematic uncertainty related to the $\Wjets$ background estimation.
  The ``$\ZDYll$'' entry corresponds to uncertainties on the $\frecoil$ selection efficiency for the $\NjetLEone$ $\SFchan$ category.
  The ``muons and electrons'' entry includes uncertainties on the lepton energy scale, lepton momentum corrections,
  lepton trigger efficiencies, and lepton isolation efficiencies.
  The ``jets'' uncertainties include the jet energy scale, jet energy resolution, and the $b$-tagging efficiency.
  Values are quoted assuming $\mH{\EQ}125.36\GeV$.
  The plot for $\VX$ (third column) has a different scale than the the other
  columns to show the relative uncertainties per column.
  The entries marked with a dash are smaller than $0.01$ or do not apply.
}
\label{tab:syst_mu}
{\small
  \centering
\begin{tabular*}{1\textwidth}{l p{0.0015\textwidth}l p{0.0015\textwidth}l}
\dbline
\begin{tabular*}{0.480\textwidth}{l cc l}
& \multicolumn{3}{c}{Observed $\sigmu{\EQ}1.09$}
\\
\clineskip
\cline{2-4}
\clineskip
\multicolumn{1}{p{0.235\textwidth}}{Source}
& \multicolumn{2}{p{0.090\textwidth}}{~~~Error}
& \multicolumn{1}{p{0.120\textwidth}}{~~Plot of error}
\\
& \multicolumn{1}{c}{$+$}
& \multicolumn{1}{c}{$-$}
& (scaled by 100)
\\
\sgline
\balkenscale{32}{-29.3}
Data statistics                             &0.16 &0.15 &{\myr\Balkenx{0}{16}{15}{0}{0}}\\
\quad Signal regions                        &0.12 &0.12 &{\myb\Balkenx{0}{12}{12}{0}{0}} \\
\quad Profiled control regions              &0.10 &0.10 &{\myb\Balkenx{0}{10}{10}{0}{0}} \\
\quad Profiled signal regions               &\multicolumn{1}{c}{-} &\multicolumn{1}{c}{-} & ~~~~~~~~-\\
\clineskip\clineskip
MC statistics                               &0.04 &0.04 &{\myr\Balkenx{0}{04}{04}{0}{0}} \\
\clineskip\clineskip
Theoretical systematics                     &0.15 &0.12 &{\myr\Balkenx{0}{14}{12}{0}{0}}\\
\quad Signal $\HWW$ $\BF$                   &0.05 &0.04 &{\myb\Balkenx{0}{05}{04}{0}{0}} \\
\quad Signal ggF cross section              &0.09 &0.07 &{\myb\Balkenx{0}{09}{07}{0}{0}} \\
\quad Signal ggF acceptance                 &0.05 &0.04 &{\myb\Balkenx{0}{05}{04}{0}{0}} \\
\quad Signal VBF cross section              &0.01 &0.01 &{\myb\Balkenx{0}{01}{01}{0}{0}} \\
\quad Signal VBF acceptance                 &0.02 &0.01 &{\myb\Balkenx{0}{02}{01}{0}{0}} \\
\quad Background $\WW$                      &0.06 &0.06 &{\myb\Balkenx{0}{06}{06}{0}{0}} \\
\quad Background top quark                  &0.03 &0.03 &{\myb\Balkenx{0}{03}{03}{0}{0}} \\
\quad Background misid.\ factor             &0.05 &0.05 &{\myb\Balkenx{0}{05}{05}{0}{0}} \\
\quad Others                                &0.02 &0.02 &{\myb\Balkenx{0}{02}{02}{0}{0}} \\
\clineskip\clineskip
Experimental systematics                    &0.07 &0.06 &{\myr\Balkenx{0}{07}{06}{0}{0}}\\
\quad Background misid.\ factor             &0.03 &0.03 &{\myb\Balkenx{0}{03}{03}{0}{0}} \\
\quad Bkg.\ $\ZDYll$                        &0.02 &0.02 &{\myb\Balkenx{0}{02}{02}{0}{0}} \\
\quad Muons and electrons                   &0.04 &0.04 &{\myb\Balkenx{0}{04}{04}{0}{0}} \\
\quad Missing transv.\ momentum             &0.02 &0.02 &{\myb\Balkenx{0}{02}{02}{0}{0}} \\
\quad Jets                                  &0.03 &0.02 &{\myb\Balkenx{0}{03}{02}{0}{0}} \\
\quad Others                                &0.03 &0.02 &{\myb\Balkenx{0}{03}{02}{0}{0}} \\
\clineskip\clineskip
Integrated luminosity                       &0.03 &0.03 &{\myr\Balkenx{0}{03}{03}{0}{0}} \\
\clineskip\clineskip
Total                                       &0.23 &0.21 &{\myr\Balkenx{0}{23}{21}{0}{0}} \vspace{1.0mm}\\
&&&
\renewcommand{\bcfontstyle}{\bfseries}%
\renewcommand{\bcfontstyle}{}%
\hspace{-13.5pt}%
\begin{bchart}[step=15,min=-30,max=30,width=0.109\textwidth,scale=1.0]\end{bchart}%
\hspace{-2.0mm}%
\\
\end{tabular*}
&
&
\begin{tabular*}{0.238\textwidth}{lll}
\multicolumn{3}{c}{Observed $\sigmu_{\ggF}{\EQ}1.02$}
\\
\sgline
\multicolumn{2}{p{0.090\textwidth}}{~~~Error}
& \multicolumn{1}{p{0.120\textwidth}}{~~Plot of error}
\\
\multicolumn{1}{c}{$+$}
& \multicolumn{1}{c}{$-$}
& (scaled by 100)
\\
\sgline
\balkenscale{32}{-29.3}
0.19 &0.19 &{\myr\Balkenx{0}{19}{19}{0}{0}}\\
0.14 &0.14 &{\myb\Balkenx{0}{14}{14}{0}{0}} \\
0.12 &0.12 &{\myb\Balkenx{0}{12}{12}{0}{0}} \\
0.03 &0.03 &{\myb\Balkenx{0}{03}{03}{0}{0}} \\
\clineskip\clineskip
0.06 &0.06 &{\myr\Balkenx{0}{06}{06}{0}{0}} \\
\clineskip\clineskip
0.19 &0.16 &{\myr\Balkenx{0}{19}{16}{0}{0}} \\
0.05 &0.03 &{\myb\Balkenx{0}{05}{03}{0}{0}} \\
0.13 &0.09 &{\myb\Balkenx{0}{13}{09}{0}{0}} \\
0.06 &0.05 &{\myb\Balkenx{0}{06}{05}{0}{0}} \\
\multicolumn{1}{c}{-} &\multicolumn{1}{c}{-} & ~~~~~~~~-\\
\multicolumn{1}{c}{-} &\multicolumn{1}{c}{-} & ~~~~~~~~-\\
0.08 &0.08 &{\myb\Balkenx{0}{08}{08}{0}{0}} \\
0.04 &0.04 &{\myb\Balkenx{0}{04}{04}{0}{0}} \\
0.06 &0.06 &{\myb\Balkenx{0}{06}{06}{0}{0}} \\
0.02 &0.02 &{\myb\Balkenx{0}{02}{02}{0}{0}} \\
\clineskip\clineskip
0.08 &0.08 &{\myr\Balkenx{0}{08}{08}{0}{0}} \\
0.04 &0.04 &{\myb\Balkenx{0}{04}{04}{0}{0}} \\
0.03 &0.03 &{\myb\Balkenx{0}{03}{03}{0}{0}} \\
0.05 &0.04 &{\myb\Balkenx{0}{05}{04}{0}{0}} \\
0.02 &0.01 &{\myb\Balkenx{0}{02}{01}{0}{0}} \\
0.03 &0.03 &{\myb\Balkenx{0}{03}{03}{0}{0}} \\
0.03 &0.03 &{\myb\Balkenx{0}{03}{03}{0}{0}} \\
\clineskip\clineskip
0.03 &0.02 &{\myr\Balkenx{0}{03}{02}{0}{0}} \\
\clineskip\clineskip
0.29 &0.26 &{\myr\Balkenx{0}{29}{26}{0}{0}} \vspace{1.0mm}\\
&&
\renewcommand{\bcfontstyle}{\bfseries}%
\renewcommand{\bcfontstyle}{}%
\hspace{-13.5pt}%
\begin{bchart}[step=15,min=-30,max=30,width=0.109\textwidth,scale=1.0]\end{bchart}%
\hspace{-2.0mm}%
\end{tabular*}
&
&
\begin{tabular*}{0.238\textwidth}{lll}
\multicolumn{3}{c}{Observed $\sigmu_{\scVX}{\EQ}1.27$}
\\
\sgline
\multicolumn{2}{p{0.090\textwidth}}{~~~Error}
& \multicolumn{1}{p{0.120\textwidth}}{~~Plot of error}
\\
\multicolumn{1}{c}{$+$}
& \multicolumn{1}{c}{$-$}
& (scaled by 100)
\\
\sgline
\balkenscale{16}{-58.3}
0.44 &0.40 &{\myr\Balkenx{0}{44}{40}{0}{0}} \\
0.38 &0.35 &{\myb\Balkenx{0}{38}{35}{0}{0}} \\
0.21 &0.18 &{\myb\Balkenx{0}{21}{18}{0}{0}} \\
0.09 &0.08 &{\myb\Balkenx{0}{09}{08}{0}{0}} \\
\clineskip\clineskip
0.05 &0.05 &{\myr\Balkenx{0}{05}{05}{0}{0}} \\
\clineskip\clineskip
0.22 &0.15 &{\myr\Balkenx{0}{22}{15}{0}{0}} \\
0.07 &0.04 &{\myb\Balkenx{0}{07}{04}{0}{0}} \\
0.03 &0.03 &{\myb\Balkenx{0}{03}{03}{0}{0}} \\
0.07 &0.07 &{\myb\Balkenx{0}{07}{07}{0}{0}} \\
0.07 &0.04 &{\myb\Balkenx{0}{07}{04}{0}{0}} \\
0.15 &0.08 &{\myb\Balkenx{0}{15}{08}{0}{0}} \\
0.07 &0.07 &{\myb\Balkenx{0}{07}{07}{0}{0}} \\
0.06 &0.06 &{\myb\Balkenx{0}{06}{06}{0}{0}} \\
0.02 &0.02 &{\myb\Balkenx{0}{02}{02}{0}{0}} \\
0.03 &0.03 &{\myb\Balkenx{0}{03}{03}{0}{0}} \\
\clineskip\clineskip
0.18 &0.14 &{\myr\Balkenx{0}{18}{14}{0}{0}} \\
0.02 &0.01 &{\myb\Balkenx{0}{02}{01}{0}{0}} \\
0.01 &0.01 &{\myb\Balkenx{0}{01}{01}{0}{0}} \\
0.03 &0.02 &{\myb\Balkenx{0}{03}{02}{0}{0}} \\
0.05 &0.05 &{\myb\Balkenx{0}{05}{05}{0}{0}} \\
0.15 &0.11 &{\myb\Balkenx{0}{15}{11}{0}{0}} \\
0.06 &0.06 &{\myb\Balkenx{0}{06}{06}{0}{0}} \\
\clineskip\clineskip
0.05 &0.03 &{\myr\Balkenx{0}{05}{03}{0}{0}} \\
\clineskip\clineskip
0.53 &0.45 &{\myr\Balkenx{0}{53}{45}{0}{0}} \vspace{1.0mm}\\
&&
\renewcommand{\bcfontstyle}{\bfseries}%
\renewcommand{\bcfontstyle}{}%
\hspace{-13.5pt}%
\begin{bchart}[step=30,min=-60,max=60,width=0.109\textwidth,scale=1.0]\end{bchart}%
\hspace{-2.0mm}%
\end{tabular*}
\\
\dbline
\end{tabular*}
\\
}
\end{table*}

The details of the uncertainties on $\sigmu$, $\sigmu_{\scggF}$, and
$\sigmu_{\scVX}$ are shown in Table \ref{tab:syst_mu}.
The statistical uncertainty is the largest single source of uncertainty
on the signal strength results, although theoretical uncertainties also play a substantial
role, especially for $\sigmu_{\scggF}$.

The signal strength results are shown in Table \ref{tab:mu} for $\mH{\EQ}\HwwHiggsMass{ATLAS}\GeV$.
The table includes inclusive results as well as results for individual categories and production modes.
The expected and observed significance for each category and production mode is also shown.

\begin{table*}[t]
\caption{
  Signal significance $Z_0$ and signal strength $\sigmu$.
  The expected (Exp)\ and observed (Obs)\ values are given;
  $\sigmu_{\exp}$ is unity by assumption.
  For each group separated by a horizontal line, the highlighted first line gives the combined result.
  The plots correspond to the values in the table as indicated.
  For the $\sigmu$ plot, the thick line represents the statistical uncertainty
  (Stat) in the signal region, the thin line represents the
  total uncertainty (Tot), which includes the uncertainty from systematic sources (Syst).
  The uncertainty due to background sample statistics is included in the latter.
  The last two rows report the results when considering ggF and VBF production modes separately.
  The values are given assuming $\mH{\EQ}125.36\GeV$.
}
\label{tab:mu}
{\small
\begin{tabular*}{1\textwidth}{ l p{0.001\textwidth} l }
\dbline
\begin{tabular*}{0.400\textwidth}{l rc l}
& \multicolumn{3}{c}{Signal significance}
\\
\clineskip
\cline{2-4}
\clineskip
  \multicolumn{1}{p{0.130\textwidth}}{Sample}
& \multicolumn{1}{p{0.040\textwidth}}{~~Exp.}
& \multicolumn{1}{p{0.050\textwidth}}{~~Obs.}
& \multicolumn{1}{p{0.128\textwidth}}{Bar graph of}
\\
& \multicolumn{1}{l}{\,~~$Z_0$}
& \multicolumn{1}{l}{\,~~$Z_0$}
& \multicolumn{1}{l}{observed $Z_0$}
\\
\sgline
\balkenscale{400}{-0.1}
$\NjetEQzero$                           &$3.70$ &$4.08$  &{\myr\Balkenx{0}{4.08}{0}{0}{0}} \\
\quad $\DFchan$, $\ell_2{\EQ}\mu$       &$2.89$ &$3.07$  &{\myb\Balkenx{0}{3.07}{0}{0}{0}} \\
\quad $\DFchan$, $\ell_2{\EQ}e$         &$2.36$ &$3.12$  &{\myb\Balkenx{0}{3.12}{0}{0}{0}} \\
\quad $\SFchan$ category                &$1.43$ &$0.71$  &{\myb\Balkenx{0}{0.71}{0}{0}{0}} \\
\clineskip\clineskip
$\NjetEQone$                            &$2.60$ &$2.49$  &{\myr\Balkenx{0}{2.49}{0}{0}{0}} \\
\quad $\DFchan$ category                &$2.56$ &$2.83$  &{\myb\Balkenx{0}{2.83}{0}{0}{0}} \\
\quad $\SFchan$ category                &$1.02$ &$0.21$  &{\myb\Balkenx{0}{0.21}{0}{0}{0}} \\
\clineskip\clineskip
$\NjetGEtwo$, ggF, $\emu$               &$1.21$ &$1.44$  &{\myr\Balkenx{0}{1.44}{0}{0}{0}} \\
\clineskip\clineskip
$\NjetGEtwo$, VBF-enr.                  &$3.38$ &$3.84$  &{\myr\Balkenx{0}{3.84}{0}{0}{0}} \\
\quad $\DFchan$ category                &$3.01$ &$3.02$  &{\myb\Balkenx{0}{3.02}{0}{0}{0}} \\
\quad $\SFchan$ category                &$1.58$ &$2.96$  &{\myb\Balkenx{0}{2.96}{0}{0}{0}} \\
\clineskip\clineskip
All $\Njet$, all signal                 &$5.76$ &$6.06$  &{\myr\Balkenx{0}{6.06}{0}{0}{0}} \\
\quad ggF as signal                     &$4.34$ &$4.28$  &{\myb\Balkenx{0}{4.28}{0}{0}{0}} \\
\quad VBF as signal                     &$2.67$ &$3.24$  &{\myb\Balkenx{0}{3.24}{0}{0}{0}} \vspace{1.0mm} \\
                                        &       &        &
\renewcommand{\bcfontstyle}{\bfseries}%
\renewcommand{\bcfontstyle}{}%
\hspace{-7.5pt}%
\begin{bchart}[step=1,max=6,width=0.130\textwidth,scale=1.03]\end{bchart}%
\vspace{-2.5mm}%
\end{tabular*}
&
&
\vspace{3.0mm}
\begin{tabular*}{0.570\textwidth}{lp{0.002\textwidth} rrr rrrrrc ll}
\multicolumn{2}{c}{Expected}
&& \multicolumn{6}{c}{Observed uncertainty}
&& \multicolumn{2}{c}{Observed central value}
\\
\clineskip
\cline{1-2}
\cline{4-9}
\cline{11-12}
\clineskip
\multicolumn{2}{c}{Tot\,err}
&&\multicolumn{2}{c}{Tot\,err}
& \multicolumn{2}{c}{Stat\,err}
& \multicolumn{2}{c}{Syst\,err}
&&\multicolumn{1}{c}{$\sigmu_{\obs}$}
&\multicolumn{1}{p{0.190\textwidth}}{~~~~$\sigmu_{\obs}{\PM}\textrm{stat\ (thick)}$}
\\
\multicolumn{1}{c}{$+$} & \multicolumn{1}{c}{$-$}
&
& \multicolumn{1}{c}{$+$} & \multicolumn{1}{c}{$-$}
& \multicolumn{1}{c}{$+$} & \multicolumn{1}{c}{$-$}
& \multicolumn{1}{c}{$+$} & \multicolumn{1}{c}{$-$}
&&
& \multicolumn{1}{l}{\,~~~~~~~~~${\PM}\textrm{total (thin)}$}
\\
\sgline
\balkenscale{700}{-1.2}
$0.35$ &$0.30$ &&$0.37$ &$0.32$ &$0.22$ &$0.22$ &$0.30$ &$0.23$ &&$1.15$ &{\myr\Balkenx{1.15}{0.22}{0.22}{0.37}{0.32}} \\
$0.41$ &$0.36$ &&$0.43$ &$0.38$ &$0.30$ &$0.29$ &$0.32$ &$0.24$ &&$1.08$ &{\myb\Balkenx{1.08}{0.30}{0.29}{0.43}{0.38}} \\
$0.49$ &$0.44$ &&$0.54$ &$0.48$ &$0.38$ &$0.37$ &$0.39$ &$0.30$ &&$1.40$ &{\myb\Balkenx{1.40}{0.38}{0.37}{0.54}{0.48}} \\
$0.74$ &$0.70$ &&$0.68$ &$0.66$ &$0.45$ &$0.44$ &$0.51$ &$0.50$ &&$0.47$ &{\myb\Balkenx{0.47}{0.45}{0.44}{0.68}{0.66}} \\
\clineskip\clineskip
$0.51$ &$0.41$ &&$0.50$ &$0.41$ &$0.33$ &$0.32$ &$0.38$ &$0.26$ &&$0.96$ &{\myr\Balkenx{0.96}{0.33}{0.32}{0.50}{0.41}} \\
$0.51$ &$0.42$ &&$0.56$ &$0.45$ &$0.35$ &$0.35$ &$0.43$ &$0.29$ &&$1.16$ &{\myb\Balkenx{1.16}{0.35}{0.35}{0.56}{0.45}} \\
$1.12$ &$0.98$ &&$1.02$ &$0.97$ &$0.80$ &$0.76$ &$0.63$ &$0.61$ &&$0.19$ &{\myb\Balkenx{0.19}{0.80}{0.76}{1.02}{0.97}} \\
\clineskip\clineskip
$0.96$ &$0.83$ &&$0.91$ &$0.84$ &$0.70$ &$0.68$ &$0.70$ &$0.49$ &&$1.20$ &{\myr\Balkenx{1.20}{0.70}{0.68}{0.99}{0.84}} \\
\clineskip\clineskip
$0.42$ &$0.36$ &&$0.45$ &$0.38$ &$0.36$ &$0.33$ &$0.27$ &$0.19$ &&$1.20$ &{\myr\Balkenx{1.20}{0.36}{0.33}{0.45}{0.38}} \\
$0.48$ &$0.40$ &&$0.47$ &$0.39$ &$0.40$ &$0.35$ &$0.24$ &$0.16$ &&$0.98$ &{\myb\Balkenx{0.98}{0.40}{0.35}{0.47}{0.39}} \\
$0.84$ &$0.67$ &&$0.97$ &$0.78$ &$0.83$ &$0.71$ &$0.51$ &$0.33$ &&$1.98$ &{\myb\Balkenx{1.98}{0.83}{0.71}{0.97}{0.78}} \\
\clineskip\clineskip
$0.23$ &$0.20$ &&$0.23$ &$0.21$ &$0.16$ &$0.15$ &$0.17$ &$0.14$ &&$1.09$ &{\myr\Balkenx{1.09}{0.16}{0.15}{0.23}{0.21}} \\
$0.30$ &$0.24$ &&$0.29$ &$0.26$ &$0.19$ &$0.19$ &$0.22$ &$0.18$ &&$1.02$ &{\myb\Balkenx{1.02}{0.19}{0.19}{0.29}{0.26}} \\
$0.50$ &$0.43$ &&$0.53$ &$0.45$ &$0.44$ &$0.40$ &$0.30$ &$0.21$ &&$1.27$ &{\myb\Balkenx{1.27}{0.44}{0.40}{0.53}{0.45}} \vspace{1.0mm} \\
       &       &&       &       &       &       &       &       &&       &
\renewcommand{\bcfontstyle}{\bfseries}%
\renewcommand{\bcfontstyle}{}%
\hspace{-6.5pt}%
\begin{bchart}[step=1.0,min=-1.0,max=3.0,width=0.158\textwidth,scale=1.0]\end{bchart}%
\vspace{-5.0mm}%
$\nq$%
\end{tabular*}
\\
\dbline
\end{tabular*}
}
\end{table*}

The $\sigmu$ values are consistent with each other and with unity within the assigned uncertainties.
In addition to serving as a consistency check, these results illustrate the
sensitivity of the different categories.  For the overall signal strength, the contribution from the
$\NjetGEtwo$ VBF category is second only to the $\NjetEQzero$ ggF category, and the $\NjetGEtwo$
ggF category contribution is comparable to those in the $\NjetEQzeroone$ $\SFchan$ categories.

For all of these results, the signal acceptance for all production modes is evaluated assuming
a SM Higgs boson.
The $\VH$ production process contributes a small number of events, amounting to about $1\%$ of the expected
signal from the $\VBF$ process.  It is included in the predicted signal yield, and where relevant, is grouped
with the $\VBF$ signal assuming the SM value of the ratio $\sigma_{\scVBF}/\sigma_{\scVH}$.
The small (${<}\,1\%$) contribution of $\Htt$ to the signal regions
is treated as signal, assuming the branching fractions as predicted by the SM.

\subsection{\boldmath Higgs couplings to fermions and vector bosons \label{sec:results_scans}}

The values of $\sigmu_{\scggF}$ and $\sigmu_{\scVX}$ can be used to test the compatibility of the fermionic
and bosonic couplings of the Higgs boson with the SM prediction using a framework motivated by
the leading-order interactions~\cite{Heinemeyer:2013tqa}.
The parametrization uses the scale factors $\kF$, applied to all fermionic couplings,
and $\kV$, applied to all bosonic couplings; these parameters are unity for the SM.

In particular, the $\ggF$ production cross section is proportional to
$\kF^2$ through the top-quark or bottom-quark loops at the production vertex, and
the $\VBF$ production cross section is proportional to $\kV^2$.
The branching fraction $\BF_{\HWW}$ is proportional to $\kV^2$ and inversely
proportional to a linear combination of $\kF^2$ and $\kV^2$.
This model assumes that there are no non-SM decay modes, so the denominator
corresponds to the total decay width in terms of the fermionic and bosonic decay amplitudes.
The formulae, following Ref.~\cite{Heinemeyer:2013tqa}, are
\begin{equation}
\np
\begin{array}{lc l}
  \sigmu_{\scggF} &\propto &{\displaystyle\frac{\kF^2 \cdot \kV^2}{(\BF_{\Hff}+\BF_{\Hgg})\,\kF^2+(\BF_{\HVV})\,\kV^2}},\\ \clineskip\clineskip
  \sigmu_{\scVBF} &\propto &{\displaystyle\frac{\kV^4            }{(\BF_{\Hff}+\BF_{\Hgg})\,\kF^2+(\BF_{\HVV})\,\kV^2}}.
\end{array}
\label{eqn:musNkappas}
\end{equation}
The small contribution from $\BF_{\Hgamgam}$ depends on both $\kF$ and $\kV$ and is not explicitly
shown.  Because $(\BF_{\Hff}+\BF_{\Hgg}) {\APPROX} 0.75$, $\kF^2$ is the dominant component of the
denominator for $\kF^2 \lesssim\, 3\,\kV^2$. As a result, the $\kF^2$ dependence for the
$\ggF$ process approximately cancels, but the rate remains sensitive
to $\kV$.  Similarly, the $\VBF$ rate scales approximately with $\kV^4 / \kF^2$ and the $\VBF$ channel
provides more sensitivity to $\kF$ than the $\ggF$ channel does in this model.
Because Eq.~(\ref{eqn:musNkappas}) contains only $\kF^2$ and $\kV^2$, this channel
is not sensitive to the sign of $\kF$ or $\kV$.

The likelihood scan as a function of $\kV$ and $\kF$ is shown in Fig.~\ref{fig:couplings}.
Both the observed and expected contours are shown, and are in good agreement.
The relatively low discrimination among high values of $\kF$ in the plot
is due to the functional behavior of the
total $\ggF$ yield. The product $\sigma_{\scggF}{\CDOT}\BF$ does not depend on $\kF$
in the limit where $\kF \gg \kV$, so the sensitivity at
high $\kF$ values is driven by the value of $\sigmu_{\scVX}$. The $\VBF$
process rapidly vanishes in the limit where $\kF \gg \kV$ due to the increase of
the Higgs boson total width and the consequent reduction of the
branching fraction to $\WW$ bosons.  Therefore, within this framework, excluding
$\sigmu_{\scVX}{\EQ}0$ excludes $\kF \gg \kV$.

\begin{figure}[b!]
\includegraphics[width=0.45\textwidth]{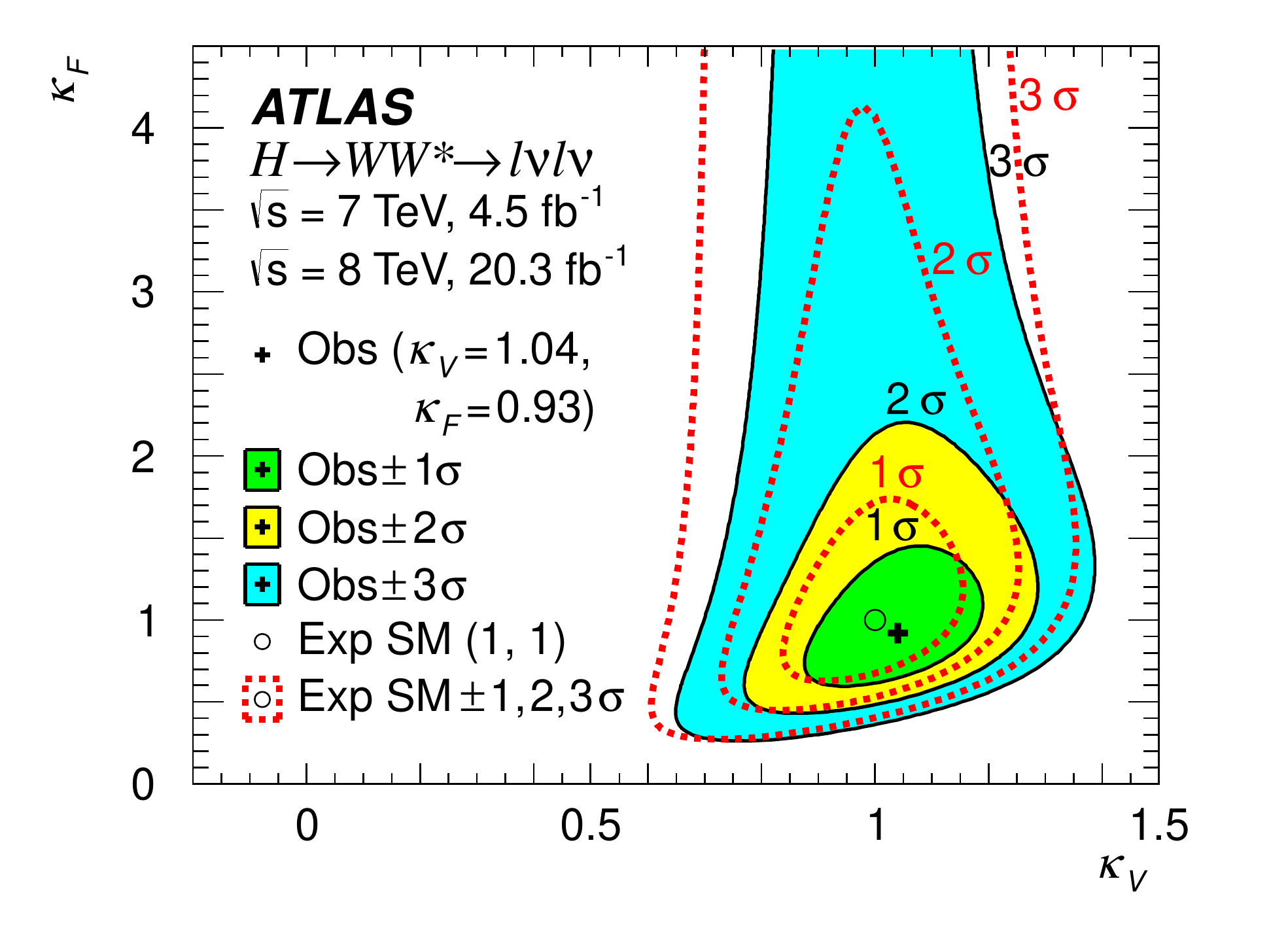}
\caption{
  Likelihood scan as a function of $\kV$ and $\kF$.
  The best-fit observed (expected SM) value is represented by the cross symbol (open circle)
  and its one, two, and three standard deviation contours are shown by solid lines surrounding the filled areas (dotted lines).
  NB. The $y$-axis spans a wider range than the $x$-axis.
}
\label{fig:couplings}
\end{figure}

The best fit values are
\begin{equation}
\begin{array}{llllll}
\no\kF\no&=0.93\no&\,\,^{+0.24}_{-0.18}\np&^{+0.21}_{-0.14}\np&=0.93&^{+0.32}_{-0.23} \\ \clineskip\clineskip
\no\kV\no&=1.04\no&\,\,^{+0.07}_{-0.08}\np&^{+0.07}_{-0.08}\np&=1.04&{\PM}0.11.\\ \clineskip\clineskip
         &        & ~\textrm{(stat)}      &\textrm{(syst)}\np
\end{array}
\end{equation}
and their correlation is $\rho{\EQ}0.47$.  The correlation is derived from the covariance matrix
constructed from the second-order mixed partial derivatives of the likelihood,
evaluated at the best-fit values of $\kF$ and $\kV$.

\subsection{\boldmath Exclusion limits \label{sec:exclusion}}

The analysis presented in this paper has been optimized for a Higgs boson of mass
$\mH{\EQ}125\GeV$, but, due to the low mass resolution of the $\lvlv$ channel, it is sensitive
to SM-like Higgs bosons of mass up to $200\GeV$ and above.
The exclusion ranges are computed using the modified frequentist method $\CLs$~\cite{CLs_2002}.
A SM Higgs boson of mass $\mH$ is considered excluded at $95\%$ C.L.~if the value $\sigmu{\EQ}1$ is
excluded at that mass.
The analysis is expected to exclude a SM Higgs boson with mass down to
$114\GeV$ at $95\%$ C.L.  The clear excess of signal over background, shown in
the previous sections, results in an observed exclusion range of
$\HwwLimit{obs}\GeV$, extending to the upper limit of the search range,
as shown in Fig.~\ref{fig:CLs}.

\begin{figure}[tb!]
  \includegraphics[width=0.45\textwidth]{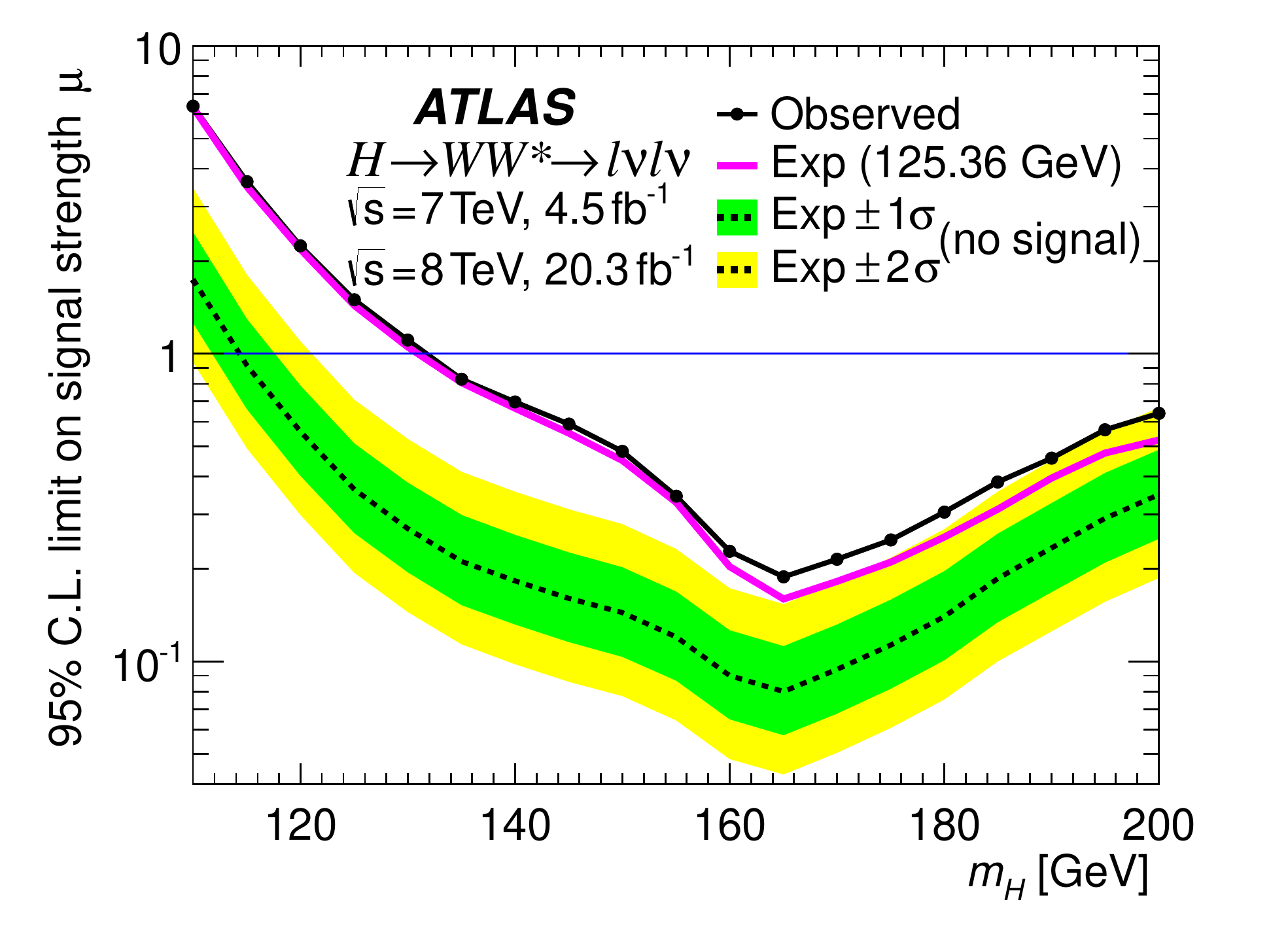}
\caption{
  $\CLs$ exclusion plot for $110{\LE}\mH{\LE}200\GeV$.
  The observed values are shown as a solid line with points where the limit is evaluated.
  The expected values for a signal at $125.36\GeV$ are given as a solid line without points.
  The expected values for scenarios without signal are given by the dotted line.
  The inner (outer) band shaded darker (lighter) represents the one (two) standard deviation uncertainty on the value for expected without signal.
  The limit of $132\GeV$ ($114\GeV$) on $\sigmu$ for the observed (expected no signal) scenario can be seen at low values of $\mH$.
}
\label{fig:CLs}
\end{figure}

\subsection{\boldmath Higgs production cross sections \label{sec:results_xsec}}

The measured signal strength can be used to evaluate the product $\sigma{\CDOT}\BF_{\HWW}$
for Higgs boson production at $\mH{\EQ}\HwwHiggsMass{ATLAS}\GeV$,
as well as for the individual $\ggF$ and $\VBF$ production modes.
The central value is simply the product of $\sigmu$ and the predicted
cross section used to define it.  The uncertainties are similarly scaled,
except for the theoretical uncertainties related to the
total production yield, which do not apply to this measurement.
These are the QCD scale and PDF uncertainties on the
total cross sections, and the uncertainty on the branching fraction for
$\HWW$, as described in Sec.~\ref{sec:signal}.
In practice, the corresponding nuisance parameters are fixed to their nominal
values in the fit, effectively removing these uncertainties from consideration.
Inclusive cross-section measurements are performed for $\ggF$ and $\VBF$ production.
The cross section is also measured for $\ggF$ production in defined fiducial
volumes; this approach minimizes the impact of theoretical uncertainties.

\subsubsection{\boldmath\textit{\textbf{Inclusive cross sections}}\label{sec:results_xsec_incl}}

Inclusive cross sections are evaluated at both $7$ and $8\TeV$ for the $\ggF$
production process and at $8\TeV$ for the $\VBF$ production
process.  The $7\TeV$ $\VBF$ cross section is not measured because of the
large statistical uncertainty.
The signal strengths used for $\ggF$ and $\VBF$ are determined
through a simultaneous fit to all categories as described in Sec.~\ref{sec:results_mu}.
The small $\VH$ contribution, corresponding to $0.9\%$, is neglected,
and its expected fractional yield is added linearly to
the total uncertainty.  The $7\TeV$ signal strength $\sigmu_{\scggF}^{7\!\TeV}$
and $8\TeV$ signal strengths $\sigmu_{\scggF}^{8\!\TeV}$ and $\sigmu_{\scVBF}^{8\!\TeV}$ are
\begin{equation}
\begin{array}{lcccc}
\sigmu_{\scggF}^{7\!\TeV}       & = 0.57 &^{+0.52}_{-0.51} &^{+0.36}_{-0.34} &^{+0.14}_{-0.004} \\ \clineskip\clineskip
\sigmu_{\scggF}^{8\!\TeV}       & = 1.09 &{\PM}0.20        &^{+0.19}_{-0.17} &^{+0.14}_{-0.09} \\ \clineskip\clineskip
\sigmu_{\scVBF}^{8\!\TeV}       & = 1.45 &^{+0.48}_{-0.44} &^{+0.38}_{-0.24} &^{+0.11}_{-0.06} \\ \clineskip\clineskip
                                &        &(\stat)          &(\syst)          &~({\rm sig})
\end{array}
\end{equation}
\noindent
where (sig)\ indicates the systematic uncertainties on the total signal yield
for the measured process, which do not affect the cross-section
measurement.  The effect of uncertainties on the signal yield for other
production modes is included in the systematic uncertainties.
In terms of the measured signal strength, the inclusive cross section is defined as
\begin{equation}
\begin{array}{ll}
\nq\big(\sigma \cdot\mathcal{B}_{\HWW}\big)_\obs
&=
\frac{\displaystyle (\Nsig)_\obs}
     {\displaystyle \Acc\cdot\Cor\cdot\mathcal{B}_{WW\rightarrow \ell\nu\ell\nu}^{\phantom{X}}}
\cdot
\frac{\displaystyle 1}{\displaystyle\Lint} \\
\clineskip
\clineskip
&=
{\displaystyle \hatsigmu \cdot (\sigma\cdot\mathcal{B}_{\HWW})_\exp}.
\end{array}
\end{equation}
\noindent
In this equation, $\Acc$ is the kinematic and geometric acceptance, and $\Cor$ is the ratio
of the number of measured events to the number of events produced in the fiducial phase
space of the detector.  The product $\Acc{\CDOT}\Cor$ is the total acceptance for reconstructed
events.  The cross sections are measured using the last line of the equation, and the results are
\begin{equation}
\begin{array}{lllllll}
\np\sigma_{\scggF}^{7\!\TeV}{\!\CDOT}\mathcal{B}_{\HWW}\np&= 2.0 \np\no&{\PM}1.7            \no\!&^{+1.2}_{-1.1}  \np\no&=2.0 \np\no&\,\,^{+2.1}_{-2.0}   \np&\pb \nq\\ \clineskip\clineskip
\np\sigma_{\scggF}^{8\!\TeV}{\!\CDOT}\mathcal{B}_{\HWW}\np&= 4.6 \np\no&{\PM}0.9            \no\!&^{+0.8}_{-0.7}  \np\no&=4.6 \np\no&\,\,^{+1.2}_{-1.1}    \np&\pb \nq\\ \clineskip\clineskip
\np\sigma_{\scVX }^{8\!\TeV}{\!\CDOT}\mathcal{B}_{\HWW}\np&= 0.51\np\no&\,\,^{+0.17}_{-0.15}\no\!&^{+0.13}_{-0.08}\np\no&=0.51\np\no&\,\,^{+0.22}_{-0.17} \np&\pb.\nq\\ \clineskip\clineskip
\np                                                       &      \np\no&\textrm{(stat)}     \no\!&\textrm{(syst)}\np\no
\end{array}
\end{equation}
\noindent
The predicted cross-section values are
$3.3{\PM}0.4\pb$,
$4.2{\PM}0.5\pb$, and
$0.35{\PM}0.02\pb$, respectively.

These are derived as described in Sec.~\ref{sec:signal}, and the acceptance
is evaluated using the standard signal MC samples.

\subsubsection{\boldmath\textit{\textbf{Fiducial cross sections}}\label{sec:results_xsec_fid}}

\begin{table}[b!]
\caption{
  Fiducial volume definitions for fiducial cross sections.
  The selection is made using only $e\mu$ events. Events in which
  one or both $W$ bosons decay to $\tau\nu$ are excluded from the fiducial volume, but are
  present in the reconstructed volume.
  Energy-related quantities are in $\!\GeV$.
}
\label{tab:fiducial}
{\small
  \centering
\begin{tabular*}{0.480\textwidth}{
    p{0.160\textwidth}
    c
    p{0.050\textwidth}
    c
}
\dbline
Type &
$\NjetEQzero$ & &
$\NjetEQone$
\\
\sgline
Preselection
& \multicolumn{3}{c}{$\pTlead{\GT}22$   } \\
& \multicolumn{3}{c}{$\pTsublead{\GT}10$} \\
& \multicolumn{3}{c}{Opposite charge $\ell$} \\
& \multicolumn{3}{c}{$\mll{\GT}10$      } \\
& \multicolumn{3}{c}{$\pTnu{\GT}20$     } \\
\sgline
$\Njet$-dependent
& $\dphi_{\ll, \nu\nu}{\GT}\pi/2$       && ~~- \\
& $\pTll{\GT}30$                        && ~~- \\
& ~~-                                   && $\mTl{\GT}50$ \\
& ~~-                                   && $\mtt{\LT}66$ \\
& $\mll{\LT}55$                         && $\mll{\LT}55$ \\
& $\dphill{\LT}1.8$                     && $\dphill{\LT}1.8$     \\
\dbline
\end{tabular*}
}
\end{table}

Fiducial cross-section measurements
enable comparisons to theoretical predictions with minimal assumptions about the
kinematics of the signal and possible associated jets in the event.
The cross sections described here are for events produced within
a fiducial volume closely corresponding to a $\ggF$ signal region.
The fiducial volume is defined using generator-level kinematic information, as specified in Table~\ref{tab:fiducial}.
In particular, the total $\pT$ of the neutrino system ($\pTnu$) replaces the
$\MPTj$, and each lepton's $\pT$ is replaced by the generated lepton $\pT$, where the lepton
four-momentum is corrected by adding the four-momenta of all photons within a cone of size $\DR{\EQ}0.1$
to account for energy loss through QED final-state radiation.  These quantities are used to compute $\mTl$.
Jets are defined at hadron level, i.e., after parton showering and hadronization but before detector simulation.
To minimize dependence on the signal model, and therefore the theoretical uncertainties, only $\emu$ events in
the $\NjetLEone$ categories are used.  Also, only the $8\TeV$ data sample is used for these measurements.

The measured fiducial cross section is defined as
\begin{equation}
\begin{array}{ll}
\sigma_{\rm fid} &= \frac{\displaystyle (\Nsig)_\obs}
                          {\displaystyle \phantom{x^X}\Cor\phantom{x^X}}
\cdot
\frac{\displaystyle 1}{\displaystyle\Lint} \\
\clineskip
\clineskip
              &= {\displaystyle \hatsigmu \cdot (\sigma \cdot\mathcal{B}_{H\rightarrow WW^* \rightarrow e\nu\mu\nu})_{\exp}}
\cdot\Acc, \\
\end{array}
\end{equation}
\noindent
with the multiplicative factor $\Acc$ being the sole difference with respect to the inclusive
cross-section calculation.  The measured fiducial cross section is not affected by the
theoretical uncertainties on the total signal yield nor by the theoretical uncertainties on
the signal acceptance. The total uncertainty is reduced compared to the value for the inclusive
cross section because the measured signal yield is not extrapolated to the total phase space.

The correction factors for $\NjetEQzero$ and $\NjetEQone$ events, $\Cor^{\scggF}_{0j}$ and $\Cor^{\scggF}_{1j}$, are
evaluated using the standard signal MC sample. The reconstructed events
include leptons from $\tau$ decays, but for simplicity, the fiducial
volume is defined without these contributions.  According to the simulation,
the fraction of measured signal events within the fiducial volume is $85\%$
for $\NjetEQzero$ and $63\%$ for $\NjetEQone$.

The values of the correction factors are
\begin{equation}
\begin{array}{ll}
\Cor^{\scggF}_{0j} &= 0.507 \PM 0.027, \\ \clineskip\clineskip
\Cor^{\scggF}_{1j} &= 0.506 \PM 0.022.
\end{array}
\end{equation}
The experimental systematic uncertainty is approximately $5\%$.
Remaining theoretical uncertainties on the $\Cor^{\scggF}$ values
were computed by comparing the $\ggF$
predictions of \POWHEG+\HERWIG, \POWHEG+\PYTHIA8 and \POWHEG+\PYTHIA6,
and are found to be approximately $2\%$ and are neglected.
The acceptance of the fiducial volume is
\begin{equation}
\begin{array}{ll}
\Acc^{\scggF}_{0j} &= 0.206 \PM 0.030, \\ \clineskip\clineskip
\Acc^{\scggF}_{1j} &= 0.075 \PM 0.017.
\end{array}
\end{equation}
The uncertainties on the acceptance are purely theoretical in origin and the
largest contributions are from the effect of the QCD scale on the jet multiplicity requirements.

The cross-section values are computed by fitting the $\sigmu$ values in
the $\NjetEQzeroone$ categories.  The $\VBF$
contribution is subtracted assuming the expected yield from the
SM instead of using the simultaneous fit to the $\VBF$ signal regions as is
done for the inclusive cross sections.
The non-negligible $\ggF$ yield in the VBF
categories would require an assumption on the $\ggF$ acceptance for different
jet multiplicities, whereas the fiducial cross-section measurement is intended to avoid this type of assumption.
The effect of the theoretical uncertainties on the $\VBF$ signal yield is included in the
systematic uncertainties on the cross sections.
The obtained signal strengths are
\begin{equation}
\begin{array}{lcccc}
\sigmu^{\scggF}_{0j,\emu} & = 1.39 &{\PM}0.27          &^{+0.21}_{-0.19}\no&^{+0.27}_{-0.17},\\ \clineskip\clineskip
\sigmu^{\scggF}_{1j,\emu} & = 1.14 &^{+0.42}_{-0.41}\no&^{+0.27}_{-0.26}\no&^{+0.42}_{-0.17},\\ \clineskip\clineskip
                 &        &(\stat)  &(\syst)        \no&~({\rm sig})
\end{array}
\end{equation}
\noindent
where (sig)\ indicates the systematic uncertainties on the signal yield and acceptance,
which do not apply to the fiducial cross-section measurements.
The corresponding cross sections, evaluated at $\mH{\EQ}\HwwHiggsMass{ATLAS}\GeV$ and
using the $8\TeV$ data, are
\begin{equation}
\begin{array}{llrlllrll}
\no\sigma_{{\rm fid},0j}^{\scggF}\no&=&\no27.6\no&\,\,^{+5.4}_{-5.3}\np&^{+4.1}_{-3.9}  \np&=&\no27.6\np&\,\,^{+6.8}_{-6.6}\np&\fb,\\ \clineskip\clineskip
\no\sigma_{{\rm fid},1j}^{\scggF}\no&=&\no 8.3\no&\,\,^{+3.1}_{-3.0}\np&^{+3.1}_{-3.0}  \np&=&\no 8.3\np&\,\,^{+3.7}_{-3.5}\np&\fb.\\ \clineskip\clineskip
                                    & &\no    \no&\textrm{(stat)}   \np&\textrm{(syst)}\np
\end{array}
\end{equation}
The predicted values are
$19.9{\PM}3.3\fb$ and $7.3{\PM}1.8\fb$, respectively.

\section{\boldmath Conclusions \label{sec:conclusions}}

An observation of the decay $\HWWlvlv$ with a significance of $\HwwSignif{obs}{all}$ standard
deviations is achieved by an analysis of ATLAS data corresponding to $\HwwLumi{0}{0}\ifb$
of integrated luminosity from $\sqrt{s}{\EQ}7$ and $8\TeV$ $pp$ collisions produced by
the Large Hadron Collider at CERN. This observation confirms the predicted decay of the
Higgs boson to $W$ bosons, at a rate consistent with that given by the Standard Model.  The
SM predictions are additionally supported by evidence for VBF production in this channel,
with an observed significance of $\HwwSignif{obs}{VX}$ standard deviations.

For a Higgs boson with a mass of $\HwwHiggsMass{ATLAS}\GeV$, the ratios of the measured
cross sections to those predicted by the Standard Model are consistent with unity for
both gluon-fusion and vector-boson-fusion production:
\begin{equation}
\begin{array}{ll}
  \sigmu          &= 1.09~^{+0.23}_{-0.21}, \\ \clineskip\clineskip
  \sigmu_{\scggF} &= 1.02~^{+0.29}_{-0.26}, \\ \clineskip\clineskip
  \sigmu_{\scVX}  &= 1.27~^{+0.53}_{-0.45}.
\end{array}
\end{equation}
\noindent
The measurement uncertainties are reduced by $30\%$ relative to the prior ATLAS $\HWWlvlv$
measurements due to improved analysis techniques.  The corresponding cross section times
branching fraction values are
\begin{equation}
\begin{array}{lllllll}
\no\sigma_{\scggF}^{7\!\TeV}\cdot\mathcal{B}_{\HWW}\no&= 2.0 \np&\,\,^{+2.1}_{-2.0}   \np&\pb,\\ \clineskip\clineskip
\no\sigma_{\scggF}^{8\!\TeV}\cdot\mathcal{B}_{\HWW}\no&= 4.6 \np&\,\,^{+1.2}_{-1.1}   \np&\pb,\\ \clineskip\clineskip
\no\sigma_{\scVX }^{8\!\TeV}\cdot\mathcal{B}_{\HWW}\no&= 0.51\np&\,\,^{+0.22}_{-0.17} \np&\pb.\\ \clineskip\clineskip
\end{array}
\end{equation}
These total cross sections, as well as the fiducial cross sections measured in the
exclusive $\NjetEQzero$ and $\NjetEQone$ categories, allow future comparisons to the
more precise cross section calculations currently under development.

The analysis strategies described in this \paper\ set the stage for more precise
measurements using future collisions at the LHC.  The larger data sets will
significantly reduce statistical uncertainties; further modeling and analysis
improvements will be required to reduce the leading systematic uncertainties.
Future precise measurements of the $\HWWlvlv$ decay will provide more stringent
tests of the detailed SM predictions of the Higgs boson properties.

\section*{\boldmath Acknowledgements \label{sec:acknowledgements}}

We thank CERN for the very successful operation of the LHC, as well as the
support staff from our institutions without whom ATLAS could not be
operated efficiently.

We acknowledge the support of ANPCyT, Argentina; YerPhI, Armenia; ARC,
Australia; BMWF and FWF, Austria; ANAS, Azerbaijan; SSTC, Belarus; CNPq and FAPESP,
Brazil; NSERC, NRC and CFI, Canada; CERN; CONICYT, Chile; CAS, MOST and NSFC,
China; COLCIENCIAS, Colombia; MSMT CR, MPO CR and VSC CR, Czech Republic;
DNRF, DNSRC and Lundbeck Foundation, Denmark; EPLANET, ERC and NSRF, European Union;
IN2P3-CNRS, CEA-DSM/IRFU, France; GNSF, Georgia; BMBF, DFG, HGF, MPG and AvH
Foundation, Germany; GSRT and NSRF, Greece; ISF, MINERVA, GIF, DIP and Benoziyo Center,
Israel; INFN, Italy; MEXT and JSPS, Japan; CNRST, Morocco; FOM and NWO,
Netherlands; BRF and RCN, Norway; MNiSW, Poland; GRICES and FCT, Portugal; MERYS
(MECTS), Romania; MES of Russia and ROSATOM, Russian Federation; JINR; MSTD,
Serbia; MSSR, Slovakia; ARRS and MIZ\v{S}, Slovenia; DST/NRF, South Africa;
MICINN, Spain; SRC and Wallenberg Foundation, Sweden; SER, SNSF and Cantons of
Bern and Geneva, Switzerland; NSC, Taiwan; TAEK, Turkey; STFC, the Royal
Society and Leverhulme Trust, United Kingdom; DOE and NSF, United States of
America.

The crucial computing support from all WLCG partners is acknowledged
gratefully, in particular from CERN and the ATLAS Tier-1 facilities at
TRIUMF (Canada), NDGF (Denmark, Norway, Sweden), CC-IN2P3 (France),
KIT/GridKA (Germany), INFN-CNAF (Italy), NL-T1 (Netherlands), PIC (Spain),
ASGC (Taiwan), RAL (UK) and BNL (USA) and in the Tier-2 facilities
worldwide.

\appendix
\section{\boldmath STATISTICAL \\ TREATMENT DETAILS \label{sec:appendix_stat}}

\subsection{\boldmath Binning of fit variables \label{sec:appendix_bins}}

\begin{figure}[b!]
  \includegraphics[width=0.40\textwidth]{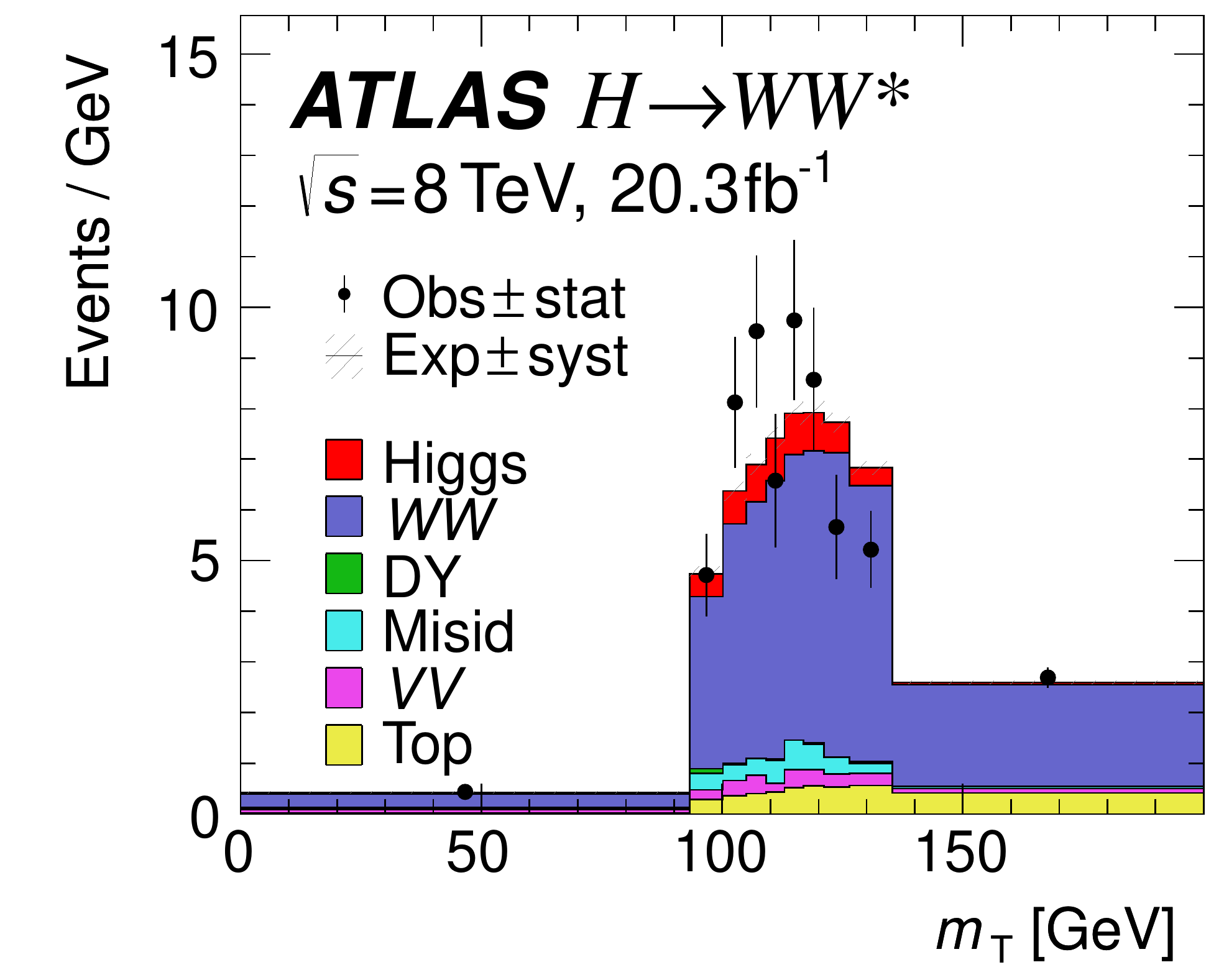}
\caption{
  Transverse mass distribution shown with variable bin widths as used in the likelihood fit.
  The most sensitive signal region is shown
  ($\NjetEQzero$, $\emu$, and $\ell_2{\EQ}\mu$ in $\mll{\GT}30\GeV$ and $\pTsublead{\GT}20\GeV$).
}
\label{fig:mTbin}
\end{figure}

The $\mTH$ distribution is used in the likelihood fit for the ggF-enriched
$\Njet$ samples (see Sec.~\ref{sec:systematics}).
Figure~\ref{fig:mTbin} shows an example of the binned $\mTH$ distribution in
the most sensitive kinematic region of $\NjetEQzero$ and $\emu$ lepton-flavor category.
The optimization procedure for the widths was discussed in Sec.~\ref{sec:systematics_region}.
Table~\ref{tab:bins} gives the details of the binning for every kinematic region.
The $\mTH$ range between the bin 1 (around $80\GeV$) and the last bin
(around $120\GeV$) is binned in variable widths.
For kinematic regions in the $\NjetEQzero$ category, the variable widths are approximately $6\GeV$;
for $\NjetEQone$, the widths are approximately $10\GeV$.
For both samples, the \RMS\ of the bin widths from the mean bin width is approximately $1\GeV$.
Lastly, the ggF-enriched $\NjetGEtwo$ and the cross-check VBF-enriched
$\NjetGEtwo$ categories use the same set of fixed $\mTH$ bin boundaries with bins of variable width.

\begin{table*}[bt!]
\caption{
  $\mTH$ bins for the likelihood fit in the $8\TeV$ analysis.
  The first bin spans $0$ to ``bin 2 left edge''; the last bin spans ``last bin
  left edge'' to $\infty$.  The bin widths $w_b$ of those between the first and
  last bins are given.  The mean of the variable width bins,
  $\overline{w}{\EQ}\sum_{b}w_b\big/(n_{\rm bins}{\MINUS}2)$,
  is given as well as the \RMS\ of the deviation with respect to the mean,
  $\sqrt{\sum_{b}(\overline{w}{\MINUS}w_b)^2/(n_{\rm bins}{\MINUS}2)}$.
  All energy-related quantities are in $\GeV$.
}
\label{tab:bins}
{\small
  \centering
\begin{tabular*}{1\textwidth}{
  lcccc p{0.001\textwidth}
  cc    p{0.001\textwidth}
  rrrr
  rrrr  p{0.001\textwidth}
  rcl
}
\dbline
\multicolumn{5}{c}{Category}
&& \multicolumn{2}{c}{Bin left edge}
&& \multicolumn{8}{c}{Bin widths $w_b$ for bin $b$}
&& \multicolumn{3}{c}{Mean width, \RMS\ of deviation}
\\
\clineskip\cline{1-5}\cline{7-8}\cline{10-17}\cline{19-21}\clineskip
\multicolumn{1}{l}{Sample}
&\multicolumn{1}{p{0.020\textwidth}}{\centering $\ell_2$      }
&\multicolumn{1}{p{0.060\textwidth}}{\centering $\mll$        }
&\multicolumn{1}{p{0.050\textwidth}}{\centering $\pTsublead$  }
&\multicolumn{1}{p{0.035\textwidth}}{\centering $n_{\rm bins}$}
&&\multicolumn{1}{c}{\centering bin\,2}
 &\multicolumn{1}{c}{\centering last\,bin}
&&\multicolumn{1}{p{0.034\textwidth}}{\centering ~2}
 &\multicolumn{1}{p{0.034\textwidth}}{\centering ~3}
 &\multicolumn{1}{p{0.034\textwidth}}{\centering ~4}
 &\multicolumn{1}{p{0.034\textwidth}}{\centering ~5}
 &\multicolumn{1}{p{0.034\textwidth}}{\centering ~6}
 &\multicolumn{1}{p{0.034\textwidth}}{\centering ~7}
 &\multicolumn{1}{p{0.034\textwidth}}{\centering ~8}
 &\multicolumn{1}{p{0.034\textwidth}}{\centering ~9}
&&$\overline{w}$~
&\RMS\
&~~Plot of $\overline{w}{\PM}{\rm \RMS}$
\\
\sgline
$\NjetEQzero$ \\
\balkenscale{160}{-0.3}
\quad $\DFchan$ & $\mu$ &10--30 &10--15       &10 &&74.5 &118.2 &&5.9  &5.0 &4.5 &4.5  &4.5  &5.0  &5.8  &8.5  &&5.5 &1.3 &{\myb\Balkenx{5.5}{1.3}{1.3}{0}{0}} \\
\quad $\DFchan$ & $\mu$ &       &15--20       &10 &&81.6 &122.1 &&6.3  &4.6 &4.1 &4.0  &4.1  &4.5  &5.3  &7.6  &&5.1 &1.2 &{\myb\Balkenx{5.1}{1.2}{1.2}{0}{0}} \\
\quad $\DFchan$ & $\mu$ &       &20--$\infty$ &10 &&93.7 &133.7 &&6.3  &4.6 &3.8 &3.9  &3.8  &4.3  &5.2  &8.1  &&5.0 &1.4 &{\myb\Balkenx{5.0}{1.4}{1.4}{0}{0}} \\
\quad $\DFchan$ & $\mu$ &30--55 &10--15       &10 &&84.1 &124.7 &&6.4  &4.6 &4.4 &4.0  &4.4  &4.4  &5.2  &7.2  &&5.1 &1.1 &{\myb\Balkenx{5.1}{1.1}{1.1}{0}{0}} \\
\quad $\DFchan$ & $\mu$ &       &15--20       &10 &&86.3 &125.8 &&6.0  &4.7 &4.4 &4.0  &4.0  &4.2  &5.1  &7.1  &&4.9 &1.0 &{\myb\Balkenx{4.9}{1.0}{1.0}{0}{0}} \\
\quad $\DFchan$ & $\mu$ &       &20--$\infty$ &10 &&93.2 &135.4 &&7.0  &4.8 &4.2 &3.8  &3.9  &4.2  &5.3  &9.0  &&5.3 &1.7 &{\myb\Balkenx{5.3}{1.7}{1.7}{0}{0}} \\
\quad $\DFchan$ & $e$   &10--30 &10--15       &10 &&76.7 &118.0 &&5.8  &4.5 &4.2 &4.0  &4.5  &4.7  &5.7  &7.9  &&5.2 &1.2 &{\myb\Balkenx{5.2}{1.2}{1.2}{0}{0}} \\
\quad $\DFchan$ & $e$   &       &15--20       &10 &&80.8 &121.4 &&5.9  &4.8 &4.4 &3.9  &4.1  &4.6  &5.3  &7.6  &&5.1 &1.1 &{\myb\Balkenx{5.1}{1.1}{1.1}{0}{0}} \\
\quad $\DFchan$ & $e$   &       &20--$\infty$ &10 &&93.1 &133.6 &&6.7  &4.9 &4.0 &3.8  &3.8  &4.2  &5.0  &8.1  &&5.1 &1.5 &{\myb\Balkenx{5.1}{1.5}{1.5}{0}{0}} \\
\quad $\DFchan$ & $e$   &30--55 &10--15       &10 &&84.9 &125.7 &&6.0  &4.7 &4.3 &3.9  &4.1  &4.4  &5.4  &8.0  &&5.1 &1.3 &{\myb\Balkenx{5.1}{1.3}{1.3}{0}{0}} \\
\quad $\DFchan$ & $e$   &       &15--20       &10 &&85.0 &125.2 &&6.6  &4.8 &4.1 &3.9  &3.9  &4.2  &5.2  &7.5  &&5.0 &1.3 &{\myb\Balkenx{5.0}{1.3}{1.3}{0}{0}} \\
\quad $\DFchan$ & $e$   &       &20--$\infty$ &10 &&93.5 &135.8 &&6.8  &4.9 &4.2 &3.8  &3.8  &4.3  &5.5  &9.0  &&5.3 &1.7 &{\myb\Balkenx{5.3}{1.7}{1.7}{0}{0}} \\
\quad $\SFchan$ &  -    &12--55 &10--$\infty$ &10 &&95.1 &128.8 &&4.9  &4.0 &3.5 &3.3  &3.4  &3.6  &4.3  &6.7  &&4.2 &1.1 &{\myr\Balkenx{4.2}{1.1}{1.1}{0}{0}} \\
                &       &       &             &   &&     &      &&     &    &    &     &     &     &     &     &&    &    &
\renewcommand{\bcfontstyle}{\bfseries}%
\renewcommand{\bcfontstyle}{}%
\hspace{-7.5pt}%
\begin{bchart}[step=5,max=15,width=0.130\textwidth,scale=1.03]\end{bchart}%
\vspace{-10pt}%
\\
$\NjetEQone$ \\
\quad $\DFchan$ & $\mu$ &10--30 &10--15       &6  &&79.0 &118.7 &&10.5 &8.5 &8.8 &11.9 &-~   &-~   &-~   &-~   &&9.9 &1.4 &{\myb\Balkenx{9.9 }{1.4}{1.4}{0}{0}} \\
\quad $\DFchan$ & $\mu$ &       &15--20       &6  &&81.6 &119.7 &&10.6 &9.6 &8.4 &9.5  &-~   &-~   &-~   &-~   &&9.5 &0.8 &{\myb\Balkenx{9.5 }{0.8}{0.8}{0}{0}} \\
\quad $\DFchan$ & $\mu$ &       &20--$\infty$ &6  &&86.7 &127.4 &&11.2 &9.1 &9.3 &11.1 &-~   &-~   &-~   &-~   &&10.2&1.0 &{\myb\Balkenx{10.2}{1.0}{1.0}{0}{0}} \\
\quad $\DFchan$ & $\mu$ &30--55 &10--15       &6  &&79.6 &116.0 &&9.1  &9.2 &8.3 &9.8  &-~   &-~   &-~   &-~   &&9.1 &0.5 &{\myb\Balkenx{9.1 }{0.5}{0.5}{0}{0}} \\
\quad $\DFchan$ & $\mu$ &       &15--20       &6  &&81.9 &120.2 &&10.3 &9.2 &8.6 &10.2 &-~   &-~   &-~   &-~   &&9.6 &0.7 &{\myb\Balkenx{9.6 }{0.7}{0.7}{0}{0}} \\
\quad $\DFchan$ & $\mu$ &       &20--$\infty$ &6  &&87.4 &127.9 &&11.1 &8.7 &9.3 &11.4 &-~   &-~   &-~   &-~   &&10.1&1.1 &{\myb\Balkenx{10.1}{1.1}{1.1}{0}{0}} \\
\quad $\DFchan$ & $e$   &10--30 &10--15       &6  &&88.1 &123.3 &&9.9  &7.9 &7.3 &10.1 &-~   &-~   &-~   &-~   &&8.8 &1.2 &{\myb\Balkenx{8.8 }{1.2}{1.2}{0}{0}} \\
\quad $\DFchan$ & $e$   &       &15--20       &6  &&88.2 &123.9 &&9.7  &7.9 &7.8 &10.3 &-~   &-~   &-~   &-~   &&8.9 &1.1 &{\myb\Balkenx{8.9 }{1.1}{1.1}{0}{0}} \\
\quad $\DFchan$ & $e$   &       &20--$\infty$ &6  &&92.0 &130.2 &&9.5  &8.2 &8.9 &11.6 &-~   &-~   &-~   &-~   &&9.6 &1.3 &{\myb\Balkenx{9.6 }{1.3}{1.3}{0}{0}} \\
\quad $\DFchan$ & $e$   &30--55 &10--15       &6  &&87.0 &121.7 &&8.9  &9.1 &7.0 &9.7  &-~   &-~   &-~   &-~   &&8.7 &1.0 &{\myb\Balkenx{8.7 }{1.0}{1.0}{0}{0}} \\
\quad $\DFchan$ & $e$   &       &15--20       &6  &&87.4 &123.2 &&9.6  &8.1 &8.6 &9.5  &-~   &-~   &-~   &-~   &&9.0 &0.6 &{\myb\Balkenx{9.0 }{0.6}{0.6}{0}{0}} \\
\quad $\DFchan$ & $e$   &       &20--$\infty$ &6  &&91.2 &129.0 &&10.1 &8.3 &8.1 &11.3 &-~   &-~   &-~   &-~   &&9.5 &1.3 &{\myb\Balkenx{9.5 }{1.3}{1.3}{0}{0}} \\
\quad $\SFchan$ &  -    &12--55 &10--$\infty$ &6  &&96.9 &126.7 &&8.3  &6.5 &6.3 &8.7  &-~   &-~   &-~   &-~   &&7.5 &1.1 &{\myr\Balkenx{7.5 }{1.1}{1.1}{0}{0}} \\
\clineskip
                &       &       &             &   &&     &      &&     &    &    &     &     &     &     &     &&    &    &
\renewcommand{\bcfontstyle}{\bfseries}%
\renewcommand{\bcfontstyle}{}%
\hspace{-7.5pt}%
\begin{bchart}[step=5,max=15,width=0.130\textwidth,scale=1.03]\end{bchart}%
\vspace{-10pt}%
\\
\multicolumn{2}{l}{$\NjetGEtwo$ ggF}\\
\quad $\DFchan$ &  -    &10--55 &10--$\infty$ &4  &&50.0 &130.0 &&30   &50  &-~  &-~   &-~   &-~   &-~   &-~   &&40  &10  &~~Not displayed \\
\clineskip\clineskip
\multicolumn{4}{l}{$\NjetGEtwo$ VBF cross-check}\\
\quad $\DFchan$ &  -    &10--55 &10--$\infty$ &4  &&50.0 &130.0 &&30   &50  &-~  &-~   &-~   &-~   &-~   &-~   &&40  &10  &~~Not displayed \\
\quad $\SFchan$ &  -    &12--55 &10--$\infty$ &4  &&50.0 &130.0 &&30   &50  &-~  &-~   &-~   &-~   &-~   &-~   &&40  &10  &~~Not displayed \\
\dbline
\end{tabular*}
}
\end{table*}

\subsection{\boldmath Drell-Yan estimate in $\SFchan$ for $\NjetLEone$ \label{sec:appendix_stat_pacman}}

The details of the treatment for the Drell-Yan estimate for the $\SFchan$ category
in the $\NjetLEone$ sample are described.

The method uses additional control regions to constrain the parameters
corresponding to the selection efficiencies of contributing processes
categorized into ``DY'' and ``non-DY''; the latter includes the signal
events.  The variable $\frecoil$ is used to separate the two categories, and
to divide the sample into ``pass'' and ``fail'' subsamples.
In the $\SFchan$ categories, the pass samples are enriched
in non-DY events and, conversely, the fail samples are enriched
in DY events.  The residual cross-contamination
is estimated using additional control regions.

Of particular interest is the data-derived efficiency of the $\frecoil$ selection for the DY
and non-DY events.  The efficiency of the applied
$\frecoil$ selection on DY events (on $\fEff_{\scDY}$) is obtained from the
$\SFchan$ sample in the $Z$ peak (in the $Z$ CR), defined by the dilepton mass range
$\ABS{\mll{\MINUS}\mZ}{\LT}15\GeV$.  Events in the $Z$ CR are relatively pure in DY.
The $\fEff_{\scDY}$ estimates the efficiency of the selection due to neutrinoless
events with $\met$ due to misreconstruction, or ``fake $\met$.''  The same
parameter appears in two terms, one for the $Z$ CR and the other
for the signal region, each composed of two Poisson functions.

The non-DY events with neutrino final states, or ``real $\met$,'' contaminate
both the $Z$ CR and the SR, and the two corresponding $\frecoil$ selection
efficiencies, $\fEff_{\scnonDY}'$ and $\fEff_{\scnonDY}$, are evaluated separately.
The non-DY efficiency in the $Z$ CR, $\fEff_{\scnonDY}'$, is evaluated using
the $Z$ CR selection except with the $\DFchan$ sample, which is pure in non-DY events.
The SR efficiency $\fEff_{\scnonDY}$ is evaluated using the $\SFchan$ SR
selection (described in Sec.~\ref{sec:bkg_dy_ll_ggf}) applied to an
$\DFchan$ sample.

The fit CR part of the likelihood function [Eq.~(\ref{eqn:likelihood})] contains
two Poisson functions that represent events---in the $Z$ mass window in the
$\SFchan$ category---that pass or fail the $\frecoil$ selection:
\begin{eqnarray}
\no f\Big(N_\pass^{\ZCR}\,&\Big|&\,\fNorm'_{\scDY}{\CDOT}\fEff_{\scDY}{\CDOT}B_{\scDY}^{\ZCR} + \fEff'_{\scnonDY}{\CDOT}B_{\nonDY}^{\ZCR}\Big)\cdot \label{eqn:pacman_stat_zsf}
\\
\no f\Big(N_\fail^{\ZCR}\,&\Big|&\,\fNorm'_{\scDY}{\,\cdot}\big(1{-}\fEff_{\scDY}\big){\cdot\,}B_{\scDY}^{\ZCR}{\,+}\big(1{-}\fEff'_{\scnonDY}\big){\cdot\,}B_{\scnonDY}^{\ZCR}\Big),\nonumber
\end{eqnarray}
where $N$ is the observed number of events and $B$ the background estimate
without applying an $\frecoil$ selection.
The superscript denotes the $Z$ CR mass window; the subscript pass (fail)
denotes the sample of events that pass (fail) the $\frecoil$ selection; and the
subscripts DY (non-DY) denotes background estimates for the Drell-Yan (all
except Drell-Yan) processes.  The non-DY estimate, $B^{\ZCR}_{\scnonDY}$, is a
sum of all contributing processes listed in Table~\ref{tab:process};
normalization factors, such as $\fNorm_{\scWW}$, that are described in
Sec.~\ref{sec:bkg} are implicitly applied to the corresponding contributions.
The Drell-Yan estimate is normalized explicitly by a common normalization
factor $\fNorm_{\scDY}'$ applied to both the
passing and failing subsamples of the $Z$ peak.

The $\fEff_{\nonDY}'$ parameter above is determined using events in the
$\DFchan$ category.  The corresponding Poisson functions are included in the
likelihood:
\begin{equation}
\begin{array}{llllll}
\nq f\Big(N_\pass^{\ZCR,\DFchan}\,\np&\Big|&\no\,\fEff'_{\scnonDY}{\CDOT}B_{\scnonDY}^{\ZCR,\DFchan}\Big)\cdot \\
\clineskip
\nq f\Big(N_\fail^{\ZCR,\DFchan}\,\np&\Big|&\no\,(1{\MINUS}\fEff'_{\scnonDY}){\CDOT}B_{\scnonDY}^{\ZCR,\DFchan}\Big),
\end{array}
\label{eqn:pacman_stat_zdf}
\end{equation}
where the $\DFchan$ in the superscript denotes the $Z$ CR mass window for
events in the $\DFchan$ category; all other notation follows the convention for
Eq.~(\ref{eqn:pacman_stat_zsf}).  The DY contamination in this
region is implicitly subtracted.

The SR part of the likelihood also contains two Poisson
functions---using the same $\fEff_{\scDY}$ above, but a
different $\fNorm_{\DY}$ and $\fEff_{\scnonDY}$ corresponding to the
SR---is
\begin{eqnarray}
\no f\Big(N_\pass^{\SR}\,&\Big|&\,\fNorm_{\scDY}{\CDOT}\fEff_{\scDY}{\CDOT}B_{\scDY}^{\SR}{\PLUS}\fEff_{\scnonDY}{\CDOT}B_{\scnonDY}^{\SR}\Big)\cdot \\
\no f\Big(N_\fail^{\SR}~\,&\Big|&\,\fNorm_{\scDY}{\CDOT}\big(1{-}\fEff_{\scDY}\big){\cdot\,}B_{\scDY}^{\SR}{\,+}\big(1{-}\fEff_{\scnonDY}\big){\cdot\,}B_{\scnonDY}^{\SR}\Big),\nonumber
\label{eqn:pacman_stat_srsf}
\end{eqnarray}
where SR denotes the signal region selection and $\fNorm_{\scDY}$ is the
common normalization factor for the Drell-Yan estimate for the pass and fail subsamples.

The parameter $\fEff_{\scnonDY}$ is constrained following the same strategy as
Eq.~(\ref{eqn:pacman_stat_zdf}) with
\begin{equation}
\begin{array}{llllll}
\no f\Big(N_\pass^{\SR,\DFchan}\no&\Big|&\no\fEff_{\scnonDY}{\CDOT}B_{\scnonDY}^{\SR,\DFchan}\Big)\cdot \\
\clineskip
\no f\Big(N_\fail^{\SR,\DFchan}\no&\Big|&\no\big(1{-}\fEff_{\scnonDY}\big){\CDOT}B_{\scnonDY}^{\SR,\DFchan}\Big),
\end{array}
\label{eqn:pacman_stat_srdf}
\end{equation}
where the $\DFchan$ in the superscript denotes the $\SFchan$ SR selection (including the
one on $\frecoil$) applied to events in the $\DFchan$ category.  As noted before, the
DY contamination in this region is implicitly subtracted.

\subsection{\boldmath Top-quark estimate for $\NjetEQone$ \label{sec:appendix_stat_jbee}}

The details of the \insitu\ treatment for the $b$-tagging efficiency for the top-quark
estimate for $\NjetEQone$ category are described.

The method uses two control regions within the $\NjetEQtwo$ sample: those with
one and two $b$-jets.  These CRs constrain the normalization parameter
for the $b$-tagging efficiency of top-quark events ($\fNorm_{\btag}$) and for
the top-quark cross section in these regions ($\fNorm_\top$).

The Poisson terms for the control regions are
\begin{eqnarray}
\no f\Big(N^{2b}_{2j}\,&\Big|&\,\fNorm_{\top}{\CDOT}\fNorm_{\btag}{\CDOT}B^{2b}_{\top}{\PLUS}B_{\rm other}\Big)\cdot \label{eqn:jbee}  \\
\no f\Big(N^{1b}_{2j}\,&\Big|&\,\fNorm_{\top}{\CDOT}B^{1b}_{\top}{\PLUS}\fNorm_{\top}{\CDOT}(1{\MINUS}\fNorm_{\btag}){\CDOT}B^{2b}_{\top}{\PLUS}B_{\rm other}\Big),\nonumber
\end{eqnarray}
where $N^{1b}_{2j}$ ($N^{2b}_{2j}$) corresponds to the number of observed
events with one (two) $b$-jets; $B^{1b}_{\top}$ ($B^{2b}_{\top}$) is the
corresponding top-quark estimates from MC samples; and $B_{\rm other}$ are the
rest of the processes contributing to the sample.

The parameter $\fNorm_{\top}$ enters only in the above terms, while
$\fNorm_{\btag}$ is applied to other regions.  In the top-quark CR, one factor
of $\fNorm_{\top}$ is applied to the expected top-quark yield.  In the SR and the $\WW$ CR, the
treatment is of the same form as the second line of Eq.~(\ref{eqn:jbee}) applied
to the $\NjetEQone$ sample, \ie, the estimated top-quark background is
$B^{0b}_{\top}{\PLUS}(1{\MINUS}\fNorm_{\btag}){\CDOT}B^{1b}_{\top}$.

In summary, the difference between the observed and the expected $b$-tagging
efficiency corrects the number of estimated untagged events in the SR.

\section{\boldmath BDT PERFORMANCE \label{sec:appendix_bdt}}

Section~\ref{sec:selection_2jvbf} motivated the choice of
variables used in the $\NjetGEtwo$ VBF-enriched category based on their
effectiveness in the cross-check analysis. Many of the variables
exploit the VBF topology with two forward jets and no activity in the
central region. The main analysis in this category is based on the
multivariate technique that uses those variables as inputs to the
training of the BDT. The training is optimized on the simulated VBF signal production and
it treats simulated ggF production as a background. Figures~\ref{fig:aux_bdtinputDF}
and~\ref{fig:aux_bdtinputSF} show
the distributions of the input variables in the $\DFchan$ and
$\SFchan$ samples, respectively. The comparison is based only on MC simulation and it shows
the separation between the VBF signal and the background processes,
motivating the use of the chosen variables.

\begin{figure*}[p!]
\includegraphics[width=0.40\textwidth]{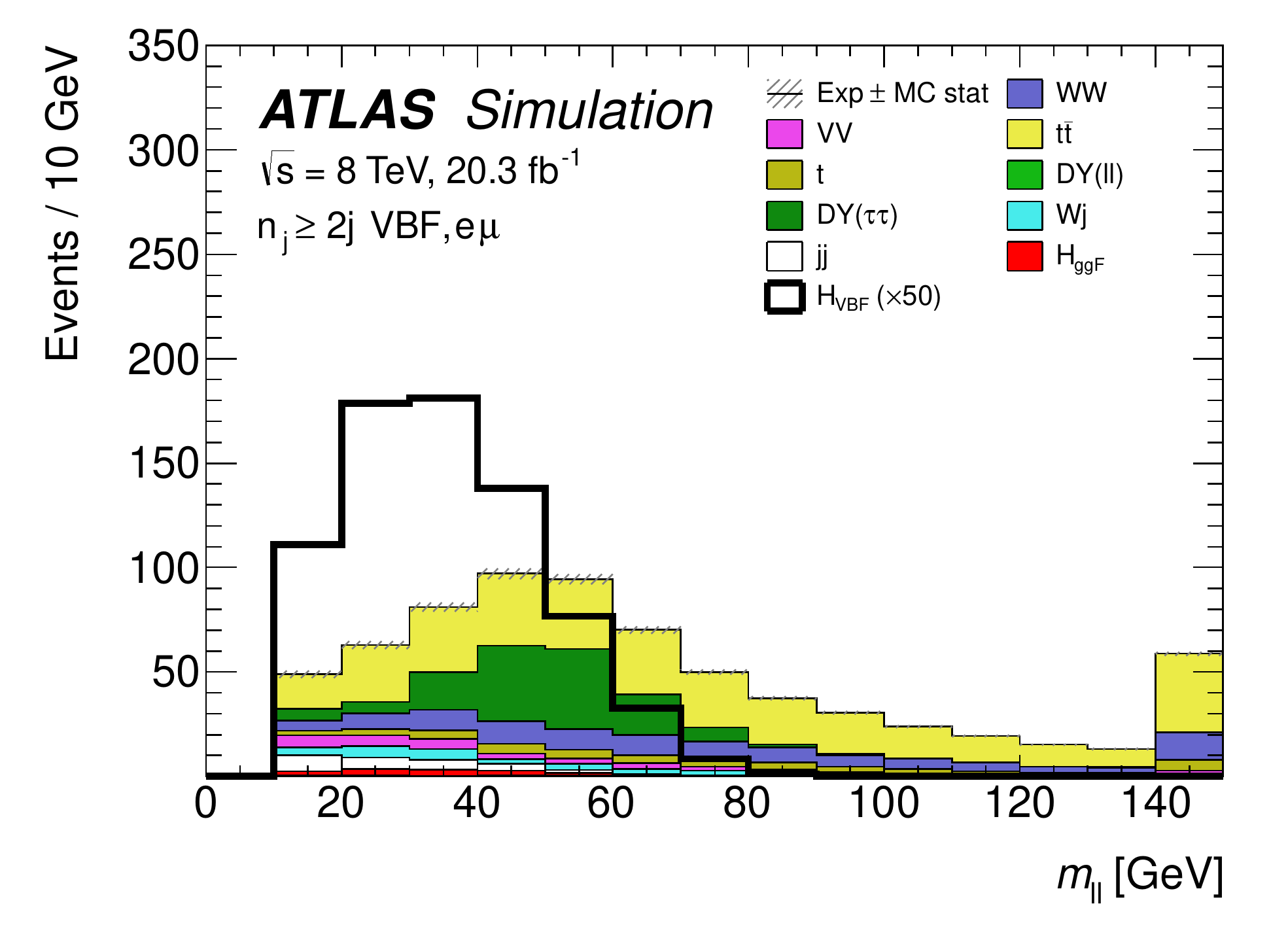}
\includegraphics[width=0.40\textwidth]{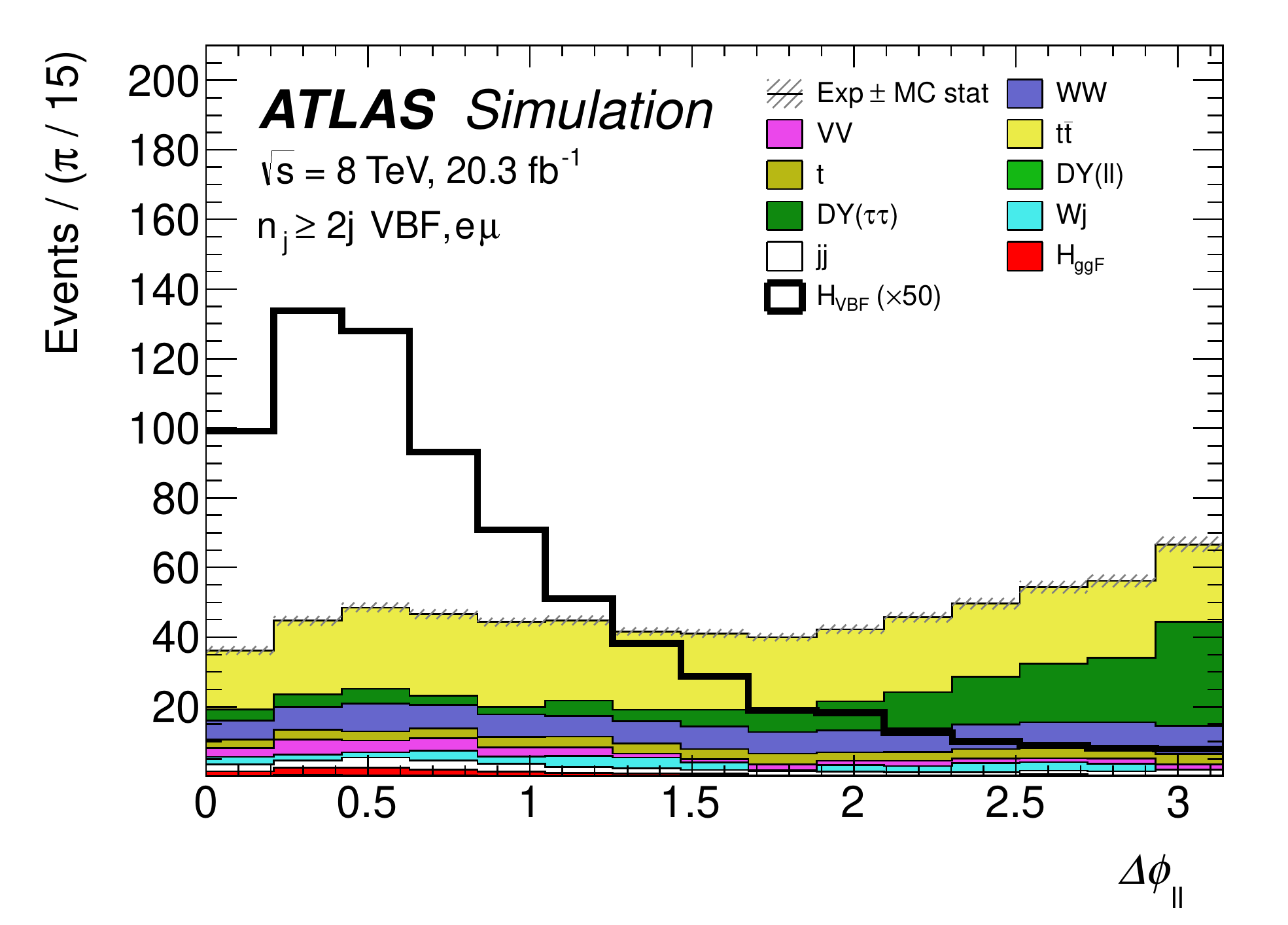}\\
\includegraphics[width=0.40\textwidth]{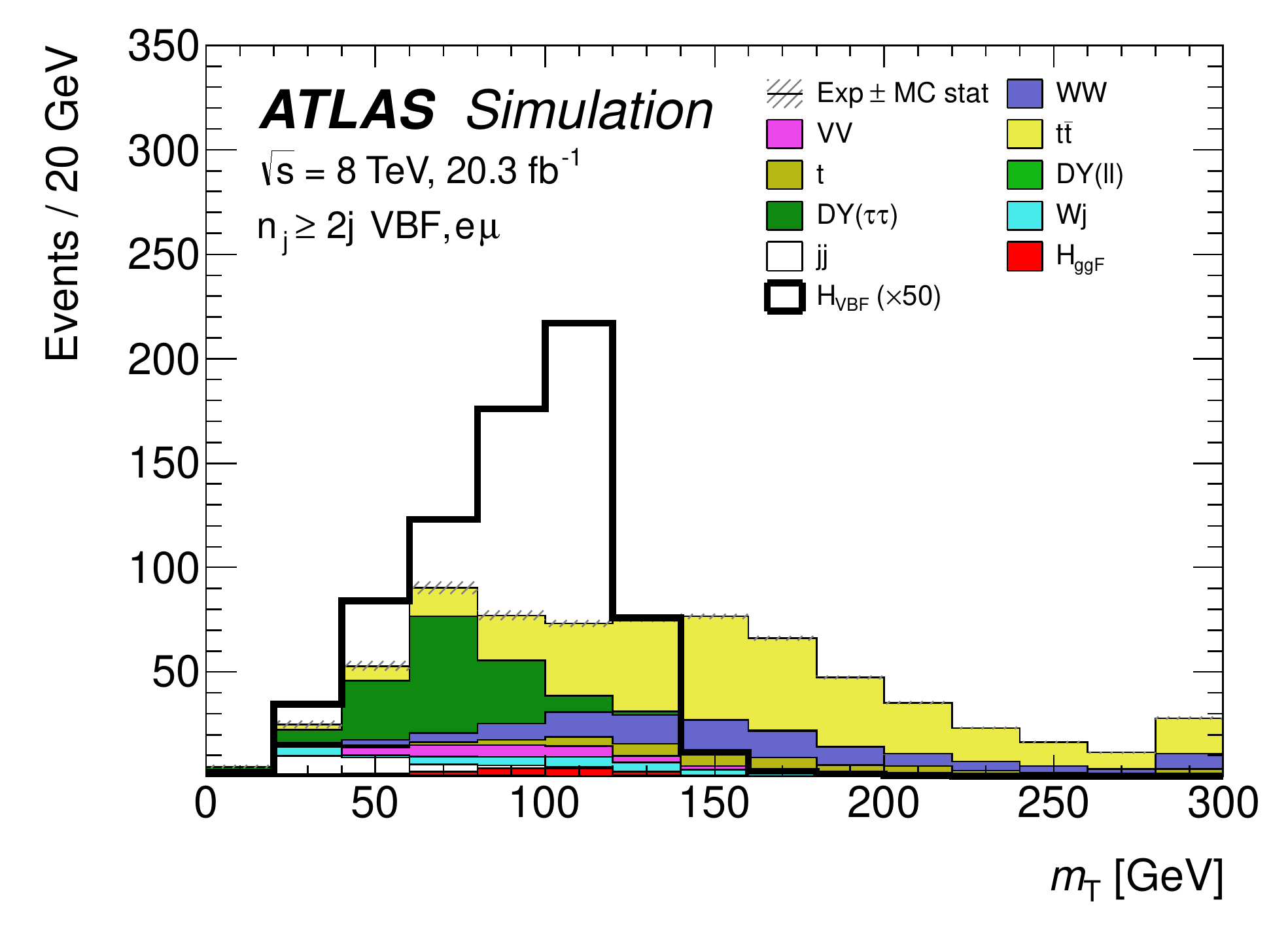}
\includegraphics[width=0.40\textwidth]{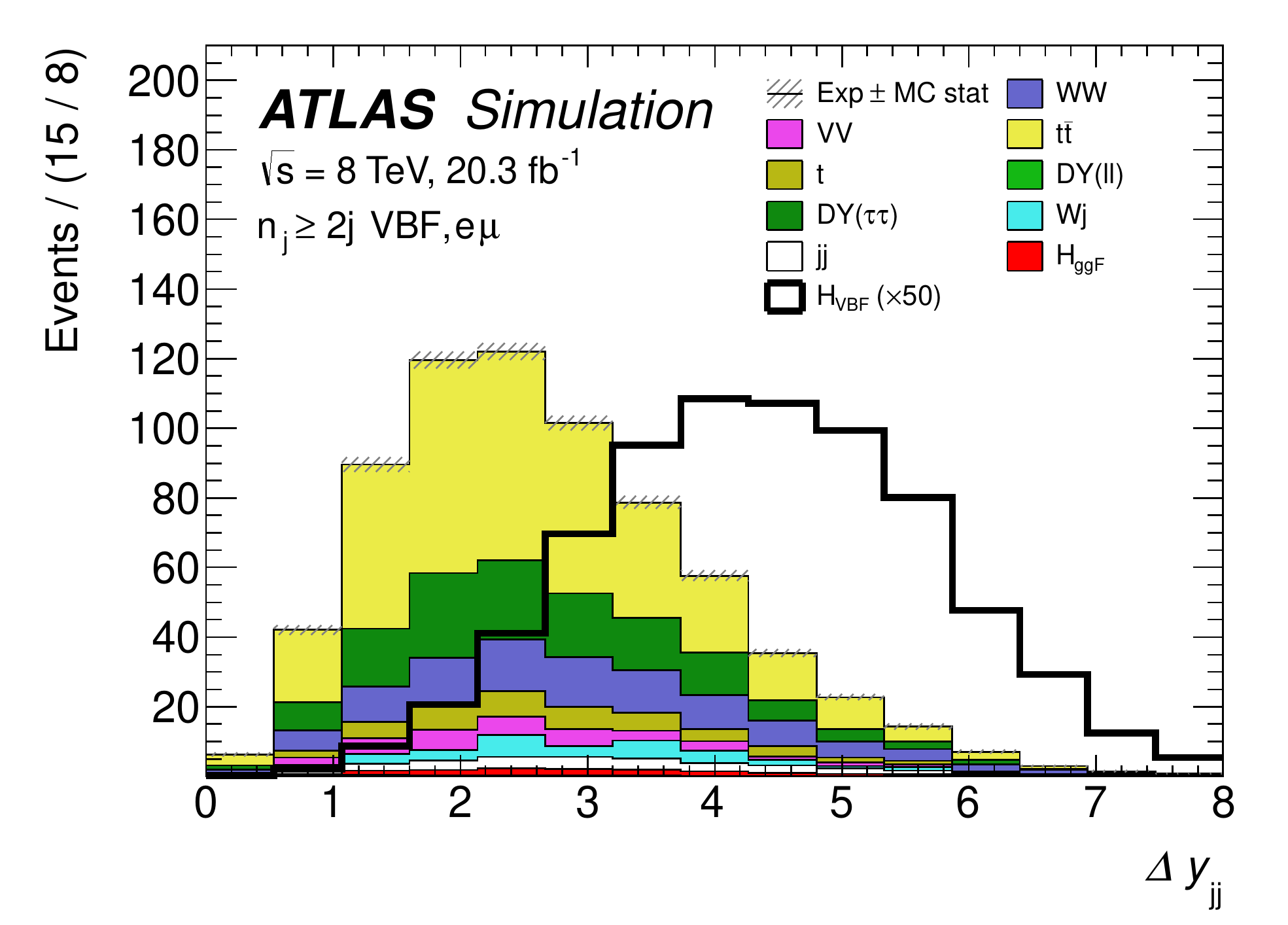}\\
\includegraphics[width=0.40\textwidth]{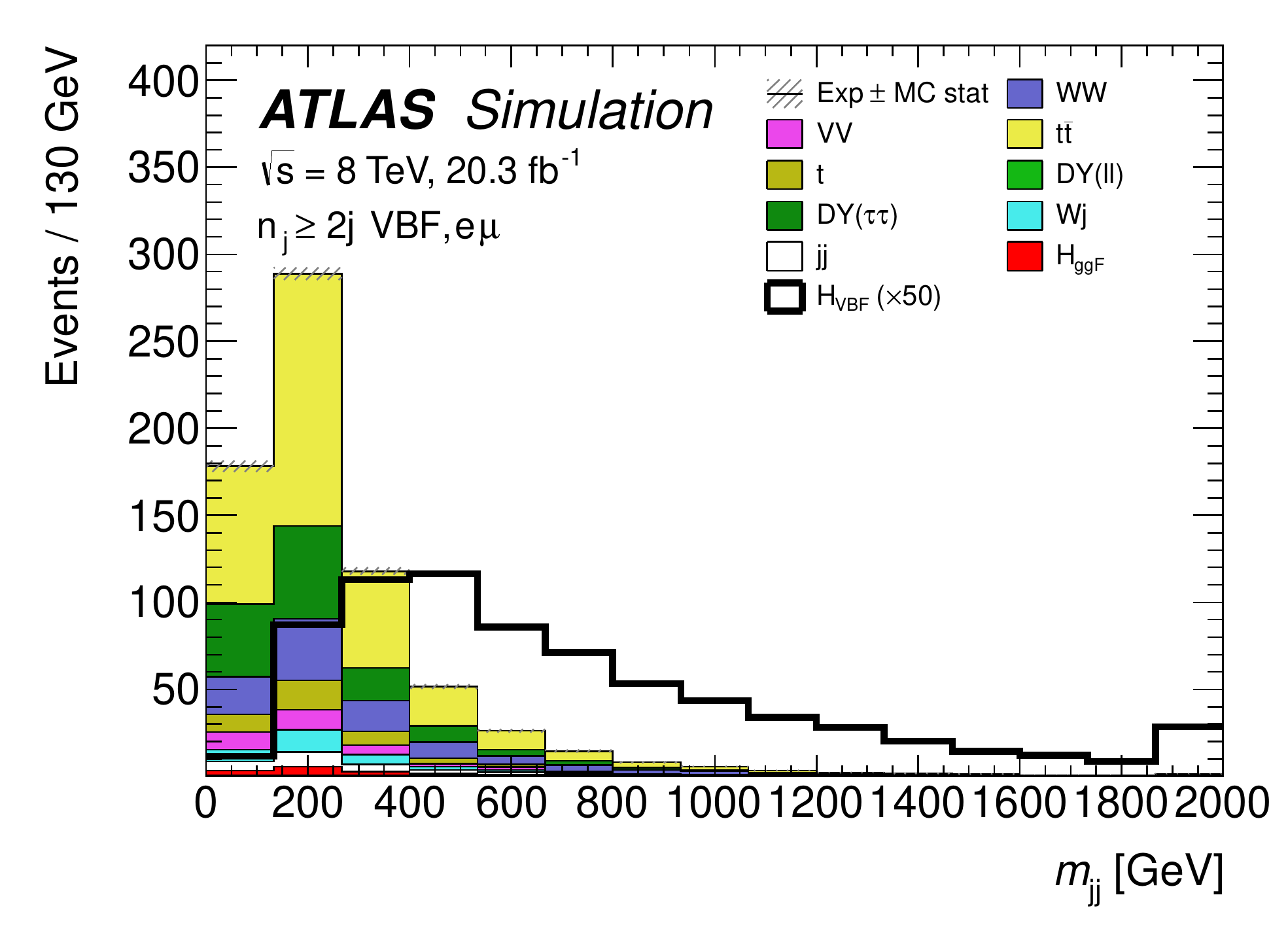}
\includegraphics[width=0.40\textwidth]{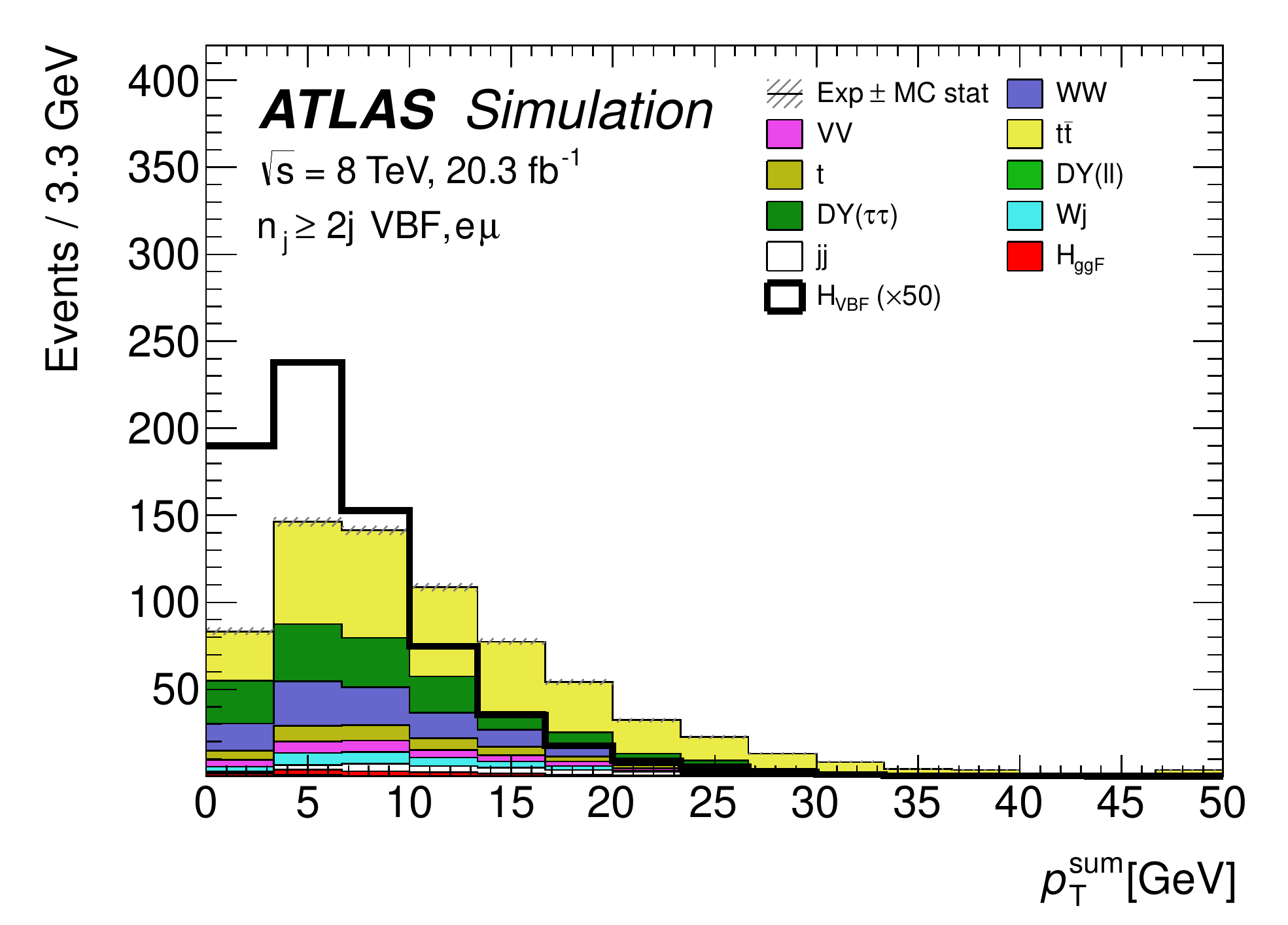}\\
\includegraphics[width=0.40\textwidth]{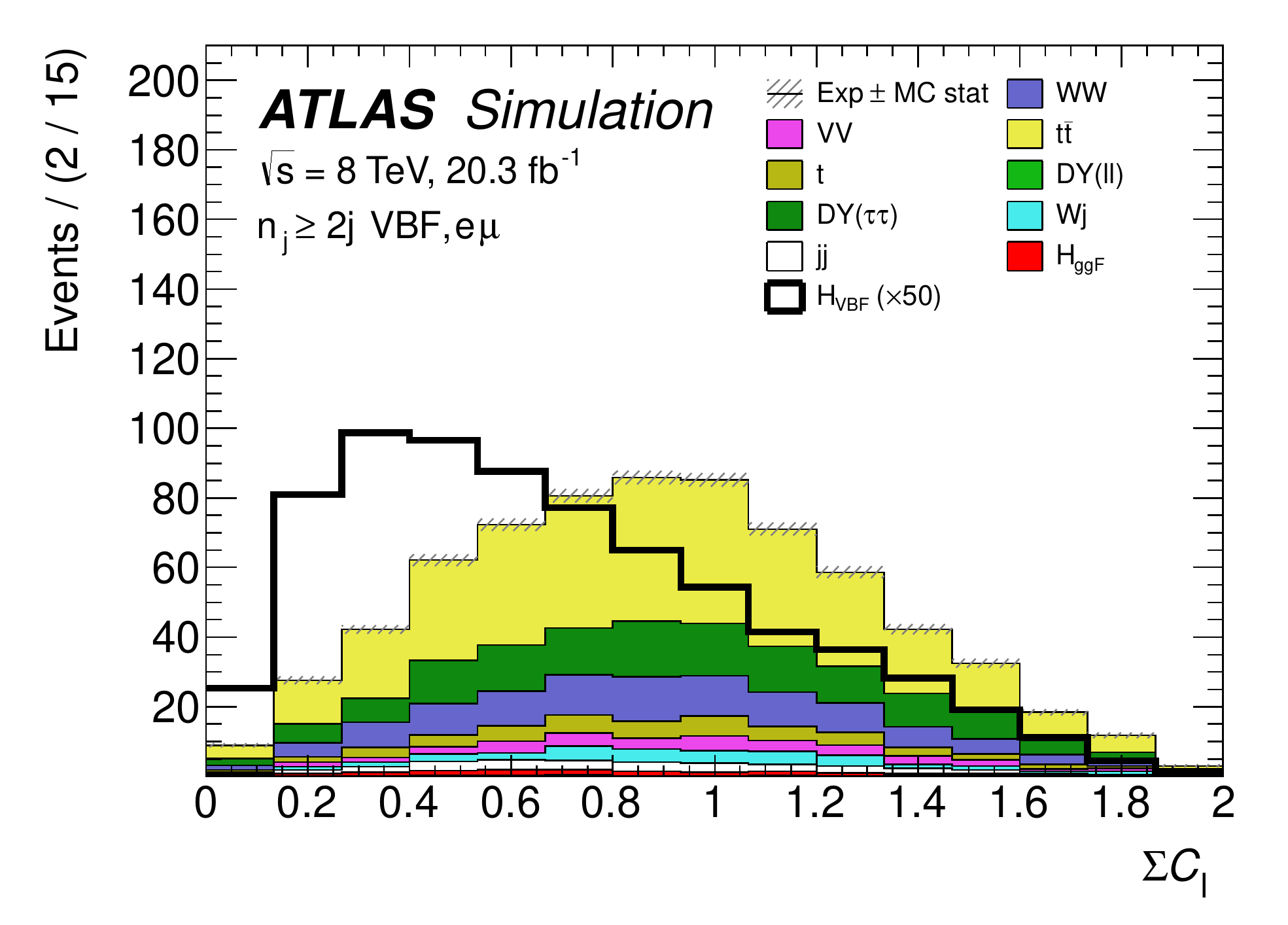}
\includegraphics[width=0.40\textwidth]{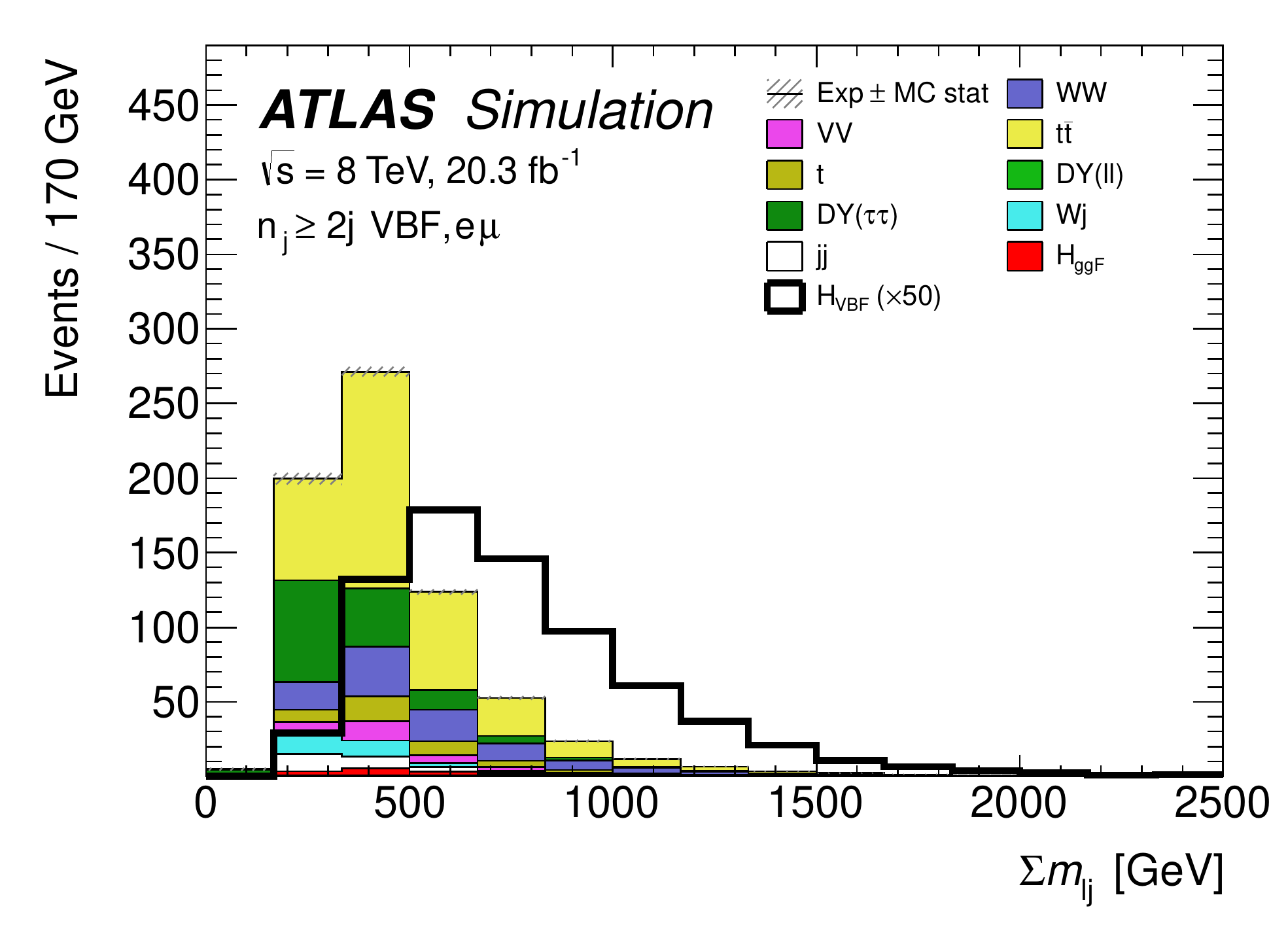}\\
\caption{Distributions of the variables used as inputs to the training
  of the BDT in the $\DFchan$ sample in the $8\TeV$ data
  analysis. The variables are shown after the common preselection and the
  additional selection requirements in the $\NjetGEtwo$ VBF-enriched
  category, and they include: $\mll$, $\dphill$,
  $\mTH$, and $\dyjj$ (top two rows); $\mjj$, $\pTtot$, $\contolv$, and
  $\mlj$ (bottom two rows). The distributions show the separation between
  the VBF signal and background processes (ggF signal
  production is treated as such). The VBF signal
  is scaled by fifty to enhance the differences in the shapes of the
  input variable distributions. The SM Higgs boson is shown at $\mH{\EQ}125\GeV$. The uncertainties on the background
  prediction are only due to MC sample size.
}
\label{fig:aux_bdtinputDF}
\end{figure*}
\begin{figure*}[p!]
\includegraphics[width=0.40\textwidth]{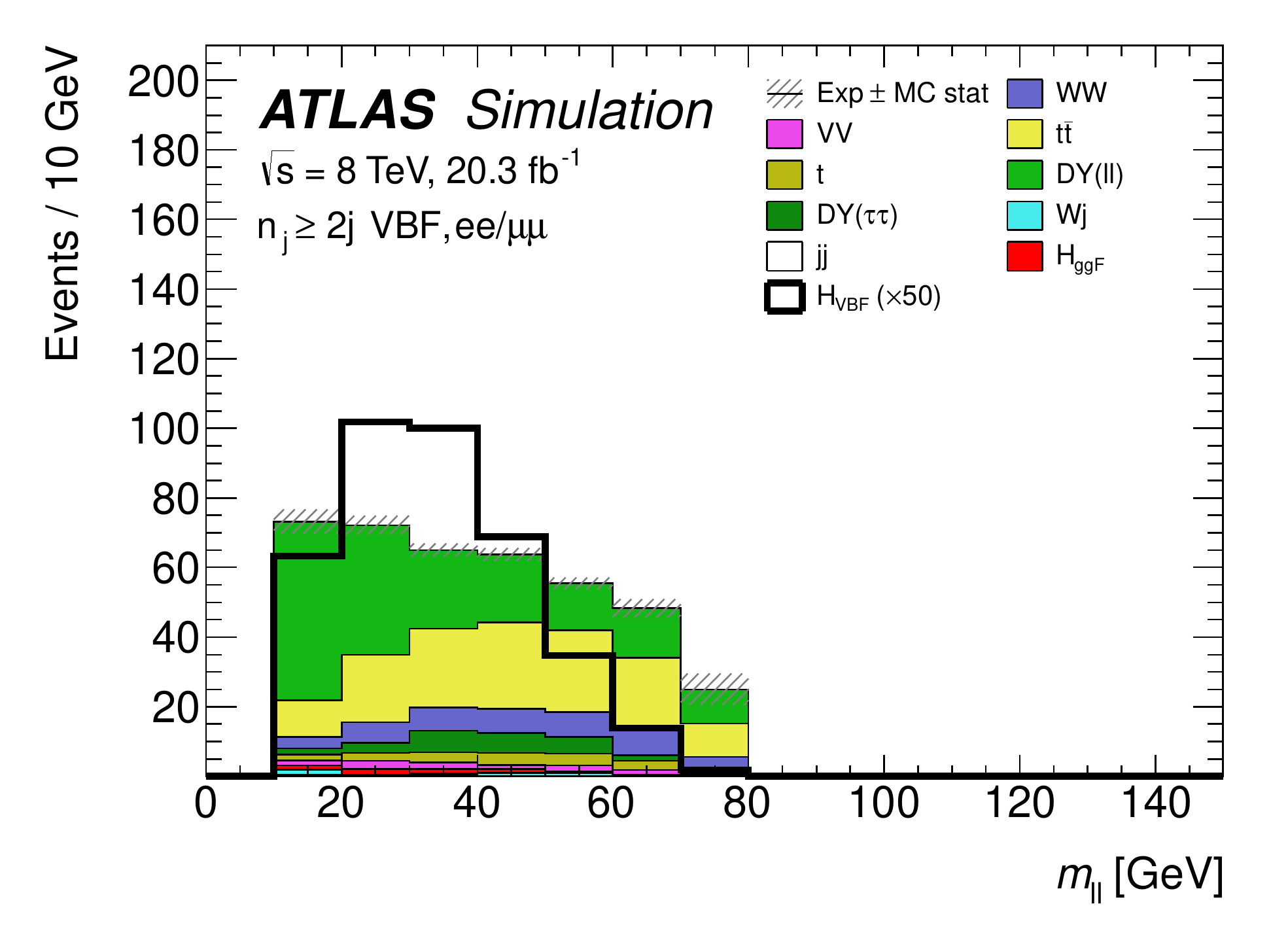}
\includegraphics[width=0.40\textwidth]{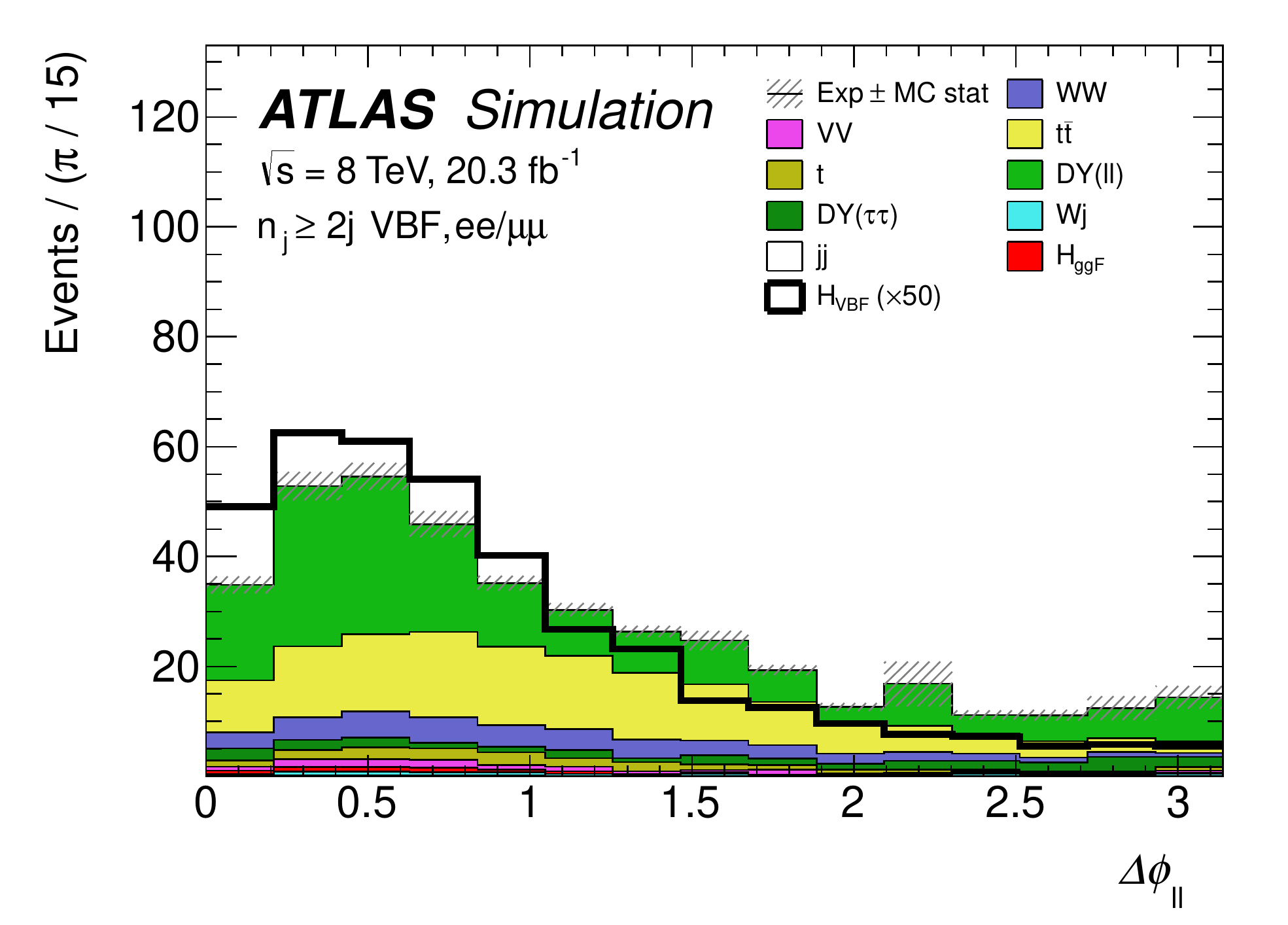}\\
\includegraphics[width=0.40\textwidth]{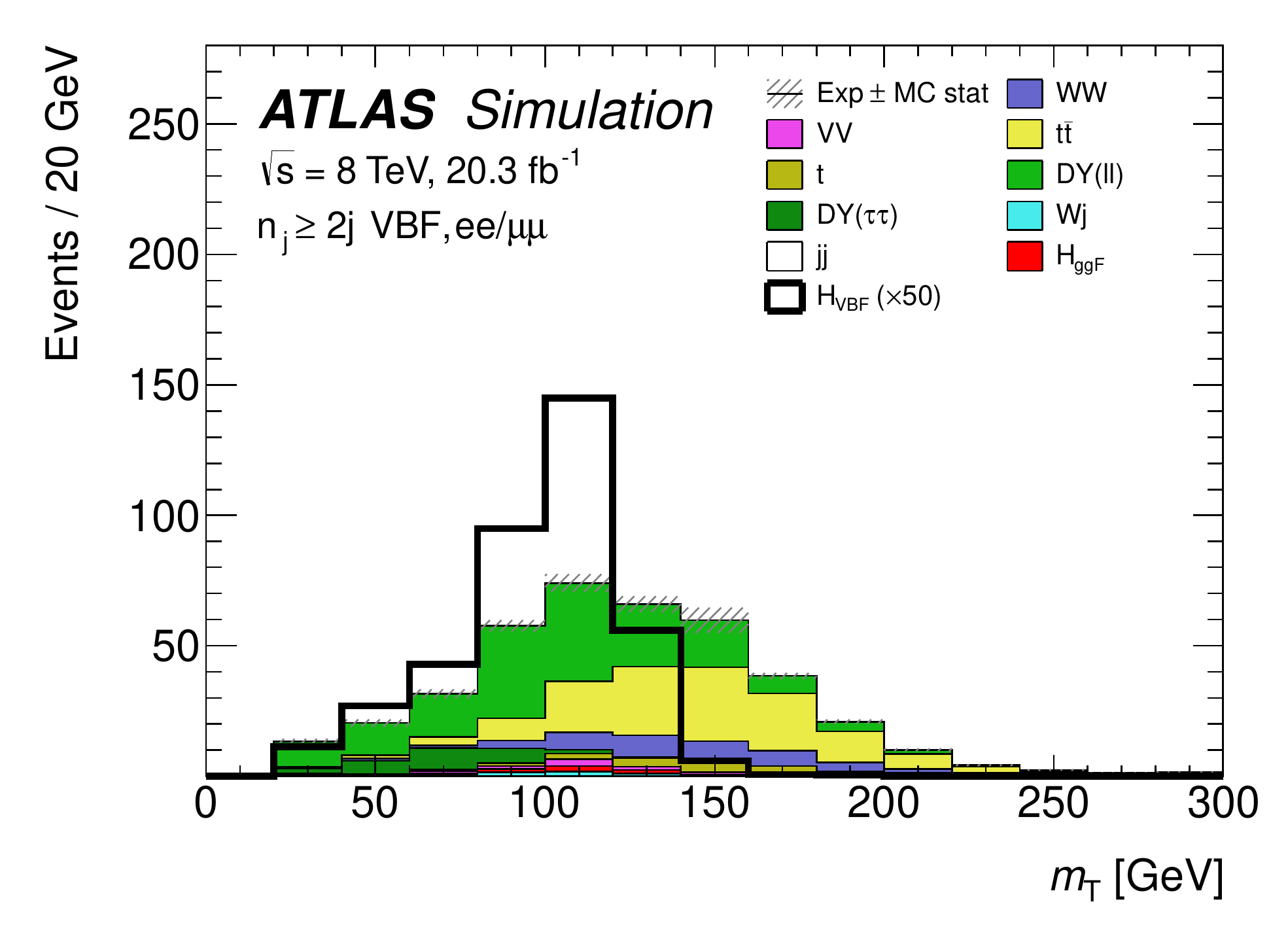}
\includegraphics[width=0.40\textwidth]{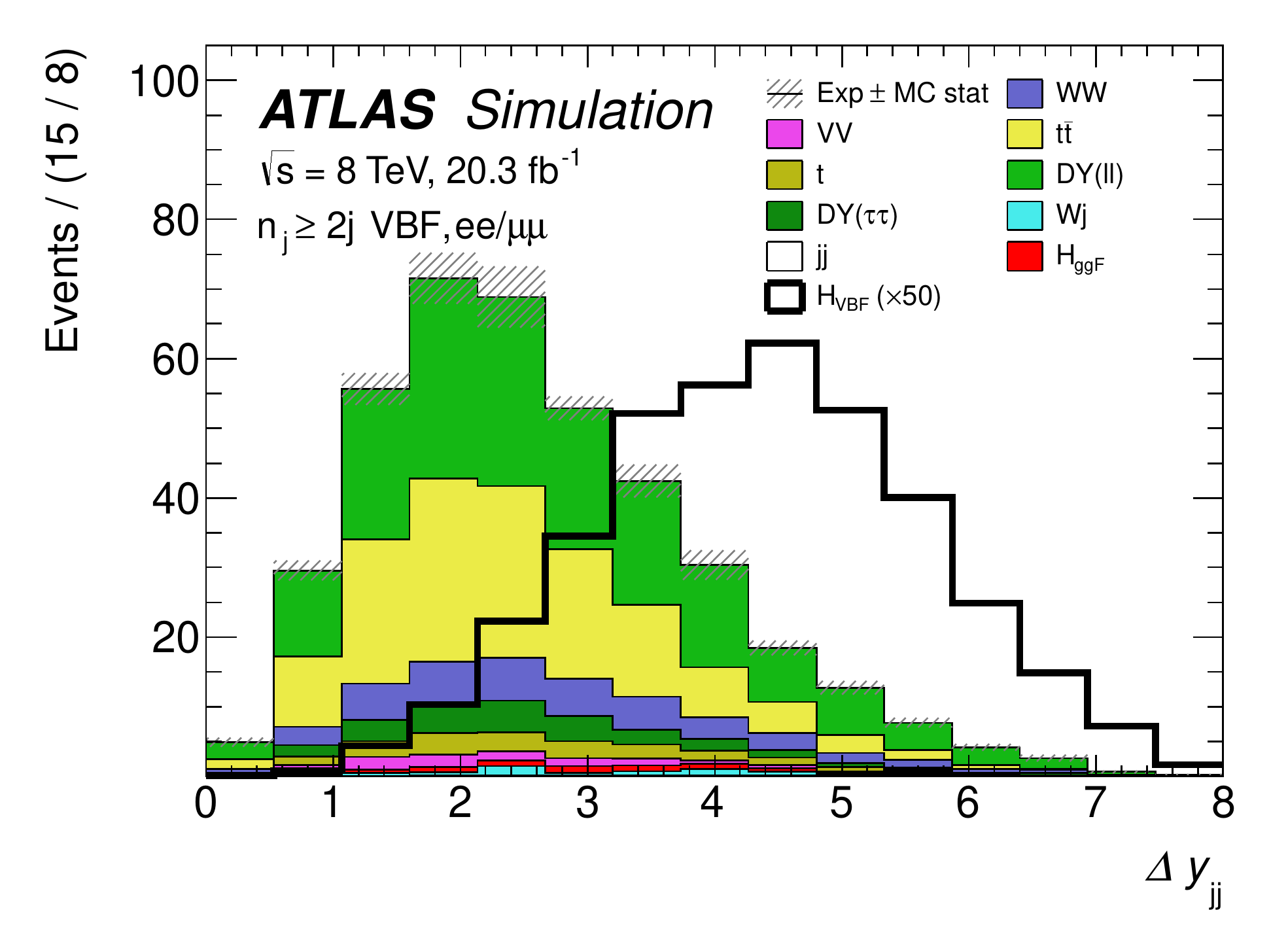}\\
\includegraphics[width=0.40\textwidth]{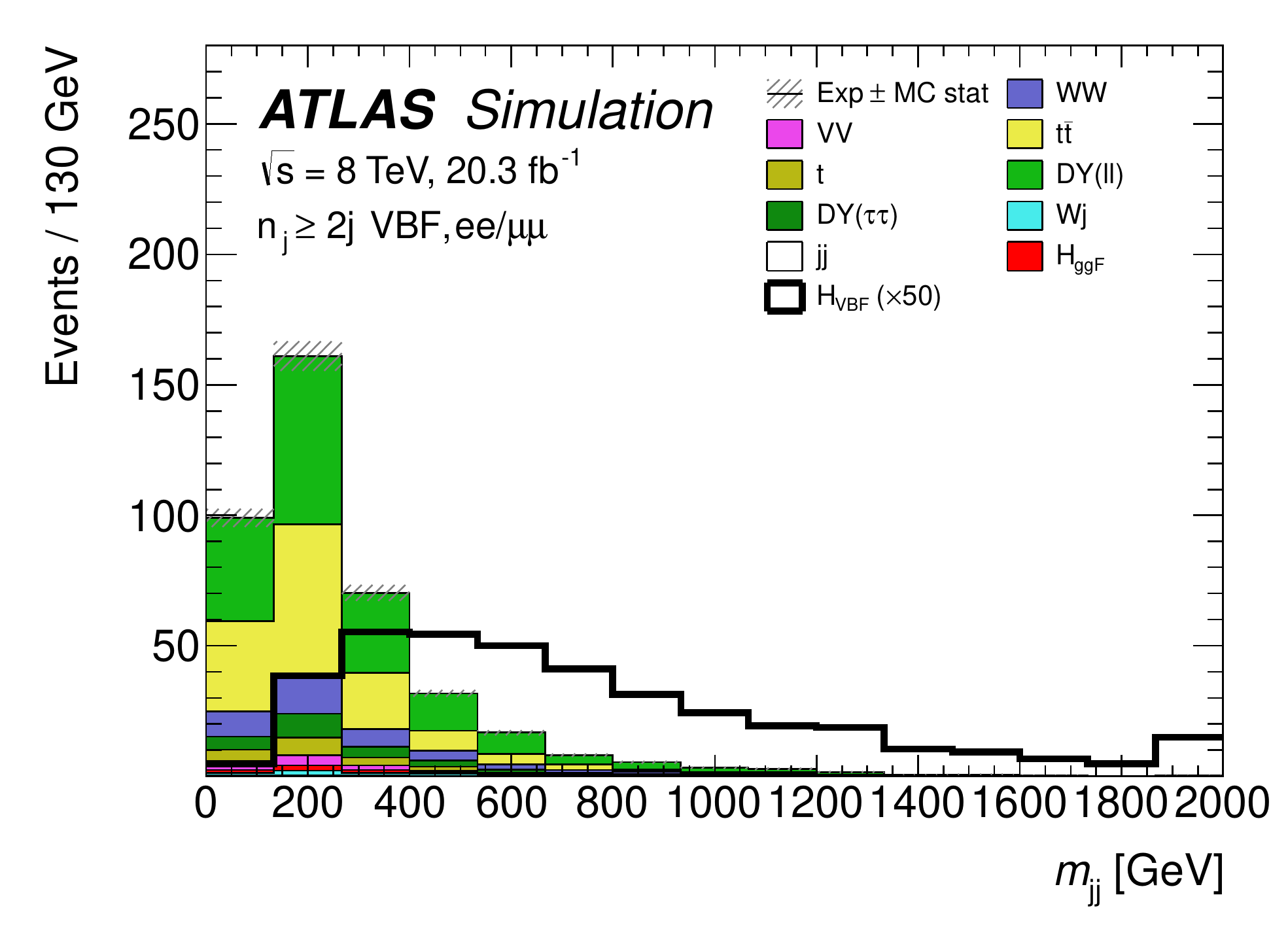}
\includegraphics[width=0.40\textwidth]{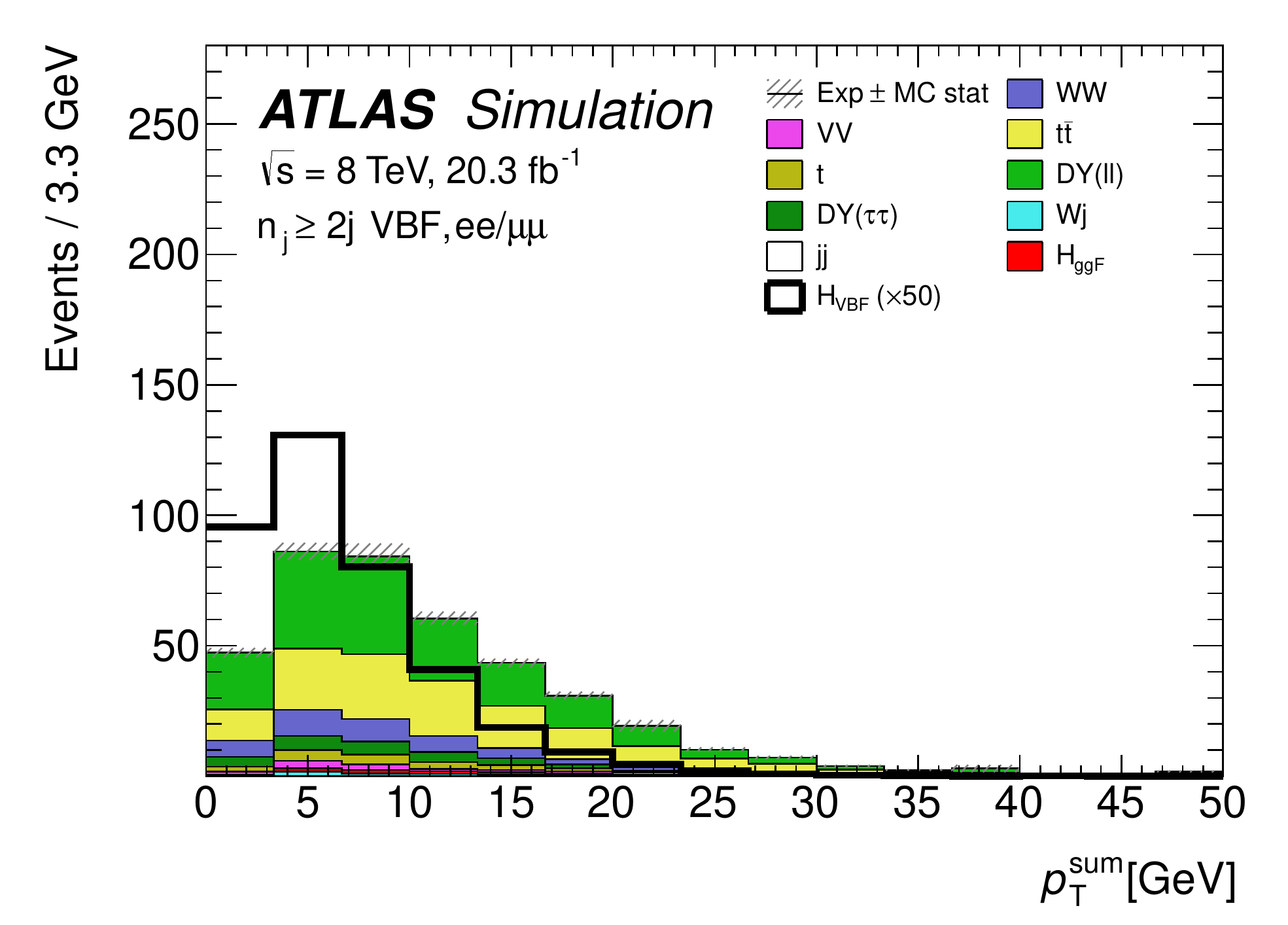}\\
\includegraphics[width=0.40\textwidth]{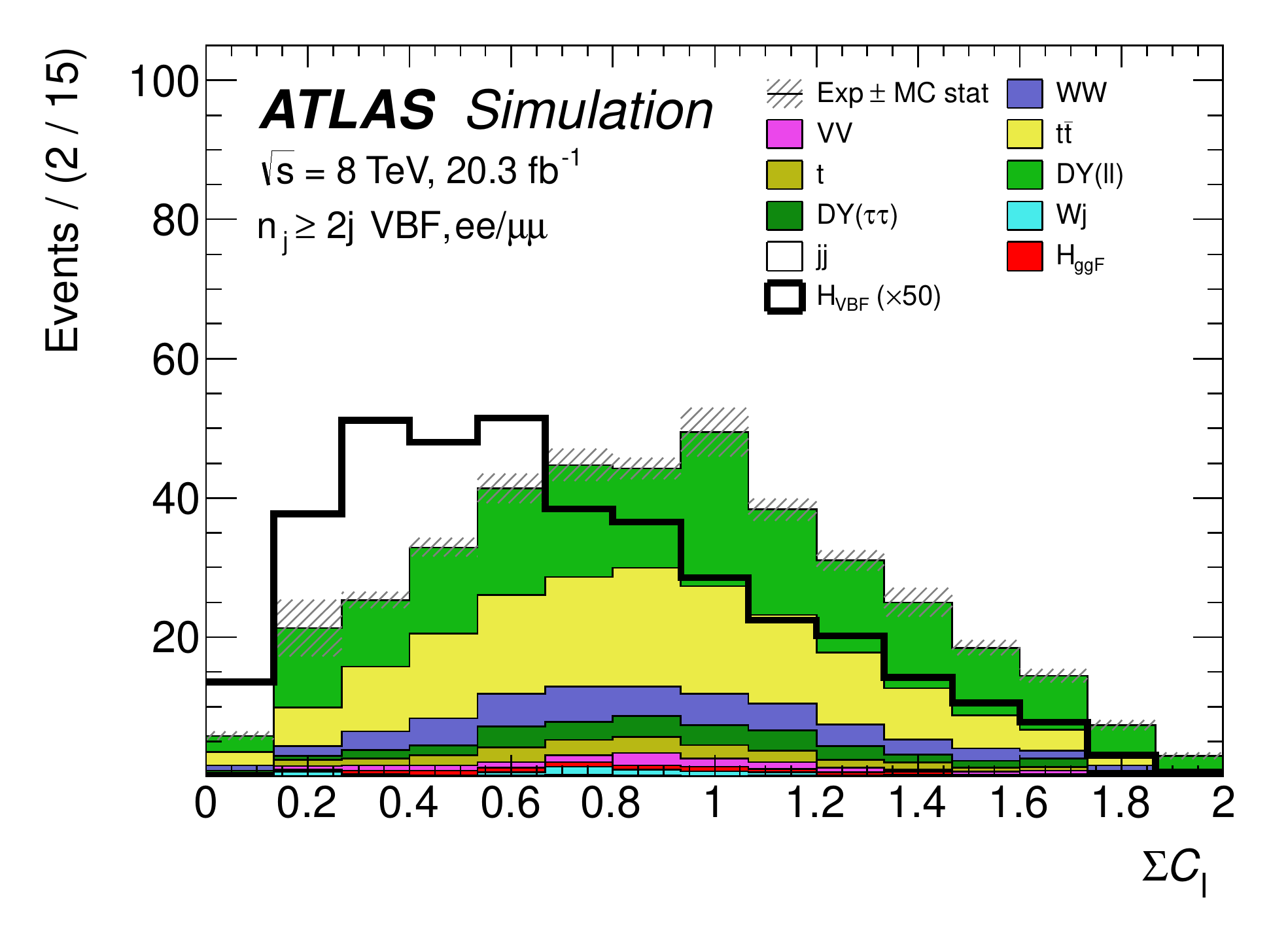}
\includegraphics[width=0.40\textwidth]{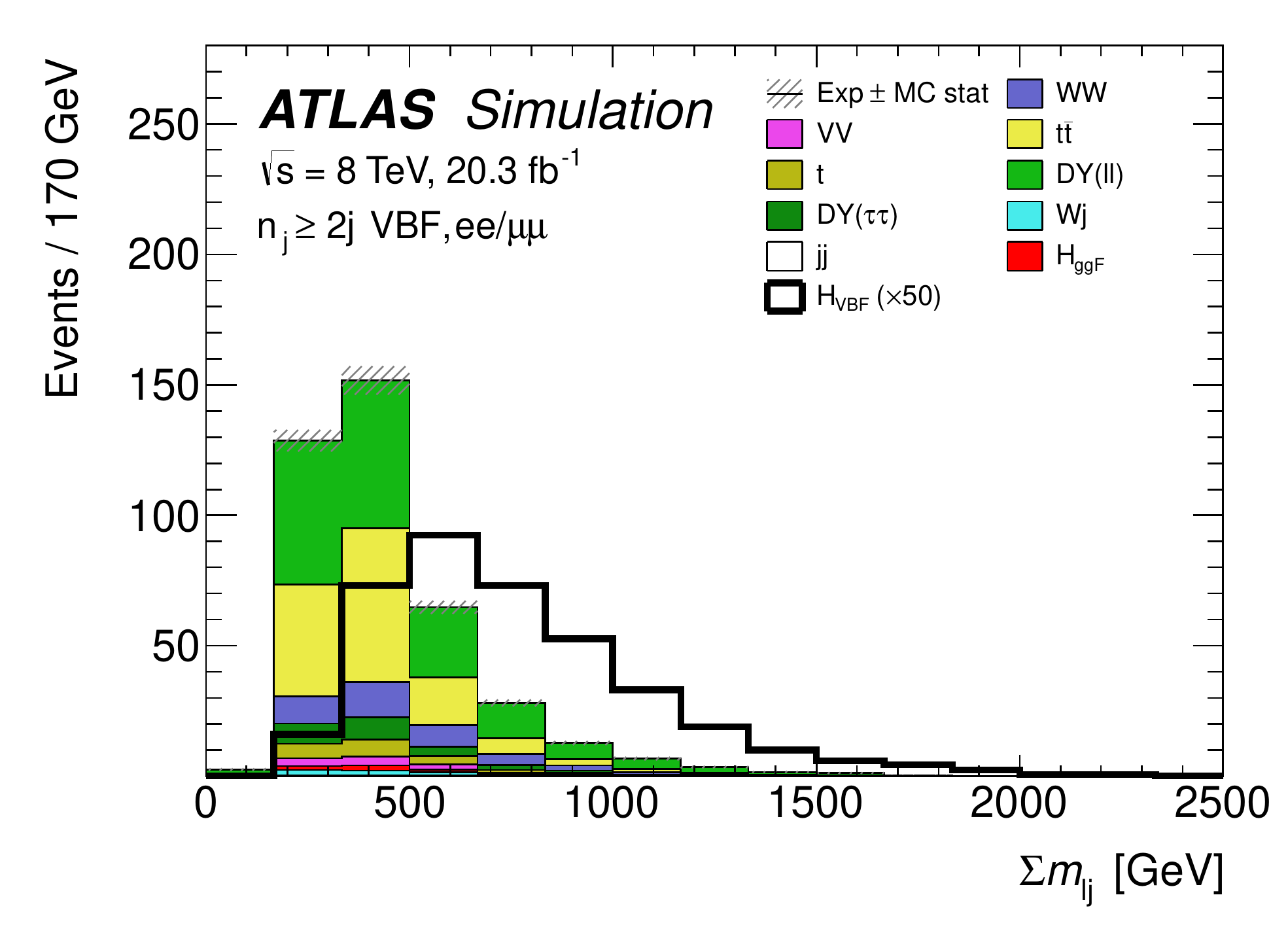}\\
\caption{Distributions of the variables used as inputs to the training
  of the BDT in the $\SFchan$ sample in the $8\TeV$ data
  analysis. The variables are shown after the common preselection and the
  additional selection requirements in the $\NjetGEtwo$ VBF-enriched
  category, and they include: $\mll$, $\dphill$,
  $\mTH$, and $\dyjj$ (top two rows); $\mjj$, $\pTtot$, $\contolv$, and
  $\mlj$ (bottom two rows). The distributions show the separation between
  the VBF signal and background processes (ggF signal
  production is treated as such). The VBF signal
  is scaled by fifty to enhance the differences in the shapes of the
  input variable distributions. The SM Higgs boson is shown at $\mH{\EQ}125\GeV$. The uncertainties on the background
  prediction are only due to MC sample size.
}
\label{fig:aux_bdtinputSF}
\end{figure*}

\begin{figure*}[tb!]
\includegraphics[width=0.45\textwidth]{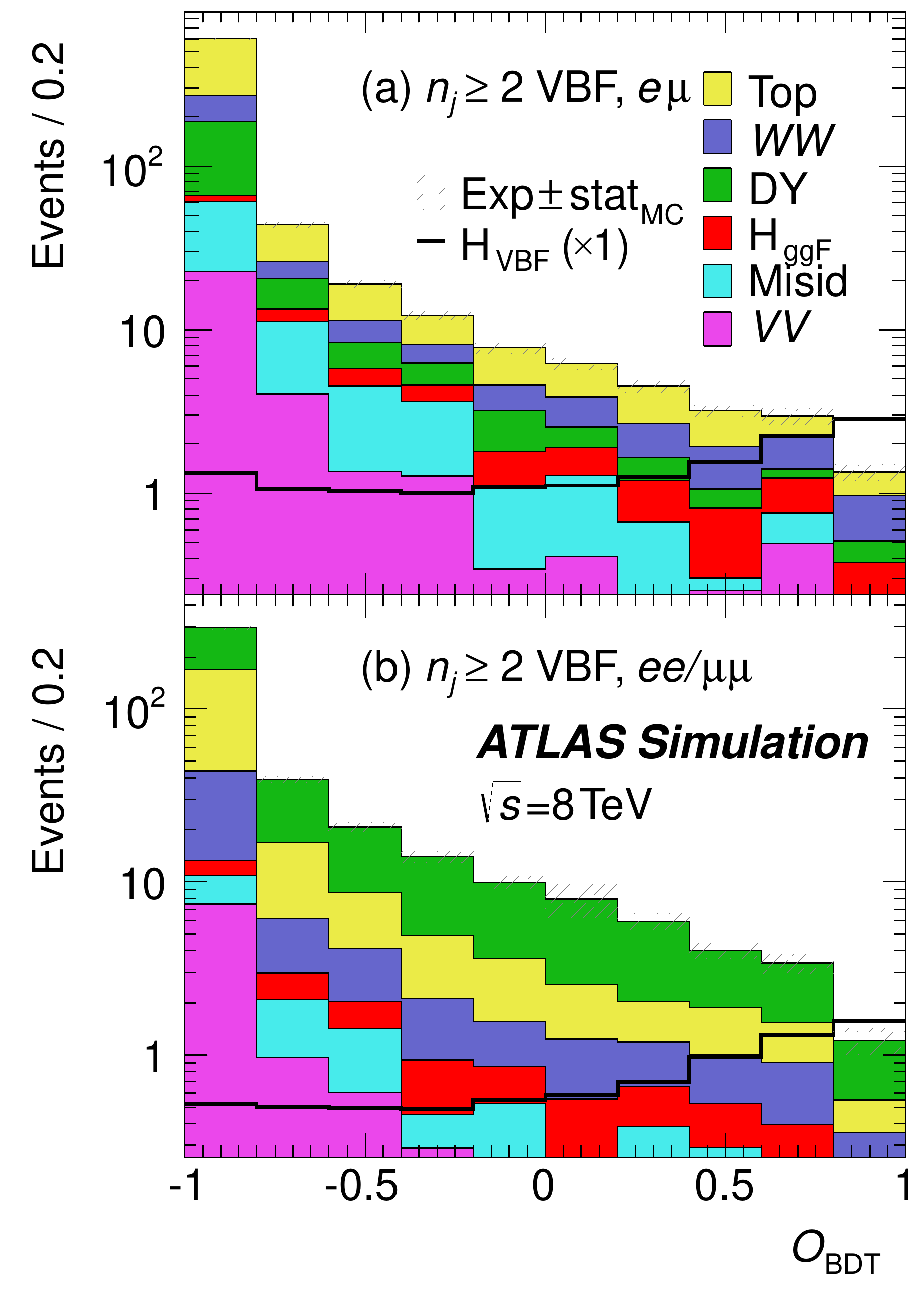}
\caption{Distributions of BDT output in the $\NjetGEtwo$ VBF-enriched category
  in the (a) $\DFchan$ and (b) $\SFchan$ samples in the $8\TeV$ data
  analysis.  The distributions show the separation between
  the VBF signal and background processes (ggF signal
  production is treated as such). The VBF signal
  is overlaid to show the differences in the shapes
  with respect to the background prediction. The DY contribution in (b) is stacked on top, unlike
  in the legend, to show the dominant contribution; the other processes
  follow the legend order. The SM Higgs boson is shown at $\mH{\EQ}125\GeV$. The uncertainties on the background
  prediction are only due to MC sample size.
}
\label{fig:aux_bdtoutputDFSF}
\end{figure*}

The $\bdt$ distributions are shown in
Fig.~\ref{fig:aux_bdtoutputDFSF}. The lowest $\bdt$ score is assigned
to the events that are classified as background, and the highest
score selects the VBF signal events. This separation can be seen in these
distributions.
The final binning configuration is four bins with
boundaries at $[-0.48,0.3,0.78,1]$, and with bin numbering from $0$ to
$3$. The background estimation and the signal extraction is then
performed in bins of $\bdt$. Figure~\ref{fig:aux_bdtinputDFSF} shows
the data-to-MC comparison of the input variables in the three highest
$\bdt$ bins. Good agreement is observed in all the distributions.
The event properties of the observed events in the highest BDT bin in
the $\NjetGEtwo$ VBF-enriched category are shown in Table~\ref{tab:eventlist}.
\vfill

\begin{figure*}[p!]
\includegraphics[width=0.40\textwidth]{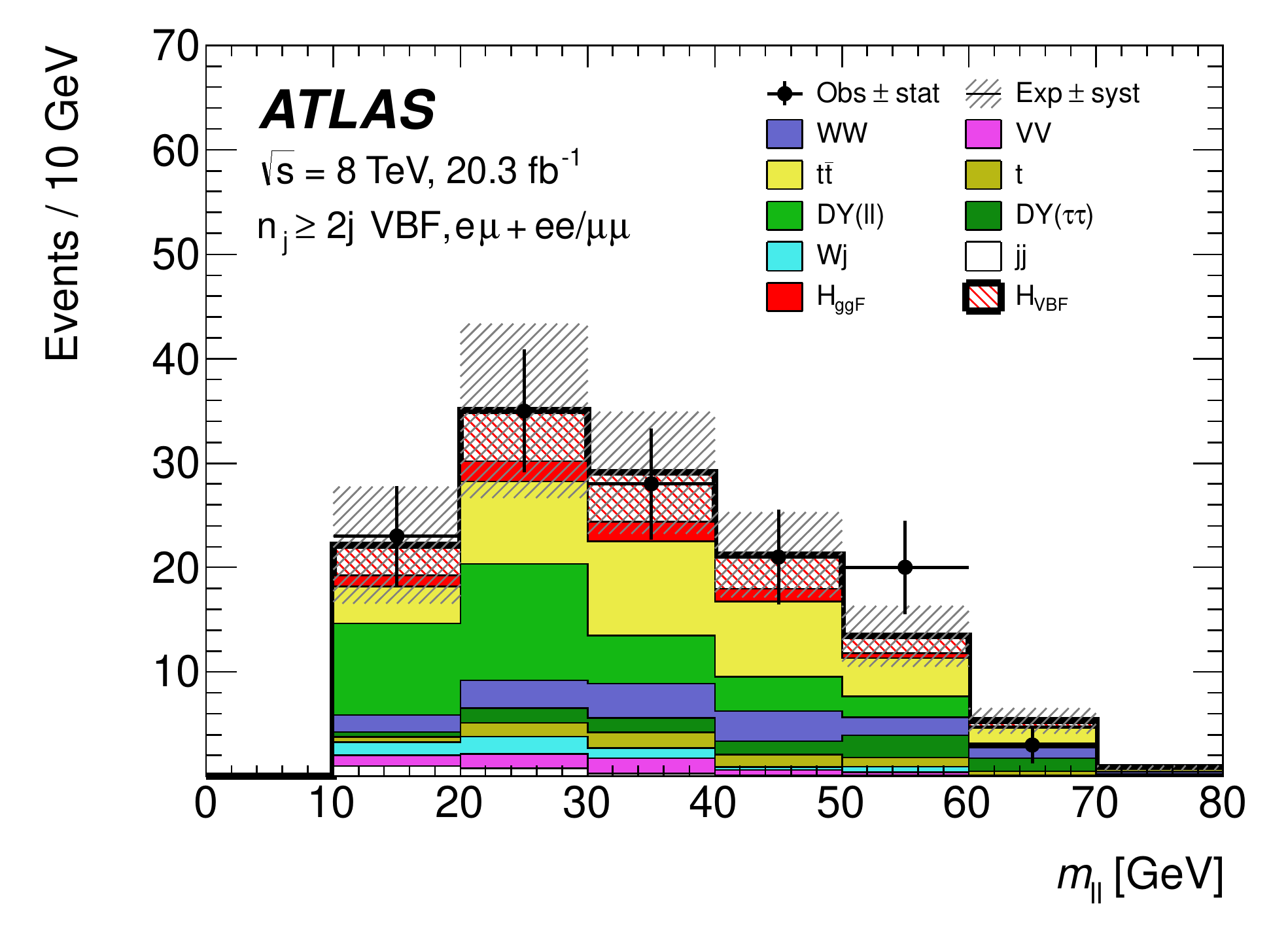}
\includegraphics[width=0.40\textwidth]{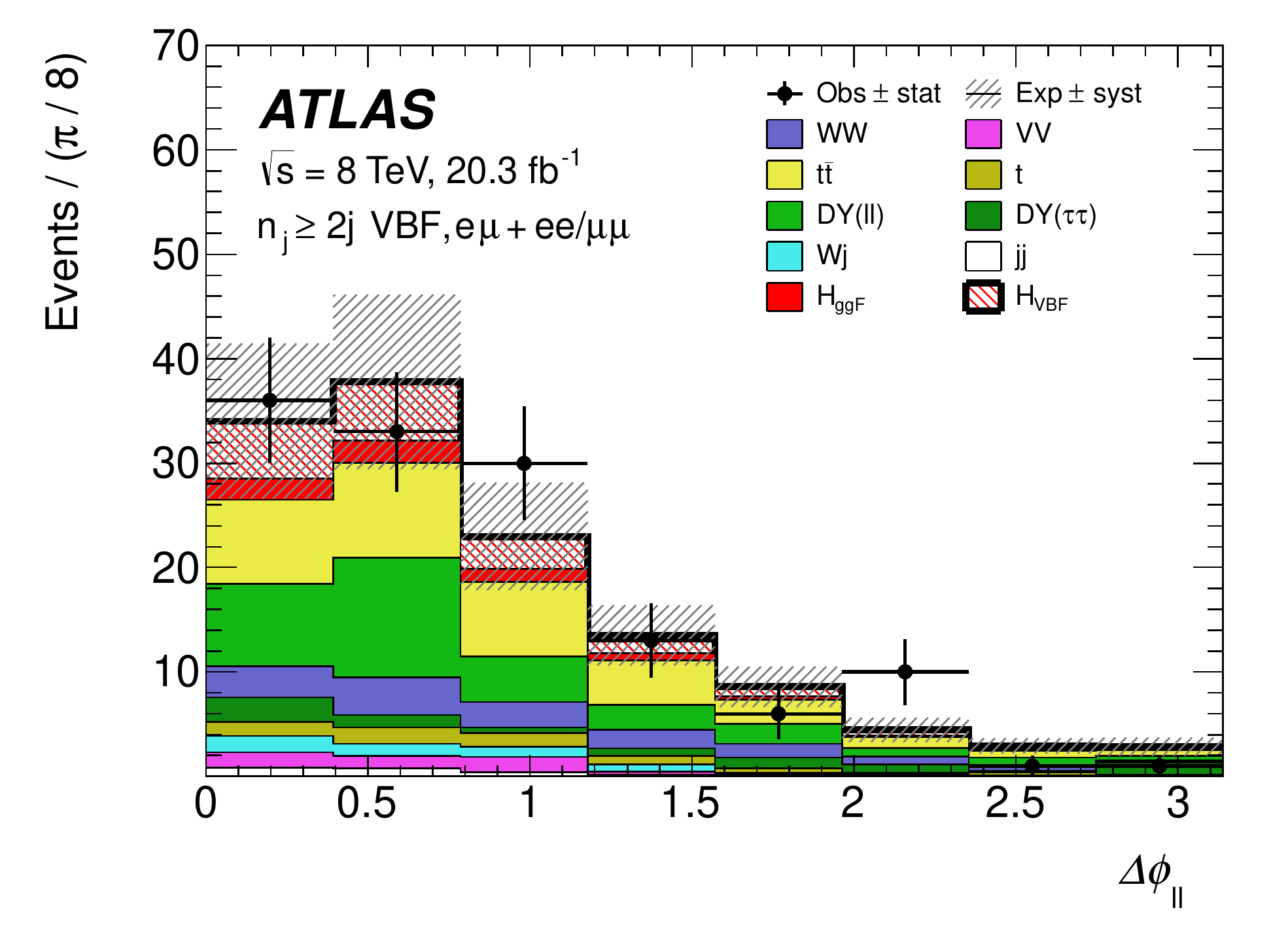}\\
\includegraphics[width=0.40\textwidth]{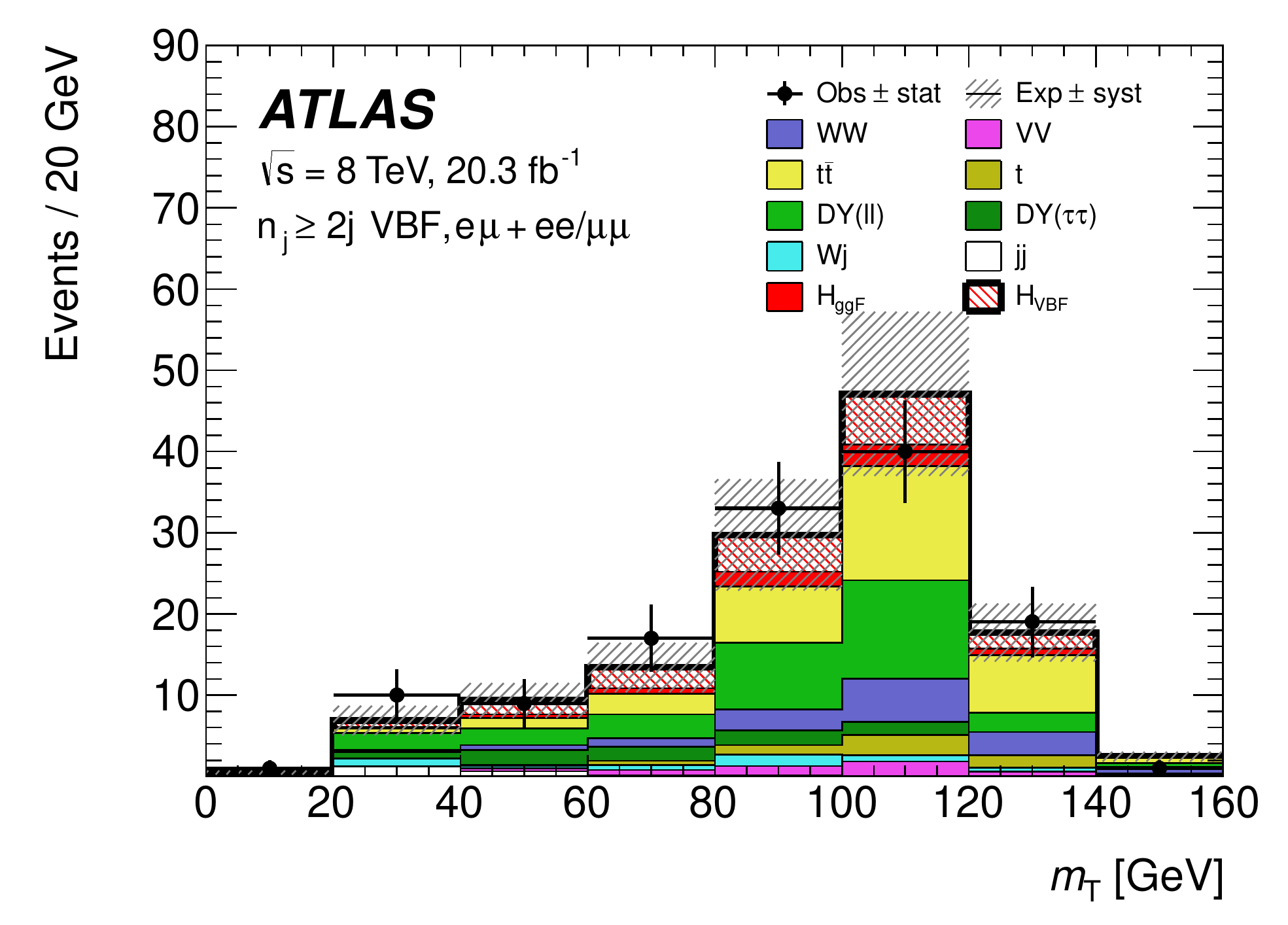}
\includegraphics[width=0.40\textwidth]{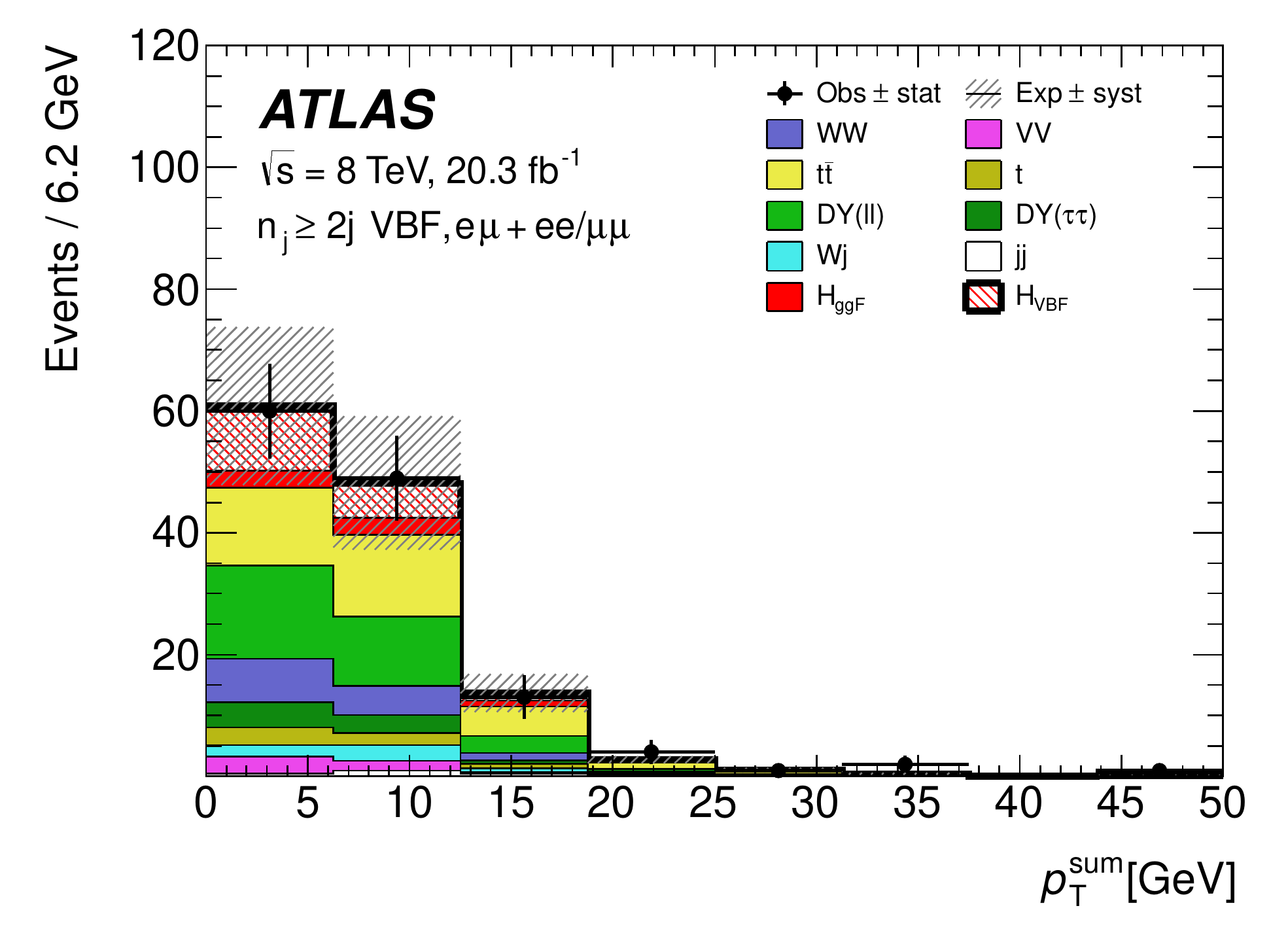}\\
\includegraphics[width=0.40\textwidth]{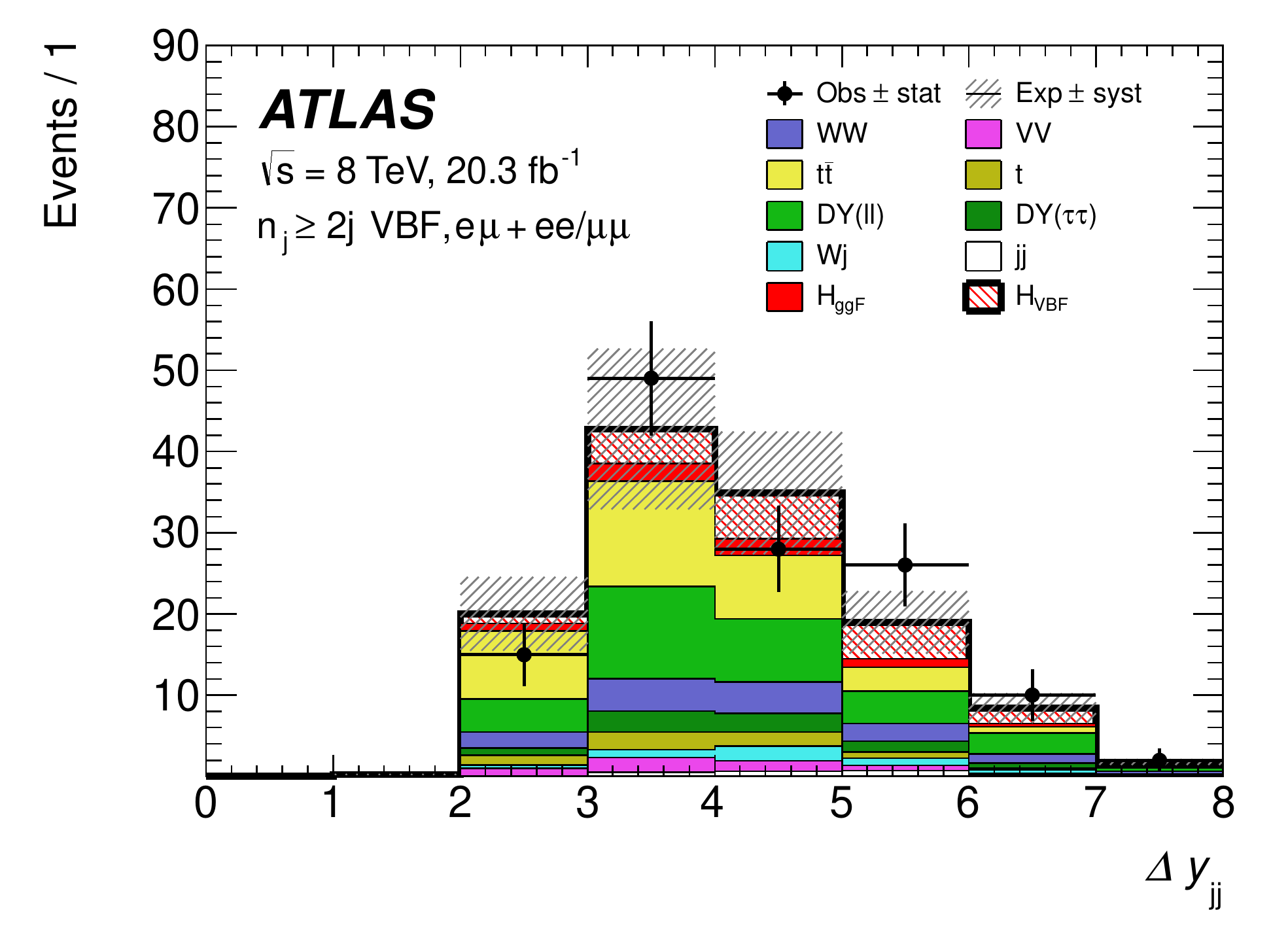}
\includegraphics[width=0.40\textwidth]{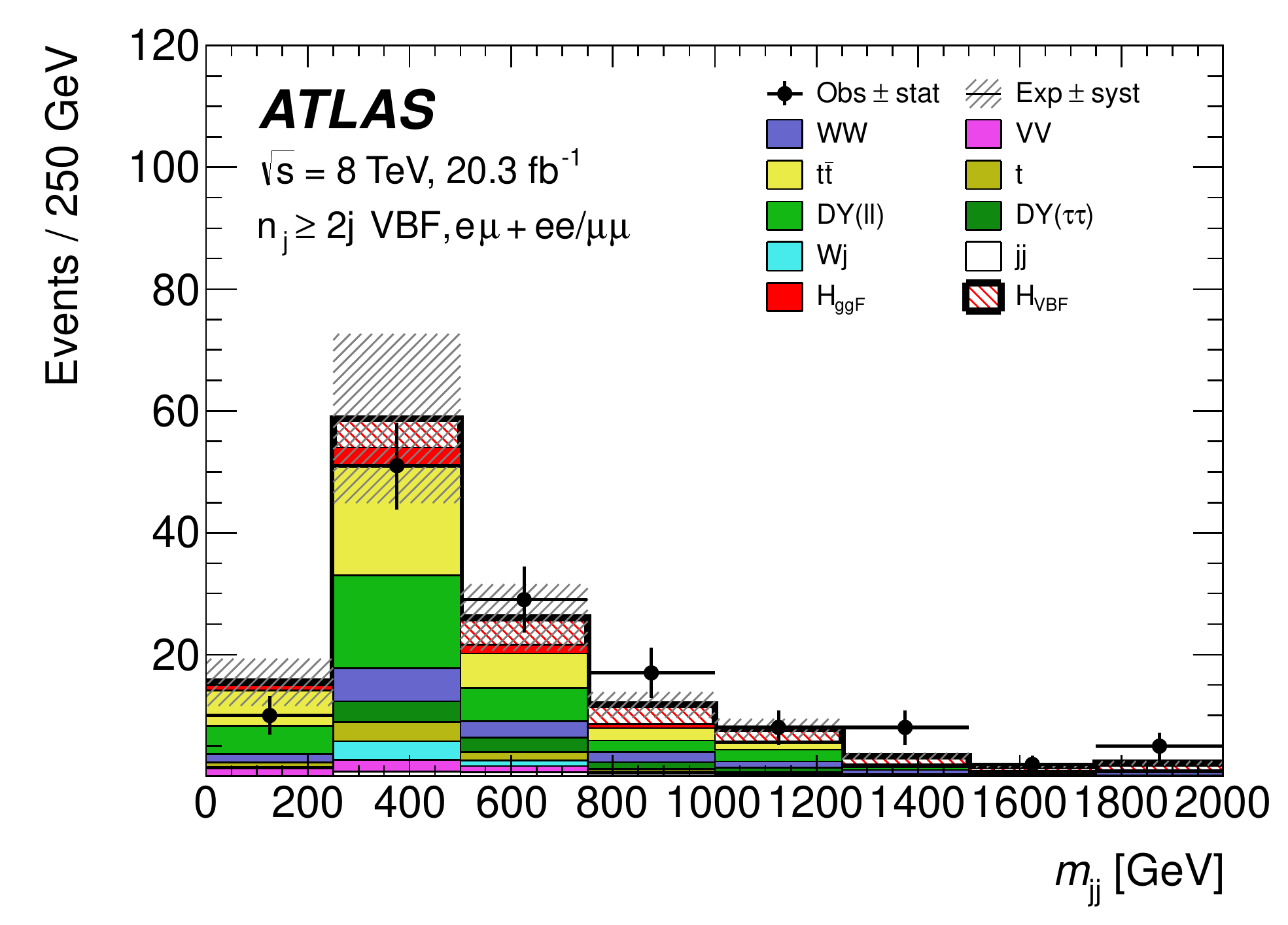}\\
\includegraphics[width=0.40\textwidth]{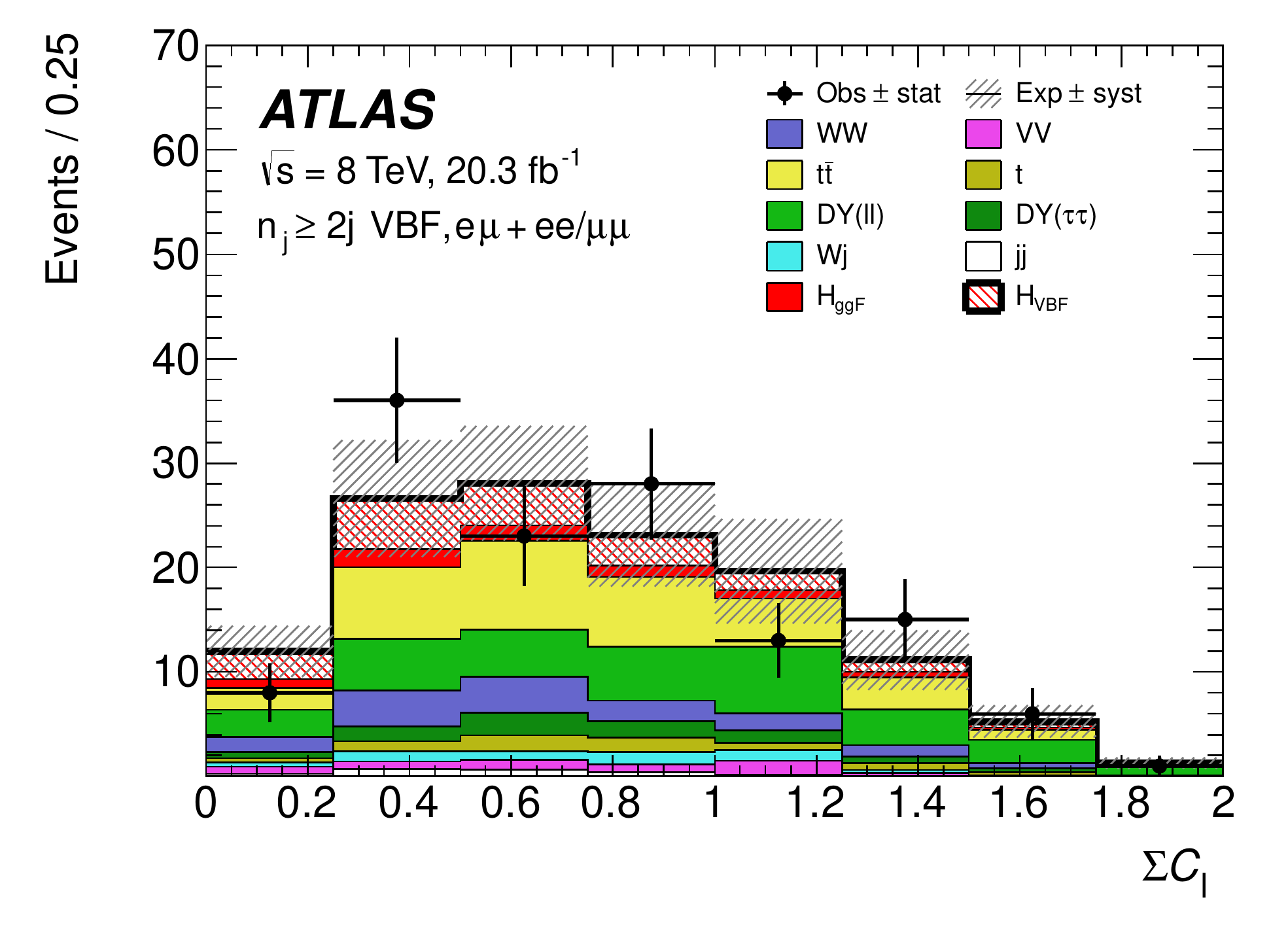}
\includegraphics[width=0.40\textwidth]{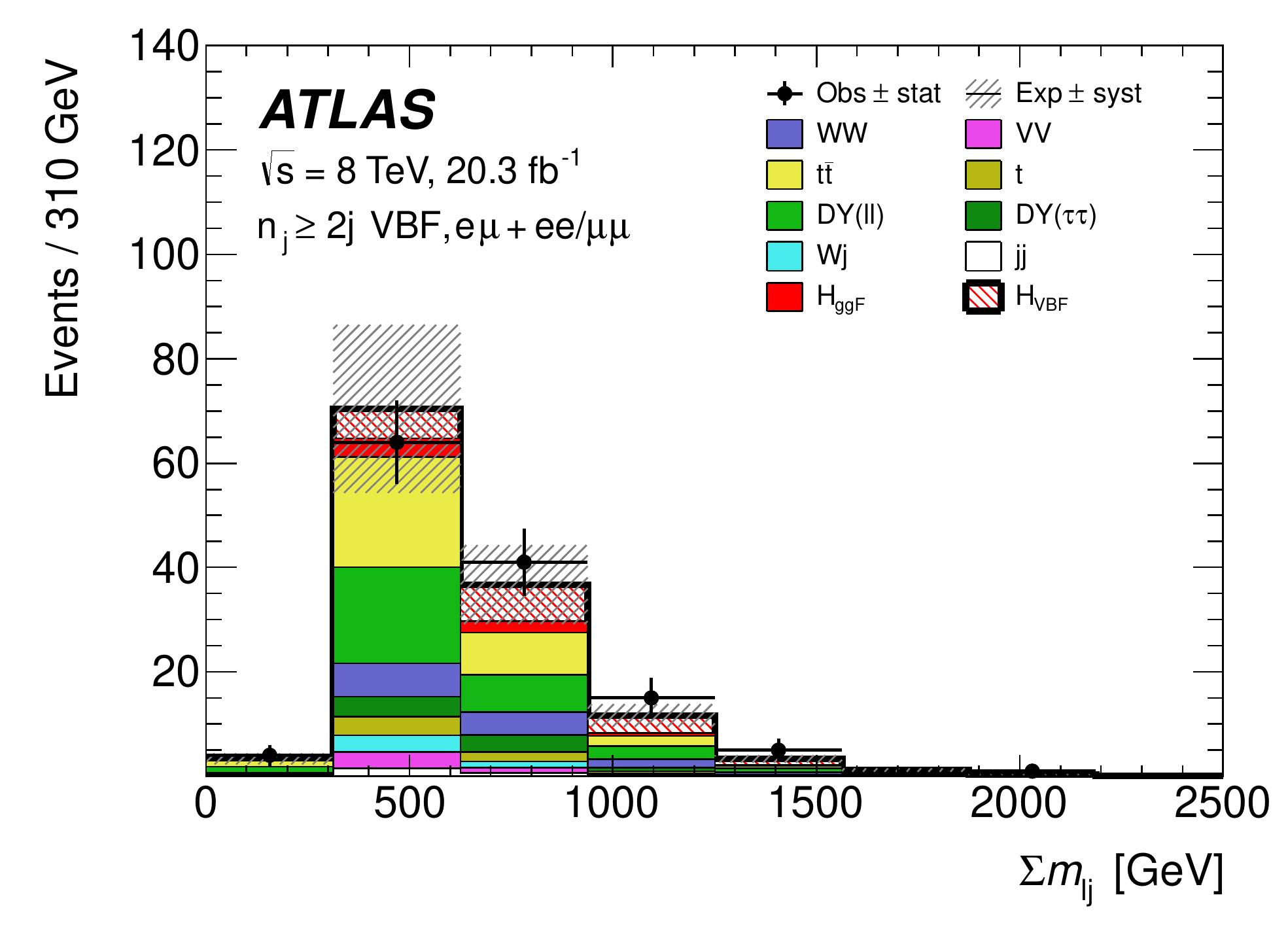}\\
\caption{Comparisons of the observed and expected distributions of the
  variables used as inputs to the training of the BDT in the $\DFchan{\PLUS}\SFchan$ samples in the $8\TeV$ data analysis.
  The variables are shown after the common preselection and the
  additional selection requirements in the $\NjetGEtwo$ VBF-enriched
  category, and after the BDT classification selecting the final signal
  region (BDT bins 1--3, $\bdt{\GT}-0.48$). The variables shown are $\mll$, $\dphill$, $\mTH$, and $\pTtot$
  (top two rows); $\dyjj$, $\mjj$, $\contolv$, and $\mlj$ (bottom two rows). The SM Higgs boson is shown at $\mH{\EQ}125\GeV$.
  Both the statistical and systematic uncertainties are included.
}
\label{fig:aux_bdtinputDFSF}
\end{figure*}
\vfill

\begin{table*}[t!]
\caption{
  Event properties of the Higgs candidates in BDT bin $3$ ($\bdt{\GT}0.78$)
  of the $\NjetGEtwo$ VBF-enriched category in the $8\TeV$ data analysis.
  The energy-related quantities are in $\!\GeV$.
}
\label{tab:eventlist}
{\small
 \centering
  \begin{tabular*}{0.99\textwidth}{ l c c
      d{0}d{1}d{2}d{2}
      d{0}d{2}d{0}d{1} p{0.005\textwidth}p{0.005\textwidth}
      d{0}d{0}d{1}d{1}
      d{0}d{0}d{1}d{1}
      l
    }
\dbline
\multicolumn{1}{p{0.090\textwidth}}{\multirow{2}{*}{Sample}}
& \multicolumn{1}{p{0.060\textwidth}}{\multirow{2}{*}{~~$\bdt$}}
&  \multicolumn{1}{p{0.000\textwidth}}{}
&  \multicolumn{9}{c}{BDT input variables}
&  \multicolumn{8}{c}{$\pT$ and $\eta$ of leptons and jets}
\vspace{-6pt}
\\
\cline{4-12}\cline{14-22}
\clineskip
&
&
& \multicolumn{1}{p{0.035\textwidth}}{~~$\mTH$ }
& \multicolumn{1}{p{0.040\textwidth}}{~~~$\mll$ }
& \multicolumn{1}{p{0.045\textwidth}}{~~$\dphill$ }
& \multicolumn{1}{p{0.050\textwidth}}{~~~$\contolv$ }
& \multicolumn{1}{p{0.050\textwidth}}{~~$\mjj$ }
& \multicolumn{1}{p{0.040\textwidth}}{~~$\dyjj$ }
& \multicolumn{1}{p{0.050\textwidth}}{~~$\mlj$ }
& \multicolumn{1}{p{0.035\textwidth}}{~~$\pTtot$ }
&&
& \multicolumn{1}{p{0.020\textwidth}}{~~$\pTlead$ }
& \multicolumn{1}{p{0.030\textwidth}}{~~$\pTsublead$ }
& \multicolumn{1}{p{0.040\textwidth}}{~~~~$\etalead$ }
& \multicolumn{1}{p{0.040\textwidth}}{~~~~$\etasublead$ }
& \multicolumn{1}{p{0.040\textwidth}}{~~~~$\pTleadjet$ }
& \multicolumn{1}{p{0.035\textwidth}}{~~~$\pTsubleadjet$ }
& \multicolumn{1}{p{0.040\textwidth}}{~~~$\etaleadjet$ }
& \multicolumn{1}{p{0.040\textwidth}}{~~~$\etasubleadjet$ }
&
\\
\sgline
$\DFchan$ sample &0.92 && 113 & 19.8 & 0.26 & 0.25 & 1290 & 5.30 &  810 & 3.2 &&& 50  & 15 & 0.6  & -0.1 & 110 & 74 & 2.9  & -2.4 \\
                 &0.88 && 102 & 27.2 & 0.62 & 0.83 &  530 & 3.75 &  540 & 2.5 &&& 39  & 16 & -0.5 & 0.4  & 80  & 79 & -2.7 & 1.1  \\
                 &0.85 &&  87 & 19.9 & 0.02 & 1.01 & 1770 & 7.55 & 1510 & 4.8 &&& 73  & 20 & 1.6  & 2.2  & 50  & 32 & 3.8  & -3.8 \\
                 &0.84 && 105 & 26.5 & 0.64 & 0.86 & 1500 & 5.58 & 1250 & 7.4 &&& 84  & 21 & 0.8  & 0.7  & 115 & 73 & 2.4  & -3.2 \\
                 &0.81 &&  43 & 31.7 & 1.08 & 0.05 &  850 & 5.44 &  710 & 2.2 &&& 48  & 20 & 0.5  & 0.4  & 84  & 37 & -2.2 & 3.2  \\
                 &0.78 &&  84 & 37.2 & 0.18 & 0.76 &  810 & 5.26 &  660 & 4.8 &&& 29  & 18 & -1.5 & 0.1  & 82  & 41 & -2.4 & 2.9  \\
\sgline
$\SFchan$ sample &0.92 &&  66 & 38.9 & 0.18 & 0.33 & 1370 & 5.15 & 1060 & 2.2 &&& 61  & 31 & 0.4  & 1.2  & 118 & 90 & -3.2 & 2.0  \\
                 &0.91 && 102 & 27.4 & 0.65 & 0.24 & 1260 & 5.97 &  780 & 2.5 &&& 34  & 23 & -0.5 & 0.2  & 101 & 40 & -3.1 & 2.9  \\
                 &0.89 &&  81 & 26.3 & 0.08 & 0.57 &  670 & 3.97 & 1210 & 8.1 &&& 147 & 44 & 1.4  & 1.1  & 168 & 49 & -1.3 & 2.7  \\
                 &0.81 && 118 & 57.6 & 2.31 & 0.28 &  700 & 3.90 &  610 & 6.6 &&& 32  & 29 & 1.3  & 0.8  & 113 & 83 & -1.2 & 2.7  \\
                 &0.81 && 116 & 18.5 & 0.33 & 0.19 &  670 & 3.97 &  650 & 2.9 &&& 54  & 25 & -1.7 & -1.3 & 119 & 68 & 0.4  & -3.6 \\
                 &0.80 && 108 & 50.1 & 0.80 & 0.87 &  740 & 4.68 &  660 & 2.1 &&& 46  & 15 & -0.4 & -2.0 & 123 & 41 & -2.5 & 2.2  \\
\dbline
\end{tabular*}
}
\end{table*}

\clearpage
\onecolumngrid
\vspace{20pt}
\begin{center}
\begin{flushleft}
{\Large The ATLAS Collaboration}

\bigskip

G.~Aad$^{\rm 85}$,
B.~Abbott$^{\rm 113}$,
J.~Abdallah$^{\rm 152}$,
S.~Abdel~Khalek$^{\rm 117}$,
O.~Abdinov$^{\rm 11}$,
R.~Aben$^{\rm 107}$,
B.~Abi$^{\rm 114}$,
M.~Abolins$^{\rm 90}$,
O.S.~AbouZeid$^{\rm 159}$,
H.~Abramowicz$^{\rm 154}$
% ...
\bigskip
\\
% ...
$^{11}$ Institute of Physics, Azerbaijan Academy of Sciences, Baku, Azerbaijan\\
% ...
$^{85}$ CPPM, Aix-Marseille Universit{\'e} and CNRS/IN2P3, Marseille, France\\
% ...
$^{90}$ Department of Physics and Astronomy, Michigan State University, East Lansing MI, United States of America\\
% ...
$^{107}$ Nikhef National Institute for Subatomic Physics and University of Amsterdam, Amsterdam, Netherlands\\
% ...
$^{113}$ Homer L. Dodge Department of Physics and Astronomy, University of Oklahoma, Norman OK, United States of America\\
$^{114}$ Department of Physics, Oklahoma State University, Stillwater OK, United States of America\\
% ...
$^{117}$ LAL, Universit{\'e} Paris-Sud and CNRS/IN2P3, Orsay, France\\
% ...
$^{152}$ Institute of Physics, Academia Sinica, Taipei, Taiwan\\
% ...
$^{154}$ Raymond and Beverly Sackler School of Physics and Astronomy, Tel Aviv University, Tel Aviv, Israel\\
% ...
$^{159}$ Department of Physics, University of Toronto, Toronto ON, Canada\\
% ...
$^{*}$ Deceased
\end{flushleft}


\end{center}

\end{document}